\documentclass[b5paper,10pt,twoside]{report}
\usepackage[utf8]{inputenc}

\usepackage{amsmath}
\usepackage{amsfonts}
\usepackage{amssymb}
\usepackage{mathrsfs}
\usepackage{graphicx}
\usepackage[retainorgcmds]{IEEEtrantools}
\usepackage{placeins}
\usepackage{enumerate}
\usepackage{mdframed}
\usepackage{hyperref}

\usepackage[font={footnotesize}]{caption}

\usepackage{fancyhdr}

\fancypagestyle{plain}{%
  \fancyhf{}%
  \fancyhead[RO]{\thepage}
  \fancyhead[LE]{\thepage}
  \fancyhead[C]{\nouppercase\rightmark}
}

\fancypagestyle{plainClearDoublePage}{%
  \fancyhf{}%
  \fancyhead[RO]{\thepage}
  \fancyhead[LE]{\thepage}
}

\usepackage{ntheorem}

\newtheorem*{theorem}{Theorem}

\newcommand{\df}{\text{d}}
\newcommand{\Lagr}{\mathscr{L}}
\newcommand{\Lie}{\mathcal{L}}

\makeatletter
\def\cleardoublepage{%
        \clearpage%
        \if@twoside%
            \ifodd\c@page%
            \else%
                \thispagestyle{plainClearDoublePage}%
                \hbox{}%
                \newpage%
            \fi%
        \fi%
    }%

\def\cleardoublepageempty{%
        \clearpage%
        \if@twoside%
            \ifodd\c@page%
            \else%
                \thispagestyle{empty}%
                \hbox{}%
                \newpage%
            \fi%
        \fi%
    }%
\makeatother

\begin{document}
\pagenumbering{gobble}
\pagestyle{empty}

\begin{titlepage}

\begin{center}
\includegraphics[width=.8\linewidth]{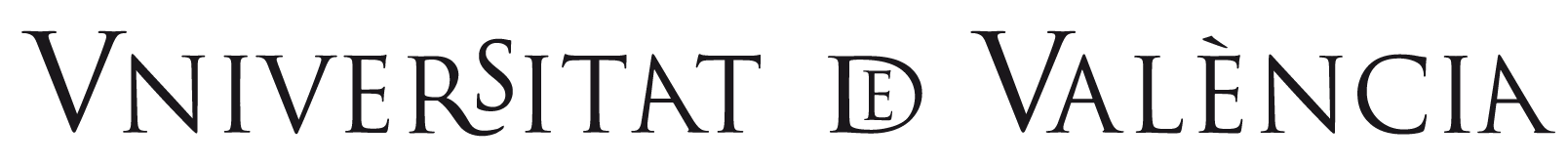}
\vspace{0.5cm}

\includegraphics[width=.4\linewidth]{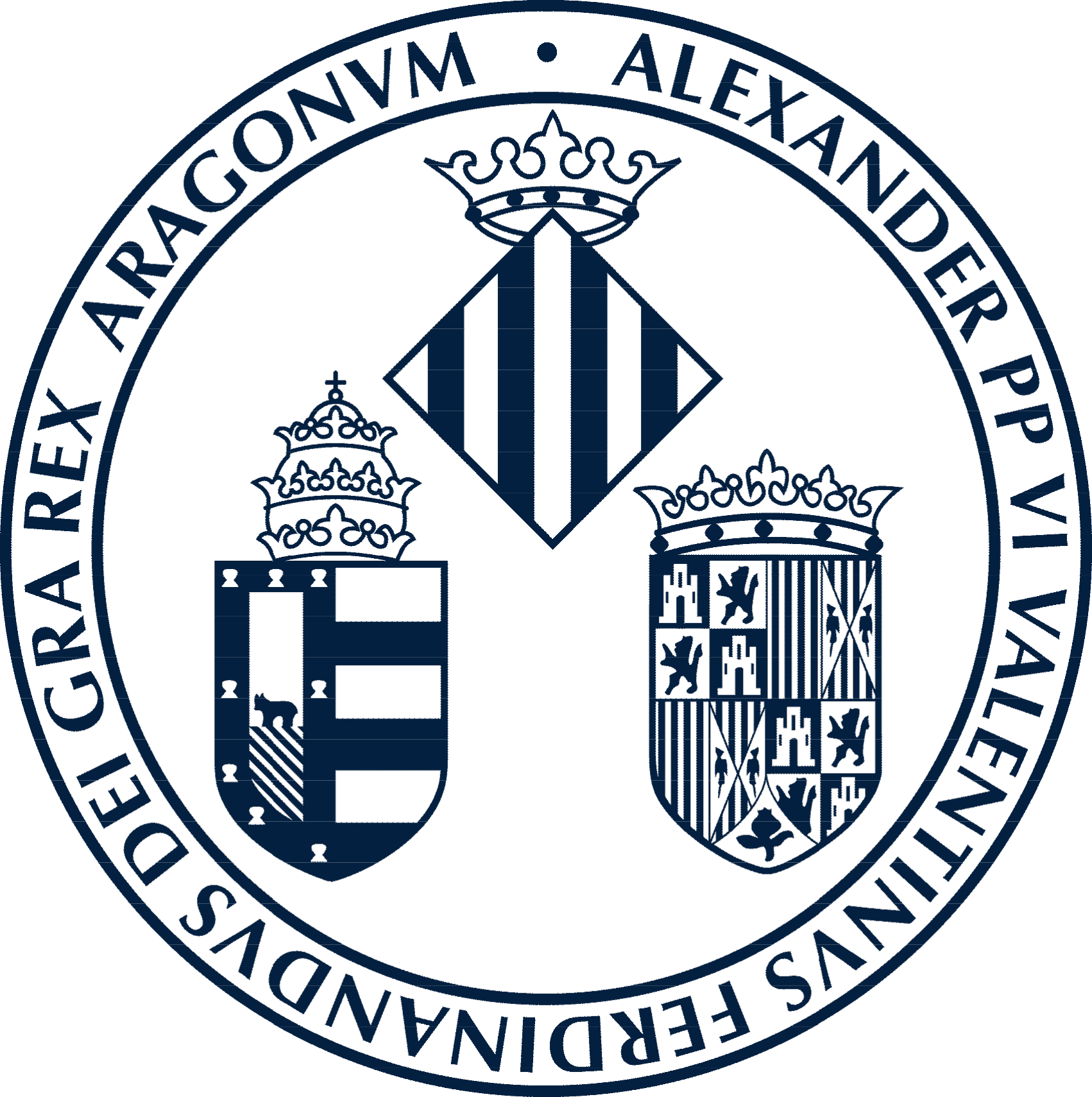}

\vspace{1cm}
\begin{huge}
\textbf{Black Holes, Geons, and Singularities in Metric-Affine Gravity}
\end{huge}
\end{center}
\vspace{1.5cm}
\begin{flushright}
 Phd Thesis by
 
 \begin{Large}
 \textbf{Antonio Sánchez Puente}
 \end{Large}
 
 \vspace{0.5cm}
 Under the supervision of
 
 \begin{Large}
\textbf{Gonzalo Olmo Alba}
 \end{Large}

\end{flushright}

\vspace{1cm}

\begin{center}
\begin{Large}
 Programa de Doctorado en Física
 
 Diciembre de 2016
\end{Large}
\end{center}

\end{titlepage}

\cleardoublepageempty
$\ $

\begin{flushright}
\textit{A mis padres \\
y hermanos}
\end{flushright}

\cleardoublepageempty

\pagenumbering{roman} 
\chapter*{List of Publications}
This PhD thesis is based on the following publications:
\begin{itemize}
 \item \emph{Classical resolution of black hole singularities via wormholes.} \cite{Olmo:2015dba}
 
 Gonzalo J. Olmo, D. Rubiera-Garcia, and A. Sanchez-Puente
 
 \emph{Eur. Phys. J.}, C76(3):143, 2016
 
 \item \emph{Classical resolution of black hole singularities in arbitrary dimension.} \cite{Bazeia:2015uia}
 
D. Bazeia, L. Losano, Gonzalo J. Olmo, D. Rubiera-Garcia, and A. Sanchez-Puente

\emph{Phys. Rev.}, D92(4):044018, 2015

 \item \emph{Geodesic completeness in a wormhole spacetime with horizons.} \cite{Olmo:2015bya}
 
 Gonzalo J. Olmo, D. Rubiera-Garcia, and A. Sanchez-Puente
 
  \emph{Phys. Rev.},D92(4):044047, 2015
  
 \item \emph{Impact of curvature divergences on physical observers in a wormhole space–time with horizons.}\cite{Olmo:2016fuc}
 
 Gonzalo J. Olmo, D. Rubiera-Garcia, and A. Sanchez-Puente
 
  \emph{Class. Quant. Grav.}, 33(11):115007, 2016
\end{itemize}

\thispagestyle{plain}
\cleardoublepage
\pagestyle{fancy}
\renewcommand{\chaptermark}[1]{\markboth{#1}{}}
\renewcommand{\sectionmark}[1]{\markright{\thesection\ #1}}
\tableofcontents
\cleardoublepage
\chapter*{Notation}
\markboth{}{}
\addcontentsline{toc}{chapter}{Notation}

In this thesis I will follow the $(-,+,+,+)$ convention for the space-time metric. I will also make use of the following abbreviations:
\newline
\newline
\begin{tabular}{ll}
 \textbf{GR} & General Relativity \\
 \textbf{EEP} & Einstein's Equivalence Principle \\
 \textbf{WEP} & Weak Equivalence Principle \\
 \textbf{SEP} & Strong Equivalence Principle \\
 \textbf{CMB} & Cosmic Microwave Background \\
 \textbf{QFT} & Quantum Field Theory \\
 \textbf{LQG} & Loop Quantum Gravity\\
 \textbf{PS} & Pfaffian System
\end{tabular}

\section*{About Planck Length and Units}

In this thesis I will work in Planck units. In these units, we make use of physical constants to relate the measurement of different magnitudes. For instance, if I have defined a unit of time, I can work with a unit of length that is the distance light travels in unit of time. In those units, the speed of light is just $c=1$, dimensionless, and any other velocity would be expressed as a fraction of the speed of light. In this way we have removed the arbitrariness of choosing units adapted to the ``human scale''. If we repeat this procedure, it would seem we still need to choose some arbitrary unit from which all the rest are derived. This is not the case: there is a preferred unit of length that we will call \emph{Planck length} or $l_P$. The Planck length appears because there are two different lengths associated to an object with mass $m$: On one hand quantum physics tells us that this object has a dual wave/particle nature and will propagate as a wave with 
Compton wavelength inversely proportional to its mass\footnote{For an observer who is not at rest respect to the object, he or she would see a wavelength inversely proportional to the momentum of the object.} $\lambda = 2 \pi \hbar c^{-1} m^{-1}$, on the other hand in General Relativity\footnote{Also Newtonian gravity can be characterized using the same radius.} this object will deform the space-time around it in a way characterized by the Schwarzschild radius $r_S=2G c^{-2} m$. These two lengths are related through the Planck length $\lambda= 4 \pi l_P^2 / r_S$, with $l_P \equiv \sqrt{G \hbar / c^3}$. 

This also gives us two different ways of relating units of length to units of mass (energy). In Planck units, they are related through the de Broglie wavelength, which makes $\hbar=1$ and $G=l_P^2$. Sometimes it is said $G=1$ which implies $l_P=1$, this can be convenient as one does not have to carry the units everywhere. However, to avoid confusion, I will write the Planck length wherever it is needed. Therefore time and length have dimensions of $l_P$, velocities are dimensionless, and mass and energy have dimensions of $l_P^{-1}$. A curious point is that $l_P$ as a unit of mass is very close to human scales ($\simeq 10^{-2} mg$), whereas as a unit of length it is really small compared to human scales ($\simeq 10^{-35} m$) which tells us that common objects are both way too big (massive) to see quantum effects, and way too small to see gravitational effects.

With this choice, if we look at Coulomb's law $F= \frac{1}{4 \pi \epsilon_0} \frac{qQ}{r^2}$, we can see that forces have dimensions of [mass] times [length${}^{-1}$], that is to say [length${}^{-2}$], and that means that the combination $qQ/(4\pi \epsilon_0)$ is dimensionless. It is possible to reabsorb the $(4 \pi \epsilon_0)$ factor into the charge definition, $q^\prime \equiv q/\sqrt{4 \pi \epsilon_0}$; this way $q=1$ is known as the \emph{Planck charge} and the elementary charge has value $e=\sqrt{\alpha} \approx \sqrt{1/137}$.

\cleardoublepage
\chapter*{Resumen en Español}
\addcontentsline{toc}{chapter}{Resumen en Español}
\markboth{Resumen en Español}{}

La Gravedad es ubicua en nuestras vidas. A diferencia de otras fuerzas de la naturaleza, para las cuales la carga de los objetos macroscópicos es prácticamente neutra, toda la materia habitual tiene masa y experimenta una aceleración hacia el resto de la materia. Observamos los efectos de la gravedad tanto en los objetos que caen al suelo, como en el movimiento de los planetas alrededor del Sol. Parece natural que fuera la primera fuerza que tuvo una descripción matemática precisa, la ley de gravitación universal, dada por sir Isaac Newton. A pesar de ello, es la interacción más misteriosa a día de hoy. Podemos estudiar el resto de fuerzas de la naturaleza construyendo experimentos en los cuales la física es testada en un amplio rango de energías, pero la gravedad es tan débil que el único experimento en el que la podemos estudiar es la observación del propio Universo, desde nuestro particular punto de vista en la Tierra. No es sorprendente por tanto, que en los últimos 100 años el avance en la comprensión de la gravedad haya sido impulsado por motivos teóricos más que por motivos experimentales.

En 1905, Albert Einstein desarrolló la teoría Especial de la Relatividad\cite{Einstein:1905ve}, que nos dice que las leyes de la física tiene que ser iguales en cualquier sistema de referencia inercial, y en particular, las ecuaciones de Maxwell para el electromagnetismo. Como consecuencia, la velocidad de la luz tiene que ser la misma para cualquier observador inercial, y este hecho transforma el ``tiempo'' en una coordenada al mismo nivel que el ``espacio''. Dos sistemas de referencia están relacionados por un conjunto de transformaciones llamadas transformaciones de Lorentz, que incluyen \emph{rotaciones} entre dos coordenadas espaciales, y \emph{boosts} entre una coordenada espacial y el tiempo. Las coordenadas que usan diferentes observadores inerciales están relacionadas por uno de estos boosts, y por lo tanto diferentes observadores medirán tiempos y distancias diferentes para los mismos eventos. Aunque estas transformaciones son anti-intuitivas, es tranquilizador saber que en el límite de observadores cuya velocidad relativa es mucho menor que la luz, estas transformaciones se aproximan a las transformaciones de Galileo, en las cuales las coordenadas espaciales cambian, pero el tiempo es universal para todos los observadores, lo cual se parece más a nuestra visión cotidiana del mundo. La teoría de la Relatividad Especial ha sido comprobada experimentalmente numerosas veces, y es tan fundamental para toda la física que se ha desarrollado posteriormente, que es prácticamente imposible imaginar un mundo en el cual no sea cierta.

La ley de la gravitación universal es invariante bajo transformaciones de Galileo, pero no bajo transformaciones de Lorentz, y por lo tanto es incompatible con la Relatividad Especial. Einstein y Hilbert (\cite{Einstein:1915ca}, \cite{Hilbert:1915tx}) desarrollaron la teoría de la Relatividad General, que describe la gravedad de una manera compatible con la Relatividad Especial. La idea fundamental de la teoría es el principio de equivalencia de Eisntein. Con el mismo espíritu que la Relatividad Especial, este principio establece que las leyes de la física no-gravitatorias son iguales localmente para cualquier sistema de referencia en caída libre. Podemos interpretar este principio como que es imposible hacer un experimento dentro de un laboratorio, sin interactuar con el exterior, que distinga si el laboratorio se encuentra estático alejado de cualquier fuente de materia, o si está orbitando alrededor de un planeta. Obviamente, si el laboratorio es realmente grande, o si está realmente cerca de la fuente gravitacional, el experimentador podría medir diferencias en la fuerza de la gravedad a lo largo del laboratorio, que se verían como fuerzas de marea. Por ello es importante el distintivo ``localmente'' en la formulación del principio de equivalencia: Es importante que el tamaño de nuestro experimento sea mucho más pequeño que el tamaño característico de la variación en la fuerza de la gravedad. Este principio, que parece muy simple e intuitivo, obliga a que la gravedad sea un fenómeno de curvatura del espacio-tiempo. Un espacio-tiempo se describe con un tensor de rango 2, la métrica, que mide las distancias y los ángulos entre elementos diferenciales de las funciones coordenadas y que en general varía a lo largo del espacio-tiempo. Un observador que no experimente ninguna aceleración proveniente de otras fuerzas seguirá una trayectoria --llamada \emph{geodésica}-- que no es recta, pero que es la trayectoria más recta posible en geometría curva\footnote{Como veremos más tarde, trayectoria geodésica tiene dos posibles significados. Uno es la curva que extremiza la longitud (tal y como lo mide la métrica), otro es la curva más recta posible (curva cuyo vector tangente es transportado de manera paralela por una conexión). En Relatividad General, la métrica y la conexión están relacionados, y ambas definiciones coinciden.}. La fuerza de la gravedad ya no es una fuerza que produce aceleraciones en los observadores, sino que es el resultado aparente de que los observadores ya no siguen líneas rectas, y en su lugar siguen geodésicas que convergen (o divergen) a lo largo de su trayectoria.

La teoría mas simple\footnote{En 1913, antes de que se formulara la Relatividad General, Gunnar Nordström propuso otra teoría que satisface el principio de equivalencia de Einstein y coincide con la Ley de Newton en el límite de campo débil. Sin embargo esta teoría no es capaz de describir la desviación gravitatoria de los rayos de luz y da una corrección incorrecta para el avance del perihelio de mercurio. Aunque se descubrió antes, la formulación de manera Lagrangiana también es más complicada que la Relatividad General, pues contiene un campo escalar adicional.} que satisface el principio de Equivalencia de Einstein y coincide con la Ley de Newton en el límite de campo débil es la Relatividad General. Esta teoría establece que la curvatura del espacio tiempo (que se mide con tensores construidos a partir de las derivadas de la métrica) es igual al contenido de materia (que se mide con el tensor de energía-momento):

\begin{equation}
R_{\mu \nu} - \frac{1}{2} R g_{\mu \nu} = 8 \pi l_P^2 T_{\mu \nu}
\end{equation}

Estas ecuaciones son aparentemente simples, pero en realidad son un sistema acoplado de ecuaciones diferenciales de segundo orden, que son difíciles de resolver en general. Sin embargo, hay determinados casos de interés para la física, en los cuales se pueden resolver de manera analítica. Por ejemplo, soluciones de vacío con rotación y carga (que describen el espacio-tiempo en el exterior de una estrella, o también un agujero negro, como veremos a continuación) o soluciones homogéneas e isótropas (que sirven para describir el universo y la cosmología).


Después de su publicación, las predicciones de la Relatividad General fueron experimentalmente comprobadas, tales como el avance del perihelio de mercurio, y la desviación de los rayos de luz al pasar cerca del Sol \cite{Dyson:1920cwa}. Desde entonces, se han propuesto muchos más experimentos gravitacionales, tales como la medida del efecto Nordtvedt \cite{Nordtvedt:1968qr} en el movimiento de la luna, la medida del tiempo de retardo de la luz \cite{Shapiro:1964uw}, la búsqueda de una ``quinta fuerza'' \cite{Fischbach:1985tk}, el decaimiento orbital de púlsares binarios \cite{Weisberg:1981mt}, la detección directa de ondas gravitacionales \cite{Abbott:2016blz}, etc. y la Relatividad General explica satisfactoriamente todos ellos \cite{Will:2014kxa}.


Entonces, si la Relatividad General da explicación a todos los experimentos que se han propuesto hasta la fecha, ¿Por qué deberían los físicos intentar buscar teorías alternativas de gravedad? En primer lugar, la mayoría de los tests a los que se ha sometido la Relatividad General tiene una escala del orden del sistema solar (incluidos los tests de gravedad fuerte, como los púlsares binarios), pero es mucho más difícil hacer esos tests a escalas cosmológicas. La expansión acelerada del universo \cite{Riess:1998cb}, y determinado número de fenómenos a escalas galácticas, tales como el aplanamiento de las curvas de rotación en galaxias espirales \cite{Rubin:1980zd}, no pueden ser explicadas en Relatividad General solo con las fuentes de materia visibles. Este hecho ha llevado a la comunidad científica a postular la existencia de energía oscura y materia oscura. Muchas extensiones de Modelo Estándar predicen partículas de materia oscura, sin embargo aún no hay evidencia directa de su existencia. Hay indicios indirectos de la existencia de materia oscura en modelos de formación de estructura en el universo primitivo, en las fluctuaciones del fondo cósmico de microondas, en la lente gravitatoria generada por clústeres de galaxias en colisión (el ``Bullet Cluster''), y otras, pero no son concluyentes, y algunos otros fenómenos, tales como la estrecha relación entre la cantidad de materia visible y materia oscura necesaria en las galaxias \cite{McGaugh:2016leg}, o la dinámica de estrellas binarias muy alejadas entre sí \cite{Hernandez:2011uf}, sugieren otro tipo de explicación. La energía oscura se puede incorporar en ambos lados de las ecuaciones de Einstein, o bien como una constante cosmológica, o bien como una fuente de energía. En cualquier caso, nuestra actual comprensión de la teoría cuántica no nos permite explicar su pequeña magnitud. Algunas teorías de gravedad modificada intentan incorporar alguno o ambos de estos efectos como una consecuencia puramente gravitacional de las ecuaciones.

En segundo lugar, la Relatividad General predice la existencia de agujeros negros. Estos objetos se forman cuando una cantidad suficiente de materia se junta en una región pequeña del espacio, de modo que la atracción gravitatoria es tan fuerte que ni la luz puede escapar hacia el exterior. La existencia de estos objetos ha sido reconocida en regiones tales como en el centro de la galaxia, en determinadas fuentes de rayos X, y ha sido confirmada por la reciente observación de ondas gravitacionales provenientes de fusiones de agujeros negros \cite{Abbott:2016blz}. La Relatividad General predice que en el interior de estos objetos existe una región llamada \emph{singularidad} donde la curvatura del espacio-tiempo diverge. El destino de cualquier observador que se adentra en un agujero negro es viajar hasta la singularidad en un plazo de tiempo finito, a partir del cual el espacio-tiempo estaría mal definido. La capacidad predictiva de la teoría se pierde en la singularidad, y es una región muy problemática desde el punto de vista matemático. La Relatividad General también predice singularidades en modelos cosmológicos, tales como la singularidad del Big Bang o singularidades cosmológicas futuras. El objetivo de determinadas extensiones de la Relatividad General es \emph{suavizar} estas singularidades de modo que ya no sean problemáticas


En tercer lugar, aunque la Relatividad General da una descripción adecuada de la gravedad, la teoría cuántica de campos es la teoría del resto de partículas y fuerzas. Se espera que a energías del orden de la escala de Planck sea necesario describir la gravedad por una teoría cuántica de la gravitación. Quizás una teoría cuántica de la gravedad resuelva el problema de las singularidades. Sin embargo, no es posible cuantizar la Relatividad General de manera perturbativa con las técnicas que conocemos. Esto ha llevado a diferentes planteamientos a la hora de afrontar la cuantización de la gravedad: Uno sería de lo fundamental hacia lo particular, buscando una teoría cuántica que tenga todas las propiedades adecuadas y que se corresponda con la Relatividad General a nuestras escalas. Este es el planteamiento de teoría de cuerdas y de Loop Quantum Gravity. La teoría de cuerdas es una teoría de unificación en la cual las partículas son descritas por cuerdas unidimensionales en un espacio de dimensión mayor, que da lugar al Modelo Estándar y a la Relatividad General tras compactificar hasta las 4 dimensiones habituales. Loop Quantum Gravity intenta cuantizar la gravedad de una manera no perturbativa, evitando así los problemas de cuantizar gravedad como una fuerza sobre un espacio-tiempo de Minkowski. Otro planteamiento sería de de lo particular hacia lo fundamental, buscando una teoría clásica (efectiva) que sea un mejor punto de partida a la hora de cuantizar. Como veremos, el problema de las singularidades también se puede resolver de una manera clásica en el formalismo Métrico-Afín, y esta teoría podría ser más apropiada que la Relatividad General para ser cuantizada.

En esta tesis estudio el problema de la singularidades para una familia de teorías de gravedad modificada en el formalismo Métrico-Afín. En el formalismo Métrico-Afín se considera que la estructura afín (que nos dice como hacer el transporte paralelo y define una derivada covariante) es independiente de la estructura métrica (que nos dice como medir tiempos y distancias, y define la estructura causal del espacio-tiempo). La Relatividad General está formulada en el formalismo Riemanniano, en el cual la estructura métrica determina la estructura afín (a través de la conexión de Levi-Civita). Esta es una extensión interesante de la Relatividad General, porque no conocemos realmente si la estructura geométrica del espacio tiempo es Riemanniana o Métrico-Afín. Además, presenta ciertas ventajas, tales como que las ecuaciones del movimiento son siempre ecuaciones diferenciales de segundo orden, incluso si consideramos correcciones cuadráticas a la curvatura. Esto hace que la teoría no sufra de inestabilidades tales como los ghosts, que serían un obstáculo a la hora de cuantizar la teoría. Además, veremos como se resuelve el problema de las singularidades en este formalismo.


\subsubsection{Geodésicas}

El principio de equivalencia de Einstein hace que cualquier teoría gravitacional que lo satisfaga trate sobre una métrica y sus geodésicas. Las geodésicas de una métrica son las curvas que extremizan la longitud entre dos puntos dados. La longitud de una curva $\gamma^\mu(\lambda)$ entre dos puntos $a$ y $b$ viene dada por:

\begin{equation}
 L = \int_a^b \sqrt{-\frac{\df \gamma^\mu}{\df \lambda} \frac{\df \gamma^\nu}{\df \lambda} g_{\mu \nu}} \df \lambda
\end{equation}

Las ecuaciones que determinan que curva es la que minimiza la longitud son:

\begin{equation}
 g_{\alpha \mu} \frac{\df^2 \gamma^\mu}{\df \lambda^2} + \frac{\df \gamma^\mu}{\df \lambda} \frac{\df \gamma^\nu}{\df \lambda} \partial_\alpha g_{\mu \nu} - \frac{1}{2} \frac{\df \gamma^\mu}{\df \lambda} \frac{\df \gamma^\nu}{\df \lambda} \partial_\mu g_{\mu \alpha}=0 \label{Aeq:Geo1}
\end{equation}

Por otro lado, estamos tratando con observadores no acelerados. La aceleración es la derivada de la velocidad (que se mide con el vector unitario tangente a la curva) a lo largo de su trayectoria. Esto se traduce en la siguiente ecuación:


\begin{equation}
 u^\mu \nabla_\mu u^\nu = 0
\end{equation}
donde $\nabla$ representa la derivada covariante que está definida por la conexión $\Gamma$. De modo que esta ecuación es equivalente a esta otra:

\begin{equation}
 \frac{\df^2 \gamma^\nu}{\df \lambda^2} + \Gamma^\nu_{\alpha \beta} \frac{\df \gamma^\alpha}{\df \lambda}\frac{\df \gamma^\beta}{\df \lambda} = 0 \label{Aeq:Geo}
\end{equation}

Comparando las ecuaciones (\ref{Aeq:Geo}) y (\ref{Aeq:Geo1}), vemos que coinciden si la conexión es la \emph{conexión de Levi-Civita}:

\begin{equation}
 \Gamma^\alpha_{\beta \gamma}\equiv \frac{1}{2} g^{\alpha \mu} \left ( \partial_\beta g_{\mu \gamma} + \partial_\gamma g_{\beta \mu} - \partial_\mu g_{\beta \gamma} \right )\label{Aeq:LCconnection}
\end{equation}

En el formalismo Riemanniano se asume que la conexión del espacio-tiempo es la de Levi-Civita, la cual satisface la propiedad de $\nabla g =0$ (donde $g$ es la métrica del espacio-tiempo). Esta conexión surge de manera natural como la conexión inducida en una subvariedad curva embebida en variedad plana (una en la que $\Gamma=0$ en toda la variedad). Por razones históricas, la Relatividad General se construyó usando este formalismo, dado que solo fue años después, que se empezó a considerar geometrías no-Riemannianas \cite{eisenhart1927non}. En esta tesis consideraremos conexiones independientes de la métrica, y veremos cuales son sus consecuencias. 

\subsubsection{Descripción de los Observadores Físicos}

Los observadores físicos se suelen describir con curvas en el espacio-tiempo, los cuales miden un tiempo propio equivalente a la longitud de dicha curva. Los observadores en caída libre (también llamados inerciales) siguen curvas geodésicas. Esta descripción, en la cual los observadores son objetos puntuales moviéndose a lo largo de una curva, no la deberíamos considerar demasiado realista. El principio de equivalencia de Einstein se preocupa exclusivamente de experimentos \emph{locales}, en los cuales el tamaño de los observadores es despreciable respecto a las variaciones de la métrica; no es de extrañar que en este caso los observadores se puedan considerar puntuales. Pero si el tamaño del observador es comparable al de las variaciones de la métrica, no solo este observador dejará de seguir trayectorias geodésicas, sino que la descripción no es la adecuada. 

En Relatividad General existen soluciones en las cuales la curvatura diverge. Si nos aproximamos a dicha divergencia de curvatura, da igual cuan pequeño sea el observador, a partir de cierto punto las variaciones en curvatura serán mayores que el tamaño del observador. Esto hace que el observador experimente fuerzas de marea que pueden cambiar su trayectoria, o al menos, deformar al observador. Por lo tanto, es importante describir a los observadores de algún modo que permita estudiar las fuerzas de marea que experimentan.

Para describir en términos matemáticos tal observador, podemos pensar en una ``nube de motas de polvo'', donde cada mota sigue una trayectoria geodésica. Según esta nube viaja, las motas de polvo que la constituyen se acercarán o alejarán entre ellas. Ahora pensemos en un objeto rígido con la forma de esta nube de polvo en su lugar: donde las motas convergen y la nube se hace más pequeña, el objeto rígido experimentaría una compresión, y donde las motas divergen y la nube de polvo se expande, el objeto rígido experimentaría una fuerza de estiramiento. Una \emph{congruencia} es el conjunto de curvas integrales de un campo vectorial no nulo. Una \emph{congruencia geodésica} es una congruencia en la cual cada curva es geodésica. La trayectoria de las motas de polvo está descrita por una congruencia geodésica, y esta herramienta matemática nos permitirá describir las fuerzas de marea que experimenta un objeto rígido.

Esta descripción de un observador, en la cual cada uno de sus constituyentes intenta seguir una trayectoria geodésica, pero no lo hacen debido a las fuerzas internas que mantienen íntegro al observador, es una mejora respecto a la sencilla descripción en base a una única geodésica. Sin embargo, según nos acercamos a una divergencia de curvatura, nos deberíamos preocupar que incluso los constituyentes elementales que forman el observador tienen un tamaño superior al de la variación en curvatura. Estos constituyentes elementales al final son electrones y protones y otras partículas fundamentales, cuyas propiedades son mejor descritas por una onda que se propaga que por una trayectoria geodésica. Por lo tanto, para entender correctamente cual es el destino de un observador que se acerca a una divergencia de curvatura, debemos estudiar también la propagación de ondas. Esta descripción es además compatible con las descripciones anteriores, puesto que en determinado límite similar a la ``óptica geométrica'', las ondas se propagan en forma de rayo, siguiendo geodésicas.

Para dar una visión completa de la física en las cercanías de una divergencia de curvatura, en esta tesis estudio la geometría del espacio-tiempo usando tanto geodésicas, como congruencias de geodésicas y ondas que se propagan.

\subsubsection{Relatividad General}

Siguiendo el principio de equivalencia, nuestro espacio-tiempo va a ser descrito por una métrica. Para construir una teoría que nos dé las ecuaciones para una métrica, podemos construir un Lagrangiano que contenga la métrica, sus derivadas, y quizás algún campo auxiliar. Podríamos intentar usar la conexión de Levi-Civita (eq. \ref{Aeq:LCconnection}), pero no es un objeto tensorial, y su valor depende de qué sistema de coordenadas estemos usando. Si que hay un objeto puramente tensorial que contiene derivadas de la métrica, que es el tensor de curvatura de Riemann:

\begin{equation}
 R^\alpha{}_{\beta \mu \nu}= \partial_\mu \Gamma^\alpha_{\nu \beta}-\partial_\nu \Gamma^\alpha_{\mu \beta} + \Gamma^\alpha_{\mu \sigma} \Gamma^\sigma_{\nu \beta}-\Gamma^\alpha_{\nu \sigma} \Gamma^\sigma_{\mu \beta}
\end{equation}

También podemos usar su traza, $R_{\alpha \beta}\equiv R^\sigma{}_{\alpha \sigma \beta}$, que es conocida como el tensor de Ricci, y el escalar de Ricci $R\equiv g^{\alpha \beta} R_{\alpha \beta}$. En principio se podrían construir más escalares de curvatura (y otros tensores) usando diversas combinaciones de estos y sus derivadas.

La Relatividad General se puede derivar usando el siguiente Lagrangiano $\Lagr_G = R$ con una constante apropiada para que reproduzca la gravedad de Newton en el límite adecuado. En la acción también debe aparecer el Lagrangiano de materia, que describe la dinámica del resto de campos de materia. La acción por lo tanto es:

\begin{equation}
 S= \frac{1}{16 \pi l_P^2} \int_\mathcal{M} R \sqrt{|g|} \df^4 x + \int_\mathcal{M} \Lagr_\text{m} \sqrt{-g} \df^4 x \label{Aeq:GRaction}
\end{equation}

La variación de esta acción respecto a la métrica da las siguientes ecuaciones.

\begin{equation}
 \underbrace{R_{\mu \nu} - \frac{1}{2} R g_{\mu \nu}}_{G_{\mu  \nu}} = 8 \pi l_P^2 T_{\mu \nu}
\end{equation}
donde $T_{\mu\nu}=\frac{2}{\sqrt{|g|}}\frac{\delta \Lagr_\text{m}\sqrt{|g|}}{g^{\mu \nu}}$ es el tensor de energía-momento. Estas ecuaciones nos dicen que la curvatura del espacio-tiempo es igual a su contenido de materia. Estas ecuaciones son difíciles de resolver en general, pero podemos resolverlas para un espacio-tiempo estático, esféricamente simétrico y vacío:

\begin{equation}
 \df s^2 = -\left ( 1-\frac{r_S}{r} \right ) \df t^2 + \frac{1}{\left ( 1-\frac{r_S}{r} \right ) } \df r^2 + r^2 \df \Omega^2
\end{equation}

Esta métrica describe el espacio exterior a una estrella, o bien un agujero negro. $r_S$ es el radio de Schwarzschild que depende de la masa $r_S \equiv 2 M l_P^2$. En este radio se encuentra el horizonte de sucesos, todo lo que atraviese el horizonte es atraído irremediablemente hasta $r=0$ y no puede escapar al exterior. En particular, si una estrella, por algún motivo, se hace más pequeña que su propio radio de Schwarzschild, colapsará hasta $r=0$ toda ella. También se conoce la solución para un espacio-tiempo estático, esféricamente simétrico y con carga $q$:

\begin{equation}
 \df s^2 = -\left ( 1-\frac{r_S}{r} + \frac{r_q^2}{r^2} \right ) \df t^2 + \frac{1}{\left ( 1-\frac{r_S}{r} + \frac{r_q^2}{r^2} \right ) } \df r^2 + r^2 \df \Omega^2
\end{equation}
donde $r_q \equiv q l_P$. Esta solución se conoce como métrica de Reissner-Nordström. La estructura de Reissner-Nordström es diferente a Schwarzschild, y depende de la relación carga-masa. Si $r_q<r_S/2$ hay dos horizontes, si $r_q=r_S/2$ solo hay uno degenerado (agujero negro \emph{extremal}), y si $r_q > r_S/2$ no hay horizontes y la región $r=0$ puede transmitir información al exterior (el caso de \emph{singularidad desnuda}).

\subsubsection{Definiendo Singularidad}

La región $r=0$ es problemática en la geometría de Schwarzschild. Cualquier observador que cruza el horizonte acaba en el origen (donde lo comprimen fuerzas infinitas) y no hay manera de continuar la evolución temporal de ese observador. Nos gustaría definir estos puntos problemáticos de la geometría como \emph{singularidades}. Sin embargo, esta definición no es tan fácil como a primera vista pudiéramos pensar. Vamos a seguir el razonamiento de Geroch (\cite{Geroch:1968ut}) para llegar hasta la definición adecuada. Un primer intento de definición sería del estilo de ``región del espacio-tiempo donde algo va mal'', y algo que va mal podría ser una magnitud geométrica que diverge. Este tipo de definición se encuentra con dos clases de problemas

\begin{itemize}
 \item El primero tiene que ver con la magnitud geométrica que diverge. Esta magnitud no pueden ser las componentes de la métrica, porque divergen en regiones que no son problemáticas, tales como el horizonte de sucesos de Schwarzschild. Las componentes del tensor de Riemann tampoco son buena elección, porque si la curvatura no es constante, siempre es posible elegir un sistema de coordenadas en el cual alguna componente diverge. Los escalares de curvatura parecen más apropiados, pues no dependen del sistema de coordenadas; pero hay un infinito número de ellos (contracciones del Riemann con la métrica, consigo mismo y con sus derivadas), y no está claro la relevancia física de todos ellos. Por otro lado, hay geometrías que no contienen divergencias de curvatura, y que consideramos singulares, tales como un espacio-tiempo de Minkowski en el cual hemos quitado un sector cilíndrico y hemos identificado los dos lados del corte. 
 
 \item El segundo problema tiene que ver con el concepto de ``región del espacio-tiempo''. En Relatividad General escogemos una variedad a la cual asignamos una métrica. Podríamos escoger una variedad en la cual hemos quitado las regiones singulares; sin embargo, nos gustaría definir ese espacio-tiempo como singular en cualquier caso. ¿Cómo nos podemos dar cuenta si hemos quitado una región del espacio-tiempo?  No es tan sencillo como a primera vista pudiéramos pensar, pues es fácil esconder dicha región en algún sistema de coordenadas. O al revés, usar un sistema de coordenadas que traiga una región inaccesible del espacio-tiempo a un valor finito de las coordenadas. Por ello debemos evitar usar coordenadas en nuestra definición. Tampoco podemos usar el concepto de distancia, pues siempre podemos encontrar una curva que une dos puntos con una longitud tan próxima a $0$ como queramos (para una métrica con signatura Lorentziana).
\end{itemize}

La solución la podemos encontrar en la física. Si quitamos una parte del espacio tiempo, existirá algún observador (en caída libre) que se encuentre con el ``fin'' del espacio-tiempo. Esta situación se corresponde con una geodésica con un punto final (o de comienzo). De este modo definimos: \emph{Un espacio-tiempo es no-singular si todas las geodésicas son completas, o están contenidas en un conjunto compacto}\footnote{El motivo por el cual la definición hace referencia a conjuntos compactos es porque existen conjuntos compactos geodésicamente incompletos \cite{Misner:1963fr}, que no pueden ser el resultado de haber quitado una región del espacio tiempo.}.

Una geodésica es completa si su parámetro afín puede tomar valores arbitrariamente grandes. Esta definición parece que le falta algo: hemos tratado el problema de qué entendemos por ``región del espacio-tiempo'', pero no hemos hecho ninguna referencia a la parte de ``donde algo va mal''. Esto es una buena propiedad, porque lo que de verdad nos importa es la existencia de observadores y de su evolución temporal. El hecho de que la curvatura diverja, y por lo tanto los observadores experimenten fuerzas infinitas, es secundario a la existencia de observadores. En la literatura, muchas veces se asocia los conceptos de divergencia de curvatura y singularidad, pues muy a menudo van de la mano en Relatividad General. En esta tesis veremos que no los debemos confundir.

\subsubsection{Teorías Métrico-Afín}

La Relatividad General trabaja en el formalismo Riemanniano. En este formalismo, la conexión --que define el transporte paralelo y la derivada covariante-- se toma como la conexión de Levi-Civita dada por la métrica. Existe otro formalismo, el formalismo Métrico-Afín, en el cual la conexión y la métrica son estructuras independientes

Parece natural trabajar en el formalismo Riemanniano, dado que la conexión de Levi-Civita garantiza que los ángulos se preservan al seguir una trayectoria no acelerada. Esto es lo que nos dice nuestra intuición, y forma parte del principio de equivalencia de Einstein: si no se preservan los ángulos a lo largo de la trayectoria, los experimentos dependerían de que trayectoria estamos siguiendo, y hasta ahora no hemos observado violaciones del principio de equivalencia. Sin embargo, podríamos argumentar que el formalismo Riemanniano restringe artificialmente la estructura geométrica del espacio-tiempo. Podríamos considerar una teoría en la cual la conexión es aproximadamente Levi-Civita excepto en las regiones donde la gravedad es más fuerte; esta teoría respetaría el principio de equivalencia dentro del rango de nuestras observaciones.

Para construir una teoría en el formalismo Métrico-Afín, sencillamente tenemos que tomar un Lagrangiano construido con escalares de curvatura, que a partir de este momento van a depender de una conexión independiente y no sólo de la métrica. Para obtener las ecuaciones del movimiento tendremos que variar la acción respecto a la métrica, y también respecto a la conexión. Análogo al tensor energía-momento, existe otro tensor llamado tensor de hipermomento que resulta de la variación de la acción de materia respecto a la conexión independiente. Por simplicidad, en esta tesis vamos a considerar el caso en que esta conexión no tiene torsión.

Una primera consideración es ver si en este formalismo es posible recuperar los resultados de la Relatividad General. A fin de cuentas, estamos interesados en hacer extensiones de la Relatividad General, pero que hagan las mismas predicciones respecto al sistema solar y otros experimentos. Tomemos la acción de la ec. \ref{Aeq:GRaction}, pero ahora considerando que la curvatura depende de la conexión independiente $R=R_{\alpha \beta}(\Gamma) g^{\alpha \beta}$, y que el Lagrangiano de materia no lo hace. Tomando variaciones respecto a la métrica y la conexión obtenemos las siguientes ecuaciones del movimiento:

\begin{IEEEeqnarray}{rCl}
 R_{\mu \nu}(\Gamma) - \frac{1}{2} R(\Gamma) g_{\mu \nu} &=& 8 \pi l_P^2 T_{\mu \nu} \\
 \nabla_\alpha (\sqrt{-g} g^{\nu \beta}) &=& 0
\end{IEEEeqnarray}

La primera ecuación es igual que la ecuación de Einstein, salvo que en este caso, la curvatura depende de la conexión independiente, en vez de la métrica. La segunda ecuación nos dice que la conexión independiente es la conexión de Levi-Civita de la métrica $g$. De modo que estas ecuaciones son las mismas que las de Relatividad General. Para obtener resultados diferentes al formalismo Riemanniano, necesitaremos elegir un Lagrangiano diferente. En esta tesis vamos a proponer el siguiente Lagrangiano $\Lagr_G=R+l_P^2 (a R^2 + R_{\mu \nu} R^{\mu \nu})$, que en el formalismo Riemanniano ha suscitado interés como un modo de incorporar correcciones cuánticas a las ecuaciones clásicas. Con este Lagrangiano las ecuaciones son:

\begin{IEEEeqnarray}{rCl}
(1+2a l_p^2 R) R_{\mu \nu}) - \frac{1}{2} (R+l_P^2 (a R^2 + R_{\alpha \beta} R^{\alpha \beta})) g_{\mu \nu} + 2 a l_p^2 R_\mu{}^\alpha R_{\alpha \nu} &=& 8 \pi l_P^2 T_{\mu \nu} \IEEEeqnarraynumspace  \label{Aeq:metricQuad} \\
 \nabla_\alpha  \underbrace{\left [\sqrt{-g}((1+2a l_p^2 R) g^{\mu \nu} + 2 a l_P^2 R^\mu{}_\alpha R^{\alpha \nu})\right ] }_{\equiv \sqrt{h} h^{\mu \nu}} &=& 0
\end{IEEEeqnarray}

En este caso, la segunda ecuación nos dice que la conexión independiente ya no es la conexión de Levi-Civita de $g$, sino que es la conexión de Levi-Civita de una métrica auxiliar $h$, diferente a $g$. Por la forma de las ecuaciones, podemos intuir que cuando el valor del tensor energía-momento es bajo --y por lo tanto la curvatura $R$ es pequeña--, $h$ y $g$ van a ser muy parecidas, y el resultado va a ser prácticamente como el formalismo Riemanniano. Además, en el caso de curvatura pequeña, el termino $R$ domina sobre los términos $R^2$ y $R_{\mu \nu} R^{\mu \nu}$, así que el Lagrangiano va a ser aproximadamente el de la Relatividad General, y recuperaremos los resultados ya conocidos en este límite. Sin embargo, cuando el valor del tensor energía-momento sea elevado y la curvatura sea grande empezaremos a ver diferencias sustanciales.

\subsubsection{Agujero de Gusano Geónico}

Para obtener soluciones para la métrica que representen agujeros negro con carga eléctrica debemos considerar un espacio-tiempo esféricamente simétrico y estático, con un tensor de energía-momento correspondiente a una carga puntual $q$:

\begin{equation}
 T_\mu{}^\nu = \frac{q^2}{8\pi r^4}\left ( \begin{array}{cccc} -1 & 0 & 0 & 0 \\ 0 & -1 & 0 & 0 \\ 0 & 0 & 1 & 0 \\ 0 & 0 & 0& 1 \end{array} \right ) \label{Aeq:EMVacuumT}
\end{equation}

La solución de las ecuaciones (\ref{Aeq:metricQuad}) con este tensor energía-momento es:

\begin{equation}
 g=-\frac{A}{\sigma_+} \df t^2 + \frac{1}{A \sigma_+} \df x^2 + r^2(x) \df \Omega^2 \label{Aeq:metricgx}
\end{equation}

Esta métrica contiene las siguientes funciones:

\begin{itemize}
 \item El radio de las 2-esferas de la geometría es función de la coordenada $x$ (ver figura \ref{4fig:x-r}):
 \begin{equation}
  r^2=\frac{x^2+\sqrt{x^4+4r_c^4}}{2}
 \end{equation}
 donde $r_c$ es un valor mínimo del radio de las 2-esferas que depende de la carga del agujero negro como $r_c = 2^\frac{1}{4} \sqrt{r_q l_P}$. Este radio mínimo es la garganta de un agujero de gusano, que separa dos regiones asintóticas diferentes $x\rightarrow +\infty$ y $x \rightarrow -\infty$.
 \item La función $\sigma_+$ vale $1+r_c^4/r^4$, y es aproximadamente $1$ para valores altos del radio, y aproximadamente $2$ para valores cercanos a $r_c$.
 \item $A$ es una función que para valores altos del radio se comporta como $A\simeq1-\frac{r_s}{r}+\frac{r_q^2}{r^2}-\frac{r_c^4}{r^4}$, que es el mismo comportamiento que Reissner-Nördstrom, con correcciones de cuarto orden. 
 \item La función $A$ para valores del radio cercanos a $r_c$ tiene un comportamiento muy diferente. Si la combinación de masa y carga $\delta_1 \equiv r_q^2/(r_S r_c)$ es igual un valor crítico $\delta_c \simeq 0.57207$, $A$ toma un valor constante, y los escalares de curvatura de esta geometría son finitos. Si $\delta_1$ es diferente a $\delta_c$, entonces $A$ diverge como $\pm 1/|x|$, de manera positiva si $\delta_1 > \delta_c$, y de manera negativa si $\delta_1 < \delta_c$. En estos dos casos los escalares de curvatura divergen.
\end{itemize}

La carga eléctrica se define como una integral de las líneas de campo eléctrico que atraviesan una superficie que encierra las fuentes. En este caso vemos que la carga en un producto de la topología del espacio-tiempo: Las líneas de campo eléctrico comienzan en una región asintótica, atraviesan el agujero de gusano, y salen por la otra región asintótica; creando la ilusión de un objeto positivamente cargado en un lado del agujero de gusano, y un objeto negativamente cargado en el otro lado. 

Hemos construido soluciones de agujero de gusano sin la necesidad de materia exótica, en contraste con otras soluciones de la literatura (\cite{Morris:1988tu}, \cite{Lobo:2009ip}, \cite{Visser:1995cc}). Las ecuaciones para el campo electromagnético y para el campo gravitacional están bien definidas en todo el espacio, y no contienen ninguna fuente. Esta solución responde a las características que Wheeler propuso para un \emph{geón} (\cite{Wheeler:1955zz}, \cite{Misner:1957mt}).

Los escalares de curvatura divergen en la garganta del agujero de gusano, pero eso no quiere decir que la geometría sea singular, tenemos que estudiar las geodésicas. En función del valor de $\delta_1$ tenemos tres casos diferentes:

\begin{itemize}
 \item $\delta_1 < \delta_c$: Este caso tiene una estructura causal similar a Schwarzschild. Todas las geodésicas, temporales o luminosas, que atraviesen el horizonte tiene como destino alcanzar la garganta del agujero gusano. No hay problema en integrar la ecuación de las geodésicas más allá, y se pueden extender de manera natural.
 \item $\delta_1 = \delta_c$: En este caso, la geometría puede tener un horizonte a cada lado del agujero de gusano, un horizonte en la garganta, o bien ningún horizonte, dependiendo del número de cargas. En cualquier caso, la métrica es suave en la garante, y no hay ningún problema en extender las geodésicas a través de la garganta. 
 \item $\delta_1 > \delta_c$: Este caso tiene una estructura causal similar a Reissner-Nordström, y la geometría puede tener dos horizontes, uno, o ninguno a cada lado del agujero de gusano, dependiendo de la relación carga-masa. En el caso con dos horizontes, las geodésicas que atraviesen el horizonte exterior, se verán obligadas a atravesar también el horizonte interior. Sin embargo, la garganta del horizonte de gusano supone una barrera de potencial, de modo que todas las geodésicas salvo las luminosas sin momento angular (que atraviesan sin problemas), son repelidas y no alcanzan la garganta.
\end{itemize}

En ninguno de los casos hay geodésicas incompletas, de modo que esta geometría es regular. Para entender lo que le sucede a un observador físico que atraviesa la garganta del agujero de gusano, también hemos estudiado las congruencias de geodésicas y la propagación de ondas. La congruencia la estudiamos con la evolución de un elemento infinitesimal de volumen transportado. Al acercarse a la garganta del agujero de gusano, este elemento de volumen se estira en la dirección radial como $1/\sqrt{|x|}$, y mantiene un valor finito en las direcciones angulares; mientras que en Schwarzschild, el elemento de volumen al acercarse a $r=0$, se estira en la dirección radial como $1/\sqrt{r}$, y se comprime en las direcciones angulares como $\sqrt{r}$. De modo que en el agujero de gusano geónico el volumen infinitesimal transportado diverge como $1/\sqrt{|x|}$, mientras que en Schwarzschild se comprime hasta volumen $0$. La compresión en Schwarzschild hace que cualquier objeto físico que caiga sea irremediablemente destruido, y se conoce en la literatura como \emph{singularidad fuerte} (\cite{Ellis:1977pj}, \cite{CLARKE1985127}); cabe preguntarse si un estiramiento infinito es igual de destructivo, pero ¿que significa realmente que un volumen infinitesimal se vuelve infinitamente grande? Para ello he estudiado la transmisión de señales entre dos elementos de un volumen finito (ver figs. \ref{5fig:LightRayWH} y \ref{5fig:LightRayTimeWH}), y llegamos a la conclusión de que en ningún momento los elementos pierden el contacto causal: la distancia espacial entre dos geodésicas (separadas de manera no-infinitesimal) siempre es finita.

La propagación de ondas la estudiamos con un campo escalar, en términos de la coordenada radial tortuga $y^*$. Usando esta coordenada, la ecuación queda: 

\begin{equation}
 \partial_{y^*} \partial_{y^*} \psi_{\omega l} + \left [ \omega^2 - \underbrace{\left ( \frac{1}{r} \partial_{y^*} \partial_{y^*} r + \left ( m^2 + \frac{l(l+1)}{r^2} \right ) \frac{A}{\sigma_+} \right )}_{V_{\text{eff}}} \right ] \psi_{\omega l}=0
\end{equation}

El comportamiento a orden más bajo del potencial es $V_{\text{eff}} \approx \kappa/|y^*|^\frac{1}{2}$. Las soluciones de esta ecuación tienen un comportamiento regular en la garganta del agujero de gusano ($y^*=0$), que bien puede ser lineal o constante. Con esto podemos realizar cálculos como el de la sección eficaz de transmisión a través del agujero de gusano para un caso sin horizontes (ver fig. \ref{6fig:CrossSect}).

\subsubsection{Conclusiones}

A la luz del trabajo realizado en esta tesis, deberíamos reconsiderar nuestro enfoque habitual a la hora de tratar las singularidades. Es posible que no sea un problema que debe resolver gravedad cuántica, sino que quizás es necesario obtener una teoría clásica que no tenga ese problema antes de intentar una cuantización. Después de todo, en esta tesis hemos introducido conceptos geométricos que están artificialmente restringidos en Relatividad General. En física del estado sólido, estas herramientas geométricas son necesarias para describir un cristal con defectos, mientras que el formalismo Riemanniano sólo es capaz de describir cristales perfectos; y estos defectos son imprescindibles a la hora de describir las propiedades globales del cristal. Esto hace plantearnos cual es la verdadera estructura geométrica del Universo. En esta tesis, hemos considerado que la materia no se acopla a estas nuevas estructuras; pero si lo hiciera, podríamos buscar violaciones del principio de equivalencia de Einstein cerca de las regiones más curvadas del Universo. Cúal es la geometría subyacente del Universo y cómo se acopla a la materia queda como una pregunta abierta que debemos resolver.

\cleardoublepage

\pagenumbering{arabic}
\chapter{Introduction: General Relativity and the Schwarzschild Geometry}

Gravity is ubiquitous in our lives. Unlike the other forces of nature, for which macroscopic objects are mostly neutrally charged, all common matter has mass and is accelerated towards all the other matter content. We see the effects of gravity as objects fall to the ground as well as in the motion of planets around the sun. It feels only natural that it was the first force for which we had an accurate mathematical description, given by sir Isaac Newton. However, it remains the most mysterious interaction. We can study other forces by building experiments in which physics is tested at a wide range of energies, but gravity is much weaker and we cannot do the same. However, we have an experiment set up for us, the Universe, which we can study from a very particular point of view, the Earth. Then, it is not surprising that the driving force behind the modern view of gravity has been theoretical in nature rather than experimental.

In 1905, Albert Einstein developed the theory of Special Relativity\cite{Einstein:1905ve}, which tells us that the laws of physics must be the same in every inertial reference frame, and in particular Maxwell's equations for electromagnetism. As a consequence, the speed of light is the same for every inertial reference frame, and this fact turns ``time'' into a coordinate in the same vein as ``space''. Two reference frames are related by a set of transformations called Lorentz transformations, which include \emph{rotations} between two spatial coordinates and also \emph{boosts} that mix the time coordinate with a spatial one. The coordinates of observers in different reference frames are related by one of these 'boosts', and consequently, different observers will measure different distances and times for the same events. This is counter-intuitive, but it is comforting that in the limit of observers with relative speed much slower than the speed of light, Lorentz transformations tend to Galilean transformations, in which spatial coordinates change with time but time remains universal, and agrees with our intuitive view of the world. Special Relativity has been extensively tested, and it is so fundamental for all the physics that came after it, that it is almost impossible to imagine a world in which it does not hold. 

Newton's law of universal gravitation is invariant under Galilean transformations, but is not invariant under Lorentz transformations. Therefore, physicists looked for a description of gravity that was compatible with Special Relativity. The theory of General Relativity was born in 1915, developed by Einstein and Hilbert (\cite{Einstein:1915ca}, \cite{Hilbert:1915tx}), to describe gravity. The fundamental idea of the theory is the Einstein Equivalence Principle (EEP). In the same spirit as Special Relativity, it states that the non-gravitational laws of physics are locally the same in every free-falling reference frame. We could think of it as that it is impossible to make an experiment in a laboratory that distinguishes if that laboratory is static far from gravitational sources, or in free-fall near a star. Obviously, if the laboratory is really large, or if it is really close to a strong gravitational source, there will be differences in the force of gravity along the laboratory. These differences would be seen as tidal forces in our experiments, hence the word ``locally'' has importance: the laboratory must be small enough for the tidal forces to be negligible. This principle, that seems very simple and intuitive, forces gravity to be a curved space-time phenomenon. Space-time must be described with a metric that will change from point to point. Observers unaffected by other forces will not follow straight lines any more, but will follow geodesics of the metric, which are the straightest possible lines in curved geometry\footnote{As we will see, geodesics can have two different meanings. One will be that of curves that extremize the length as measured by a metric; the other is that of straightest possible path in the sense that the tangent vector is parallel transported along the curve according to some connection. In GR -- and in the Riemannian formalism in general --, the connection and the metric are related, and both definitions agree.}. Then, the force of gravity is the apparent result of those observers not following straight lines any more, but geodesics that converge or diverge along their path.


The simplest theory\footnote{In 1913, before GR was formulated, Gunnar Nordström proposed another theory (\cite{ANDP:ANDP19133450503}, \cite{ANDP:ANDP19133471303}) that agreed with EEP and Newton's law in the weak field limit. This theory is not able to describe the gravitational bending of light, and gives a wrong correction to the perihelion advance of Mercury. I call GR the simplest in terms that it will be the simplest when written as a Lagrangian, as we will see.} that agrees with the EEP and agrees with Newton's law in the weak field limit is General Relativity. It equates the curvature of the geometry (which is made of derivatives of the metric tensor that describes our geometry) with the matter content of the space-time (given by the energy-momentum tensor):

\begin{equation}
 R_{\mu \nu} - \frac{1}{2} R g_{\mu \nu} = 8 \pi l_P^2 T_{\mu \nu}
\end{equation}

These equations look deceptively simple, but actually are a system of coupled second-order differential equations on the metric which is very difficult to solve. However, it can be analytically solved for a number of different scenarios of physical interest, like vacuum solutions with rotation and charge, or a homogeneous and isotropic universe.

Soon after its publication, GR predictions were successfully tested, such as the measurement of the perihelion advance of Mercury, and the deflection of light by the Sun\cite{Dyson:1920cwa}. Since then, many more tests have been proposed to test gravity, such as measurement of the Nordtvedt effect\cite{Nordtvedt:1968qr} in lunar motion, time delay of light\cite{Shapiro:1964uw}, fifth-force searches\cite{Fischbach:1985tk}, orbital decay of binary pulsars\cite{Weisberg:1981mt}, direct detection of gravitational waves\cite{Abbott:2016blz}, etc. and GR has been successful in all of them \cite{Will:2014kxa}.

If GR has passed every test it has faced, why should physicists search for modified gravity theories? First, although GR has passed many tests at the scales of the size of the solar system (including the dynamics of binary systems with strong gravity, like binary pulsars), it is more difficult to make such tests at the cosmological level. The accelerated expansion of the universe \cite{Riess:1998cb} and a number of phenomena at galactic scales, such as the flattening of rotation curves in spiral galaxies \cite{Rubin:1980zd},  cannot be explained in GR with just the visible matter sources. This has led the scientific community to postulate the existence of Dark Energy \cite{Peebles:2002gy} and Dark Matter \cite{Bertone:2004pz}. Dark Matter particles are predicted in many extensions of the Standard Model, however, direct searches have been unsuccessful so far. There is indirect evidence on the existence of Dark Matter through gravitational lensing in colliding cluster of galaxies (the ``Bullet Cluster''), models of formation of large structures in the early universe, fluctuations on the Cosmic Microwave Background (CMB), and others, but none are conclusive and some other observed phenomena such as the tight relation between baryonic matter and dark matter in galaxies \cite{McGaugh:2016leg}, or the motion of widely separated binaries \cite{Hernandez:2011uf}, suggest a different kind of explanation. Dark Energy can be incorporated on either side of Einstein's equations, as a cosmological constant or as an energy source. In either case, its magnitude cannot be explained with our current understanding of quantum theory. Some modified gravity theories try to incorporate one or both of these effects as a purely gravitational consequence of the equations.

Second, GR predicts the existence of black holes. These objects are formed when enough mass lies in a small region of space, so that gravity becomes so strong that not even light can scape. The existence of such objects has been established in regions such as the centre of our galaxy, for some X-ray sources (such as Cyg X-1), and has been confirmed by the recent observation of gravitational waves coming from mergers of black holes\cite{Abbott:2016blz}. In the interior of these objects (as predicted by GR), there is a region called \emph{singularity} where the curvature of space-time diverges. The fate of every observer that goes inside the black hole is to travel to the singularity in a finite amount of time, at which space-time becomes ill-defined. The predictive power of the theory is lost at the singularity and it is a very problematic region from a mathematical point of view. Singularities are also found in cosmological models, as the Big Bang singularity and future singularities. One aim of some modified gravity theories is to ``smoothen'' these singularities so that they are no longer problematic.

Third, although GR describes gravity accurately, Quantum Field Theory (QFT) is the theory that describes the rest of particles and forces. It is expected that gravity at energies of the order of the Planck scale should be described by a quantum theory of gravitation. A quantum theory of gravitation could possibly solve the problem of singularities. However, GR cannot be properly quantized using the perturbative techniques of QFT. This has led to different approaches to search for this quantum gravity theory: A top-down approach would be to find a theory that has all the right quantum properties and breaks down to GR at the right scales. This is the approach of string theory and loop quantum gravity (LQG). String theory is a unification theory in which particles are represented by strings in a higher dimensional space, which gives rise to the Standard Model and GR after compactifying it to 4 dimensions. Loop quantum gravity tries to quantize gravity in a non-perturbative way, bypassing the problems of quantizing gravity like other forces in a Minkowski background. 
Another approach would be bottom-up, looking for classical (effective) theories which might be a better starting point to quantize gravity. As we will see, singularities can be cured in a classical way for certain modified gravity theories; such a theory might also be more appropriate to develop a quantization of gravity.

This thesis deals with the problem of singularities for a family of modified gravity theories in the metric-affine framework. A metric-affine theory is one in which the affine structure (which tells us how to parallel transport and define a covariant derivative) is independent of the metric structure (which tells us how to measure times, distances, and the causal structure of space-time); meanwhile GR is formulated in the Riemannian formalism, in which the metric structure determines the affine structure. This is an interesting extension of GR as we do not really know if the structure of the space-time is Riemannian or Metric-Affine. It also contains certain advantages, such as that the resulting equations of motion are at most second-order differential equations, even if we consider corrections that are quadratic or higher in the curvature scalars (Unlike in the Riemannian formalism, in which quadratic corrections to the curvature lead to fourth-order differential equations). This eliminates ghost instabilities that would happen if we try to quantize the theory. We will use black holes as a test for these theories, and see how in this scenario, the problem of singularities can be solved. 

This thesis is divided in seven chapters: In Chapter 1, I will introduce General Relativity and the most simple black hole solution. We will see that this black hole solution has a curvature divergence at the center, and we will study what happens to observers that fall into it. Also, we will describe charged black hole solutions in GR, which will serve as a comparison in the next chapters. In Chapter 2, we will examine the definition of singularity, which does not make mention of curvature divergences. We will also study one of the singularity theorems and how black hole solutions in GR are singular anyway. We will also see a different way of describing geodesics near a curvature divergence and  will study an interesting modified gravity theory called quadratic gravity. In Chapter 3, the fundamentals of metric affine theories will be explained, and we will see condensed-matter scenarios in which this formalism is necessary, such as to describe crystals with a density of defects. These systems need new tools to describe the underlying geometry, which gives rise to properties that is not possible to describe using only Riemannian geometry. In Chapter 4, we will obtain charged black hole solutions in the metric-affine formalism for a quadratic Lagrangian, and will see how these solutions have a wormhole structure with a curvature divergence at their throat. In Chapter 5, we will analyse geodesics and congruences of geodesics of these wormoles, and will determine that these new solutions are non-singular. The curvature divergence at the wormhole throat is not an impediment for an observer to cross it in a well-defined way. In Chapter 6, we will study the propagation of waves in these geometries, and see how this propagation is also non-singular. In Chapter 7 we will study charged black holes in d-dimensions for a different gravity Lagrangian to see that the absence of singularities is a generic feature of these theories. The last chapter contains a summary and some conclusions.


\section{Einstein Equivalence Principle}

The Einstein Equivalence Principle (EEP) is the fundamental principle in which GR is constructed. It tells us that\cite{Will:2014kxa}:
\begin{itemize}
 \item The Weak Equivalence Principle (WEP) holds true. This principle states that the trajectory of a freely falling test body is independent of its internal structure and composition. 
 \item The outcome of any local non-gravitational experiment is independent of the velocity of the freely-falling reference frame in which it is performed.
 \item The outcome of any local non-gravitational experiment is independent of where and when in the universe it is performed.
\end{itemize}

The first point states that the ``gravitational mass'' of a test body (as in the charge of the gravitational potential) is the same as its ``inertial mass'' (as in the resistance to acceleration caused by a force). The classic experiment is that two objects made of different material and shape dropped from the same place fall with the same speed and acceleration because of the force of gravity (in vacuum). The second and third points are called local Lorentz invariance and local position invariance, and tell us that Special Relativity holds in the coordinates of the freely-falling reference frame, no matter the velocity or the position of the freely-falling frame. In other words, there is not a special place or reference frame in the universe where physics behaves differently. This restricts gravitational theories to theories where:

\begin{itemize}
 \item The space-time has a metric
 \item The world lines of test bodies are geodesics of that metric
 \item In local freely falling frames, the non-gravitational laws of physics are those of special relativity.
\end{itemize}

Any theory that satisfies these three properties, satisfies the EEP. The converse is also true \cite{Will:Libro81}. As in special relativity, massive test bodies will follow time-like geodesics (whose length as measured by the metric is negative), while light rays will follow null geodesics (whose length as measured by the metric is zero). The metric will no longer be Minkowski everywhere as in Special Relativity (although it will be locally Minkowski in a freely falling reference frame), but in general, will change from point to point, describing a curved geometry.

We should note that there is a stronger version of the EEP, which is called Strong Equivalence Principle (SEP), that states the same conditions hold true also for gravitational experiments. This has a big implication with respect to self-gravitating bodies, such as stars: If our gravitational theory satisfies the SEP, self-gravitating bodies should follow the same trajectories as test bodies, but if it only satisfies the EEP, this is not the case. GR satisfies the SEP in addition to EEP, but other modified theories of gravity only satisfy EEP. Throughout this thesis, we will concern ourselves only about the EEP, and consequently, we will consider test bodies that are not massive enough to modify the space-time around them.


\subsection{Geodesics}

The EEP makes any gravitational theory to be about a metric an its geodesics.  The metric is a rank 2 tensor that describes the distance between two neighbouring points, and the angles between the different directions. A geodesic is the path that extremizes the length between two points, as measured by that metric. 

We can use the variational principle to obtain the geodesics of a metric. If we consider two points, $a$ and $b$, and a curve $\gamma^\mu(\lambda)$ between those two points, the length between the points $a$ and $b$ along the curve is:

\begin{equation}
 L = \int_a^b \sqrt{-\frac{\df \gamma^\mu}{\df \lambda} \frac{\df \gamma^\nu}{\df \lambda} g_{\mu \nu}} \df \lambda
\end{equation}

To obtain the geodesic, we have to minimize this length. It can be shown that this is the same problem as minimizing this other integral:

\begin{equation}
 E = \frac{1}{2} \int_a^b \frac{\df \gamma^\mu}{\df \lambda} \frac{\df \gamma^\nu}{\df \lambda} g_{\mu \nu} \df \lambda
\end{equation}

Let us note that minimizing this other functional selects a particular parametrization of the curve $\gamma^\mu(\lambda)$. The freedom of parametrizations would have been a problem minimizing the length, as the different parametrizations of the same curve gives the same length. The Euler-Lagrange equations of this other functional are:

\begin{equation}
 g_{\alpha \mu} \frac{\df^2 \gamma^\mu}{\df \lambda^2} + \frac{\df \gamma^\mu}{\df \lambda} \frac{\df \gamma^\nu}{\df \lambda} \partial_\alpha g_{\mu \nu} - \frac{1}{2} \frac{\df \gamma^\mu}{\df \lambda} \frac{\df \gamma^\nu}{\df \lambda} \partial_\mu g_{\mu \alpha}=0 \label{1eq:Geo1}
\end{equation}

Given the metric and initial conditions $\gamma^\nu(0)$, $\frac{\df \gamma^\nu}{\df \lambda}(0)$, we can solve these equations and obtain the geodesics. On the other hand, EEP identifies geodesics with the paths of unaccelerated observers, so it should also be possible to obtain the geodesic paths using this property. Acceleration is the derivative of the velocity along its path, but in curved space-time there is a problem defining directional derivatives, because partial derivatives give a result that is not tensorial. Using partial derivatives to calculate the acceleration would give a result that would depend on which coordinates we are using. To define a good directional derivative, called \emph{covariant derivative} $\nabla$, we need to introduce an additional affine structure, the connection $\Gamma$. The covariant derivative of a vector $u$ along the direction of $v$ is:

\begin{equation}
 \nabla_v u= v^\alpha \partial_\alpha u^\beta + v^\alpha \Gamma_{\alpha \gamma}^\beta u^\gamma
\end{equation}

The result of the covariant derivative is a tensor; however, the connection is not, because it has to balance that the partial derivative is not either. The covariant derivative can be generalized to act on tensors of any kind:

\begin{IEEEeqnarray}{rCl}
 (\nabla_v T)^{\alpha_1...\alpha_p}{}_{\beta_1...\beta_q} &=& v^\gamma \partial_\gamma T^{\alpha_1...\alpha_p}{}_{\beta_1...\beta_q} + \sum_{i=1}^p v^\gamma \Gamma^{\alpha_i}_{\gamma \delta} T^{\alpha_1...\delta...\alpha_p}{}_{\beta_1...\beta_q}\nonumber\\
 &&- \sum_{i=1}^q v^\gamma \Gamma^{\delta}_{\gamma \beta_i} T^{\alpha_1...\alpha_p}{}_{\beta_1...\delta...\beta_q}
\end{IEEEeqnarray}

It is common to understand the covariant derivative like a gradient that takes a $(p,q)$ type tensor $T$ and gives a $(p,q+1)$ tensor $\nabla T$ with components $(\nabla T)_\mu{}^{\alpha_1...\alpha_p}{}_{\beta_1...\delta...\beta_q} \equiv \nabla_\mu T^{\alpha_1...\alpha_p}{}_{\beta_1...\delta...\beta_q} = (\nabla_{e_\mu}T)^{\alpha_1...\alpha_p}{}_{\beta_1...\delta...\beta_q}$ where $e_\mu$ are the basis vectors. With this notation, $\nabla_v u = v^\alpha \nabla_\alpha u$. We will study the covariant derivative and the affine structure in more detail in chapter \ref{M-AChap}.

Now that we have defined correctly the directional derivative, we can work with the acceleration as the derivative of the velocity along its path (given a connection $\Gamma$). The velocity of a test body is the unitary tangent vector to its trajectory. If we are parametrizing the curve of the test body as $\gamma^\mu(\lambda)$, then the tangent vector $u^\mu = \frac{\df \gamma^\mu}{\df \lambda}$ will not be unitary in general, and its modulus might vary along the path of the curve. However, if the derivative of the tangent vector is directed along the tangent vector itself, the variation happens only in the modulus of the tangent vector, and will leave the unitary part unchanged. Therefore, $\gamma^\mu(\lambda)$ will be unaccelerated if
\begin{equation}
 u^\mu \nabla_\mu u^\nu = f\  u^\nu
\end{equation}
for some function $f$. However, no matter which $f$ we choose, this equation describes the same curve, but parametrized in a different way. It is common to choose $f=0$, that corresponds to a reparametrization of the curve $\gamma^\mu(\lambda)$, which is called \emph{affine parametrization}. In that case, the tangent vector of the geodesic satisfies:

\begin{equation}
 u^\mu \nabla_\mu u^\nu = 0 \label{1eq:GeoParallel}
\end{equation}

This equation is equivalent to:

\begin{equation}
 \frac{\df^2 \gamma^\nu}{\df \lambda^2} + \Gamma^\nu_{\alpha \beta} \frac{\df \gamma^\alpha}{\df \lambda}\frac{\df \gamma^\beta}{\df \lambda} = 0 \label{1eq:Geo}
\end{equation}
which is called the \emph{geodesic equation}. This is a set of second-order differential equations which have a unique solution given the initial conditions $\gamma^\nu(0)$, $\frac{\df \gamma^\nu}{\df \lambda}(0)$, and a connection $\Gamma^\nu_{\alpha \beta}$. If our space-time has a metric, it induces a special connection called the \emph{Levi-Civita connection} which is defined as:

\begin{equation}
 \Gamma^\alpha_{\beta \gamma}\equiv \frac{1}{2} g^{\alpha \mu} \left ( \partial_\beta g_{\mu \gamma} + \partial_\gamma g_{\beta \mu} - \partial_\mu g_{\beta \gamma} \right )\label{1eq:LCconnection}
\end{equation}

Comparing with eq. \ref{1eq:Geo1}, the equation of an unaccelerated test body is the same as the equation of a curve that extremizes the length between two points if the connection is Levi-Civita\footnote{Actually, if the symmetric part of the connection is.}. This connection defines a covariant derivative with the property that $\nabla g =0$. It is easy to check that the affine parametrization corresponds to one in which the tangent vector is normalized $\tilde{u}^\mu = u^\mu/\sqrt{|u^\alpha u^\beta g_{\alpha \beta}|}$, and that the affine parameter measures the length (time) of the curve. The Levi-Civita connection also corresponds to the connection that a flat manifold (one in which $\Gamma=0$ everywhere and partial derivatives constitute a good directional derivative) induces in a (possibly curved) embedded submanifold; for example, in classical theory of curves and surfaces, the induced derivative of a surface in euclidean space is the Levi-Civita one.

For historical reasons, GR was constructed assuming that the connection was compatible with the metric (Levi-Civita). This is the basic characteristic of Riemannian geometry, the only non-Euclidean geometry known at the time GR was proposed. Mathematicians developed the theory of connections, and non-Riemannian geometries, a few years after the publication of GR (see \cite{eisenhart1927non}). In Chapter \ref{M-AChap} we will relax this condition and see what are its consequences.


\subsection{Description of Physical Observers}\label{1sec:PhysObs}

Physical observers are usually described by curves in the space-time, who measure time with the proper length of the curve. Free-falling observers (also called inertial observers) follow geodesics as any other test body. However, we should think that this description, in which observers are point-like and move along some curve, is not very reallistic. The conditions for EEP explicitly concern themselves only with local experiments, in which the observers (and observed phenomena) size can be disregarded with respect to the characteristic size of the variations in the metric; therefore, the geodesic description is fitting in this scenario. But if the size of a test body is comparable to the change in curvature, not only the test bodies will not necessarily follow a geodesic path, but also the description is not suitable.


Furthermore, we will see that there are geometries where the curvature diverges. If we approach such a curvature divergence, no matter how small the test body, there will be a point from which there will be differences in the curvature of space-time along its extension. This will cause any test body to experience tidal forces that might change its trajectory or deform it. Therefore, it is important to describe observers with non-zero size in such a way we can  study the tidal forces that act upon them.


To describe in mathematical terms such an object we can consider a ``cloud of dust'', where each of its components follow a geodesic path. As we follow the path of the elements of this cloud, its components may converge or may drift apart. Now consider a rigid object with the shape of the cloud of dust in its place: where the cloud of dust converges, the rigid object would feel a compression force, and where the elements of the cloud of dust drift apart, the rigid object would feel a force trying to stretch it. A \emph{congruence} is the set of integral curves of a non-vanishing vector field. A \emph{geodesic congruence} is a congruence where every curve is geodesic. The path of the particles of the ``cloud of dust'' is described by a geodesic congruence, and this is the mathematical tool we need to understand the tidal forces that a rigid object experiences.

This description of a physical observer, in which every one of its elements try to follow a geodesic path but are unable to do so due to the internal forces that keep the rigid observer together, is an improvement over the simplistic geodesic description. However, we should worry that at some point near a curvature divergence, even the elements that constitute the rigid body will be too big with respect to the change in curvature. Ultimately, these elements are electrons and protons or other fundamental particles, whose quantum properties are more aptly described by a propagating wave than a geodesic path. Therefore, to describe correctly the fate of a physical observer near a curvature divergence, it is necessary to study the wave propagation. This wave description is consistent with the geodesic description, because in a certain limit akin to ``geometrical optics'' waves propagate like a ray, following geodesics.

To give a complete picture of the physics near curvature divergences, in this thesis I will study the geometry of the space-time using geodesics, congruences of geodesics, and the propagation of waves.

\section{General Relativity}

%
%

We want to construct a theory that gives a metric $g$ to the space-time $\mathcal{M}$ in a covariant way. In order to do that, we can write a gravity Lagrangian made up of scalars of the geometry and perhaps some auxiliary fields. As we have seen, the connection is not a tensorial object, so it cannot enter the Lagrangian directly, but we can construct a truly tensorial object, called the Riemann curvature tensor, from it:

\begin{equation}
 R^\alpha{}_{\beta \mu \nu}= \partial_\mu \Gamma^\alpha_{\nu \beta}-\partial_\nu \Gamma^\alpha_{\mu \beta} + \Gamma^\alpha_{\mu \sigma} \Gamma^\sigma_{\nu \beta}-\Gamma^\alpha_{\nu \sigma} \Gamma^\sigma_{\mu \beta}
\end{equation}

We can take the trace of the Riemann tensor to construct the Ricci tensor, $R_{\alpha \beta}\equiv R^\sigma{}_{\alpha \sigma \beta}$. And contracting the Ricci tensor with the metric we obtain the scalar curvature $R\equiv g^{\alpha \beta} R_{\alpha \beta}$. It is possible to construct more scalars of the geometry with the different contractions between these tensors and their derivatives.

The simplest gravity Lagrangian we can write is just the scalar curvature $\Lagr_G=R$. The full action also contains the matter Lagrangian, which describes the equations of motion of the different matter fields, but that now will act as the source of gravity, too. This matter Lagrangian must be written in a covariant way, and must agree with the usual Lagrangian when the space-time is flat. To write this matter Lagrangian, we can take the matter Lagrangian of special relativity and substitute partial derivatives with covariant derivatives, although this is not the only possibility (for example, we could include terms multiplied by the scalar curvature which would vanish for a flat space-time). Finally, we have to include a factor $\frac{1}{16 \pi l_P^2}$ to provide the correct dimensions and so that the theory agrees with Newton's law in the linear limit. The Einstein-Hilbert action for GR is finally:

\begin{equation}
 S= \frac{1}{16 \pi l_P^2} \int_\mathcal{M} R \sqrt{|g|} \df^4 x + \int_\mathcal{M} \Lagr_\text{m} \sqrt{-g} \df^4 x
\end{equation}

Taking the variation of the action with respect to changes in the metric gives us the equations of motion:

\begin{equation}
 \underbrace{R_{\mu \nu} - \frac{1}{2} R g_{\mu \nu}}_{G_{\mu  \nu}} = 8 \pi l_P^2 T_{\mu \nu}
\end{equation}
where $T_{\mu\nu}=\frac{2}{\sqrt{|g|}}\frac{\delta \Lagr_\text{m}\sqrt{|g|}}{g^{\mu \nu}}$ is the energy-momentum tensor, and the left hand side is $G_{\mu\nu}$, the Einstein tensor, which is a sum of curvature terms. These equations tell us that the curvature of the space-time is equal to its matter content. If we think in terms of the metric, the curvature terms contain first derivatives of the connection, which in turn contains first derivatives of the metric. So it is a second-order differential equation for the metric. These equations are difficult to solve, and will usually require the use of symmetries to be solved analytically, or to deal with them numerically.

%

\section{The Schwarzschild Black Hole}

One of the simplest solutions of GR is the Schwarzschild metric. This metric describes a static, spherically symmetric, vacuum solution of the Einstein's equations. This is the case of the space-time outside a spherical source, such as a star. The equations of motion in this case are:

\begin{equation}
 G_{\mu \nu}=0
\end{equation}

The metric will have the symmetries of the space-time. A general spherically symmetric and static metric can be written as\footnote{To be precise, we should have used a radial coordinate $x$ different from the area function $r^2(x)$ that goes into the angular part of the metric. Only if we stablish that the relation between $x$ and $r$ is monotonic, $r$ can be used as a coordinate, this is the case in the Schwarzschild geometry.}:

\begin{equation}
 \df s^2 = -A(r) e^{B(r)} \df t^2 + \frac{1}{A(r)} \df r^2 + r^2 \df \Omega^2 \label{1eq:MetricSpheric1}
\end{equation}
where $A$ and $B$ are two arbitrary functions, that will be determined when we solve Einstein's equations. We can calculate each of the Einstein tensor components that corresponds with the above metric. Each of those components must be equal to 0 if we are in vacuum. This gives us the following equations:

\begin{IEEEeqnarray}{rCl}
 G^t{}_t &=& \frac{r \partial_r A + A -1}{r^2}=0 \label{1eq:Gtt}\\
 G^r{}_r &=& \frac{r A \partial_r B + r  \partial_r A + A -1}{r^2}=0 \label{1eq:Grr}\\
 G^\theta{}_\theta &=& G^\phi{}_\phi = \nonumber \\&=& \frac{2 r A \partial_r \partial_r B + r A (\partial_r B)^2 + (3 r \partial_r A + 2 A) (\partial_r B)+2 r \partial_r \partial_r A+4\partial_r A}{4 r}\nonumber \\
 &=&0
\end{IEEEeqnarray}

Comparing eq. \ref{1eq:Gtt} and eq. \ref{1eq:Grr} we see that $\partial_r B=0$, which means that $B(r)=B_0$ is constant. It is possible to absorb this factor into a redefinition of the time coordinate $t^\prime = e^{B_0/2} t$. Then Eq. \ref{1eq:Gtt} has solution:

\begin{equation}
 A(r) = 1-\frac{2 M l_P^2}{r}
\end{equation}
where $M$ is an integration constant that corresponds to the mass of the black hole\footnote{It is possible to define a mass in a ``Gauss's law form'' for space-times which have time translation symmetry. It is called Komar mass and it is defined as:
\begin{equation}
 M_K = -\frac{1}{8\pi} \int_S \epsilon_{\alpha \beta \mu \nu} \nabla^\mu \xi^\nu
\end{equation}
Where $\xi^\mu$ is the killing vector associated to the time translation symmetry, $S$ is a surface that encloses the sources and $\epsilon_{\alpha \beta \mu \nu} = \sqrt{|g|} \df x^4$ is the volume form of the space-time. The integral curves of the killing vector $\xi^\mu$ are the paths of observers that remain static respect to the black hole. Therefore, $\nabla^\mu \xi^\nu$ is the acceleration that they need to remain static with respect to the black hole. Integrating it over a surface that encloses the sources gives us a definition of mass analogue to the Gauss's Law for Newtonian gravity.}. This mass gives us a characteristic length called the \emph{Schwarzschild radius} $r_S\equiv 2 M l_P^2$. The resulting metric looks like:

\begin{equation}
 \df s^2 = -\left (1- \frac{r_S}{r} \right ) \df t^2 + \frac{1}{ \left (1- \frac{r_S}{r} \right ) } \df r^2 + r^2\df \Omega^2
\end{equation}

We could have dropped the condition of staticity, and the result would have been the same. This is known as Birkhoff's theorem: any spherically symmetric space-time with its sources confined behind a certain radius $r<R$ must match the Schwarzschild metric for $r>R$ (see \cite{HawkingEllisLandshoffNelsonSciamaWeinberg197503}). Therefore, this metric is an accurate representation of the space-time deformation that a planet or a star generates around it, even if their structure change over time, depending only on the mass $M$.

When we are far from the sources, $r \rightarrow \infty$, this metric tends to the Minkowski metric, as expected. We will call this region far from the sources \emph{asymptotic flat infinity}.

From the expression of the metric, we can see that $r=r_S$ is a special hypersurface, called \emph{event horizon}, where some of the components of the metric diverge. What would happen if the sources are found behind that radius? Is it possible to reach it? What would happen to an observer that does? The components of the metric also diverge for $r=0$. As we have discussed, the Schwarzschild  solution is an accurate representation of the space-time outside a star, but a star has a finite radius. Does the full Schwarzschild solution, from $r=\infty$ to $r=0$ represent a conceivable physical scenario? If it does, what happens if an observer reaches $r=0$?


An important observation is that inside the Schwarzschild radius, the $r$ coordinate becomes time-like while the $t$ coordinate becomes space-like. This forces any observer inside $r_S$ to move in the $r$ coordinate, either falling into $r=0$ or escaping outside, without the possibility of changing directions. As we will see in the next section, there are two disjoint regions inside the Schwarzschild radius, called \emph{black hole} and \emph{white hole}. In the black hole, every observer must travel towards $r=0$ and nothing, not even light, can escape, hence the name. In the white hole, it happens the other way round, and everything must escape this region, and not even light can remain inside. 

In a realistic process of gravitational collapse\footnote{The Oppenheimer-Snyder model describes the collapse of a spherically symmetric, pressureless and homogeneous ball of dust beyond its own horizon into a Schwarzschild solution, which can serve as an approximate description of these events, for more detail see Chapter 1 of \cite{fabbri2005modeling}, for example. }, for example a massive star which has burnt up all its fuel and can no longer stand the gravitational force of its own mass, the description of the region outside the collapsing star is accurately given by the Schwarzschild metric. If the star collapses beyond its Schwarzschild radius, a black hole will form, and all the matter of the star will be forced to travel to $r=0$, no matter the increase in the interior pressure of the star. After a finite amount of time, all the matter will have travelled to $r=0$ leaving all the space between $r=0$ and $r=r_S$ empty. Therefore, the Schwarzschild geometry is what GR predicts as the final state of a gravitational collapse that collapses beyond $r=r_S$, and therefore we need to understand the $r=0$ region in order to understand black holes.


%
%

\subsection{Geodesics of a Spherically Symmetric and Static Space-time}\label{1sec:Geo}

To study the $r=0$ and $r=r_S$ regions of the Schwarzschild geometry, which are seemingly problematic, we can study the fate of observers that reach them. We could possibly study general observers, but for simplicity, we will consider free-falling observers. Not only it is simpler, but accelerated observers might have some unexpected behaviours (such as reaching regions of the space-time for which you need infinite energy to arrive).

Let $\gamma^\mu(\lambda)$ be the path that describes an unaccelerated observer, affinely parametrized such  that the tangent vector $u^\mu = \frac{\df \gamma^\mu}{\df \lambda}$ is unitary $u^\alpha u_\alpha = -1$. Therefore, if the observer experiences no acceleration it will satisfy the geodesic equation \ref{1eq:Geo}.
Let us first consider the geodesics of a generic spherically symmetric and static space-time. In the next section we will consider the Schwarzschild case in particular. The general metric with those symmetries is\footnote{This is just like eq. \ref{1eq:MetricSpheric1}, with $\df x = e^{B(r)/2} \df r$ and $F(x)=A(r)e^{B(r)}$}:

\begin{equation}
 \df s^2 = -F(x) \df t^2 + \frac{1}{F(x)} \df x^2 + r^2(x) \df \Omega^2
\end{equation}

Instead of integrating directly the geodesic equation to obtain the paths of test particles and light rays, it is more convenient to take advantage of the symmetries of the geometry to obtain conserved quantities that simplify the analysis. First of all, because of spherical symmetry the geodesics lie on a plane, and we can rotate our coordinate system so that the plane is $\theta= \frac{\pi}{2}$ without loss of generality, and therefore $\frac{\df \gamma^\theta}{\df \lambda}=0$. Second, if the geodesics are time-like, we can normalize their tangent vector to $-1$; if they are null, the norm of the tangent vector is $0$. Third, the symmetries under rotations and temporal translations give us two conserved quantities\footnote{Each symmetry has an associated Killing vector $\chi^\mu = \frac{\partial}{\partial t}$, $\frac{\partial}{\partial \phi}$ such that $\Lie_\chi g =0$, which in turn implies $\nabla_\mu \chi_\nu = - \nabla_\nu \chi_\mu$. Then, the quantity $g_{\mu \nu} \chi^\mu u^\nu$ is conserved along the geodesic: $u^\alpha \nabla_\alpha (g_{\mu \nu} \chi^\mu u^\nu)= \chi_\nu \underbrace{u^\alpha \nabla_\alpha u^\nu}_{=0}+u^\nu u^\alpha \nabla_\alpha \chi_\nu =0$ }: 

\begin{equation}
 E=F(x) \frac{\df \gamma^t}{\df \lambda}\qquad L=r^2 \frac{\df \gamma^\phi}{\df \lambda}\label{1eq:GeoEL}
\end{equation}

For time-like geodesics, $E$ can be interpreted as the total energy per unit mass, and $L$ as angular momentum per unit mass. In the case of light rays, it is not possible to normalize the tangent vector and consequently, $E$ and $L$ lack meaning by themselves; but the quotient $L/E$ can be interpreted as the apparent impact parameter as seen from the asymptotically flat infinity. The condition that the tangent vector to the geodesics has to be normalized to $\kappa = 0$ or, $\kappa = -1$ gives us another equation:

\begin{equation}
 -\kappa = -F(x) \left ( \frac{\df \gamma^t}{\df \lambda} \right )^2 + \frac{1}{F(x)} \left ( \frac{\df \gamma^x}{\df \lambda} \right )^2 + r^2(x) \left ( \frac{\df \gamma^\phi}{\df \lambda} \right )^2
\end{equation}

Substituting the value of the conserved quantities, this equation gives us the final component of the tangent vector:

\begin{equation}
 \frac{\df \gamma^x}{\df \lambda}=\pm \sqrt{E^2-F(x)\left ( \kappa + \frac{L^2}{r^2(x)} \right ) } \label{1eq:geosss}
\end{equation}

Knowing the tangent vector of the geodesic for every $x$, now it is possible to integrate the geodesic paths. 


\subsection{Trajectory of Infalling Radial Light Rays in the Schwarzschild Geometry}

Now that we know how to obtain the geodesic paths, we can study the path of a light ray sent towards the centre of the Schwarzschild geometry. A light ray follows null geodesics, and if it is sent towards the centre, it has no angular momentum, $L=0$. For the Schwarzschild metric, we just have to substitute $r=x$, $F(x)=1-\frac{r_S}{r}$, in the formulas of the previous section. Then, the solution for the trajectory of a radial null geodesic $\gamma^\mu(\lambda)$ and its tangent vector $\dot{\gamma}^\mu(\lambda)$ is:

\begin{IEEEeqnarray}{rCl}
 \dot{\gamma}^\mu (\lambda) &=& \left ( \frac{E}{1-\frac{r_S}{r(\lambda)}}, \pm E, 0 , 0 \right ) \\
 \gamma^\mu(\lambda) &=& \left (t_0 + E \lambda \pm r_S \log \left ( \frac{r_0 \pm E\lambda -r_S}{r_0 -r_S} \right ),r_0 \pm E \lambda, \frac{\pi}{2}, \phi_0 \right )
\end{IEEEeqnarray}

where we consider $E$ to be positive, and the choice of sign depends on whether the geodesic is outgoing/ingoing. Looking at the expression for the geodesic, it seems that the hypersurface $r=r_S$ cannot be reached from outside, because as we get closer to it, $\gamma^t$ goes to infinity. A closer inspection reveals that the affine parameter of the geodesic reaches a finite value at $r=r_S$ (The value of the affine parameter at $r_S$ satisfies $r_S=r_0-E \lambda_S$, if we substitute its value into the time component of the geodesic, it gives a $0$ inside the logarithm, which implies that $t$ goes to infinity). It would be really strange that an observer would cross the entire space-time in a finite amount of proper time. This is a hint that $t$ is not a good coordinate to describe the $r=r_S$ region, and there is something beyond $t=\infty$.

Since we suspect that $t$ is not a good coordinate, we can try to use a different one that works better. The $t$ coordinate of the light-ray trajectory can be written as:
\begin{equation}
 \gamma^t(\lambda)= t_0 +r_0+r_S \log \left (\frac{r_0}{r_S-1} \right ) -\gamma^r(\lambda) - r_S \log \left (\frac{\gamma^r(\lambda)}{r_S-1} \right )
\end{equation}

It is easy to see that given a ray of light, the quantity $\gamma^t(\lambda)+\gamma^r(\lambda)+r_S \log(\gamma^r(\lambda)/r_S-1)$ is constant along the curve. Then, if we change the $t$ coordinate to $v=t+r+r_S \ln \left | \frac{r}{r_S} - 1 \right |$, the $v$ component of the light trajectory should be constant (and finite). Let us write the metric with this coordinate instead of $t$:

\begin{equation}
 -\left ( 1 - \frac{r_S}{r} \right ) \df v^2 + 2 \df v \df r + r^2 \df \Omega^2
\end{equation}

The infalling null geodesics in these coordinates are:

\begin{IEEEeqnarray}{rCl}
 \dot{\gamma}^\mu (\lambda) &=& (0, -E, 0 , 0) \\
 \gamma^\mu(\lambda) &=& (v_0,r_0 -E \lambda, \frac{\pi}{2}, \phi_0)
\end{IEEEeqnarray}
where $v_0 = t_0 + r_S \log \left ( \frac{r_0 - r_S}{r_0} \right )$. In these coordinates we can see that the geodesics show no problems crossing the event horizon, and the metric is regular at that point. The Schwarzschild metric expressed in $(t,r,\theta,\phi)$ is said to have a \emph{coordinate singularity} at $r=r_S$. These coordinates are simply a bad choice for the region $r=r_S$, because $(t,r,\theta,\phi)$ were chosen so that the metric looks like the Minkowski metric for the asymptotic infinity ($r\rightarrow \infty$). Let us think of a free-falling observer A moving towards the black hole who sends periodically pulses of light to a static observer B at infinity. B would receive those pulses more and more spaced in time as A approaches the black hole, and actually, B would never see A entering into the black hole. Therefore, the static observer B cannot assign a ``time'' value to the moment when the free-falling observer A crosses the event horizon, and that is the reason why these coordinates are unsuitable to describe the event horizon and whatever lies beyond it.

%
%
%
%
%

In $(v,r,\theta,\phi)$ coordinates, infalling geodesics show no problems crossing the $r=r_S$ boundary. Let us now take a look at the behaviour of outgoing geodesics:

\begin{IEEEeqnarray}{rCl}
 \dot{\gamma}^\mu (\lambda) &=& \left (\frac{2E}{1-\frac{r_S}{r}}, E, 0 , 0 \right ) \\
 \gamma^\mu(\lambda) &=& \left (v_0 + 2 E \lambda + 2 r_S \log \left ( \frac{r_0+E \lambda - r_S}{r_0 - r_S} \right ),r_0 -E \lambda, \frac{\pi}{2}, \phi_0 \right )
\end{IEEEeqnarray}

If we trace back the outgoing geodesic to the past $t\rightarrow -\infty$ we see that the outgoing geodesic was never inside $r=r_S$. This seems to agree with our intuition that nothing can escape through the event horizon. But if we look at what happens with the affine parameter, it does not go back to $\lambda=-\infty$, but to a finite value. This looks exactly like a coordinate singularity as happened with the $(t,r,\theta,\phi)$ coordinates. This is because analogue to the black hole and its horizon, there is a white hole region with a horizon, where all the trajectories must head outwards, and no signal from infinity can enter. If instead of the coordinate $v$ we use the coordinate $u=t-r-r_S \ln \left | \frac{r}{r_S} - 1 \right |$, the white hole region would be properly mapped, but then we would encounter the coordinate singularity when approaching the black hole region. Both the black hole and the white hole region lie behind $r=r_S$, but are actually disjoint, as the region $r<r_S$ mapped by the $(u,r,\theta, \phi)$ coordinates is different from the one mapped by the $(v,r,\theta, \phi)$ coordinates.

White holes are a consequence of the symmetries of this geometry, as any solution must remain invariant under the change $t\rightarrow -t$. In a realistic scenario of the formation of a black hole, like the collapse of a massive star, there is no white hole in the past, as the time symmetry has been broken.

Finally, there is a set of coordinates that map both the black hole and the white hole region in a proper way. These coordinates are the Kruskal-Szekeres coordinates:

\begin{equation}
 U=-e^{-\frac{u}{2 r_S}} \qquad V = e^\frac{v}{2 r_S} \qquad r=r(U,V)
\end{equation}

The range of the coordinates $U,V$ is $(-\infty, \infty)$. Now $r$ is not a coordinate, but it still has the meaning of the area associated to the 2-spheres symmetric by rotations in the $\theta, \phi$ angles. It is a function that depends on $U,V$ and can be solved from the equation:

\begin{equation}
 -UV=\left ( \frac{r}{r_S}-1 \right ) e^\frac{r}{r_S} \label{1eq:req}
\end{equation}

Finally, the metric is:

\begin{equation}
 \df s^2 = -\frac{4 r_S^3 e^{-\frac{r}{r_S}}}{r} \df U \df V +r^2 \df \Omega^2
\end{equation}

\begin{figure}[h!]
\centering
  \includegraphics[width=.5\linewidth]{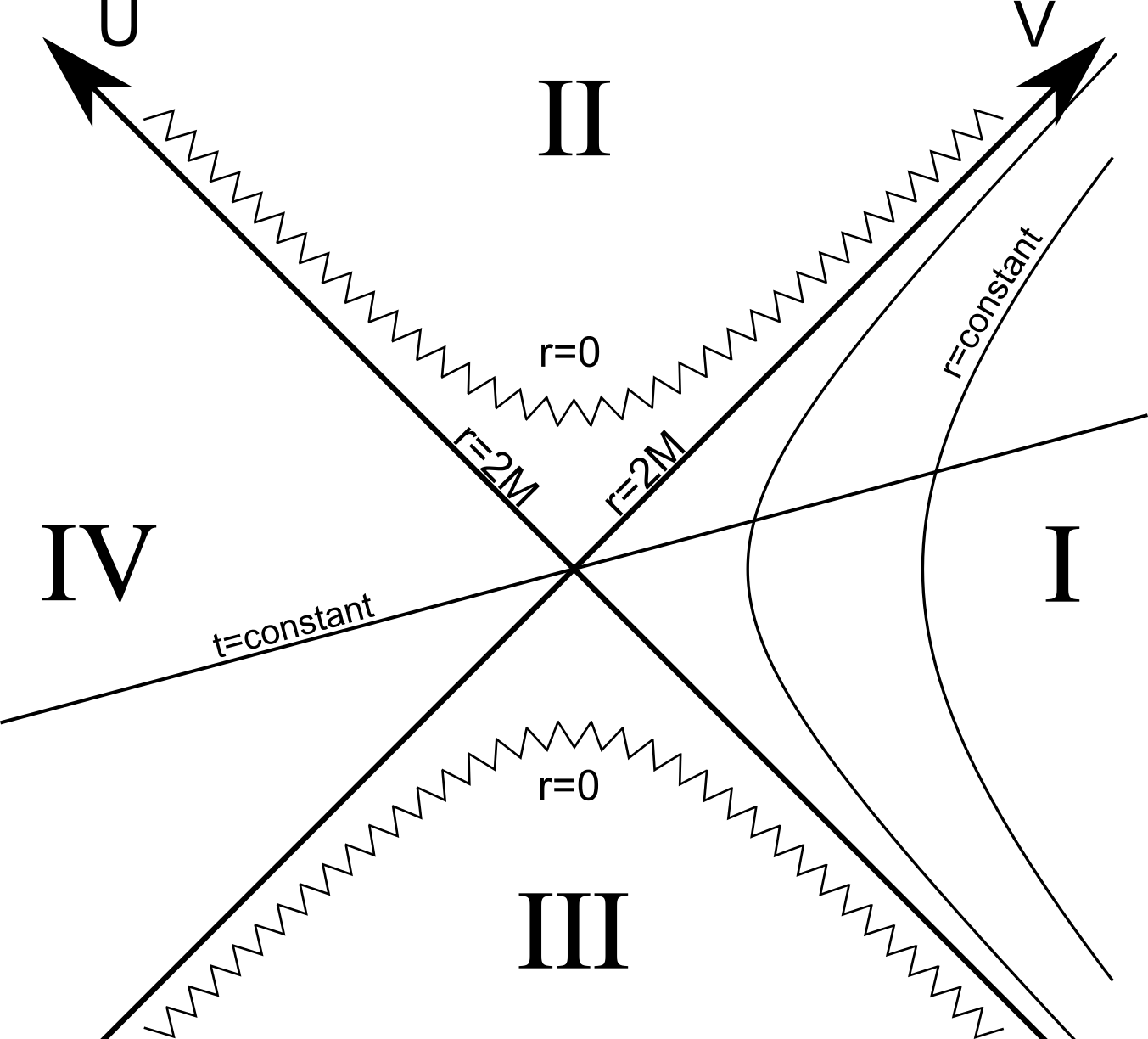}
\caption{Representation of the Schwarzschild space-time in Kruskal-Szekeres coordinates. Region I is the exterior region. Region II is the black hole region. Region III is the white hole region. Region IV is a parallel exterior region, unreachable from Region I. The $(v,r)$ coordinates map regions I and II, meanwhile $(u,r)$ coordinates map regions I and III.}\label{1fig:Kruskal}
\end{figure}

Although this metric is defined beyond the scope of our original solution, it satisfies Einstein's equations for vacuum in all the range of its coordinates. In these coordinates, it is possible to represent all of the Schwarzschild space-time (see Fig. \ref{1fig:Kruskal}). The horizon located at $r=r_S$ corresponds to the surface $U V=0$. We can distinguish four regions in the solution. Region I ($U<0$, $V>0$) corresponds the exterior of the black hole. Region II ($U>0$, $V>0$) is the black hole region. Region III ($U<0$, $V<0$) is the white hole region. Region IV ($U>0$, $V<0$) is a region we have not yet described, that corresponds to a parallel exterior region, unreachable from region I. 

The light cones in Fig. \ref{1fig:Kruskal} can be seen as lines of $U$ or $V$ constant. The trajectories of physical observers must go forward in the quadrant of increasing $U$ and $V$. Therefore, the future of any physical observer inside region III (the white hole) is to exit it. The future of any observer inside region II (the black hole) is to remain there and reach $r=0$. And region IV is not in the future of any observer in region I and the other way around.

In these coordinates, $r=0$ corresponds to the condition $U V =1$, and the components of the metric are still divergent at that point. Actually, it is impossible to find a coordinate transformation that smooths the metric at $r=0$. This can be seen from the fact that the Kretschmann curvature scalar diverges, $K \equiv R^\alpha{}_{\beta \mu \nu} R_\alpha{}^{\beta \mu \nu} = \frac{24 r_S}{r^6}$, so some of the components of the metric or its derivatives must diverge, too. Any observer inside the black hole will encounter $r=0$ in its future. What happens there?

First of all, eq. \ref{1eq:Geo} will not be well defined at $r=0$, because if the metric is divergent, so will its Levi-Civita connection. Also, if we tried to continue our trajectory to values for which $UV>1$ (even though the geodesic equation is ill-defined), eq. \ref{1eq:req} that gives us the value of $r(U,V)$ would stop having real solutions. There is no natural way to extend the geometry beyond it, although we could hypothetically extend it by hand connecting it to some other space-time and giving a prescription on how to continue the geodesics. 

\FloatBarrier

%
%
%

\section{Geodesic Congruences In The Schwarzschild Geometry}

%

In the last section, we have studied the geodesics of the geometry, as they are the paths of freely falling observers according to the EEP. But the EEP concerned itself with \emph{local} experiments, i.e., experiments in which the variation in gravity along the size of our laboratory is negligible. As we discussed in section \ref{1sec:PhysObs}, this description is inadequate to capture the physics near a curvature divergence (such as $r=0$ in the Schwarzschild geometry). For that reason, we will study congruences of geodesics, which will be more appropiate to describe observers with non-negligible physical size.

\subsection{Evolution of a Geodesic Congruence}\label{1sec:EvoGeoC}

Let $\gamma^\mu = \gamma^\mu (\lambda, \xi)$ be an uniparametric family of geodesics of the congruence with affine parameter $\lambda$ and family parameter $\xi$. Its tangent vector is $u^\mu = u^\mu (\lambda, \xi) = \frac{\df \gamma^\mu}{\df \lambda}(\lambda, \xi)$, normalized to $-1$ or $0$ for all the geodesics of the uniparametric family. The deviation vector is $Z^\mu = Z^\mu (\lambda, \xi) = \frac{\df \gamma^\mu}{\df \xi}(\lambda, \xi)$.

\begin{figure}[h!]
 \centering
 \includegraphics[width=.35\linewidth]{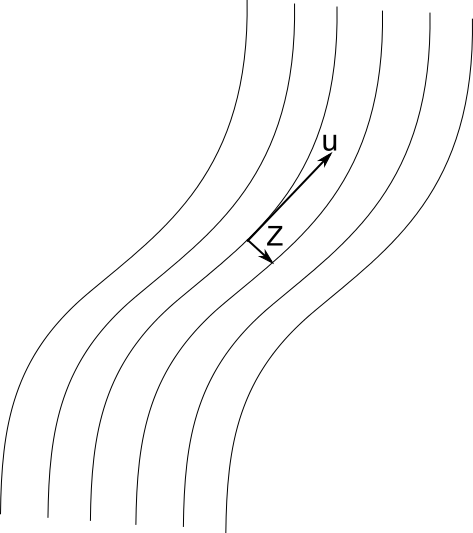}
 \caption{An uniparametric family of geodesics belonging to a congruence with tangent vector $u$ and deviation vector $Z$}
\end{figure}
\FloatBarrier

This deviation vector describes the infinitesimal displacement from one geodesic to a nearby geodesic from the same uniparametric family. As partial derivatives commute, we have that $u^\alpha \nabla_\alpha Z^\beta = Z^\alpha \nabla_\alpha u^\beta$. Also, the tangent part of the deviation vector is conserved along the geodesics\footnote{\begin{eqnarray}
 u^\alpha \nabla_\alpha (u_\beta Z^\beta) &=& Z_\beta (u^\alpha \nabla_\alpha u^\beta) + u^\beta u^\alpha \nabla_\alpha Z_\beta \\
 &=& Z_\beta (u^\alpha \nabla_\alpha u^\beta) + u^\beta Z^\alpha \nabla_\alpha u_\beta \\
 &=& Z_\beta (u^\alpha \nabla_\alpha u^\beta) + \frac{1}{2} Z^\alpha \nabla_\alpha (u^\beta u_\beta) =0
\end{eqnarray}
the first term of the last line of the equation vanishes because of the geodesic equation (eq. \ref{1eq:Geo}), and the last term is $0$ because we have normalized the tangent vector of the geodesics.
}. This last property makes it possible to reparametrize $\tilde{\gamma}^\mu(\lambda,\xi) = \gamma^\mu(\lambda+f(\xi),\xi)$ so that the deviation vector is perpendicular to the tangent vector of the geodesics along all the family of geodesics. The vector $Z^\alpha$ satisfies the \emph{geodesic deviation equation}\footnote{
\begin{eqnarray}
 u^\rho \nabla_\rho (u^\sigma \nabla_\sigma Z^\alpha) &=& u^\rho \nabla_\rho (Z^\sigma \nabla_\sigma u^\alpha) \\
 &=& (u^\rho \nabla_\rho Z^\sigma)(\nabla_\sigma u^\alpha) + u^\rho Z^\sigma \nabla_\rho \nabla_\sigma u^\alpha \\
 &=& (u^\rho \nabla_\rho Z^\sigma)(\nabla_\sigma u^\alpha) + u^\rho Z^\sigma \nabla_\sigma \nabla_\rho u^\alpha + u^\rho Z^\sigma R^\alpha{}_{\nu \rho \sigma} u^\nu \\
 &=& (Z^\rho \nabla_\rho u^\sigma)(\nabla_\sigma u^\alpha) + u^\rho Z^\sigma \nabla_\sigma \nabla_\rho u^\alpha - u^\rho Z^\sigma R^\alpha{}_{\nu \sigma \rho} u^\nu \\
 &=& Z^\rho \nabla_\rho (u^\sigma \nabla_\sigma u^\alpha) - u^\rho Z^\sigma R^\alpha{}_{\nu \sigma \rho} u^\nu \\
 &=&   -R^\alpha{}_{\nu \sigma \rho} u^\nu Z^\sigma u^\rho
\end{eqnarray}
}:

\begin{equation}
 u^\rho \nabla_\rho (u^\sigma \nabla_\sigma Z^\alpha) = - R^\alpha{}_{\beta \mu \nu} u^\beta Z^\mu u^\nu \label{1eq:GeoDevEq}
\end{equation}

This is a set of 3 second-order differential equations (because we have set the perpendicular part of $Z^\alpha$ to 0). The equation becomes clearer if we use an adapted coordinate system such that the $e_1, e_2, e_3$ base vectors are orthogonal to $u^\mu$ and are parallel transported along the geodesic. In this base, $Z^\mu$ can be written as $Z^a e_a$. We will use latin letters for the components 1,2,3 of the vectors written in this basis. Then it is possible to write the geodesic deviation equation as an ordinary differential equation:

\begin{equation}
 \frac{\df^2 Z^a}{\df \lambda^2} = - R^a{}_{\mu b \nu} u^\mu Z^b u^\nu
\end{equation}

Given initial conditions $Z^a$ and $\frac{\df Z^a}{\df \lambda}$ at some point $\lambda_i$, the equation can be integrated along the geodesic, which gives raise to 6 independent solutions. A vector field that satisfies this equation everywhere is called a \emph{Jacobi field}. The linearity of Eq. \ref{1eq:GeoDevEq} allows us to express the components of $Z^\mu(\lambda)$ everywhere along the geodesic in terms of a linear combination of the initial conditions $Z^\mu(\lambda_i)$ as:

\begin{equation}
 Z^a(\lambda) = A^a{}_b(\lambda) Z^b(\lambda_i)\label{1eq:matrixA}
\end{equation}

Where at the initial time, $A^a{}_b(\lambda_i)=\delta^a{}_b$. If the deviation vector vanishes at the initial time $\lambda_i$ (all the geodesics in the uniparametric family start from the same point $Z^a(\lambda_i)=0$), we can write the solution in terms of the initial value of the first derivative of the deviation vector, $\left. \frac{\df Z^b}{\df \lambda}\right |_{\lambda = \lambda_i}$ as:

\begin{equation}
 Z^a(\lambda) = \tilde{A}^a{}_{b}(\lambda) \left. \frac{\df Z^b}{\df \lambda}\right |_{\lambda = \lambda_i}\label{1eq:tildeA}
\end{equation}

And in this case at the initial time, $\tilde{A}^a{}_b(\lambda_i)=0$. If we are given 3 independent Jacobi fields, we could think of them as representing the edges of an infinitesimal cube transported by the congruence. The volume of this cube would be:

\begin{equation}
 V(\lambda) = \det (Z^a_{(1)},Z^b_{(2)},Z^c_{(3)})
\end{equation}

Or in terms of the matrix $A$ (or $\tilde{A}$ as appropriate):

\begin{equation}
 V(\lambda)=\det \left ( A(\lambda) \right ) V(\lambda_i)
\end{equation}

Changes in this volume would correspond to compression or expansion tidal forces. We are specially interested in the points where this volume goes to zero for every congruence, as it would correspond to infinite compression. That would be an extreme case where any finite body, no matter their internal interactions, would be unavoidably destroyed and has been catalogued as \emph{strong singularity} in the literature (\cite{Ellis:1977pj}, \cite{CLARKE1985127}). In the next chapter we will talk more about the concept of singularity. Another interesting case is when this infinitesimal volume diverges, we will study that case more in section \ref{5sec:congruences}.

%
%
%
%
%

\subsection{Congruence Around A Time-like Radial Geodesic For A Spherically Symmetric And Static Space-time}\label{1sec:CongSph}

As we did with the study of the geodesic, we will consider first a generic spherically symmetric and static space-time, and then we will particularize for the Schwarzschild case. Such space-time has a metric that can be written as:

\begin{equation}
 \df s^2 = -F(x) \df t^2 + \frac{1}{F(x)} \df x^2 + r^2(x) \df \Omega^2 \label{1eq:sphmetric}
\end{equation}

We will focus on a time-like radial geodesic, which is the simplest case. The components of the tangent vector to a time-like radial geodesic are:

\begin{equation}
 u^\mu = (E/F(x), \sqrt{E^2-F(x)},0,0)
\end{equation}

We choose three deviation vectors orthogonal to this tangent vector, which can be written as:
\begin{IEEEeqnarray}{rCl}
 Z_{(1)}&=&B(\lambda) \left (\frac{\sqrt{E^2-F(x)}}{F(x)},E,0,0 \right ) \\
 Z_{(2)}&=&P(\lambda) (0,0,1,0) \\
 Z_{(3)}&=&Q(\lambda) (0,0,0,\frac{1}{\sin \theta})
\end{IEEEeqnarray}

Applying eq. \ref{1eq:GeoDevEq} we obtain the functions $B(\lambda)$, $P(\lambda)$, $Q(\lambda)$. The solutions for $P(\lambda)$ and $Q(\lambda)$ are identical and of the form\footnote{
 First, let us calculate de second derivative of the modulus of the deviation vector along the geodesic:
\begin{IEEEeqnarray}{rCl}
 u^\mu \nabla_\mu (u^\nu \nabla_\nu |Z^a|) &=& \frac{Z_a}{|Z|} u^\mu \nabla_\mu ( u^\nu \nabla_\nu Z^a) 
 \\&&+ \frac{1}{|Z|}(u^\mu \nabla_\mu Z^a)(u^\nu \nabla_\nu Z_a) \nonumber
 \\&&- \frac{Z_a Z_b}{|Z|^3}(u^\mu \nabla_\mu Z^a)(u^\nu \nabla_\nu Z^b) \nonumber 
\end{IEEEeqnarray}
Given that, in this case, the derivative of the deviation vector is parallel to the deviation vector itself, the expression simplifies to:
\begin{equation}
 u^\mu \nabla_\mu (u^\nu \nabla_\nu |Z^a|) = \frac{Z_a}{|Z|} u^\mu \nabla_\mu ( u^\nu \nabla_\nu Z^a) \label{1eq:ModGeoDev}
\end{equation}
On the one hand, $|Z_{(2)}|=r P(\lambda)$, so the left hand side is $u^\mu \nabla_\mu (u^\nu \nabla_\nu |Z^a|) = \ddot{P} r + 2\dot{P} \dot{r} + P \ddot{r}$. On the other hand, and using eq. \ref{1eq:GeoDevEq}, the right hand side of the equation transforms to $\frac{Z_a}{|Z|} u^\mu \nabla_\mu ( u^\nu \nabla_\nu Z^a)=\frac{Z_a}{|Z|} R^a{}_{\mu b \nu} u^\mu Z^b u^\nu $. The relevant components of the Riemann tensor are:
\begin{equation}
 R^\theta{}_{t\theta t} = \frac{F(x) \partial_x F \partial_x r}{2 r} \qquad R^\theta{}_{x\theta x} = -\frac{\partial_x F \partial_x r}{2 F(x) r} - \frac{\partial_x \partial_x r}{r}
\end{equation}
Then, the right hand side of eq. \ref{1eq:ModGeoDev} becomes:
\begin{equation}
\frac{Z_a}{|Z|} R^a{}_{\mu b \nu} u^\mu Z^b u^\nu = P \left \{ (E^2-F(x)) \partial_x\partial_x r - \frac{(\partial_x r) (\partial_x F)}{2} \right \}=P\ddot{r} 
\end{equation}
Putting everything together, it results the simple equation:
\begin{equation}
 \ddot{P} r + 2\dot{P} \dot{r} =0
\end{equation}
Integrating this expression twice gives the result of the text. An alternative derivation can be found in \cite{Nolan:1999tw}.
}:

\begin{equation}
 P(\lambda) = P_0 + C \int \frac{\df \lambda}{r^2(\lambda)}\label{1eq:PQeq}
\end{equation}

For $B(\lambda)$, the solution does not simplify as much. Substituting in eq. \ref{1eq:GeoDevEq} and using that $R^x{}_{txt}=F(x) \partial_x\partial_x F(x)/2$ we find:

\begin{equation}
 \ddot{B}(\lambda)+\frac{\partial_x \partial_x F(x)}{2}B(\lambda)=0\label{1eq:Beq}
\end{equation}

The volume transported by this congruence of geodesics is $V=|Z_{(1)}||Z_{(2)}||Z_{(3)}|=r^2 B(\lambda)P(\lambda)Q(\lambda)$.

\subsection{Evolution of the congruence near the singularity of a Schwarzschild black hole}\label{1sec:CongSchw}

In the case of the Schwarzschild black hole we have $F(x)=1-\frac{r_S}{r}$. We can approximate the radial component of the tangent vector to the geodesic near $r=0$ as:

\begin{equation}
 \frac{\df r}{\df \lambda}= \pm \sqrt{E^2-\left (1-\frac{r_S}{r} \right ) } \simeq \sqrt{\frac{r_S}{r}}
\end{equation}

Inside the black hole, the sign must always be negative, otherwise the observer lies in the white hole region. We can integrate this last expression to know the radial part of the trajectory in terms of the affine parameter:

\begin{equation}
 \lambda \simeq  -\frac{2}{3\sqrt{r_S}} r^{3/2} \qquad r(\lambda) = \left ( -\frac{3\sqrt{r_S}}{2} \lambda \right )^{2/3}
\end{equation}

We have chosen to parametrize the proper time such that the geodesic has negative value of its parameter in its trajectory and reaches $r=0$ at $\lambda=0$, and the value of the proper time is negative at the initial point. Let us consider a congruence that starts at the initial point $\lambda_i$, that is $B(\lambda_i),P(\lambda_i),Q(\lambda_i)=0$. Integration of eqs. \ref{1eq:PQeq} and \ref{1eq:Beq} give:

\begin{IEEEeqnarray}{rCl}
 P(\lambda) &\simeq& K_P (|\lambda_i|^{-\frac{1}{3}}-|\lambda|^{-\frac{1}{3}})\\
 Q(\lambda) &\simeq& K_Q (|\lambda_i|^{-\frac{1}{3}}-|\lambda|^{-\frac{1}{3}})\\
 B(\lambda) &\simeq& K_B \left (\frac{1}{|\lambda|^\frac{1}{3}} - \frac{|\lambda|^\frac{4}{3}}{|\lambda_i|^\frac{5}{3}} \right ) \label{1eq:Z1}
\end{IEEEeqnarray}

Where $K_P,K_Q,K_B$ are integration constants. Then, the volume near the singularity behaves as:

\begin{equation}
 V(\lambda) \propto |\lambda|^\frac{1}{3} \propto r^\frac{1}{2}
\end{equation}

No matter where the congruence started, it is crushed to zero volume when it reaches the $r=0$, and therefore, this point is a strong singularity. Any object following a radial time-like geodesic would be crushed by tidal forces of infinite magnitude. The tidal forces do not work in a homogeneous way, and the object would be crushed in the angular directions, but would actually be stretched in the radial direction, in a process usually called \emph{spaghettization}. This is a catastrophic event that would destroy anything sent towards the singularity. Not only the geodesics reach $r=0$ in a finite time and there is no natural way to extend the space-time, but even if we extended the space-time 'by hand' it may have no physically reasonable extension as all the information sent is crushed to a single point.

\section{Charged Black Holes in GR}

In the next chapters we will be interested in charged black hole solutions, because the Metric-Affine theories we are going to study, only depart from the Riemannian formalism when the energy-momentum tensor is different from zero. In GR, these solutions are described by the Reissner-Nordström metric and present a causal structure very different from the one of the Schwarzschild black hole.

\subsection{Spherically Symmetric Electrovacuum Field}

In order to obtain charged black hole solutions, we must introduce the electromagnetic sector in the matter action, and consider no sources\footnote{Except perhaps in the central region of the geometry.}. The action for the electromagnetic sector and its energy-momentum tensor are:

\begin{equation}
 S_m = -\frac{1}{16 \pi l_P^2} \int F_{\alpha \beta} F^{\alpha \beta} \sqrt{-g} \df^4 x  \label{1eq:EM-action}
\end{equation}
\begin{equation}
 T_\mu{}^\nu = -\frac{1}{4\pi} \left ( F_\mu{}^\alpha F_\alpha{}^\nu-\frac{F_\alpha{}^\beta F_\beta{}^\alpha}{4}\delta_\mu{}^\nu \right )
\end{equation}

where $F_{\mu \nu}=(\df A)_{\mu \nu} = \partial_\mu A_\nu - \partial_\nu A_\mu$ is the electromagnetic field and $A$ is the potential. The sourceless equations of motion are:

\begin{eqnarray}
 \df F&=&0 \label{1eq:EMmov1}\\
 \df (*F)&=&0 \qquad \Rightarrow \qquad \nabla_\mu F^{\mu \nu}=0 \label{1eq:EMmov2}
\end{eqnarray}

Since we are considering a static and spherically symmetric solution, the metric can be written as $\df s^2 = g_{tt}(r) \df t^2 + g_{rr}(r) \df r^2 + r^2 \df \Omega^2$, and the components of the field $F_{\mu \nu}$ depend only on the coordinate $r$. With this information, eq. \ref{1eq:EMmov2} can be integrated and the only non-zero component of the field strength tensor is $F^{tr}$: 

\begin{equation}
 F^{tr} = \frac{q}{r^2} \frac{1}{\sqrt{-g_{tt} g_{rr}}} = -F^{rt} \label{1eq:EMflux}
\end{equation}

Where $q$ is an integration constant that corresponds to the charge measured by computing the electric flux that passes through a surface that encloses the centre of the geometry $\int_S *F = 4 \pi q$. Now, the corresponding energy-momentum tensor in spherical coordinates $(t,r,\theta,\phi)$ has this simple form:

\begin{equation}
 T_\mu{}^\nu = \frac{q^2}{8\pi r^4}\left ( \begin{array}{cccc} -1 & 0 & 0 & 0 \\ 0 & -1 & 0 & 0 \\ 0 & 0 & 1 & 0 \\ 0 & 0 & 0& 1 \end{array} \right ) \label{1eq:EMVacuumT}
\end{equation}

\subsection{The Reissner-Nordström Metric and its Geometry}

The Reissner-Nordström metric (\cite{ANDP:ANDP19163550905}, \cite{Nordstrom:1918}) satisfies the Einstein equations with the energy-momentum tensor of a spherically symmetric electrovacuum field:

\begin{equation}
 \df s^2=-\left ( 1-\frac{r_S}{r}+\frac{r_q^2}{r^2} \right )\df t^2 + \left ( 1-\frac{r_S}{r}+\frac{r_q^2}{r^2} \right )^{-1} \df r^2 + r^2\df \Omega^2
\end{equation}

\begin{figure}[h!]
 \centering
 \includegraphics[width=.7\linewidth]{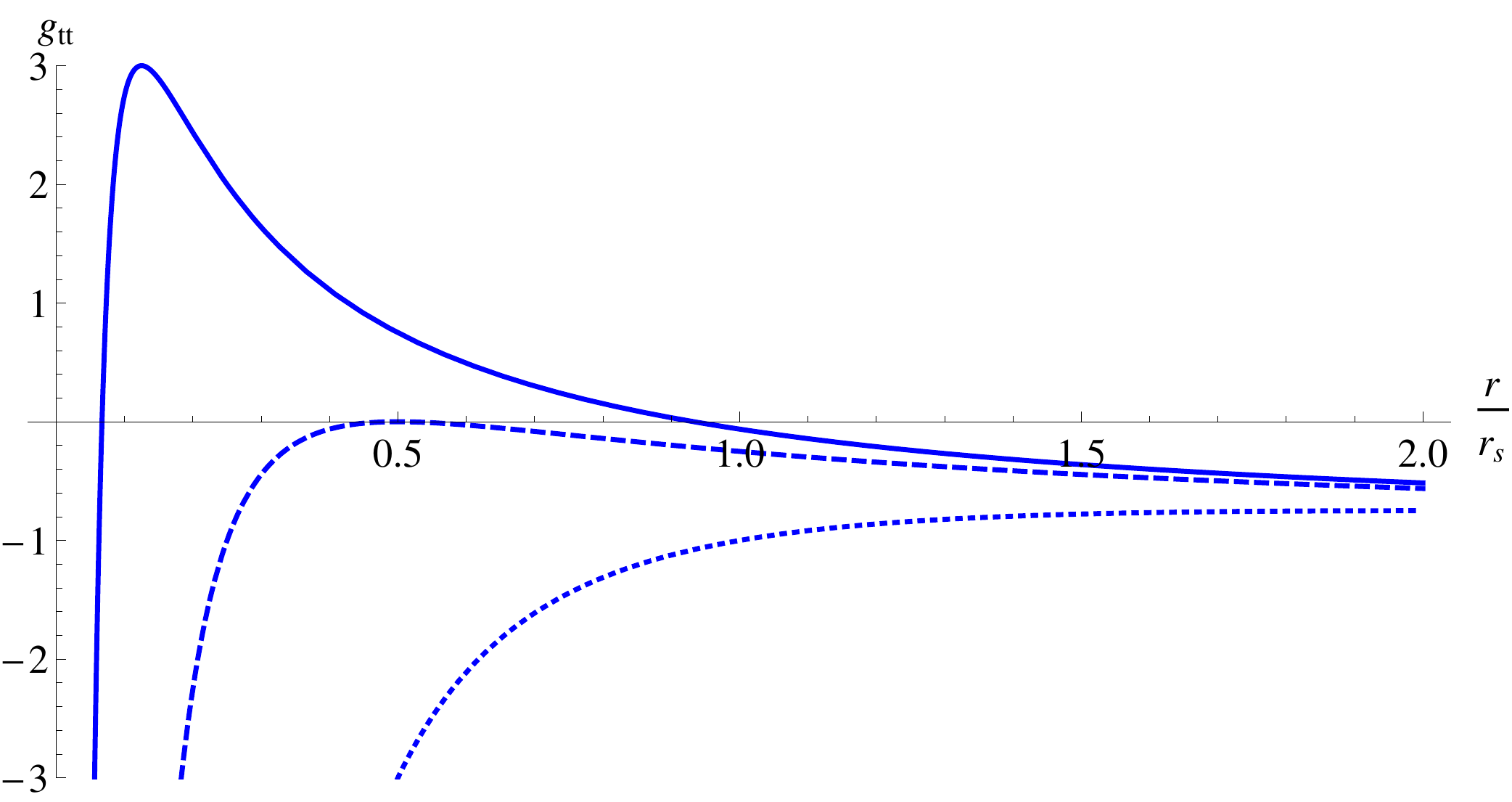}
 \caption{Function $g_{tt}$ in terms of the coordinate $r$, for a charged black hole with $r_q=r_S/4$ (continuous), $r_q=r_S/2$ (dashed) and $r_q=r_S$ (dotted).}\label{1fig:gttRN}
\end{figure}

Where $r_q \equiv q l_P$ and $r_S$ is an integration constant that corresponds to the mass of the black hole ($M =r_S/(2 l_P^2)$) as felt at the asymptotic flat infinity ($r\rightarrow \infty$). This metric has a different structure than Schwarzschild. At $r=0$ the components of the metric diverge, and also the Kretschmann curvature scalar $K=\frac{12 r_S^2}{r^6} - \frac{48 r_S r_q^2}{r^7} + \frac{56 r_q^4}{r^8}$, so the divergence is not a consequence of a coordinate singularity. Let us note that at $r=0$ the energy-momentum tensor is ill-defined and that the Einstein's and Maxwell's equations are not solved in that region
. Depending on the value of $r_S$ and $r_q$ the metric components will also diverge at $r_\pm=(r_S \pm \sqrt{r_S^2-4 r_q^2})/2$ (see fig. \ref{1fig:gttRN}), these are coordinate singularities that signal the horizons of the geometry: If $r_q<r_S/2$, there will be two horizons, if $r_q > r_S/2$ there will be no horizon (known as naked case), and if $r_q = r_S/2$ there will be a single degenerated horizon (known as extremal case). 

\begin{figure}[h!]
\centering
 \begin{tabular}{ccc}
  \includegraphics[height=.35 \paperheight]{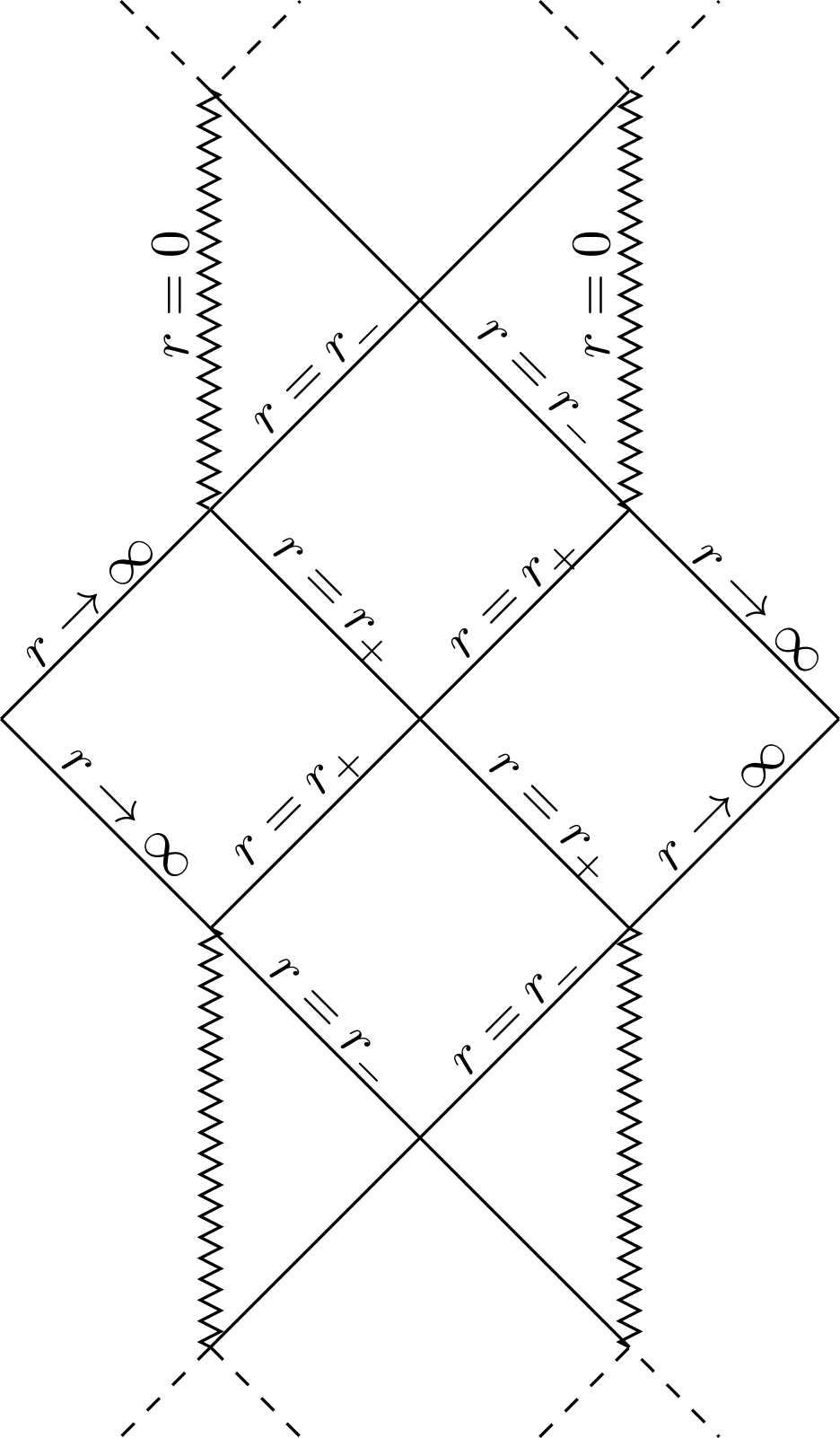} \ \ \ \ \ \ \  &
  \includegraphics[height=.3 \paperheight]{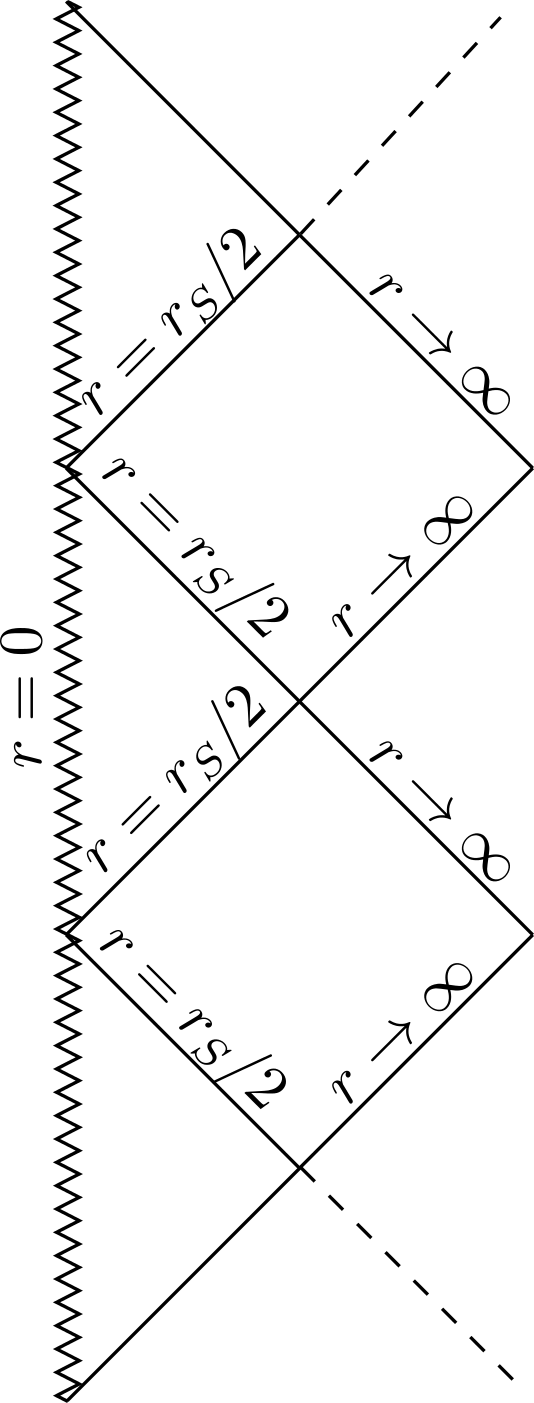} \ \ \ \ \ \ \ &
  \includegraphics[height=.1 \paperheight]{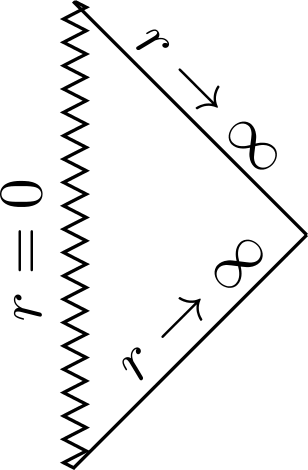} \\
  (a) \ \ \ \ \ \ \  & (b) \ \ \ \ \ \ \  & (c)
 \end{tabular}
 \caption{Penrose diagrams of the different configuration of the Reissner-Nordström geometry, (a) is $r_q<r_S/2$, (b) is $r_q=r_S/2$, (c) is $r_q>r_S/2$. In these diagrams, we represent the radial and time coordinates of the space-time (every point would correspond to a two-sphere) in compactified way, such that light rays are straight lines at 45 degrees. These diagrams are useful to illustrate the causal structure of the space-time.}\label{1fig:PenRN}

\end{figure}

The causal structure of this regions is represented in Penrose diagrams in fig. \ref{1fig:PenRN}. In all cases, $r=0$ is a timelike region. In the naked case, the region $r=0$ is not hidden behind any horizon, and there would be information travelling to and coming from this region to observers at infinity. In the extremal case, there is a single horizon at $r=r_S/2$ separating the asymptotic flat infinity from $r=0$. Inside this horizon the $t$ coordinate still has a time-like nature, and the $r$ coordinate is still space-like, so it is possible for an observer to cross back the horizon outside; but the asymptotic region you reach in this way, is different from the one you started with (There are infinite regions like this one, and an observer could travel between all of them).

The subextremal case is a bit more complicated, there are two horizons, $r_+$ and $r_-$. Outside $r_+$ we have an asymptotic region. Between $r_+$ and $r_-$ there is a region where $t$ becomes space-like and $r$ time-like, which forces any observer that crosses from $r_+$ to travel to $r_-$, and any observer that comes from $r_-$ to travel to $r_+$. The $r=0$ region lies inside $r_-$, and $t$ is again time-like and $r$ space-like,  so it is possible to enter inside $r_-$ and then exit again. Therefore it is possible for an observer to cross $r_+$ (which would be seen as the black hole horizon from infinity), $r_-$, interact with $r=0$, exit throught $r_-$, and then $r_+$ (which now would be seen as a white hole horizon from infinity), to an asymptotic region which would be different from the one the observer started its trip. Besides this ``parallel'' asymptotic region (infinite of them), there is also another asymptotic region, akin to the region IV of the Schwarzschild geometry, which is unreachable from the starting one.

If we take a look at the geodesics through eq. \ref{1eq:geosss}, where we have to substitute $F=\left ( 1-\frac{r_S}{r}+\frac{r_q^2}{r^2} \right )$, we can see that there will be geodesics able to cross the horizons of the geometry. But since $F$ diverges to positive infinity at $r=0$, only null geodesics with no angular momentum will be able to reach that region, and the rest will be repelled by the potential.


\cleardoublepage
\chapter{Introduction: Singularities and Quadratic Gravity}\label{Int2}

In this chapter, we will start discussing singularities. First of all, we will formalize what it means to have a singular space-time, and we will see it is characterized by the extendibility of geodesics. We will study geodesics in this context, and will check that the emergence of singularities is not a consequence of the high degree of symmetry of the problem (such as the gravitational collapse of spherical symmetric star). After that, we will see a new approach to the extendibility of geodesics that will be relevant for future chapters.

Finally, I will introduce an alternative gravity theory called \emph{quadratic gravity}. This theory was born after the first attempts to quantize gravity. We will treat this theory in the Metric-Affine formalism in the next chapters.

\section{Defining a Singular Space-time}

As we have seen in the previous chapter, $r=0$ is a problematic point in the Schwarzschild geometry. Every observer that crosses the event horizon must end up at $r=0$ where it will be crushed by infinite forces, and there is no natural way to extend the geometry beyond it. One would like to define such point as a \emph{singularity}.
Finding a suitable definition of singularity is a difficult task, and there have been different attempts in the literature each with its advantages and disadvantages. In this section we will follow the work of Geroch (\cite{Geroch:1968ut}), which we consider stays closer to the physics behind the geometry. A first attempt would be to try a definition of the type ``the region of space-time where something goes wrong'', where something going wrong would be a geometric quantity that diverges. This type of definition faces two kinds of problems:

\begin{itemize}
 \item The first one has to do with the quantity that diverges. The components of the metric tensor are clearly a bad choice, since they also diverge at $r=r_S$ --the event horizon--, which is not a problematic region of space-time. The components of the Riemann tensor are also a bad choice, because if the curvature is not constant, it is always possible to choose a coordinate system where some component diverges. The curvature scalars are a more interesting choice, since they are independent of coordinates, and do not diverge at the horizon. However, there are infinitely many of them (contractions of the Riemann with itself and with derivatives of itself), and it is not clear that all of them are physically relevant.
 
 On the other hand, there are space-times that show no curvature divergences but we would want to consider singular. For example, let us take Minkowski with cylindrical coordinates, remove the portion of space between $\phi = 0$ and $\phi=\phi_0$, and identify both edges. The flat metric can be continued everywhere but at $r=0$, where it is undefined. At that point, there is a \emph{conical singularity}, even though the curvature tensor vanishes everywhere.
 \item The second problem has to do with the concept of ``region of space-time''. In GR, we can choose whatever manifold to define a geometry with a metric tensor. In particular, we could choose to use a manifold with the singular points removed; however, we would clearly want to define such space-time as singular. How can we tell if a region has been removed? It is not easy: First of all, it is possible to choose a coordinate system that hides the removed region, taking it to infinity (the black hole region of Schwarzschild expressed in $(t,r)$ coordinates lies beyond $t=\infty$). On the other hand, we could also bring an inaccessible region of the space-time to a finite coordinate value through some coordinate redefinition ($t^\prime = \arctan t$ in Minkowski space-time). Whatever the definition, it cannot make use of coordinates. Second, since the metric has Lorentzian signature, we do not have a good concept of distance between two points: we can always find a curve that joins two points with length as close to 0 as we want. This makes complicated to detect whether a region of space has been artificially removed, as every point can be connected by a curve of 0 length.
\end{itemize}

We can find a solution to the second problem if we think in physical terms. If we remove a part of the space-time, a physical observer that passed through the removed region before, now would meet the ``end'' of the space-time. Free-falling observers follow geodesics, so mathematically, this scenario corresponds to a geodesic with and endpoint. This lead us to define singularity as:

\begin{definition*}
 A space-time is non-singular if every half-geodesic is either complete or else is contained in a compact set.
\end{definition*}


A geodesic is complete if its affine parameter can attain arbitrarily large values. If a region of the space-time is removed, geodesics that reached that region would be incomplete now. Moreover, if a problematic point is not reached by any geodesic, the space-time would be considered non-singular. This could happen with a space-time that have a curvature that diverges as we approach some unreachable asymptotic region. We would not wish to call that space-time singular, so this definition works well in that regard. This definition is independent on the coordinate system chosen, and works with conical singularities. The reason the definition makes mention of compact sets is that there are geodesically incomplete compact sets \cite{Misner:1963fr}, but such set can not have resulted from the removal of a part of larger (connected) space-time. In that sense, having incomplete geodesics in a compact set would not signal that a portion of the space-time has been removed. 

This definition of singularity is tied with the physical concept of free-falling observers. An observer that follows an incomplete geodesic would reach the end of it. What would happen afterwards? Would the observer simply disappear? The equations that govern the evolution of the physical system clearly have to break at that point, and most likely we would lose properties such as unitarity. But, should not we care about accelerated observers too? In Minkowski space-time, an observer with acceleration that grows to infinity can cross the entire space-time in finite amount of proper time. In this sense, it would be an incomplete curve. However, such observer would need an infinite amount of energy to accelerate that much, which would be physically impossible. Therefore, it seems much more reasonable to concern ourselves only with observers with bounded acceleration. Geroch \cite{Geroch:1968ut} showed an example of a geometry where geodesics are complete but an observer with bounded acceleration crosses the entire space-time in finite proper time. So it may be interesting to study the completeness of curves of bounded acceleration after we have established if geodesics are complete. 

This definition does not concern itself with any geometric quantity diverging. On one hand, this is good, because the physical implications of the curvature divergences are not clear. The Riemann tensor is associated to the tidal forces, but even if tidal forces were able to rip apart any infalling observer, they do not pose the same problems as an observer simply disappearing into nothingness. On the other hand, it seems that we are missing half the definition. Our first attempt at a singularity definition was the intuitive idea of ``the region of space-time where something goes wrong''. We have dealt with the problems of the idea of ``region of space-time, but we have forgotten about the part of ``something going wrong''. This is arguably a good property of the definition, the real problematic  space-times are the ones where the trajectories of physical observers are not well-defined, and this should not concern us with diverging quantities: The existence of observers is more important (and takes precedence over) than them suffering infinite forces. However, it is often implicitly assumed in the literature that if a metric is not $\mathcal{C}^2$ (or perhaps $\mathcal{C}^1$) something goes wrong and the space-time is singular. I will discuss this point in the following sections.
%

\section{Extension of Geodesics}

As we have seen, the definition we have given in the previous section has many good properties. But is this the definition we were looking for? We wanted specifically to consider the central point of the Schwarzschild space-time as a singularity, since it crushes every observer that falls into it to zero volume. This is a very pathological behaviour, but it is not clear what it has to do with geodesic completeness. The Schwarzschild metric does not extend in a natural way through $r=0$, and the geodesics would meet their endpoint there. But one could try to extend it ``by hand'': for example, gluing together the black hole side of the Schwarzschild metric with the white hole side, or perhaps simply allowing the coordinate $r$ to take negative values (which would be equivalent to a Schwarzschild metric with $r$ positive, but negative mass $M<0$). Then one could possibly give a prescription on how to extend the geodesics from one side to the other. Would this space-time be non-singular?

The usual answer to this question is no. The reason is that the metric is no longer $\mathcal{C}^2$ at the origin. This is important because it guarantees that the coefficients of the connection are $\mathcal{C}^1$, and therefore, the standard existence and uniqueness theorems of ODE guarantee a solution for the components of the tangent vector of the geodesics in eq. \ref{1eq:Geo}. If they were not, the solution may not exist or may not be unique.


For this section we will consider a $\mathcal{C}^2$ metric, we will study conjugate points and its relation to Jacobi fields, and then we will introduce one of the singularity theorems that will tell us a minimum set of conditions for which the space-time will develop a singularity. In this case, the space-time cannot be extended, not even ``by hand''. In the next section the requirement of a $\mathcal{C}^2$ metric will be relaxed and an alternative representation of the geodesic equation will be introduced. 
%

%

\subsection{Conjugated Points}

Geodesics are the curves that maximize proper time, or minimize the distance between two points\footnote{Also, in the Riemannian formalism, the tangent vector is parallel transported along the curve, see eq. \ref{1eq:GeoParallel}.}. However, the geodesic equation (eq. \ref{1eq:Geo}) gives us curves that maximize proper (minimize distance) time \emph{locally}, but that does not imply that those curves maximize the proper time (minimize distance) \emph{globally}. A simple example would be the following: In a 2-sphere (like the surface of the Earth), the geodesic curves that minimize the distance are the great circles. Now, consider a traveller going from the south pole to the north pole following a great circle. Once the traveller reaches the north pole and continues beyond it, he is no longer following the shortest path, as the shortest path would be the one that goes from the south pole to his current location by the other side of the world.

\begin{figure}
\centering
\includegraphics[width=.35\linewidth]{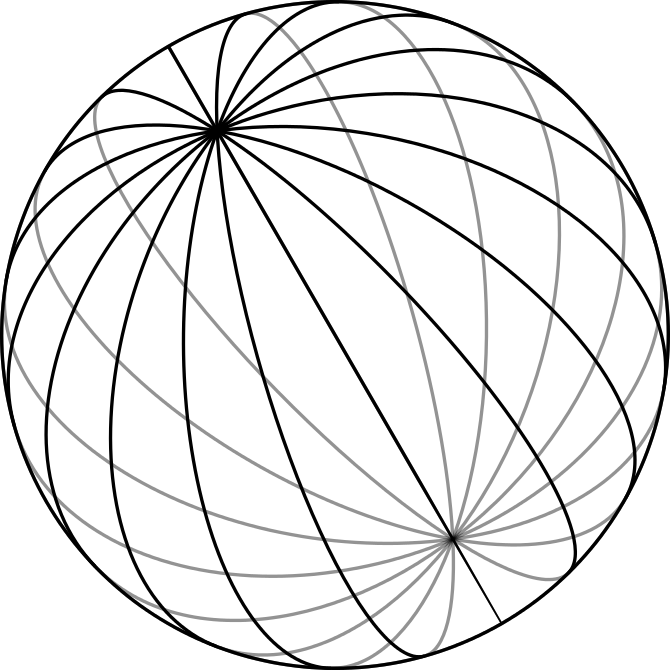}
 \caption{A 2-sphere. The great circles are geodesics, and the north pole and the south pole are conjugated points.}
\end{figure}

All geodesics that start from the south pole, coalesce again in the north pole. These two points are called \emph{conjugated points}. A more formal definition would be this:
\emph{Two points $p,q$ of a geodesic $\gamma$ are said to be conjugated if there is a Jacobi field not identically zero that vanishes at $p$ and $q$}. Let us recall that a Jacobi field is a vector field that satisfies the geodesic deviation equation eq. \ref{1eq:GeoDevEq}. Conjugated points are important because a segment of a geodesic that contains two conjugated points no longer extremizes the path between its endpoints\footnote{See \cite{HawkingEllisLandshoffNelsonSciamaWeinberg197503}, section 4.5}.

It is also possible to define points conjugated to a hypersurface. \emph{A point $p$ of a geodesic $\gamma$ belonging to a congruence of geodesics orthogonal to the hypersurface $\Sigma$ if there exists a Jacobi field that vanishes at $p$ and non-zero and tangent to $\Sigma$ in $\Sigma$}. In a similar fashion, the path between a hypersurface $\Sigma$ and a point $q$ will be extremized by a geodesic orthogonal to $\Sigma$ if it has no conjugate points between $\Sigma$ and $q$.

If a Jacobi field vanishes, then the volume transported by the congruence will go to $0$, too (unless some other deviation vector that diverges to infinity). So we start seeing the relation between the problematic points from a physical point of view (the volume going to zero) and possibly problematic global properties of the geometry (geodesics no longer extremizing length globally).

\subsubsection{Raychaudhuri Equation and Focalization Theorem}

To understand when a conjugate point forms, we have to look again at the evolution of a geodesic congruence. In the last chapter, we looked at the second derivatives of the deviation vectors that describe the congruence along the geodesic, and we obtained the geodesic deviation equation. Now, we are going to look at the first derivative of a deviation vector $Z$ along a null geodesic with tangent vector $u$:

\begin{equation}
 u^\mu \nabla_\mu Z^\nu = Z^\mu \underbrace{\nabla_\mu u^\nu}_{\equiv B^\nu{}_\mu}
\end{equation}

From the definition, $B^\nu{}_\mu$ is a matrix that describes the infinitesimal deformation of the deviation vector along the geodesic. As we know that the tangent part of the deviation vector is conserved along the geodesic, the $B^\nu{}_\mu$ must contain redundant information. We could try to decompose the deviation vector into a tangent part $t^\mu$ and an orthogonal part $o^\mu$, but as $u^\mu$ is null, this decomposition is not unique, as $t^\mu+f u^\mu$ and $o^\mu-f u^\mu$ are also tangent and orthogonal. To remove this freedom we can use a projector:

\begin{equation}
 h_{\mu \nu} = g_{\mu \nu} + u_\mu m_\nu + m_\mu u_\nu
\end{equation}
where $m$ is a null auxiliary vector such that $u^\mu m_\mu =-1$. The choice of $m$ gives raise to different projectors, but the results of this section will be independent of the choice. Now we can project $B$ and $Z$.

\begin{equation}
 \bar{B}_{\mu \nu} = h^\alpha{}_\mu h^\beta{}_\nu B_{\alpha \beta} \qquad \bar{Z}^\mu = h^\mu{}_\alpha Z^\alpha \label{1eq:bbarb}
\end{equation}

The projected matrix $\bar{B}$ can be expressed in terms of the old one:

\begin{equation}
 \bar{B}_{\mu \nu} = B_{\mu \nu} + u_\mu m^\alpha B_{\alpha \nu} + u_\nu m^\alpha B_{\mu \alpha} + u_\mu u_\nu m^\alpha m^\beta B_{\alpha \beta}
\end{equation}

$\bar{B}_{\mu \nu}$, as any rank 2 tensor, can be decomposed into an antisymmetric part, a traceless symmetric part and a trace:

\begin{equation}
 \bar{B}_{\mu \nu}=\omega_{\mu \nu} + \sigma_{\mu \nu} + \frac{1}{2} \theta h_{\mu \nu}
\end{equation}

$\omega$, $\sigma$, $\theta$ are called the twist, the shear, and the expansion of the congruence. Let us study how these objects evolve along the geodesic:

\begin{IEEEeqnarray}{rCl}
 u^\alpha \nabla_\alpha B_{\mu \nu} &=& u^\alpha \nabla_\alpha \nabla_\nu u_\mu = u^\alpha \nabla_\nu \nabla_\alpha u_\mu + R_{\alpha \nu \mu \beta} u^\alpha u^\beta \\
 &=& \nabla_\nu (u^\alpha \nabla_\alpha u_\mu) -(\nabla_\nu u^\alpha) (\nabla_\alpha u_\mu)+ R_{\alpha \nu \mu \beta} u^\alpha u^\beta\\
 &=& - B^\alpha{}_\nu B_{\mu \alpha} + R_{\alpha \nu \mu \beta} u^\alpha u^\beta\label{1eq:evw}
\end{IEEEeqnarray}

Now, taking the trace of this equation and realizing that $B^{\alpha \beta}B_{\alpha \beta}=\bar{B}^{\alpha \beta}\bar{B}_{\alpha \beta}$ and that $B^\mu{}_\mu = \bar{B}^\mu{}_\mu = \theta$ we have:

\begin{equation}
 u^\mu \nabla_\mu \theta = -\frac{1}{2} \theta^2-\sigma^{\alpha \beta} \sigma_{\alpha \beta} + \omega^{\alpha \beta} \omega_{\alpha \beta} - R_{\alpha \beta} u^\alpha u^\beta 
\end{equation}

This equation is known as the \emph{Raychaudhuri equation}, and tells us the evolution of the expansion of the congruence along the geodesic. Now, in this equation $\sigma^{\alpha \beta} \sigma_{\alpha \beta}$ is a positive quantity, $R_{\alpha \beta} u^\alpha u^\beta$ will also be positive in GR if matter holds the weak energy condition: $T_{\alpha \beta} u^\alpha u^\beta\geq0$ for $u^\alpha$ non-space-like. Also, $\omega^{\alpha \beta} \omega_{\alpha \beta}$ will be $0$ if the geodesics are orthogonal to a hypersurface\footnote{Frobenius theorem says that if a vector $u^\mu$ is orthogonal to hypersurface then $u_{[\alpha} \nabla_\beta u_{\gamma]} =0$. Contracting this expression with the projector gives:

\begin{equation}
 h_\mu{}^\alpha h_\nu{}^\beta u_\gamma \nabla_\alpha u_\beta - h_\mu{}^\alpha h_\nu{}^\beta u_\gamma \nabla_\beta u_\alpha \qquad \Rightarrow \qquad \bar{B}_{[\mu \nu]}=0 
\end{equation}
Which in turn implies $\omega_{\mu \nu}=0$.}. We are left with this equation:

\begin{equation}
 u^\alpha \nabla_\alpha \theta + \frac{1}{2} \theta^2 \leq 0
\end{equation}

That can be integrated in terms of the affine parameter of the geodesic:

\begin{equation}
 \frac{1}{\theta} \geq \frac{1}{\theta_0} + \frac{\lambda}{2}
\end{equation}
with $\theta_0$ being the expansion at $\lambda=0$. This equation tells us that if the null congruence orthogonal to a surface has negative expansion, it will become infinitely negative for a finite value of the affine parameter, bounded by $\lambda \leq 2/|\theta_0|$. This will be a conjugate point to the surface. To see how it is so, let us relate the expansion $\theta$ with the matrix $A(\lambda)$ found in eq. \ref{1eq:matrixA}, that gave us the evolution of the deviation vectors from the initial values at some point. As in section \ref{1sec:EvoGeoC}, latin indices is an adapted coordinate system parallel transported to the geodesic. Then differentiating eq. \ref{1eq:matrixA} with respect to $\lambda$ we have:

\begin{equation}
 \frac{\df Z^a(\lambda)}{\df \lambda}=u^\mu \nabla_\mu Z^a(\lambda) = \frac{\df A^a{}_b(\lambda)}{\df \lambda} Z^b(\lambda_i)
\end{equation}

From which we have:

\begin{equation}
 B^a{}_b Z^b(\lambda) = \frac{\df A^a{}_b(\lambda)}{\df \lambda} Z^b(\lambda_i) \qquad \Rightarrow \qquad B^a{}_c A^c{}_b(\lambda) Z^b(\lambda_i) = \frac{\df A^a{}_b(\lambda)}{\df \lambda} Z^b(\lambda_i)
\end{equation}

Since this equation must be valid for all sets of initial conditions of the deviation vector (as long as the deviation vector is different form $0$, in which a similar construction with $\tilde{A}$ from eq. \ref{1eq:tildeA} could be made), we simply have (in matrix notation):

\begin{equation}
  \frac{\df A}{\df \lambda} = B A 
\end{equation}

which gives us the following relation between $\theta$ and $A$:

\begin{equation}
\theta = \text{tr}(B) = \text{tr} \left (A^{-1} \frac{\df A}{\df \lambda} \right ) = \frac{1}{\det(A)}\frac{\df (\det(A))}{\df \lambda}
\end{equation}

We see that when the expansion becomes infinitely negative, the volume (given by $\det(A)$) goes to 0. Therefore, a point with infinite negative expansion is a conjugate point of any hypersurface orthogonal to the congruence.

\subsection{Singularity Theorems}\label{2sec:SingTh}

One might think that the singularity at $r=0$ in the Schwarzschild geometry is a consequence of the spherical symmetry of the problem. After all, in Newtonian gravity, if a sphere of dust collapses under its own gravity, it would form a point of infinite density at the centre. But if the symmetry is broken, for example if the sphere of dust had a little angular momentum, no singular points would appear.

Penrose and Hawking gave several theorems that state basic conditions under which, a singularity will be formed. In these theorems, no symmetries are assumed. I am going to show one of them, given by Penrose in 1965 \cite{Penrose:1964wq} (see \cite{HawkingEllisLandshoffNelsonSciamaWeinberg197503} for a thorough review of this theorem and others).

\begin{theorem}
 Space-time ($\mathcal{M}$, $g$) cannot be null geodesically complete if:
 \begin{enumerate}[(i)]
  \item $R_{\mu \nu} k^\mu k^\nu \geq 0$ for all null vector $k^\mu$.
  \item There is a non-compact Cauchy surface $\mathcal{H}$ in $\mathcal{M}$.
  \item There is a closed trapped surface $\mathcal{T}$ in $\mathcal{M}$
 \end{enumerate}
\end{theorem}

Let us understand what these three conditions mean. We saw the first one when we talked about the Raychaudhuri equation, and it is a condition that will hold in GR as long as the matter satisfies the weak energy condition, which is satisfied by known sources.

The second one refers to a \emph{Cauchy surface}  $\mathcal{H}$. A (global) Cauchy surface is a space-like hypersurface in which every non-space-like curve intersects exactly once. In Minkowski space-time, the hypersurfaces of constant $t$ would be Cauchy surfaces. If the space-time admits a Cauchy surface $\mathcal{H}$, and we know the relevant data there, then we can evolve the equations of motion (of the metric and the matter fields) forwards or backwards in time, and know the state of the universe at any point. This is a really nice property to have, as it is tied with the concept of causality: We can foliate the space-time with these surfaces, and each of them is either in the past or the future of the rest. However, Einstein's equations do not imply the existence of a Cauchy surface, and there are many known solutions that do not admit one. But the Schwarzschild metric does admit them, and a slight deformation of its geometry should also admit a Cauchy surface.

The third one refers to a \emph{closed trapped surface} $\mathcal{T}$. This is a closed surface where any congruence of null geodesics orthogonal to $\mathcal{T}$ has negative expansion. We can think of it as that $\mathcal{T}$ is in a region where gravity is so strong that even outgoing light rays are pushed back and made to travel inside the trapped surface. The surface of constant $r$ and $t$ inside the event horizon of the Schwarzschild metric are trapped surfaces. Moreover, slight deformations of a trapped surface will not change its nature. So we can break the spherical symmetry of the Schwarzschild geometry, but still have a trapped surface.

The idea of the proof is the following: Let $J_+(\mathcal{T})$ be the set of points that can be joined from $\mathcal{T}$ with a non-space-like curve orientated to the future. Let $\dot{J_+}(\mathcal{T})$ be the boundary of $J_+(\mathcal{T})$. $\dot{J}_+(\mathcal{T})$ is a three dimensional submanifold of $\mathcal{M}$ without boundary, and must be \emph{achronal} (it contains no two points that can be joined by a time-like curve). $\dot{J}_+(\mathcal{T})$ is generated by two families of null geodesics orthogonal to $\mathcal{T}$ (akin to the light cone of special relativity). Because the expansion of this null geodesics is negative, and $R_{\mu \nu} k^\mu k^\nu \geq 0$, they will encounter a conjugate point after a finite amount of proper time. After this conjugate point, null geodesics no longer extremize proper time, so every point in the geodesic beyond the conjugate point can be joined to $\mathcal{T}$ through a time-like geodesic instead of a null one, and therefore are inside $J_+(\mathcal{T})$. As this happens for a finite value of the affine parameter for every null geodesic, which we are assuming complete, then $\dot{J}_+(\mathcal{T})$ must be a compact hypersurface. Can $\dot{J}_+(\mathcal{T})$ be compact without boundary, $\mathcal{H}$ non-compact, and all the geodesics complete at the same time?

Let us consider a time-like vector field, defined in all $\mathcal{M}$. Its integral curves will intersect $\mathcal{H}$ exactly once, because it is a Cauchy surface, and will intersect $\dot{J}_+(\mathcal{T})$ at most once because it is an achronal surface. We can define a one-to-one map of $\dot{J}_+(\mathcal{T})$ into $\mathcal{H}$, making a correspondence between the points intersected by the same curve. The image of $\dot{J}_+(\mathcal{T})$ is homeomorphic to $\dot{J}_+(\mathcal{T})$, and so, if $\dot{J}_+(\mathcal{T})$ is compact and three dimensional, then its image must be too. The image is a three dimensional compact subset of a three dimensional non-compact set. Therefore it must have a boundary, and so will $\dot{J}_+(\mathcal{T})$. Then we reach to a contradiction, since we stated that $\dot{J}_+(\mathcal{T})$ has no boundary. 

\begin{figure}[h!]
 \centering
  \includegraphics[width=.5\linewidth]{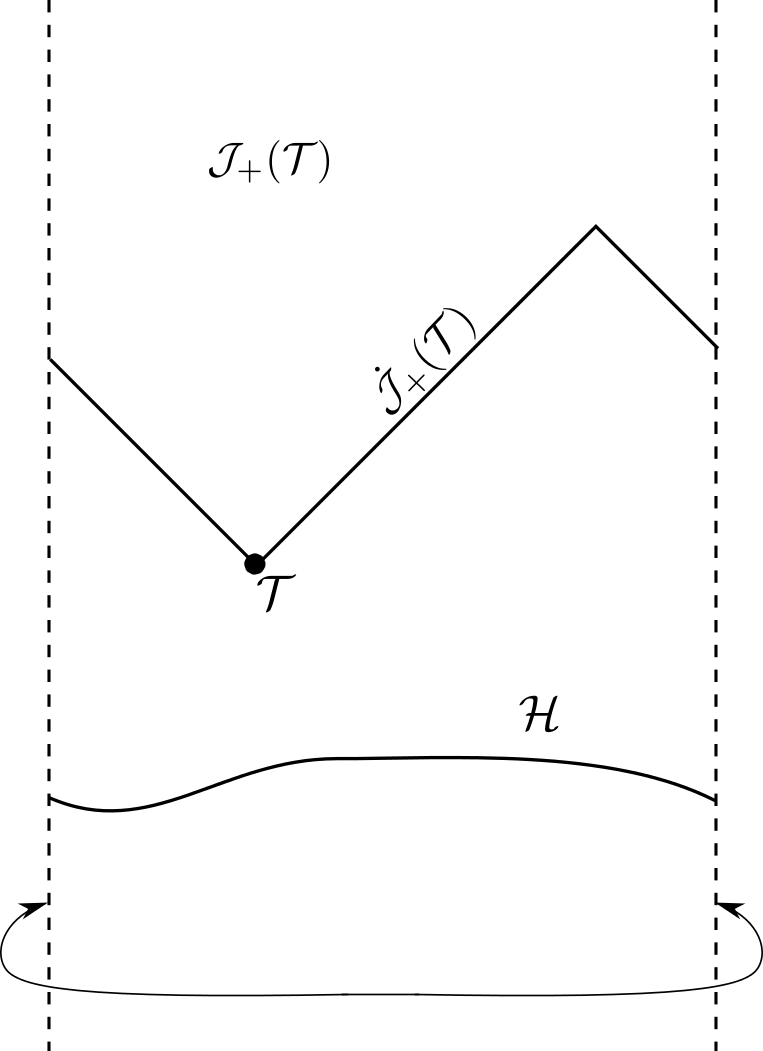}
 \caption{Diagram showing how if the Cauchy surface $\mathcal{H}$ is compact, the image of $\dot{J}_+(\mathcal{T})$ can be compact too. The spacial part of the space-time consists of a cube with identified edges. In this diagram we have suppressed 2 spatial dimensions, which makes $\mathcal{T}$ a point, and the left and right edges are identified; time increases in the vertical direction. We can see that light rays that start from $\mathcal{T}$ in one direction, cross the edge and meet the light-rays sent in the other direction, making $\dot{J}_+(\mathcal{T})$ compact. The meeting point of geodesics would be a conjugated point, and any point beyond that can be joined by a time-like curve. If we were to draw a time-like congruence in all the space-time, it would define a one to one map from $\dot{J}_+(\mathcal{T})$ to $\mathcal{H}$ which is also compact. }\label{1fig:Hcompact}
\end{figure}

If $\mathcal{H}$ was compact, then the image of $\dot{J}_+(\mathcal{T})$ could be the whole $\mathcal{H}$, and there would be no contradiction (see fig. \ref{1fig:Hcompact}). Also, if $\dot{J}_+(\mathcal{T})$ did not need to be achronal, it would be impossible to define a one-to-one map on $\mathcal{H}$, and $\dot{J}_+(\mathcal{T})$ could be compact without boundary . 

We assumed null geodesic completeness to prove $\dot{J}_+(\mathcal{T})$ is compact. Then, this assumption must be incorrect, some null geodesics will not be complete and the space-time will develop a singularity. Since it is possible to break spherical symmetry while keeping $\mathcal{T}$ a trapped surface and $\mathcal{H}$ a Cauchy surface, we prove that singularities are not a consequence of the symmetries of the geometry. It is also clear that since the Schwarzschild geometry satisfies the 3 conditions, there is nothing we can do to extend the geometry in a non-singular way.

However, in a general scenario with a trapped surface, it is also possible that one of the other conditions does not hold. Condition (i) $R_{\mu \nu} k^\mu k^\nu \geq 0$ will hold in GR as long as the matter holds the null energy condition, $T_{\mu \nu} k^\mu k^\nu \geq 0$ for every null vector $k^\mu$, which is true for every known source. But extensions of GR that change Einstein equations might break this correspondence, allowing for $R_{\mu \nu} k^\mu k^\nu < 0$ even if matter holds the null energy condition. The other possibility is that condition (ii) does not hold. In order to know if $\mathcal{H}$ is a Cauchy surface, we should know all the future of $\mathcal{H}$, which is not possible in a realistic scenario. So it would be possible that the space-time develops a \emph{Cauchy horizon}, and beyond that horizon, there are time-like curves that do not intersect $\mathcal{H}$. This is the case of the Kerr and Reissner-Nordström black holes (which also do contain a singularity). There are other theorems that also give conditions for which the space-time will develop a singularity and do not require a Cauchy surface \cite{Hawking:1969sw}.

In the last chapter, we said that if the volume transported by the congruences of geodesics goes to zero when approaching a certain point, there is a \emph{strong singularity}. The choice of name becomes clear now, as a strong singularity implies that there is a point conjugate to every geodesic that reaches it, and signals the existence of a trapped surface. Not only those points are bad from the perspective that every physical object would be destroyed, but are also problematic from a geometric perspective.

I want to emphasize that in all this section we have assumed a $\mathcal{C}^2$ metric. In particular, two half-geodesics joined at a point where the metric is not $\mathcal{C}^2$ might maximize proper time even if there is a conjugate point at the union. The converse could also be true, they might not maximize proper time even if there is not a conjugate point. But such a curve is not a geodesic under our assumptions.

\section{Extension of Geodesics for Discontinuous Metrics}\label{1sec:ExtDisc}

In this section, I want to relax the condition that the metric is $\mathcal{C}^2$ and study whether it is possible to define geodesics in a unique way. The geodesic equation (eq. \ref{1eq:Geo}) requires $\mathcal{C}^1$ coefficients of the connection to satisfy the conditions of the standard theorems of existence and uniqueness of solutions for ordinary differential equations\footnote{Actually, it requires them to be Lipschitz continuous}. If they are not, the geodesic equation may have more than one solution. However, this may be a consequence of the freedom we have in choosing both the dependent and independent variables of the differential equations, and not a problem of the geometry. The independent variable is the parameter $\lambda$, but the geodesic curves do not change under reparametrizations of $\lambda$. The dependent variables are coordinate functions, and we can choose whatever coordinates to work with.

One should realize that the geodesic equation is not an equation for a curve, but an equation for a vector (field) whose integral curves are geodesics. It would be better to find an equation whose solutions are curves directly, and therefore, is independent of reparametrizations. One possible solution would be to use a \emph{Pfaff system} (PS)\footnote{See (\cite{libermann1987symplectic},\cite{awane2013pfaffian}) for a rigorous treatment of Pfaffian systems.}. Such a system is determined by $k$ 1-forms $\alpha^k$ defined in a $d$-dimensional manifold. An \emph{integral manifold} of the PS is a $(d-k)$-dimensional submanifold such that all the 1-forms restricted to it vanish\footnote{If all the $k$ 1-forms are linearly independent.}. A Pfaff system is said to be \emph{closed} if $\df \alpha^i \wedge \alpha^1 \wedge ... \wedge \alpha^k = 0$ for $i=1,...,k$. A Pfaff system is completely integrable in the neighbourhood of a generic point if and only if it is closed in this neighbourhood.


Let us consider geodesics in a 2-dimensional manifold. We need to find just one differential 1-form such that the integral manifolds of the system are geodesic curves. Such system is always closed. Let us think of an uniparametric family of geodesics $\gamma_\xi(\lambda)$ with family parameter $\xi$ through the geodesic equation. This family of geodesics gives coordinates to the manifold by assigning to a point $p$ the coordinates $(\xi,\lambda)$, which are the value of the family parameter of the geodesic that passes through $p$, and the value of the affine parameter of that geodesic when it passes through $p$.  It is easy to see that the 1-form we need in the Pfaff system is $\df \xi$, the geodesic curves are just the curves of constant $\xi$. In other words, the differential form $\df \xi$ restricted to the geodesic curves must vanish:

\begin{equation}
 \text{``}\df \xi \big |_\gamma=0\text{''}\label{1eq:pfaff}
\end{equation}

As long as $\df \xi$ exists, this system will have solutions. 
We have to be careful that the components of the 1-form might not be well-defined in certain coordinate systems, but be regular in others. We have to realize that when we coordinate our space-time $\mathcal{M}$, we are giving it a differential structure, but we do not know if it is an appropriate one a priori. However, there are 1-forms whose components will diverge no matter the coordinate system. For example, in 2 dimensional euclidean space, the 1-form $\text{``}\df \theta\text{''}=\frac{y\df x - x\df y}{x^2+y^2}$ is not well defined at the origin, and as such, its components in any coordinate system will not be $\mathcal{C}^1$. For our Pfaff system, this type of divergence will happen at points where the geodesics originate or converge.

The existence of solutions for this equation is independent of the coordinate system or the particular parametrization of the geodesic; however, we have only changed the problem to finding a function $\xi$ whose curves of constant value are geodesics. But this equation does let us know if it makes sense to even try to extend the geodesic beyond the discontinuity. If we know how a family of geodesics reaches the region where the metric is discontinuous, we can check whether the equation $\df \xi|_\gamma=0$ still works there. First of all, we must relate $\df \xi$ with the deviation vector $Z^\alpha=\left .\frac{\partial \gamma^\alpha}{\partial \xi} \right |_{\lambda=\text{const.}}$. If $Z^\alpha$ and the tangent vector of the geodesics $u^\beta$ are orthogonal\footnote{This is always possible to get through a reparametrization of $\lambda$.}, then we simply have $(\df \xi)_\mu =g_{\mu \nu} Z^\nu /(Z^\alpha Z_\alpha)$. Thus calculating $Z^\nu$ through eq. \ref{1eq:GeoDevEq} we can obtain $\df \xi$ and check whether it will be possible to extend the geodesics beyond the discontinuity.


\subsection{Two Dimensional Study}

Let us start looking at two dimensions, and study some geometries that will be relevant:

\begin{enumerate}[i)]
 \item $\df s^2 = \frac{1}{|r|} \df t^2 - |r| \df r^2$
 \item $\df s^2 = -|r| \df r^2 + r^2 \df \phi^2$
 \item $\df s^2 = -|r| \df r^2 + (r^2+1) \df \phi^2$
 \item $\df s^2 = |r| \df t^2 - |r| \df r^2$
\end{enumerate}

With $r\in (-\infty, \infty)$, $t\in (-\infty, \infty)$, $\phi\in [-\pi, \pi)$. Each of these geometries has a curvature divergence at $r=0$. If we look at the region $r>0$, the (i) and (ii) geometries are like the ($t,r$), and  ($r,\phi$) part of the Schwarzschild black hole near the origin. Geometry (iii) would correspond to the ($r, \phi$) part of a geometry somewhat similar to Schwarzschild near the origin, but where the curvature divergence happens at a 2-sphere, instead of at a point. We will see an example of a geometry like this in Chapter \ref{SolChap}. Geometry (iv) will appear as the ($t,r$) part of a particular case of $d$-dimensional black hole in Chapter \ref{DDim}. The region $r<0$ of these geometries would correspond to a naive attempt to extend them beyond the divergence.

\begin{figure}[h!]
\centering
 \includegraphics[width=.6\linewidth]{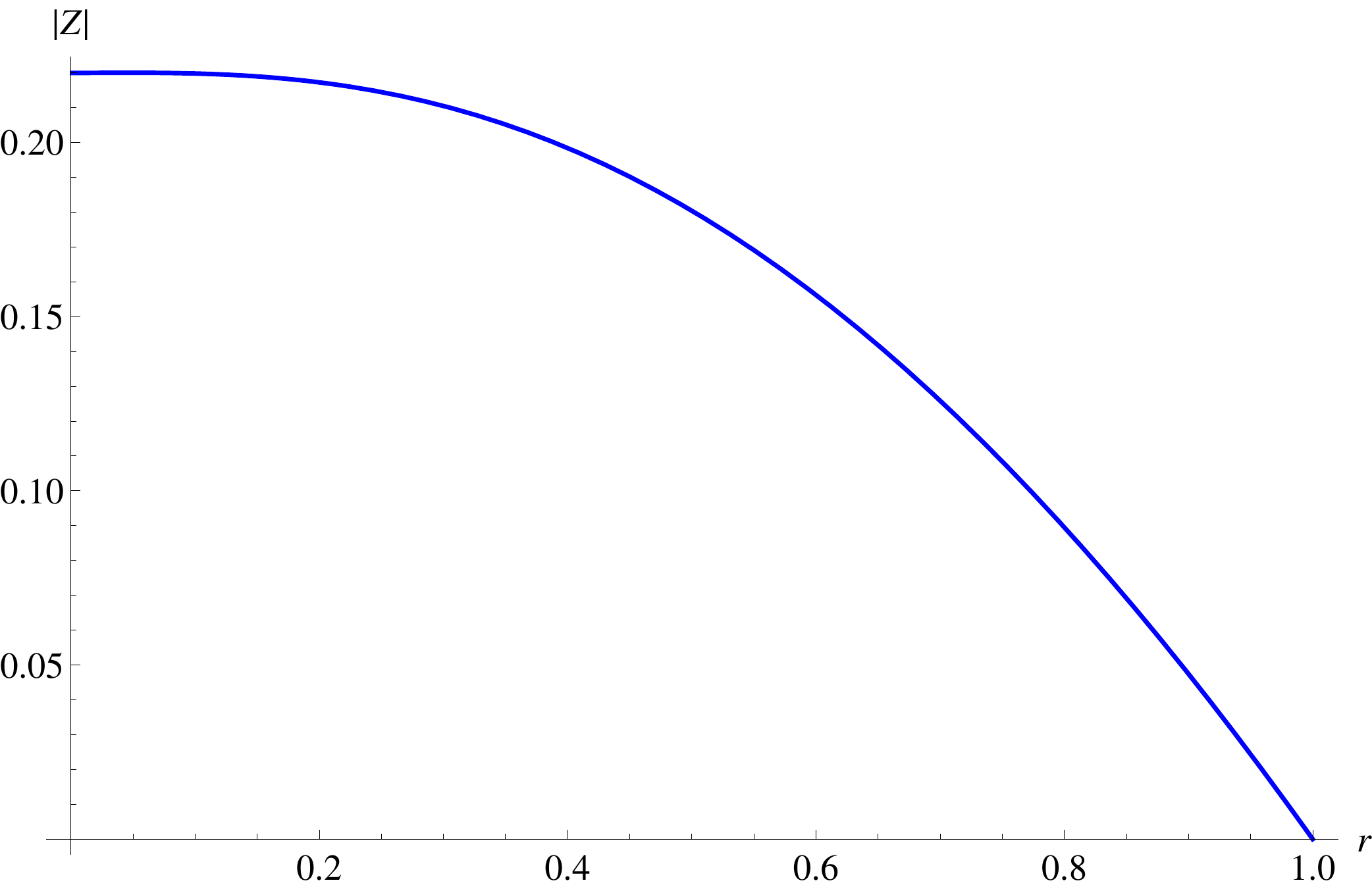}
 \caption{Modulus of the deviation vector (Jacobi field) $Z^\alpha$, for an uniparametric family of geodesics in geometry (ii) with common origin $r_0=1$, $\phi_0=0$ and different angular momentum, as calculated through the geodesic deviation equation (eq. \ref{1eq:GeoDevEq}).}\label{1fig:(ii)deviation}
\end{figure}

\begin{figure}[h!]
 \centering
 \includegraphics[width=.9\linewidth]{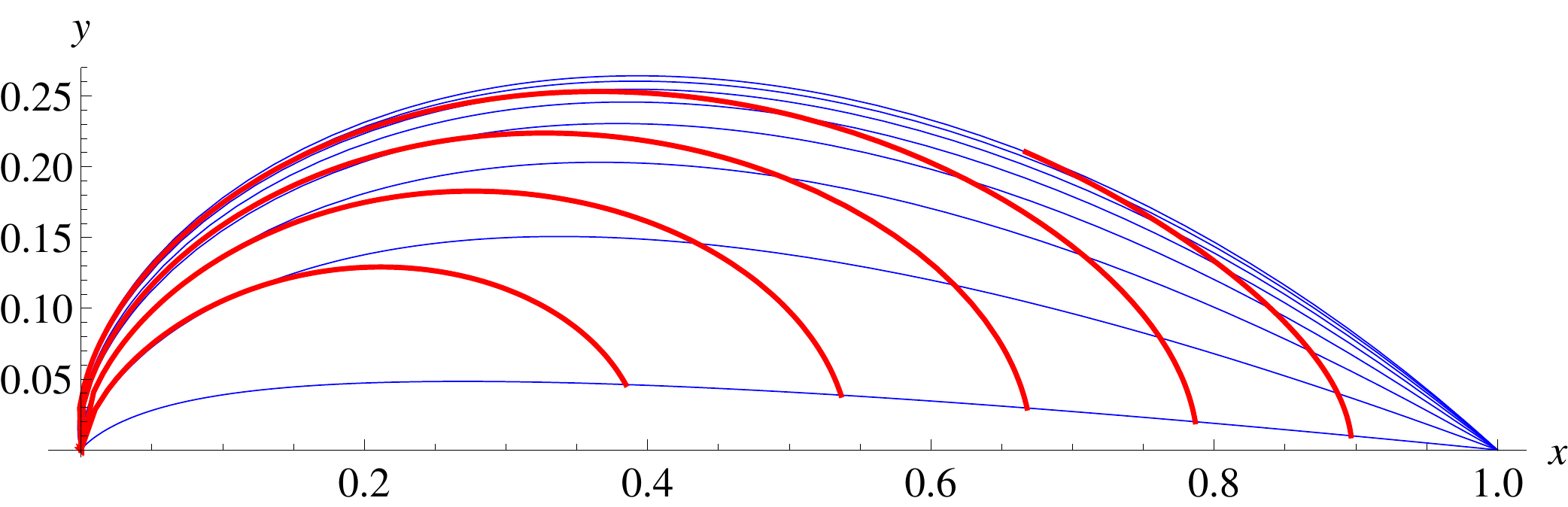}
 \caption{In blue, different geodesics of an uniparametric family in geometry (ii) with common origin $r_0=1$, $\phi_0=0$ and different angular momentum, shown in polar coordinates $(x,y)=(r \cos \phi, r \sin \phi)$. In red, curves of constant affine parameter $\lambda$. We can see how the geodesic with higher angular momentum reach the origin faster than the ones with less angular momentum. One would expect the deviation vector of this uniparametric family of geodesics to vanish at the endpoints $r=1, \phi=0$ and $r=0$, but actually, the deviation vector is different from $0$ at $r=0$ as a consequence of the divergence of the metric.}\label{1fig:(ii)constantlambda}
\end{figure}

In all these geometries, any time-like geodesic starting at $r>0$ and heading towards the origin must reach $r=0$ in finite proper time, and cannot change direction to $r=\infty$. It is possible to check that, because of the discontinuous nature of the metric, a geodesic might fail to maximize proper time even though there is not a pair of conjugate points. For example, let us take geometry (ii) and think of a congruence of geodesics where each geodesic parts from ($r=1$, $\phi=0$) with different angular velocity towards the origin. Each of the geodesics will reach the point $r=0$, but only one of them is the one that actually maximizes the proper time between the starting point and $r=0$. However, the deviation vector shows no conjugate point (see fig. \ref{1fig:(ii)deviation}). This seems counterintuitive, because all the geodesics reach the same point: How can the deviation vector be anything but $0$? The answer is that the deviation vector compares geodesics at $\lambda$ constant, but the geodesics reach the point $r=0$ with different values of $\lambda$ (see fig. \ref{1fig:(ii)constantlambda}). This is impossible if the metric is not discontinuous.

\begin{figure}[h!]
\centering
\begin{tabular}{cc}
 \includegraphics[width=.5\linewidth]{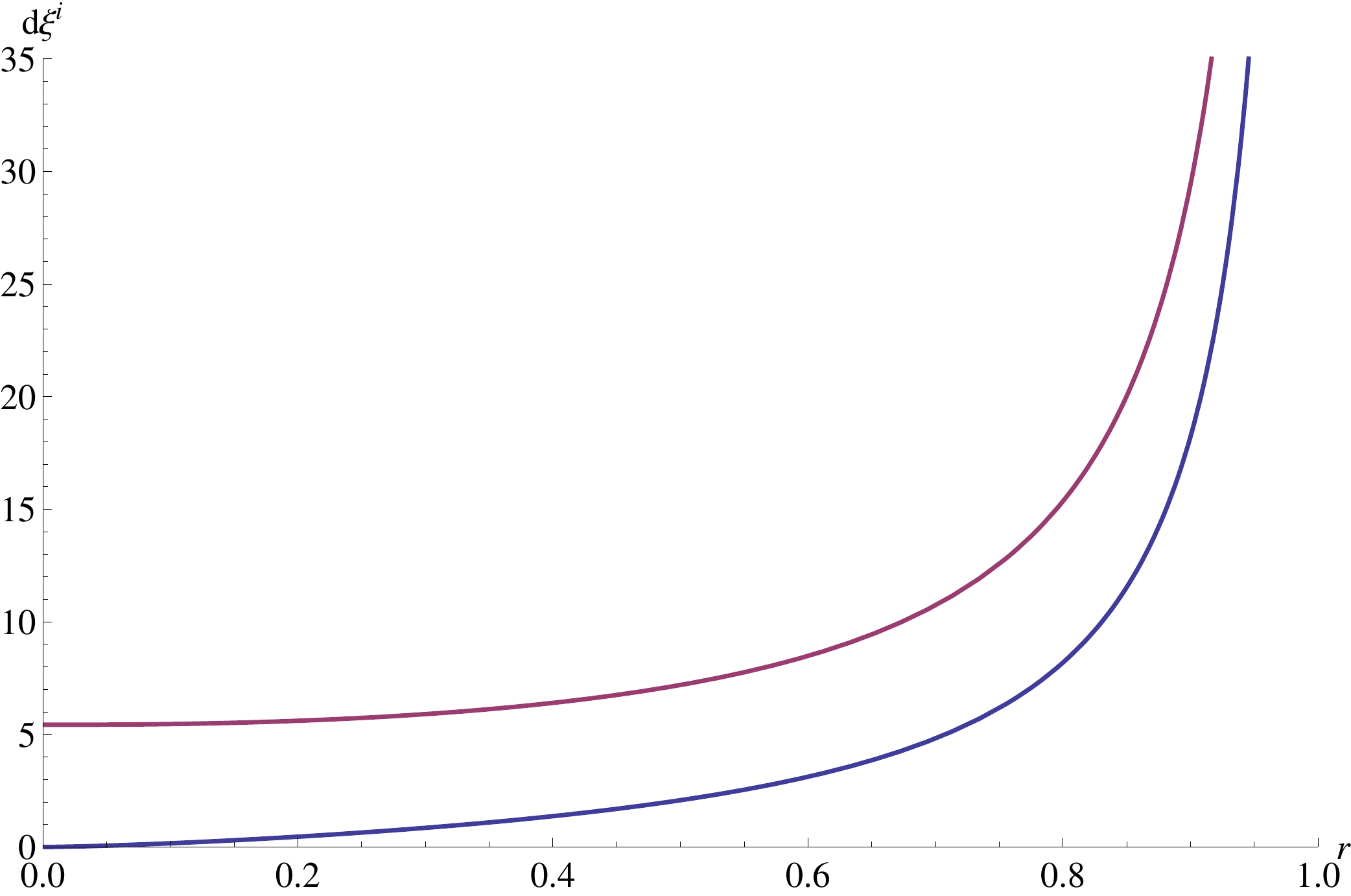} & 
 \includegraphics[width=.5\linewidth]{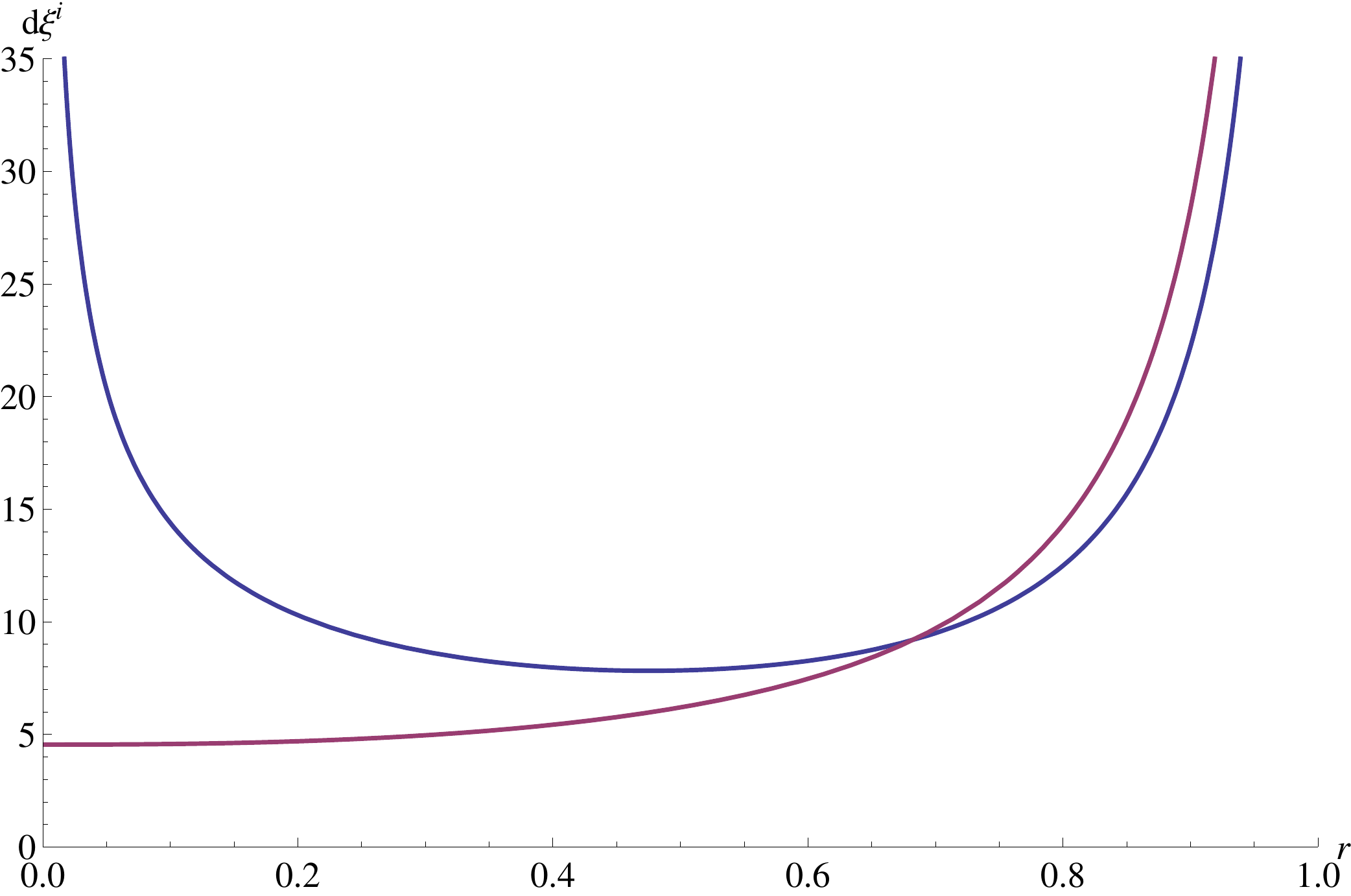} \\
 (i) & (ii) \\
 \includegraphics[width=.5\linewidth]{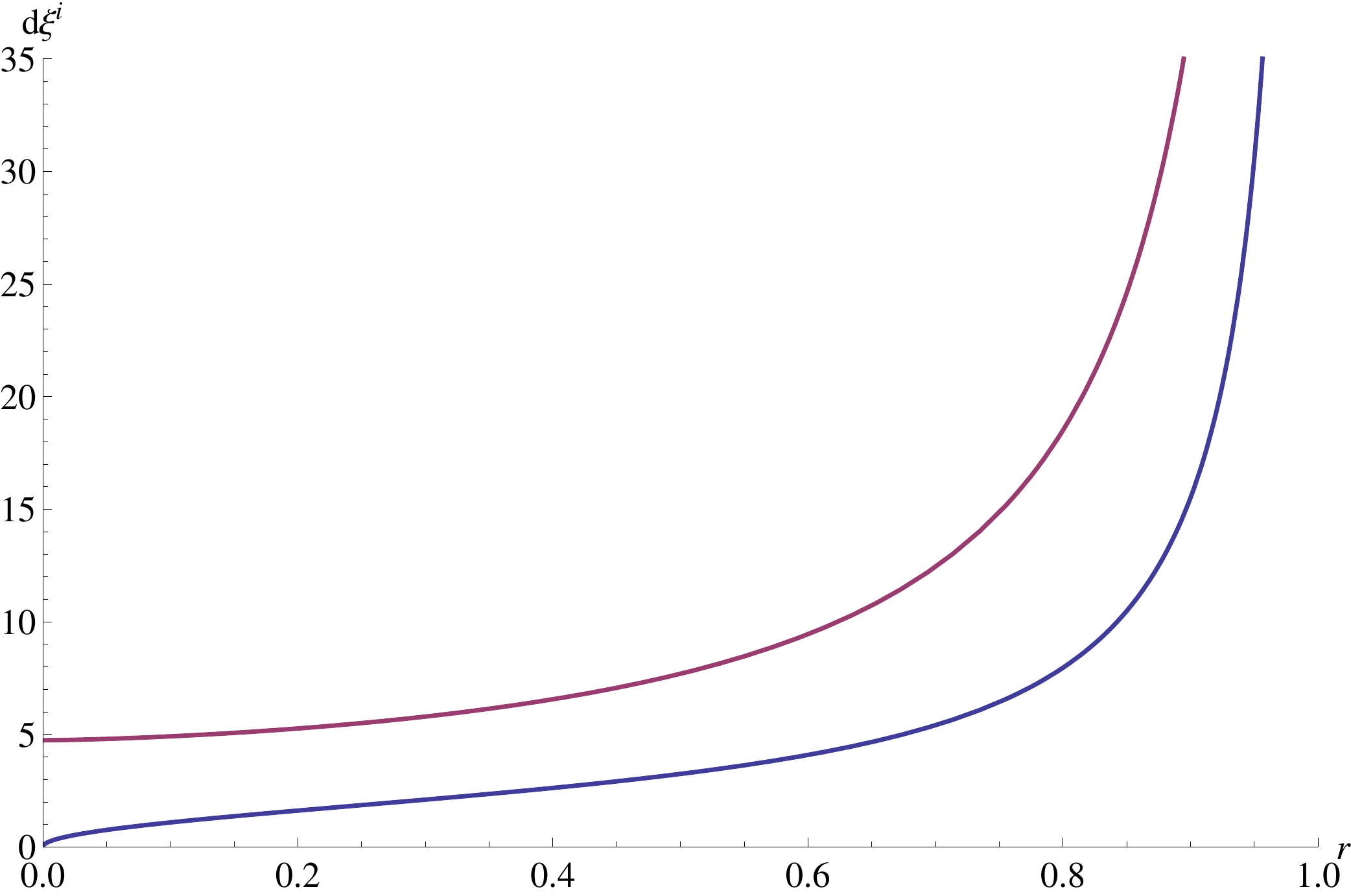}&
 \includegraphics[width=.5\linewidth]{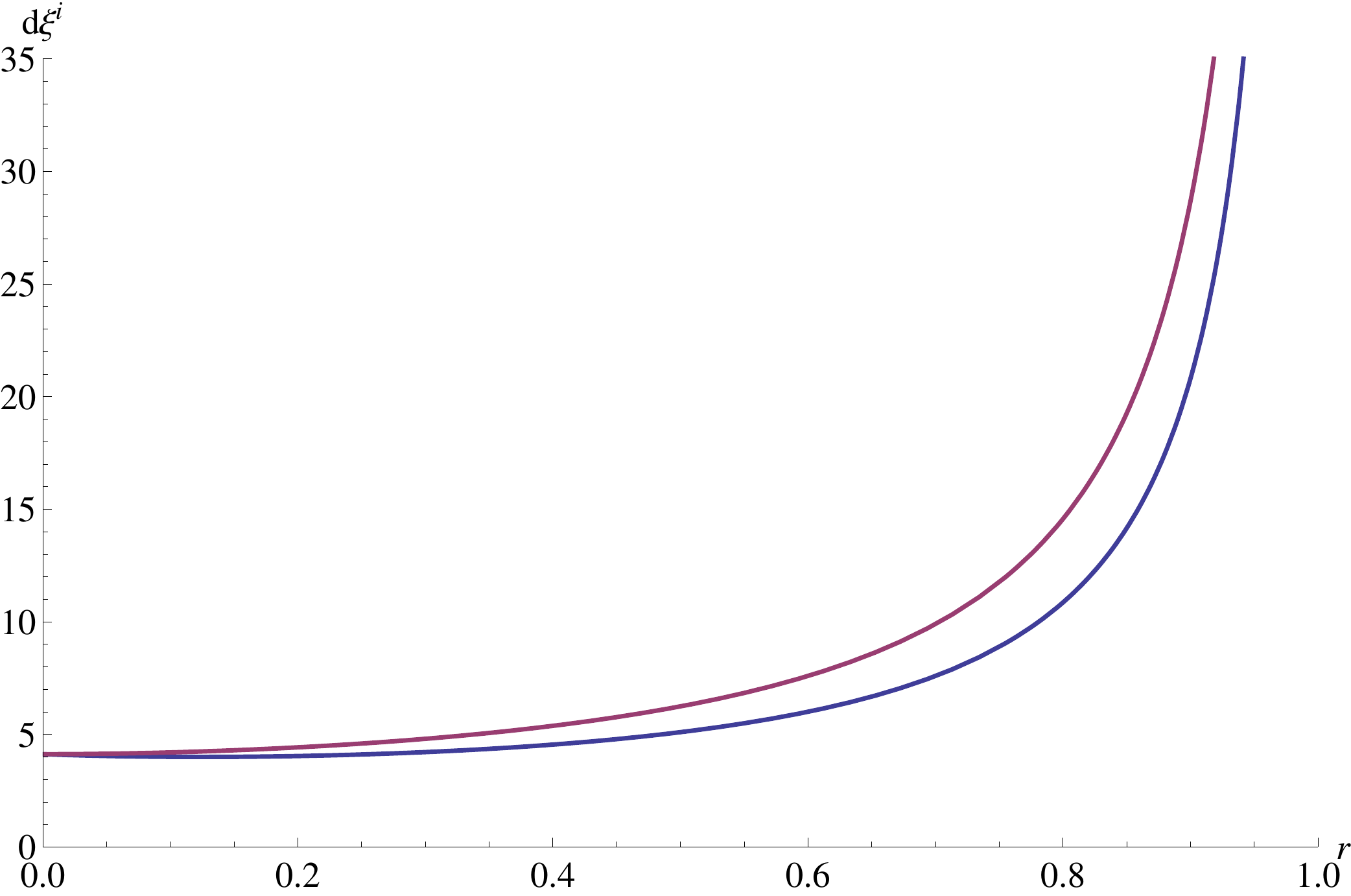} \\
 (iii) & (iv)
\end{tabular}

 \caption{Components of the $\df \xi$ form in the (i), (ii), (iii) and (iv) geometries along a geodesic corresponding to a family of geodesics with common origin $r=1$, $t \backslash \phi = 0$. In blue, the $r$ component, and in violet the $t \backslash \phi$ component. We can see that the $\df \xi$ components only diverge at $r=0$ in geometry (ii), and thus, the Pfaff equation is ill-defined there.}\label{1fig:dxicomponents}
\end{figure}

In certain sense, the 1-form $\df \xi$ is like an inverse of the deviation vector: when its modulus grows, the geodesics converge. In this sense, it agrees with our intuition, and in geometry (ii), the components of $\df \xi$ become infinite as we approach $r=0$. $\df \xi$ is not well defined at that point because in this geometry each of the geodesics reach $r=0$ with different value of $\xi$; and consequently, the geodesics cannot be extended. In figure \ref{1fig:dxicomponents} we represent the value of the components of the 1-form $\df \xi$ for a family of geodesics that reach the discontinuity $r=0$ in the geometries (i)-(ii)-(iii)-(iv).

\begin{figure}[h!]
\begin{tabular}{cc}
 \includegraphics[width=.5\linewidth]{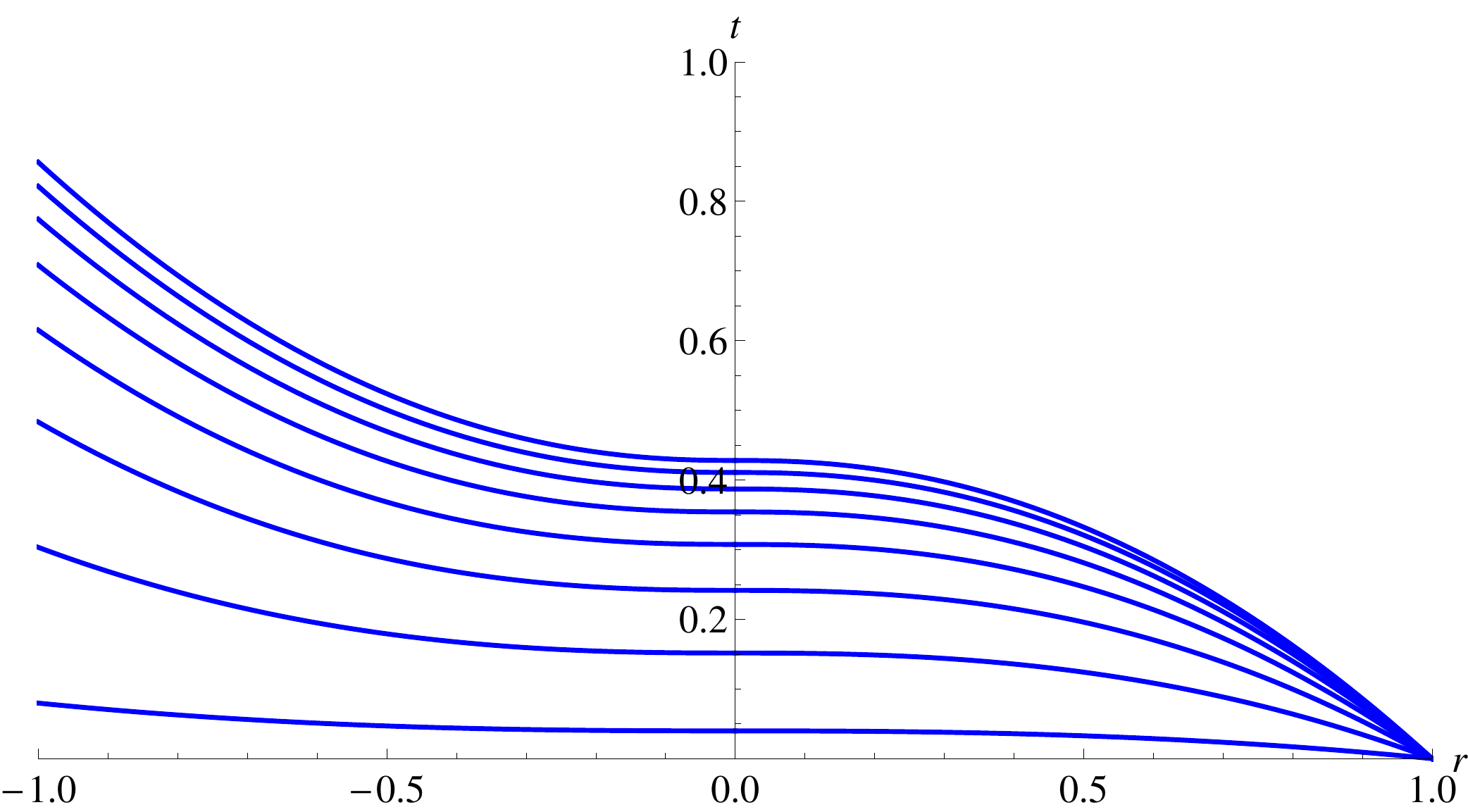} & 
 \includegraphics[width=.5\linewidth]{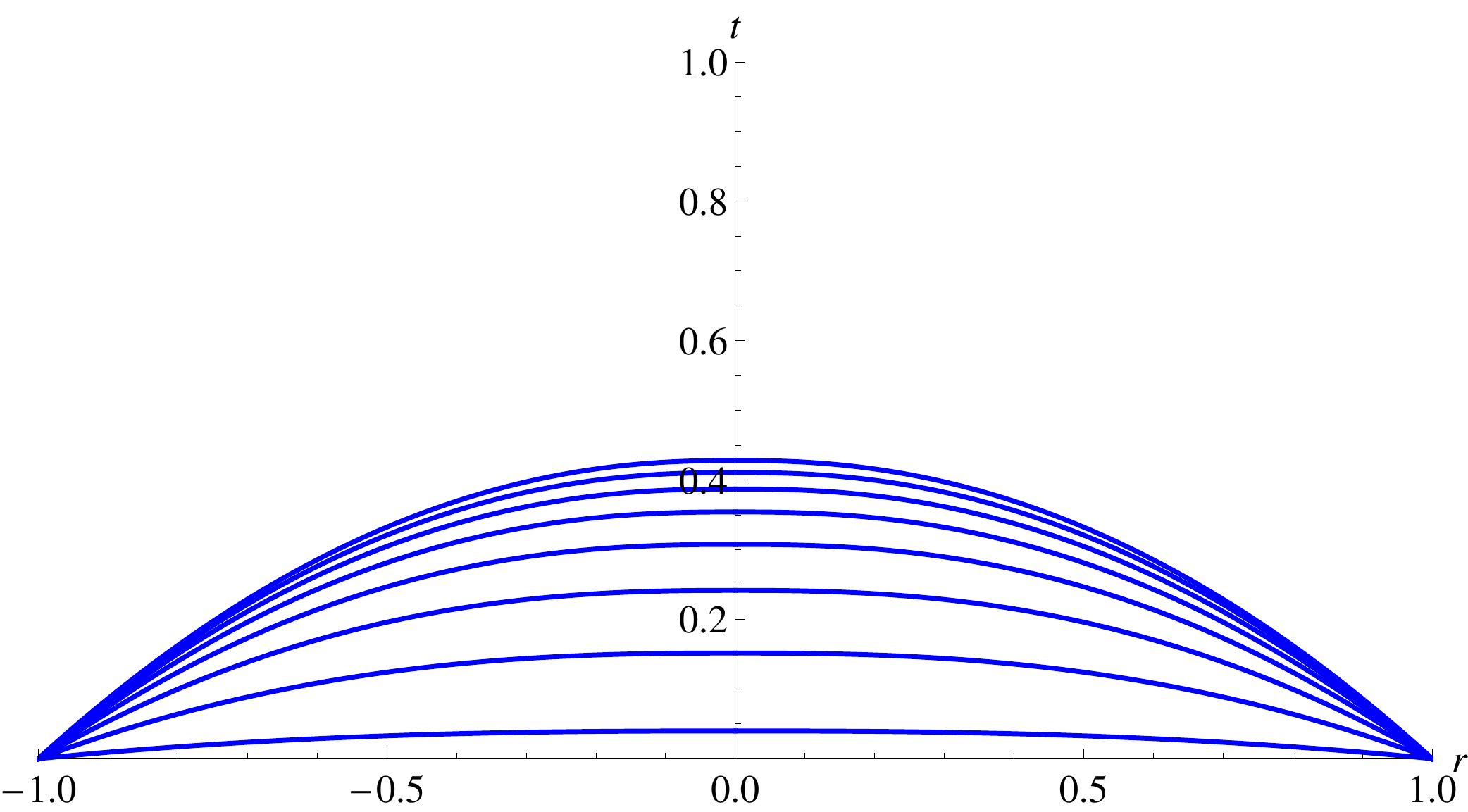}
\end{tabular}
 \caption{Two different ways of extending the geodesics with origin $r=1, t=0$ that reach $r=0$ in the (i) geometry. On the left one, the geodesic is matched with another geodesic with the same conserved energy as the incoming one. On the right, the geodesic is matched with another geodesic with \emph{minus} the conserved energy as the incoming geodesic. In both cases, $\df \xi$ matches correctly from both sides. Since the geodesic equation is singular at $r=0$, it does not preclude us from joining the geodesics either way. However, only the curves in the left diagram extremize the proper time between points at each side of $r=0$.}\label{1fig:(i)matching}
\end{figure}

For geometries (i), (iii) and (iv), it is possible to extend the geodesics beyond the discontinuity, but we have yet to know if the extension is unique. Perhaps it would be possible to find different ways of matching geodesics on one side to the other. See fig. \ref{1fig:(i)matching} for a naive attempt at matching different geodesics through $r=0$ in geometry (i). However, in geometry (ii) it will be impossible to extend the geodesics.

\FloatBarrier
\subsubsection{Extremizing the Proper Time}

Let us consider geometry (i) and study the extremization of proper time for both cases of figure \ref{1fig:(i)matching}. Obtaining the geodesics for this geometry is easy, since there is time translation symmetry, generated by the killing vector $\tau^\alpha = ( \partial/ \partial t  )^\alpha$. Let us think of a geodesic described by its unitary tangent vector $u$, then the ``energy'' $E=g_{\mu \nu} u^\mu \tau^\nu$ is conserved along the geodesic:

\begin{equation}
 \Lie_u E= \tau^\mu g_{\mu \nu} u^\alpha \nabla_\alpha u^\nu + u^\nu u^\alpha g_{\mu \nu} \nabla_\alpha \tau^\mu=0
\end{equation}

The first term vanishes because $u$ satisfies the geodesic equation, and the second term vanishes because $\tau$ is a killing vector and therefore $\nabla_\alpha \tau_\nu$ is antisymmetric in $(\alpha, \nu)$. Consequently, $E$ is conserved along the geodesic as long as the geodesic equation holds. One might wonder if crossing $r=0$, where the coefficients of the geodesic equation are not $\mathcal{C}^1$ might allow for a sudden change in the energy of the geodesics. 
We will see that in that case, the curve does no longer extremize the proper time. Let us consider a curve $\gamma(\lambda)$ that connects two points $a$ and $b$ of our space-time, one at each side of $r=0$. This curve is going to be geodesic from $a$ to $r=0$, and from $r=0$ to $b$. Now let us consider a uniparametric congruence of curves $\gamma_\xi (\lambda)$ with family parameter $\xi$ such that $\gamma_0(\lambda)= \gamma(\lambda)$. Let the tangent vector of these curves be $k^\alpha = \left. \frac{\partial (\gamma_\xi)^\alpha}{\partial \lambda} \right |_{\xi=\text{const.}}$, then the proper time from $a$ to $b$ along each curve is\footnote{By abuse of notation, I have identified the value of the parameter at the endpoints of the curve $a,b$ with its images $\gamma_\xi(a),\gamma_\xi(b)$.}:

\begin{equation}
 L(\gamma_\xi)=\int_a^b \sqrt{-g_{\alpha \beta} k^\alpha k^\beta} \df \lambda
\end{equation}

If the curve $\gamma(\lambda)$ extremizes the proper time from $a$ to $b$, then the derivative of the proper time with respect to the family parameter must vanish $(\df L / \df \xi)_{\xi =0}=0$. Let $u^\alpha$ be the unitary part of $k^\alpha$ and let $Z^\alpha= \left. \frac{\partial \gamma^\alpha}{\partial \xi} \right |_{\lambda=\text{const.}}$ be the deviation vector of this congruence of curves. The variation of the proper time with respect to the parameter $\xi$ is\footnote{Let us note a couple of things. First we are considering that all the curves in the congruence take values in $\lambda$ in the same range $[a,b]$ (which can be done in general), and therefore, it is not possible to reparametrize every curve in the congruence such that $k$ is unitary. Second, $k$ and $Z$ commute because they are partial derivatives.}:

\begin{IEEEeqnarray}{rCl}
 \frac{\df}{\df \xi} L &=& \int_a^b Z^\mu \nabla_\mu \sqrt{-g_{\alpha \beta} k^\alpha k^\beta} \df \lambda \\
 &=& - \int_a^b \frac{1}{\sqrt{-g_{\alpha \beta} k^\alpha k^\beta}}k^\nu Z^\mu \nabla_\mu k_\nu \df \lambda \nonumber \\
 &=& - \int_a^b \frac{1}{\sqrt{-g_{\alpha \beta} k^\alpha k^\beta}}k^\nu k^\mu \nabla_\mu Z_\nu \df \lambda \nonumber \\
 &=& - \int_a^b \underbrace{k^\mu \nabla_\mu}_{\frac{\partial}{\partial \lambda}} \left ( \frac{k^\nu}{\sqrt{-g_{\alpha \beta} k^\alpha k^\beta}} Z_\nu \right ) \df \lambda + \int_a^b Z^\nu k^\mu \nabla_\mu \left ( \frac{k^\nu}{\sqrt{-g_{\alpha \beta} k^\alpha k^\beta}} \right ) \df \lambda \nonumber
\end{IEEEeqnarray}

The second term vanishes if $u$ satisfies the geodesic equation (except perhaps for $r=0$ which has null measure). The first term is a total derivative that can be evaluated at the boundary. As the integrand might be discontinuous at $r=0$, we can split the integral in two pieces, one from $a$ to $r=0$, and other from $r=0$ to $b$. At $a$ and $b$ the deviation vector vanishes, so we are left with:

\begin{equation}
 \frac{\df}{\df \xi} L = \left . u^\mu Z_\mu \right |_{r\rightarrow 0^+} - \left. u^\mu Z_\mu \right |_{r\rightarrow 0^-}
\end{equation}

The deviation vector and the tangent vector to the geodesic can be written in components as $Z^\mu = Z^t \partial_t + Z^r \partial_r$, $u^\mu = u^t \partial_t + u^r \partial_r$. From the conservation of the energy we can see that the term $\left. u^t Z^t g_{tt}\right |_{r\rightarrow 0^\pm}=Z^t E_\pm$, where $E_\pm$ is the energy of the geodesics at each side of $r=0$. On the other hand, the term  $\left. u^r Z^r g_{rr}\right |_{r\rightarrow 0^\pm}=0$ because $u^r = \sqrt{(1+E^2)/|r|}$ and $g_{rr} =-|r|$. Then we are left with:

\begin{equation}
 \frac{\df}{\df \xi} L = (E_+ - E_-) Z^t\big |_{r=0}
\end{equation}

Only in the case $E_+=E_-$, the geodesic will extremize the proper time. An equivalent construction can be done for geometry (iii), in which the conserved quantity is the ``angular momentum'', $u^\mu (\partial/\partial \phi)^\nu g_{\mu \nu}$, that must be the same at both sides of $r=0$ for the geodesic to extremize proper time.

\subsubsection{Evolution Equation for $\xi$}

The Pfaff equation (eq. \ref{1eq:pfaff}) is a good way to write an equation for a curve, independent of parametrizations or coordinate choices. However, we have just limited ourselves to construct a family of geodesics through the geodesic equation, obtain $\xi$, write the Pfaff equation and check whether it is well defined. In order to do something more useful, we should be able to check when a function $\xi$ gives us a geodesic congruence without the need of using the tangent vector and the geodesic equation. Then that way, we could really check if the geodesics can be extended through $r=0$ in an unique way.

Let us take a time-like congruence, defined by an unitary vector field $u^\mu$. Let us construct a projector $h_{\mu \nu}=g_{\mu \nu} + u_\mu u_\nu$. The covariant derivative of $u$ is: 

\begin{equation}
 \nabla_\mu u_\nu = \underbrace{h_\mu{}^\alpha h_\nu{}^\beta \nabla_\alpha u_\beta}_{B_{\mu \nu}} - u_\mu \underbrace{u^\alpha \nabla_\alpha u_\nu}_{a_\nu}
\end{equation}

In a similar fashion as we did for null geodesic congruence when discussing the Raychaudhuri equation, $B_{\mu \nu}$ can be separated into an antisymmetric part, a traceless symmetric part and a trace,  $B_{\mu \nu}=\omega_{\mu \nu} +\sigma_{\mu \nu} + \frac{1}{d-1} \theta h_{\mu \nu}$. Antisymmetrizing last equation we obtain the exterior derivative of $u$ (with lowered index):

\begin{equation}
 (\df u)_{\mu \nu}=\nabla_{[\mu} u_{\nu]}= \omega_{\mu \nu} - (u\wedge a)_{\mu \nu}
\end{equation}

In two dimensions, or for a congruence orthogonal to a hypersurface, the term $\omega_{\mu \nu}$ vanishes. In those cases, the geodesic equation is equivalent to:

\begin{equation}
 \df u=0\label{1eq:ueq}
\end{equation}

In two dimensions we can construct another vector field, $\eta$ with the hodge star operator $\eta=*u$. This vector is orthogonal to $u$, and it must be unitary because $u$ is unitary. So it must be related to $\xi$ by simply $\eta= \df \xi / \sqrt{(\df \xi)^\alpha (\df \xi)_\alpha}$. Then, eq. \ref{1eq:ueq} will be satisfied if:

\begin{equation}
 *\ \df * \eta = \nabla^\mu \eta_\mu = \frac{1}{\sqrt{-g}}\partial_\mu (\sqrt{-g} g^{\mu \nu} \eta_\nu) = 0 \label{1eq:etaeq}
\end{equation}

This can be seen as the equations of motion of the action:
\begin{equation}
 S=\int\sqrt{(\df \xi)^\alpha (\df \xi)_\alpha}\sqrt{-g}\df^4 x
\end{equation}

\begin{figure}
 \begin{tabular}{cc}
 \includegraphics[width=.5\linewidth]{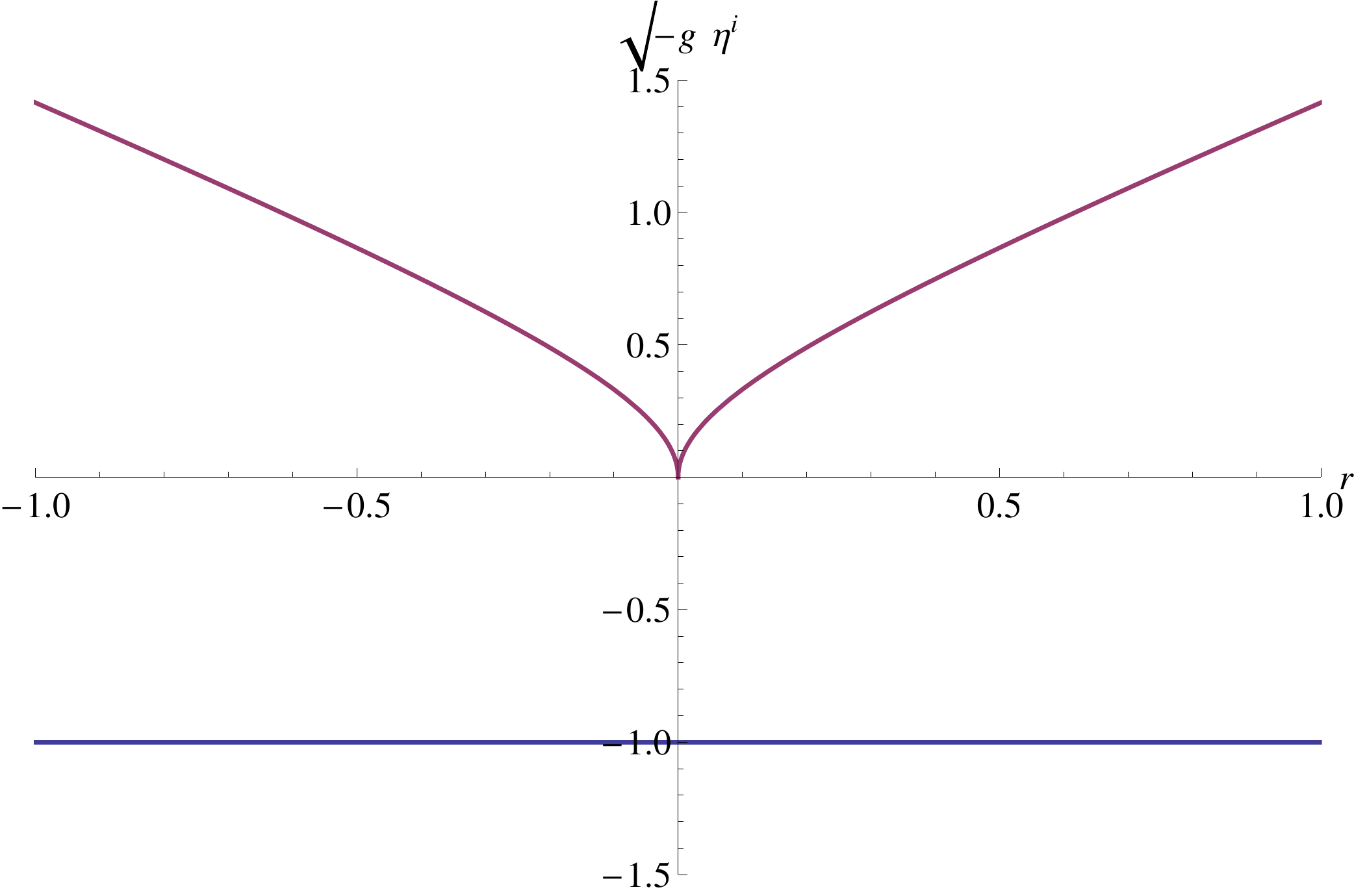} & 
 \includegraphics[width=.5\linewidth]{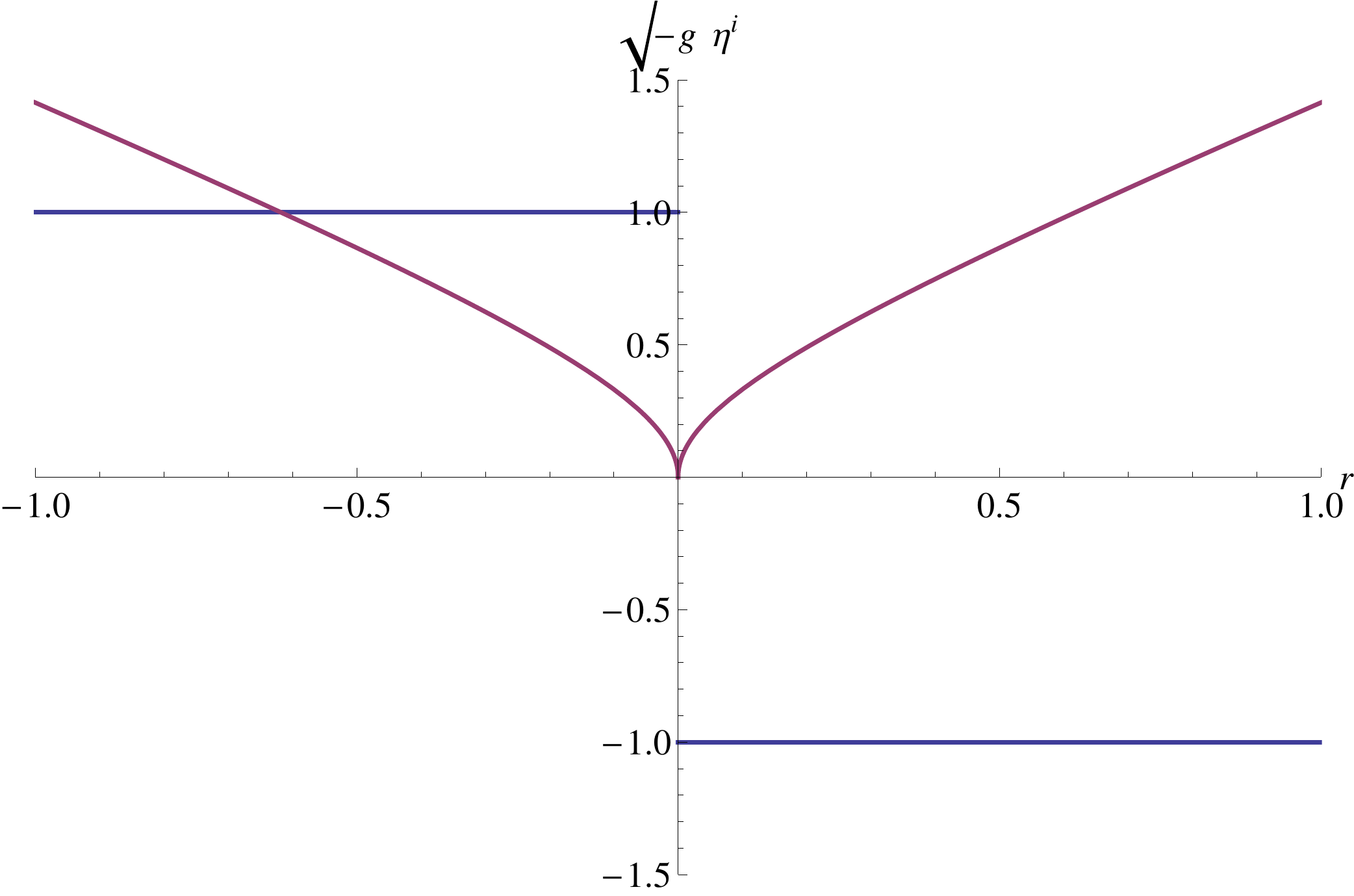}
\end{tabular}
 \caption{Components of $\sqrt{-g} g^{\mu \nu} \eta_\nu$ along a geodesic for the two scenarios depicted in fig. \ref{1fig:(i)matching}, the $r$ component in blue, and the $t$ component in violet. The left scenario is the one that actually solves eq. \ref{1eq:etaeq}. }\label{1fig:(i)eta}
\end{figure}

Now we would like to know which one of the two scenarios of fig. \ref{1fig:(i)matching} satisfies eq. \ref{1eq:etaeq} (although we already have argued that only the one with the same energy on both sides of $r=0$ corresponds to curves that extremize the length). In fig. \ref{1fig:(i)eta} we show the components of $\sqrt{-g} g^{\mu \nu} \eta_\nu$ along a geodesic in geometry (i) for these two scenarios. The relevant quantities in eq. \ref{1eq:etaeq} are $\partial_t \eta^t$ and $\partial_r \eta^r$, and as we can see from the figure,  $\partial_r \eta^r$ is not well defined at $r=0$ (meanwhile $\partial_t \eta^t=0$ at $r=0$, because all geodesics from the congruence have $\eta^t=0$ at $r=0$). This might seem surprising because $\df \xi$ is matched on both sides of the geometry for the two scenarios, but actually the $r$ component of $\df \xi$ (that tends to $0$ as we approach to $r=0$) multiplied by $g_{rr}^{-1}$ (that tends to $\infty$) gives a different limit at $r=0$ depending on each geodesic.

In the case of a congruence in a $d$-dimensional space-time, the congruence would be defined by $(d-1)$ 1-forms. The situation is more complicated as there is a lot of freedom choosing said forms, but we can check that the congruence is made of geodesic curves, if every uniparametric family of the congruence satisfies eq. \ref{1eq:etaeq}.


\FloatBarrier

%

\section{Quadratic Gravity}

In this section I will introduce a modified theory of gravity which received much interest after the first attempts to quantize gravity, adding new terms to the GR Lagrangian that would make the theory renormalizable. This Lagrangian has been understood as an effective approximation to the new gravitational physics at high energies, and might contain information relevant to the internal structure of black holes or the propagation of gravitational waves in the early universe. A reduced version of this Lagrangian is the so called Starobinsky model \cite{Starobinsky:1980te}, which has received wide interest in the context of cosmic inflation. As we will see, the resulting theory suffers from ghost instabilities which render the theory inadequate as a fundamental theory; however the theory can still be used as an effective theory. 

Our interest in this theory comes from its treatment in the Metric-Affine formalism. As we will see in the next chapter, in the Metric-Affine formalism the connection is no longer required to be the Levi-Civita connection of the metric. As a consequence, the resulting equations of motion in this formalism do not have the higher-order derivatives that cause ghost instabilities in the Riemannian formalism. Therefore, the theory in the Metric-Affine formalism represents an interesting playground for modified gravity theories that has not been sufficiently explored. The results already obtained in the literature in the usual (Riemannian) formalism thus will serve as a reference point for comparison. In particular, charged black hole solutions have been found (with some approximation scheme) \cite{Economou:1993va}, and even ``regular'' black hole solutions for some non-linear electrodynamic theory \cite{Matyjasek:2004vr} .

\subsection{Linearised GR}


Quadratic gravity was born out of the non-renormalizability of the perturbative quantization of GR. The starting point of this quantization procedure is the linearised theory of GR. The metric is splitted into a Minkowski background $\eta_{\mu \nu} = \text{diag}(-1,1,1,1)$ plus some perturbation $\gamma_{\mu \nu}$. This perturbation will be gravitational waves from a classical point of view, and gravitons from a quantum point of view: 

\begin{equation}
 g_{\mu \nu} = \eta_{\mu \nu} + \gamma_{\mu \nu} \qquad (g^{-1})^{\mu \nu} = (\eta^{-1})^{\mu \nu}-\gamma^{\mu \nu} + O(\gamma^2) \qquad \gamma^{\mu \nu} \equiv (\eta^{-1})^{\mu \alpha}(\eta^{-1})^{\nu \beta} \gamma_{\alpha \beta}
\end{equation}

With this decomposition, the volume form and the connection take the form:

\begin{IEEEeqnarray}{rCl}
 \sqrt{\det(g)}&=&1+\frac{1}{2} \gamma^\alpha{}_\alpha + O(\gamma^2)\\
 \Gamma^\alpha_{\beta \sigma} &=&  \frac{1}{2} \left \{ \partial_\sigma \gamma_\beta{}^\alpha + \partial_\beta \gamma^\alpha{}_\sigma-\partial^\alpha \gamma_{\beta \sigma} \right \} + O(\gamma^2)
\end{IEEEeqnarray}

Substituting into the Lagrangian we have:

\begin{equation}
 \sqrt{-g} R = \left ( 1+\frac{1}{2} \gamma^\delta{}_\delta \right )(\eta^{\lambda \sigma} - \gamma^{\lambda \sigma}) (\partial_\alpha \Gamma^\alpha_{\lambda \sigma} - \partial_\lambda \Gamma^\alpha_{\alpha \sigma} + \Gamma^\alpha_{\alpha \beta} \Gamma^\beta_{\lambda \sigma} - \Gamma^\alpha_{\lambda \beta} \Gamma^\beta_{\alpha \sigma} )
\end{equation}

Terms of the type $\eta^{\lambda \sigma} \partial_\alpha \Gamma^\alpha_{\lambda \sigma}$ give terms which are first order on $\gamma$ but are a total derivative inside the lagrangian and can be rewritten as surface terms that do not contribute to the action. The second-order terms are:

%

\begin{equation}
 \Lagr_E(\gamma_{\mu \nu}) = - \frac{1}{4} \gamma^\alpha{}_\alpha \Box \gamma^\beta{}_\beta + \frac{1}{4} \gamma^{\alpha \beta} \Box \gamma_{\alpha \beta} + \frac{1}{2} \gamma^\alpha{}_\alpha \partial_\lambda \partial_\sigma \gamma^{\lambda \sigma}-\frac{1}{2} \gamma^{\alpha \sigma} \partial_\sigma \partial_\beta \gamma^\beta{}_\alpha
\end{equation}
where $\Box \equiv \partial^\alpha \partial_\alpha$. This is the linearised Einstein Lagrangian that describes a massless graviton. This theory is non-renormalizable (\cite{Goroff:1985th}, \cite{vandeVen:1991gw}). Naively, GR has a dimensional coupling constant which would make loop diagrams at each loop level have a higher degree of divergence, each of them needing its own counterterm, i.e., the theory would need an infinite number of counterterms (unlike renormalizable theories, in which the cancellation of the divergences at each order of the perturbation theory can be done with a finite number of counterterms). If quadratic terms are added to the gravity action, the theory might be renormalizable, because the behaviour of the propagator of the graviton for high momenta would be dominated by this quadratic terms (as $q^{-4}$, with $q$ being the momentum of the graviton), and as a consequence, power-counting shows that all divergences are of degree four or less \cite{Stelle:1976gc}.

\subsection{Fourth Order Derivatives and Ghosts in Quadratic Gravity}\label{2subsec:ghosts}

The Lagrangian with quadratic terms in the curvature looks like:

\begin{equation}
 S= \frac{1}{16\pi l_P^2} \int \left \{ R + a R^2 + b R^{\alpha \beta} R_{\alpha \beta} \right \} \sqrt{-g} \df^4  x + S_m
\end{equation}

Where $a$ and $b$ are two constants with dimensions of length squared (which makes $a/8\pi l_P^2$ dimensionless). We have not considered the term $R^{\mu \nu \lambda \sigma} R_{\mu \nu \lambda \sigma}$ because it can be absorbed using the Gauss-Bonnet term $R^2-4R_{\mu \nu} R^{\mu \nu} + R^{\mu \nu \lambda \sigma} R_{\mu \nu \lambda \sigma}$, whose integral is a topological invariant in 4 dimensions. Let us study its equations of motion:

\begin{IEEEeqnarray}{l}
 (\alpha -2\beta) \nabla_\mu \nabla_\nu R - \alpha \nabla^\lambda \nabla_\lambda R_{\mu \nu} - (\frac{1}{2} \alpha - 2\beta) g_{\mu \nu} \nabla^\lambda \nabla_\lambda R + 2\alpha R^{\lambda \sigma} R_{\mu \lambda \nu \sigma} \nonumber \\
 -  2\beta R R_{\mu \nu} - \frac{1}{2} g_{\mu \nu} (\alpha R^{\lambda \sigma} R_{\lambda \sigma} - \beta R^2) + R_{\mu \nu} - \frac{1}{2} R g_{\mu \nu} = 8 \pi l_P^2 T_{\mu \nu}
\end{IEEEeqnarray}

These equations of motion contain terms that with four derivatives of the metric tensor. For example, the terms of the type $\nabla_\mu \nabla_\nu R$ have two derivatives of the scalar curvature, which itself contains two derivatives of the metric tensor. Let us study the linearised theory. The expansion of $R^2$ and $R^{\alpha \beta} R_{\alpha \beta}$ is:


\begin{IEEEeqnarray}{rCl}
 R^2 &=& (\partial_\alpha \partial_\beta \gamma^{\alpha \beta} - \Box \gamma^\lambda{}_\lambda)^2+O(\gamma^3) \\
 R^{\alpha \beta} R_{\alpha \beta} &=& -\frac{1}{2} (\partial_\alpha \partial_\lambda \gamma^\lambda{}_\beta) (\partial^\alpha \partial_\sigma \gamma^{\sigma \beta} ) + \frac{1}{2} ( \partial_\beta \partial_\lambda \gamma^{\lambda \beta}) (\partial_\alpha \partial_\sigma \gamma^{\sigma \alpha} ) - \frac{1}{2} (\Box \gamma^\alpha{}_\alpha)(\partial_\beta \partial_\lambda \gamma^{\lambda \beta}) \nonumber \\
 && + \frac{1}{4} (\Box \gamma^{\alpha \beta})(\Box \gamma_{\alpha \beta} ) + \frac{1}{4} (\Box \gamma^\alpha{}_\alpha) (\Box \gamma^\beta{}_\beta) + O(\gamma^3)
\end{IEEEeqnarray}

With this, the linearised action looks like:

\begin{IEEEeqnarray}{rCl}
 S= \frac{1}{16\pi l_P^2} \int && \Big \{   \Lagr_E(\gamma_{\mu \nu})+ a(\partial_\alpha \partial \beta \gamma^{\alpha \beta} - \Box \gamma^\lambda{}_\lambda)^2\\
 && + b \Big ( -\frac{1}{2} (\partial_\alpha \partial_\lambda \gamma^\lambda{}_\beta) (\partial^\alpha \partial_\sigma \gamma^{\sigma \beta} ) + \frac{1}{2} ( \partial_\beta \partial_\lambda \gamma^{\lambda \beta}) (\partial_\alpha \partial_\sigma \gamma^{\sigma \alpha} ) \nonumber \\
 &&- \frac{1}{2} (\Box \gamma^\alpha{}_\alpha)(\partial_\beta \partial_\lambda \gamma^{\lambda \beta}) + \frac{1}{4} (\Box \gamma^{\alpha \beta})(\Box \gamma_{\alpha \beta} ) + \frac{1}{4} (\Box \gamma^\alpha{}_\alpha) (\Box \gamma^\beta{}_\beta) \Big ) \nonumber \\
 && + 8\pi l_P^2 T^{\mu \nu} \gamma_{\mu \nu} \Big \} \df^4 x \nonumber
\end{IEEEeqnarray}


\begin{mdframed}
 \textbf{Ghosts in Theories with Higher Derivatives}
 
 Theories with higher derivative terms usually give raise to ghost-like instabilities. A very simple example (taken from \cite{Creminelli:2005qk}) of this would be a theory of a massless scalar field with a fourth order derivative term. We will see that this theory can be rewritten in terms of two scalar fields, one of them with a kinetic term of the wrong sign. This kinetic term would make the Hamiltonian unbounded from below, and it would be possible to excite both scalar fields without bound. Let us write the Lagrangian of such a theory:
 
\begin{equation}
 \Lagr = -\frac{1}{2} \nabla^\mu \phi \nabla_\mu \phi + \frac{s}{2 \Lambda^2} (\Box \phi)^2 - V_\text{int}(\phi)
\end{equation}

In this Lagrangian $\Lambda$ is some energy scale and $V_\text{int}(\phi)$ is a self-interaction term. The second term contains the D'Alambertian operator $\Box$ squared, and therefore, four derivatives. We have introduced a sign $s=\pm 1$ in order to check whether the sign of this term with four derivatives affect the existence of ghost instabilities or not.  We can rewrite the same theory using an auxiliary scalar field $\chi$:

\begin{equation}
 \Lagr=-\frac{1}{2}  \nabla^\mu \phi \nabla_\mu \phi - s \nabla_\mu \chi \nabla^\mu \phi - \frac{1}{2} s \Lambda^2 \chi^2 - V_\text{int} (\phi)
\end{equation}

Solving the Euler-Lagrange equations for the $\chi$ field  gives the following relation $\chi = -\frac{1}{\Lambda^2} \Box \phi$, and it can be seen that both Lagrangians are equivalent. Now we can diagonalize this new Lagrangian by substituting $\phi = \tilde{\phi} - s \chi$:

\begin{equation}
 \Lagr = -\frac{1}{2}  \nabla^\mu \tilde{\phi} \nabla_\mu \tilde{\phi} +\frac{1}{2}  \nabla_\mu \chi \nabla^\mu \chi - \frac{1}{2} s \Lambda^2 \chi^2 - V_\text{int} (\tilde{\phi},\chi)
\end{equation}

As we can see, this Lagrangian describes two scalar fields $\tilde{\phi}$ and $\chi$, but $\chi$ has a kinetic term of the wrong sign, no matter the sign of the fourth order derivative term in the original Lagrangian. The scalar field $\chi$ has mass $\Lambda$, and if the sign $s$ is negative, it will be tachyonic. $\chi$ and $\tilde{\phi}$ are coupled by the self-interaction term $V_\text{int}(\tilde{\phi}, \chi)$, and so, above energies of the mass $\Lambda$, these new degrees of freedom will be excited, creating infinitely many particles of the fields $\tilde{\phi}$ and $\chi$. Nevertheless, these excitations will only happen above the energy $\Lambda$ and it is possible to work with the theory  as an effective field theory up to that energy. In order to consider higher energies, we would need to complete the theory in the UV.
 
\end{mdframed}
\bigskip

Linearised quadratic gravity has a ghost field \cite{Stelle:1977ry}. This can be seen from this equivalent Lagrangian: 

\begin{IEEEeqnarray}{rCl}
 S = \frac{1}{16 \pi l_P^2} \int \Big \{ && \Lagr_E (\gamma_{\mu \nu} ) - \frac{1}{2} \Sigma_{\alpha \beta} \Box \gamma^{\alpha \beta} + \frac{1}{2} \Sigma^\alpha{}_\alpha \Box \gamma^\beta{}_\beta - \frac{1}{2} \Sigma^\sigma{}_\sigma \partial_\alpha \partial_\beta \gamma^{\alpha \beta} \IEEEeqnarraynumspace \label{1eq:LinQuaG} \\
 && -\frac{1}{2} \Sigma^{\alpha \beta} \partial_\alpha \partial_\beta \gamma^\sigma{}_\sigma + \frac{1}{2} \Sigma^{\alpha \beta} (\partial_\beta \partial_\sigma \gamma^\sigma{}_\alpha+\partial_\alpha \partial_\sigma \gamma^\sigma{}_\beta) \nonumber \\
 &&+\frac{1}{4b} \Sigma_{\alpha \beta} \Sigma^{\alpha \beta} - \frac{a}{4b(4a+b)} (\Sigma^\alpha{}_\alpha)^2 + 8\pi l_P^2 T^{\mu \nu} \gamma_{\mu \nu} \Big \} \df^4 x \nonumber
\end{IEEEeqnarray}

The Euler-Lagrange equations for $\Sigma$ give:

\begin{IEEEeqnarray}{rCl}
 \Sigma_{\mu \nu} &=& b \left (\Box \gamma_{\mu \nu} - g_{\mu \nu} \Box \gamma^\alpha{}_\alpha +\partial_\mu \partial_\nu \gamma^\alpha{}_\alpha +g_{\mu \nu} \partial_\alpha \partial_\beta \gamma^{\alpha \beta} -\partial_\mu \partial_\alpha \gamma^\alpha{}_\nu -\partial_\nu \partial_\alpha \gamma^\alpha{}_\mu \right ) \nonumber \\
 && + 2a g_{\mu \nu} \left (\partial_\alpha \partial_\beta \gamma^{\alpha \beta} - \Box \gamma^\alpha{}_\alpha \right )
\end{IEEEeqnarray}

Substituting this into the eqs. of motion obtained from \ref{1eq:LinQuaG}, we can see that both Lagrangians give the same theory. The kinetic term in this Lagrangian is not diagonal, but making the change $\gamma_{\mu \nu} \rightarrow \phi_{\mu \nu} + \Sigma_{\mu \nu}$ we obtain:

\begin{IEEEeqnarray}{rCl}
 S = \frac{1}{16 \pi l_P^2} \int && \Big \{ \Lagr_E(\phi_{\mu \nu})-\Lagr_E(\Sigma_{\mu \nu}) +\frac{1}{4b} \Sigma_{\alpha \beta} \Sigma^{\alpha \beta} - \frac{a}{4b(4a+b)} (\Sigma^\alpha{}_\alpha)^2 \nonumber \\
 &&+ 8\pi l_P^2 T^{\mu \nu} (\phi_{\mu \nu} + \Sigma_{\mu \nu})\Big \} \df x^4
\end{IEEEeqnarray}

This action describes a massless graviton with its kinetic term $\Lagr_E(\phi_{\mu \nu})$, but also another graviton with kinetic term $(-\Lagr_E(\Sigma_{\mu \nu}))$ having the wrong sign, and with a mass term. This mass term does not correspond to a spin-2 field, but the field $\Sigma$ can be further separated into a pure spin-two and a scalar part. The theory thus contains ghost instabilities, although it can be useful to study possible modifications to GR in as effective field in certain curvature regime. However, these instabilities would make the theory to break down when the field is strong, in particular in the moment of formation of curvature divergences, and therefore might not be appropriate to study the problem of singularities.

\cleardoublepage
\chapter{Metric-Affine Gravity}\label{M-AChap}

In this chapter, I will present the framework in which I am going to study new extensions to GR, the Metric-Affine formalism. For that purpose, first I will study in more detail the concept of covariant derivative, and how it is conceptually separated from the metric structure. Then I will present the Metric-Affine formalism and study how to derive the equations of motion for a variety of Lagrangians. Finally I will conclude motivating why it is a good idea to study physics in this framework, and I will provide an example of a physical system, the Bravais crystal, that can also be described in the Metric-Affine formalism.

\section{Connections and Curvature}

In order to describe Nature, it is common to compare the value of a scalar magnitude between two different points in space and time. If this comparison is done between infinitesimally close points along some direction, we can obtain a rate of change of this scalar magnitude, or as it is usually known, the derivative of that magnitude along that direction. 

However, what is a straightforward operation for scalar magnitudes, it is not so trivial for vectorial ones. A vector is defined in the tangent space of a given point. It is not possible to compare two vectors in two different points directly, because the tangent spaces of each vector are different. In order to compare them, we need to take both vectors to the same vectorial space, usually taking one of them and transporting it to the tangent space of the other. Once both vectors are in the same space, it is possible to obtain the difference of the two, which is another vector. This will also be the case if we want to compare tensors of any other type, which will need to be transported to the same tangent space. 

For a differentiable manifold, there is a way to transport vectors from one point to another, using a diffeomorphism given by a vector field $v$ whose flow lines connect one point to the other. This diffeomorphism also transports the tangent space at each point, and gives rise to a derivative called the \emph{Lie derivative} $\Lie_v$. However, this derivative depends on $v$ as a vector field. In other words, we need to know $v$ in a neighbourhood of the point we are taking the derivative, when a directional derivative should only depend on the direction of $v$ at that point. Therefore, the Lie derivative is not a satisfactory directional derivative. 

To transport the tangent space from one point to another infinitesimally close in a way that it only depends on the direction (which we will call \emph{parallel transport}), we will need to provide an additional structure that tells us how to do that. This is known as the \emph{affine structure}, and will be encoded into an object called \emph{connection}. The connection allows us to specify how to transport the basis vectors, as there is no rule that tells us how to do it. In general, if we transport the basis vectors from one point to other, the transported basis vectors will be different from the basis vectors in the transported point. Therefore, to transport any other vector, we can write it as a linear combination of the basis vectors in the original point, and the transported vector will just be the same linear combination of the transported basis vectors.
This way of transporting vectors will give raise to another derivative along some direction $v$, which we will call \emph{covariant derivative} $\nabla_v$.

%

This strikes against our intuition: It seems that in flat euclidean space, there is already a reasonable way of transporting vectors. Namely, one would naturally consider a vector to be parallel transported if its components in a Cartesian basis do not change along its path. We have to realize two things: First one is that if we transport Cartesian basis vectors from one point to another, the transported basis vectors will coincide with the basis vectors in the transported point. However, this will not be true for any other basis vector fields, in general. If we work in any other basis, we will still need a connection to describe the parallel transport. In other words, we still need a connection, it just happens to vanish for a Cartesian basis. The second thing is that the Cartesian basis is orthogonal and unitary, and to define such properties we have to make use of the metric. This means that the metric structure of the space is inducing an affine structure that tells us to do parallel transport in this way.

For a generic smooth manifold with a metric, it is possible to define parallel transport in a similar way, as a small enough neighbourhood of a given point will be approximately flat (In the sense that the metric can be described as an Euclidean metric up to first-order using appropriate coordinates). This way, the metric gives rise to a connection known as the Levi-Civita connection that will define parallel transport. This connection was already introduced in eq. \ref{1eq:LCconnection}, and was used to define the covariant derivative in GR and to calculate the geodesics (paths of longest proper time / shortest length) of a given metric.




I will leave to the next section the discussion of whether we should use the Levi-Civita connection as the affine structure of the space-time, or we should use any other. In this section I will study how the connection is needed to define a proper covariant derivative. I will present which additional properties the covariant derivative must satisfy in order to be the Levi-Civita connection of a metric, and I will study different decompositions of a generic connection. Finally, I will construct the curvature tensors and study their properties.

\subsection{Covariant Derivative}

First of all, we must define what constitutes a covariant derivative. Let $\nabla_v$ be a differential operator, characterized by vector $v$, that acts on tensors of type $(p,q)$ giving another tensor of type $(p,q)$. Let $A$, $B$, be tensor fields defined on the space-time $\mathcal{M}$; $f$ be a function on $\mathcal{M}$; and $\alpha$, $\beta$ be constants. Then $\nabla_v$ is a covariant derivative if it satisfies\footnote{As I already explained in the introduction, it is also common to understand the covariant derivative like a gradient that takes a $(p,q)$ type tensor $T$ and gives a $(p,q+1)$ tensor $\nabla T$ with components $(\nabla T)^{i_1...i_p}_{j_1...j_1 \mu} \equiv \nabla_\mu T^{i_1...i_p}_{j_1...j_1} \equiv (\nabla_{e_\mu}T)^{i_1...i_p}_{j_1...j_1}$ where $e_\mu$ are the basis vectors. With this notation, $\nabla_v u = v^\alpha \nabla_\alpha u$.}:

\begin{enumerate}[(1)]
 \item \emph{Linearity}: $\nabla_v (\alpha A + \beta B) = \alpha \nabla_v A + \beta \nabla_v B$
 \item \emph{Leibniz Rule}: $\nabla_v (A \otimes B) = (\nabla_v A) \otimes B + A \otimes (\nabla_v B)$
 \item \emph{Commutativity with contraction}: $\nabla_u(A^{\alpha i_2 ... i_p}{}_{\alpha j_2 .. j_q}) = (\nabla_u A)^{\alpha i_2 ... i_p}{}_{\alpha j_2 .. j_q}$
 \item \emph{Consistency}: $\nabla_v f = v^\alpha \partial_\alpha f$ 
 \item $\nabla_{f v} A = f \nabla_v A$, $\qquad \nabla_{u+v} A = \nabla_u A + \nabla_v A$
\end{enumerate}

These properties gather what is understood by directional derivative. Properties (1) and (2) belong to any linear differential operator. Property (3) implies that the tangent and cotangent spaces are transported the same way. Property (4) makes the covariant derivative consistent with the fact that vector fields are already a good directional derivative for functions. Property (5) tells us that this operator is indeed a directional derivative, and depends only in the direction in at the point we are taking the derivative, and not in the value of $v$ in a neighbourhood of the point.

The partial derivative operator $\partial_v = v^\alpha \partial_\alpha$, satisfies these five properties but does not produce a tensorial result. It depends on which coordinate system we are working, and therefore it lacks real meaning. We are looking for a tensorial differential operator. Let us start looking at how the covariant derivative should act on a vector field $u$. Let us work in coordinates $\{ x^\alpha \}$ with base vectors $\{ e_\alpha \equiv \frac{\partial}{\partial x^\alpha}\}$ and dual base $\{ \df x^\alpha \}$, in which the vector is expressed as $u=u^\alpha e_\alpha$. Using the properties of consistency and the Leibniz rule we have that in general:

\begin{equation}
 \nabla_v u = \nabla_{(v^\beta e_\beta)} (u^\alpha e_\alpha) = v^\beta (\partial_\beta u^\alpha) e_\alpha + u^\beta v^\gamma \nabla_{e_\gamma} e_\beta
\end{equation}

By the definition of covariant derivative, $\nabla_{e_\gamma} e_\beta$ is another vector, which measures the failure of the basis vector field to be parallel transported along a direction (given by another basis vector). This is what is understood as the connection. In particular, the connection can be defined (in a particular choice of basis vectors) as the set of functions $\Gamma^\alpha_{\beta \gamma}$ such that:

\begin{equation}
\Gamma^\alpha_{\beta \gamma} \equiv \df x^\alpha ( \nabla_{e_\gamma} e_\beta )
\end{equation}


So $\Gamma^\alpha_{\beta \gamma}$ measures the variation in the direction $e_\alpha$, of the vector $e_\beta$, when it is transported along the direction $e_\gamma$. With the connection, we can write the covariant derivative of a vector as:

\begin{equation}
 \nabla_v u = (v^\beta \partial_\beta u^\alpha + u^\beta v^\gamma \Gamma^\alpha_{\beta \gamma}) e_\alpha
\end{equation}

Now, we can use the properties of the covariant derivative to know how it acts on other tensor fields. For example, using properties (3) and (4) on the covariant derivative of the contraction of a 1-form $\omega$ and a vector $u$ along the direction $v$, we can obtain the equation for the covariant derivative of a 1-form:



\begin{equation}
\nabla_v (u \otimes \omega)^\beta{}_\beta = (u\otimes (\nabla_v \omega) + (\nabla_v u) \otimes \omega)^\beta{}_\beta = u^\alpha (\nabla_v \omega)_\alpha + (v^\alpha \partial_\alpha u^\beta + \Gamma^\beta_{\alpha \gamma} v^\alpha u^\gamma) \omega_\beta \\ 
\end{equation}

On the other hand we have:

\begin{equation}
 \nabla_v (u \otimes \omega)^\beta{}_\beta = v^\alpha \partial_\alpha (u^\beta \omega_\beta)= \omega_\beta v^\alpha \partial_\alpha u^\beta  + u^\beta v^\alpha \partial_\alpha\omega_\beta 
\end{equation}

Comparing both expressions we obtain:
\begin{equation}
 \nabla_v \omega = (v^\alpha \partial_\alpha \omega_\beta - v^\alpha \omega_\gamma \Gamma^\gamma_{\alpha \beta})\df x^\beta
\end{equation}

Following this procedure we can obtain the covariant derivative of generic $(p,q)$-tensor field (in components):

\begin{IEEEeqnarray}{rCl}
 \nabla_v T^{\alpha_1...\alpha_p}{}_{\beta_1...\beta_q} &=& v^\gamma \partial_\gamma T^{\alpha_1...\alpha_p}{}_{\beta_1...\beta_q} + \sum_{i=1}^p v^\gamma \Gamma^{\alpha_i}_{\gamma \delta} T^{\alpha_1...\delta...\alpha_p}{}_{\beta_1...\beta_q}\nonumber\\
 &&- \sum_{i=1}^q v^\gamma \Gamma^{\delta}_{\gamma \beta_i} T^{\alpha_1...\alpha_p}{}_{\beta_1...\delta...\beta_q}
\end{IEEEeqnarray}

The connection $\Gamma^\alpha_{\beta \gamma}$ is a set of functions (64 in 4 dimensions) that specifies completely a covariant derivative in a given coordinate system. As the covariant derivative is a tensorial operator, the connection will not transform tensorially, because it has to compensate that partial derivatives do not transform tensorially either. Let us figure how the connection transforms under a coordinate change. If we change from coordinates $\{x^{\alpha^\prime}\}$ to coordinates $\{x^\alpha\}$ we have:

\begin{IEEEeqnarray}{rCl}
 \nabla_v u &=& (v^{\beta^\prime} (\partial_{\beta^\prime} u^{\alpha^\prime}) + u^{\beta^\prime} v^{\gamma^\prime} \Gamma^{\alpha^\prime}_{{\beta^\prime} {\gamma^\prime}}) e_{\alpha^\prime} \\
 &=& \left (v^\beta (\partial_\beta u^\alpha) \frac{\partial x^{\alpha^\prime}} {\partial x^\alpha}+ v^\beta u^\alpha \frac{\partial^2 x^{\alpha^\prime}}{\partial x^\beta \partial x^\alpha} + \frac{\partial x^{\beta^\prime}}{\partial x^\beta} u^\beta \frac{\partial x^{\gamma^\prime}}{\partial x^\gamma} v^\gamma \Gamma^{\alpha^\prime}_{{\beta^\prime} {\gamma^\prime}} \right ) \frac{\partial x^\sigma}{\partial x^{\alpha^\prime}} e_\sigma \IEEEeqnarraynumspace 
\end{IEEEeqnarray}

In order for the derivative operator to transform as tensor, then $\Gamma^\alpha_{\beta \gamma}$ must transform as:

\begin{equation}
 \Gamma^\alpha_{\beta \gamma} = \frac{\partial x^{\beta^\prime}}{\partial x^\beta}  \frac{\partial x^{\gamma^\prime}}{\partial x^\gamma}  \Gamma^{\alpha^\prime}_{{\beta^\prime} {\gamma^\prime}}\frac{\partial x^\alpha}{\partial x^{\alpha^\prime}} + \frac{\partial^2 x^{\alpha^\prime}}{\partial x^\beta \partial x^\gamma} \frac{\partial x^\alpha}{\partial x^{\alpha^\prime}}
\end{equation}

The first term of the transformation relation is a tensorial transformation, but there is an additional second term that depends on the second derivative of the coordinate change, hence the connection is not a tensor, but a pseudo-tensorial object. However, from this relation we can see that the difference of two connections is a tensorial object, and so will be $\sum_\sigma \Gamma^\sigma_{\mu \sigma}$, and $\sum_\sigma \Gamma^\sigma_{\sigma \mu}$.


If the space-time has a metric $g$, there exists a unique connection called the Levi-Civita connection, that gives rise to a covariant derivative ${}^g\nabla$ (As notation, in case there could be confusion to which affine structure the covariant derivative is using, I will denote it next to the $\nabla$ symbol) that satisfies these two additional conditions:

\begin{enumerate}[(1)]
 \setcounter{enumi}{5}
 \item \emph{Torsion free}: ${}^g\nabla_u v - {}^g\nabla_v u - [u,v] =0$
 \item \emph{Metric-compatible with $g$}: ${}^g\nabla_u g=0$
\end{enumerate}

From these last two properties we can obtain an expression for the Levi-Civita connection. Torsion free implies that the connection is symmetric $\Gamma^\alpha_{\mu \nu} = \Gamma^\alpha_{\nu \mu}$. Writing the second condition in components, and rotating the indexes we have:

\begin{equation}
 \left \{ \begin{array}{rcccl} 
 0&=&{}^g \nabla_\alpha g_{\beta \gamma} &=& \partial_\alpha g_{\beta \gamma} - \Gamma^\lambda_{\alpha \beta} g_{\lambda \gamma} - \underbrace{\Gamma^\lambda_{\alpha \gamma}}_{\Gamma^\lambda_{\gamma \alpha}} g_{\beta \lambda} \\ 
 0&=&{}^g \nabla_\beta g_{\gamma \alpha} &=& \partial_\beta g_{\gamma \alpha} - \Gamma^\lambda_{\beta \gamma} g_{\lambda \alpha} - \underbrace{\Gamma^\lambda_{\beta \alpha}}_{\Gamma^\lambda_{\alpha \beta}} g_{\gamma \lambda} \\ 
 0&=&{}^g \nabla_\gamma g_{\alpha \beta} &=& \partial_\gamma g_{\alpha \beta} - \Gamma^\lambda_{\gamma \alpha} g_{\lambda \beta} - \underbrace{\Gamma^\lambda_{\gamma \beta}}_{\Gamma^\lambda_{\beta \gamma}} g_{\alpha \lambda}  
 \end{array} \right.
\end{equation}

Adding the first two identities, subtracting the third and multiplying by $g^{\gamma \sigma}/2$ we obtain:

\begin{equation}
 \Gamma^\sigma_{\alpha \beta}= \frac{1}{2} g^{\sigma \gamma} \left ( \partial_\alpha g_{\beta \gamma} + \partial_\beta g_{\gamma \alpha} - \partial_\gamma g_{\alpha \beta} \right )
\end{equation}

\noindent which is the Levi-Civita connection. Its components are the Christoffel symbols and are sometimes denoted as  ${ \sigma \brace \alpha \beta}$. The quantity $\Gamma^\sigma_{\mu \sigma}$ is a tensor for any connection, in particular for the Levi-Civita connection its value is $\partial_\mu \log(\sqrt{-g})$.

Any other connection can be constructed as the sum of the Levi-Civita connection and a tensor:

\begin{equation}
 \tilde{\Gamma}^\sigma_{\alpha \beta} = { \sigma \brace \alpha \beta} + W_{\alpha \beta}{}^\sigma
\end{equation}

It is possible to decompose a general connection further. A particularly interesting decomposition can be done in terms of the torsion tensor $S$ and the non-metricity tensor $Q$:

\begin{equation}
 S_{\mu \nu}{}^\alpha \equiv \frac{1}{2}\left ( \tilde{\Gamma}^\alpha_{\mu \nu} - \tilde{\Gamma}^\alpha_{\nu \mu} \right ) = \frac{1}{2}\left ( W_{\mu \nu}{}^\alpha - W_{\nu \mu}{}^\alpha \right ) \qquad Q_{\mu \nu \alpha} \equiv - {}^{\tilde{\Gamma}} \nabla_\mu g_{\nu \alpha} = W_{\mu \nu}{}^\lambda g_{\lambda \alpha}+W_{\mu \alpha}{}^\lambda g_{\lambda \nu}
\end{equation}

The torsion tensor is the antisymmetric part of the connection in the first two indices (24 independent components), and the non-metricity tensor is the symmetric part in the last two indices (40 independent components). It is not possible to write the connection directly as the sum of the torsion and non metricity, but we can construct another two tensors $K$ (called contorsion tensor) and $L$:\footnote{Here we have lowered the indexes of $S$ with the metric $g$}

\begin{IEEEeqnarray}{rCl}
 K_{\mu \nu}{}^\sigma &=& g^{\sigma \alpha} \left ( S_{\mu \nu \alpha} - S_{\mu \alpha \nu} - S_{\nu \alpha \mu} \right ) \\
 L_{\mu \nu}{}^\sigma &=& \frac{1}{2} g^{\sigma \alpha} \left ( Q_{\mu \nu \alpha} + Q_{\nu \mu \alpha} - Q_{\alpha \mu \nu} \right )
\end{IEEEeqnarray}

where $K_{\mu \nu \alpha}$ is antisymmetric in ($\nu$, $\alpha$), while $L_{\mu \nu}{}^\sigma$ is symmetric in ($\mu$, $\nu$). Then the connection can be separated as:

\begin{equation}
 \tilde{\Gamma}^\sigma_{\alpha \beta} = { \sigma \brace \alpha \beta} + K_{\alpha \beta}{}^\sigma + L_{\alpha \beta}{}^\sigma
\end{equation}

From this expression we can check that if we give the torsion and non-metricity with respect to some metric, the connection is completely determined; therefore, the Levi-Civita connection is the only connection for which the torsion and non-metricity both vanish. Let us note that this decomposition depends on the metric chosen, and a different metric will give a different decomposition with a different Levi-Civita connection, different $L$ and different $K$ (although the torsion $S$ is independent of the metric, $K$ changes with the choice of metric because the metric is introduced in its definition, raising and lowering different indices).

Given a generic connection $\tilde{\Gamma}$, an interesting proposition is to try to find a metric $h$ such that the torsionless part of the connection is given by the Levi-Civita connection of $h$. Not every connection can be obtained from a metric; for example, connections which are Levi-Civita satisfy $\tilde{\Gamma}^\sigma_{\mu \sigma}=\partial_\mu \sqrt{\det{h}}$, but this is not true in general: $\tilde{\Gamma}^\sigma_{\mu \sigma}$ is a 1-form, and it does not have to be exact. Moreover, it would be possible that $\tilde{\Gamma}^\sigma_{\mu \sigma}$ is a closed 1-form, which can be written as the gradient of a function \emph{locally}, but not exact, and consequently there does not exist a global function $\sqrt{\det h}$ such that $\tilde{\Gamma}^\sigma_{\mu \sigma}$ is the gradient of. In general, it is not a trivial task to find such a metric $h$, and the problem depends crucially on the topology of space.

%

\subsection{Curvature Tensors}

Though the connection is not itself a tensor, it is possible to construct different tensorial quantities out of it. An example is the torsion and another is the curvature (or Riemann) tensor. Curvature manifests as the failure of a vector to remain parallel to itself after being parallel transported along a closed loop. The Riemann tensor measures this failure around infinitesimal loops in space-time. This tensor takes 3 vectors (two that represent the directions in the loop, and one that is the transported vector) and gives us another one (that measures the difference of being transported along one path or another). Given a covariant derivative, the Riemann tensor is defined as:

\begin{equation}
 {}^\Gamma R(u,v)w= {}^\Gamma \nabla_u {}^\Gamma\nabla_v w -{}^\Gamma \nabla_v {}^\Gamma\nabla_u w  -{}^\Gamma \nabla_{[u,v]} w 
\end{equation}

or in components in a coordinate basis:

\begin{equation}
 R^\alpha{}_{\beta \mu \nu} (\Gamma) = \partial_\mu \Gamma^\alpha_{\nu \beta}-\partial_\nu \Gamma^\alpha_{\mu \beta} + \Gamma^\alpha_{\mu \sigma} \Gamma^\sigma_{\nu \beta}-\Gamma^\alpha_{\nu \sigma} \Gamma^\sigma_{\mu \beta}
\end{equation}

The Riemann tensor for a general connection does not have the same symmetries as for the Levi-Civita connection of a metric. In particular $R^\alpha{}_{\alpha \mu \nu} \neq 0$ in general. But it is still antisymmetric in ($\mu$, $\nu$) by construction and satisfies the Bianchi identities, in particular the first\footnote{The first Bianchi identity can be derived noting that the commutator of two vector fields can be written as:
\begin{equation}
 [u,v]^\mu = u^\alpha \nabla_\alpha v^\mu - v^\alpha \nabla_\alpha u^\mu - S_{\alpha \beta}{}^\mu u^\alpha v^\beta
\end{equation}
And applying it to the Jacobi identity:
\begin{equation}
 [u,[v,w]]+[v,[w,u]]+[w,[u,v]]=0
\end{equation}
}:

\begin{IEEEeqnarray}{rCl}
 R^\alpha{}_{\beta \mu \nu} &=& -R^\alpha{}_{\beta \nu \mu }\\
 R^\alpha{}_{\beta \mu \nu} +R^\alpha{}_{ \mu \nu \beta} +R^\alpha{}_{\nu \beta \mu } &=& S_{\mu \nu}{}^\lambda S_{\lambda \beta}{}^\alpha +S_{\nu \beta}{}^\lambda S_{\lambda \mu}{}^\alpha+S_{\beta \mu}{}^\lambda S_{\lambda \nu}{}^\alpha \\
 &&+ \nabla_\mu S_{\nu \beta}{}^\alpha + \nabla_\nu S_{\beta \mu}{}^\alpha + \nabla_\beta S_{\mu \nu}{}^\alpha \nonumber
\end{IEEEeqnarray}

Given two connections $\Gamma$, $\tilde{\Gamma}$ related by a tensor $W$ as $\tilde{\Gamma}=\Gamma+W$, their associated Riemann tensors are related as:

\begin{IEEEeqnarray}{rCl}
 R^\alpha{}_{\beta \mu \nu} (\tilde{\Gamma}) &=& R^\alpha{}_{\beta \mu \nu} (\Gamma) + {}^\Gamma \nabla_\mu W_{\nu \beta}{}^\alpha - {}^\Gamma \nabla_\nu W_{\mu \beta}{}^\alpha + (\Gamma_{\mu \nu}^\lambda - \Gamma_{\nu \mu}^\lambda ) W_{\lambda \beta}{}^\alpha \IEEEeqnarraynumspace \\
 && + W_{\mu \lambda}{}^\alpha W_{\nu \beta}{}^\lambda - W_{\nu \lambda}{}^\alpha W_{\mu \beta}{}^\lambda \nonumber
\end{IEEEeqnarray}

The Ricci tensor is defined as one of the traces of the Riemann tensor $R_{\beta \nu} = R^\alpha{}_{\beta \alpha \nu}$. Given the same two connections, their Ricci tensors will be related as:

\begin{IEEEeqnarray}{rCl}
 R_{\beta \nu}(\tilde{\Gamma}) &=& R_{\beta \nu}(\Gamma)+{}^\Gamma \nabla_\alpha W_{\nu \beta}{}^\alpha - {}^\Gamma \nabla_\nu W_{\alpha \beta}{}^\alpha + (\Gamma_{\alpha \nu}^\lambda - \Gamma_{\nu \alpha}^\lambda ) W_{\lambda \beta}{}^\alpha \label{3eq:RicciW} \\
 && + W_{\alpha \lambda}{}^\alpha W_{\nu \beta}{}^\lambda - W_{\nu \lambda}{}^\alpha W_{\alpha \beta}{}^\lambda \nonumber
\end{IEEEeqnarray}

The Ricci tensor for a general connection does not need to be symmetric, even if that connection is torsionless. In the previous equation, if $\Gamma$ is the Levi-Civita connection of a metric and $W$ is symmetric, we have that the antisymmetric part of the Ricci tensor is:

\begin{equation}
 R_{[\beta \nu]}(\tilde{\Gamma}) = \partial_\beta W_{\nu \alpha}{}^\alpha - \partial_\nu W_{\beta \alpha}{}^\alpha
\end{equation}

Which is non-zero in general. The antisymmetric part of the Ricci is in general the exterior derivative of the vector $W_{\nu \alpha}{}^\alpha$. If it is indeed different from zero, then $\tilde{\Gamma}^\sigma_{\mu \sigma}$ will not be the gradient of a function, and this means that $\tilde{\Gamma}$ cannot be written as the Levi-Civita connection of a metric $h$.

\section{Metric-Affine Formalism}

In the last section we have introduced an affine structure to our manifold in order to define a covariant derivative. We have also seen that if the manifold has a metric structure, it naturally induces an affine structure on the manifold through the Levi-Civita connection. A natural question is whether we should use the Levi-Civita connection, or a connection independent of the metric.

Taking the Levi-Civita connection of the metric as our affine structure is known as the \emph{Riemannian formalism}. General Relativity is done in this formalism. At the time GR was developed, this was the only known possibility, as the theory of affine connections had not been developed yet. Working with a connection independent of the metric is known as \emph{Metric-Affine formalism} (\cite{Hehl:1976my}, \cite{Hehl:1994ue}). 

It may feel natural to work in the Riemannian formalism, as unaccelerated observers will preserve angles along their path, which agrees with our notion that moving an object around space does not change that object. The Einstein Equivalence Principle makes use of this notion because if the rest frame did not preserve angles, non-gravitational experiments would be frame-dependent, and EEP violations have not been observed. However, the Riemannian formalism imposes artificial restrictions on the affine structure. A theory in which matter follows geodesics of a connection that is independent of the metric, but approximately equal to the Levi-Civita connection except in the regions where gravity is most strong, would still agree with EEP within the experimental range. Another thing we should note is that in order to define 'unaccelerated' we have to look into the equations of motion of the matter. The equations of motion for matter in the Riemannian formalism tell us that matter particles follow geodesics of the metric. The path of matter particles in the Metric-Affine formalism depends on how matter couples to the connection. If matter does not couple to the connection, matter particles will still follow geodesics of the metric and the EEP would not be violated. If this is the case, the geodesics of the independent connection would be accelerated paths, and not preserving angles would just translate into forces to an object following them (This is the case, for example, of some models in which dark matter follows geodesics from a different metric than the space-time one, see \cite{Nojiri:2014zqa}, \cite{vandeBruck:2016jgg}). So we should consider that the Metric-Affine formalism is a viable strategy to study modifications to gravity. 

Now, I want to know how to construct a theory of gravity in the Metric-Affine formalism. In the Riemannian formalism we construct a gravity action with a Lagrangian made of curvature scalars. Performing the variation of the action with respect to the metric gives us a set of equations of motion which can be solved, that equate the curvature of space-time (which comes from the variation of gravity action with respect to the metric) to the matter content of the universe, represented by the energy-momentum tensor (which comes from the variation of the matter action with respect to the metric). These equations are enough to describe gravity. The case of the Metric-Affine formalism is similar, we can construct a gravity Lagrangian made of curvature scalars, and perform the variation of the action with respect to both the metric and the connection, obtaining two sets of equations of motion. Analogous to the energy-momentum tensor, there will be a hypermomentum tensor that corresponds to variation of the matter action with respect to the connection. It should be possible to solve these equations to obtain both the metric and the connection.

The first question the Metric-Affine formalism faces is whether it is possible to recover the GR results. It would be good if we can find theories which are different from GR, but not too different, since GR makes correct predictions for the solar system and many other experiments. Let us consider the same Lagrangian as in GR, $L_G=R$, which now depends both on the connection and the metric $R=g^{\alpha \beta} R_{\alpha \beta}(\Gamma)$.

\begin{equation}
 S= \frac{1}{16 \pi l_P^2} \int_\mathcal{M} g^{\alpha \beta} R_{\alpha \beta}(\Gamma) \sqrt{|g|} \df^4 x + \int_\mathcal{M} \Lagr_\text{m} \sqrt{-g} \df^4 x
\end{equation}

The variation of the action with respect to the metric gives us a familiar equation:

\begin{equation}
 R_{\mu \nu}(\Gamma) - \frac{1}{2} R(\Gamma) g_{\mu \nu} = 8 \pi l_P^2 T_{\mu \nu} \label{3eq:metriceq}
\end{equation}

This equation looks exactly like Einstein's equations, but instead of the curvature tensor associated to the Levi-Civita connection, we have a curvature tensor associated to an independent connection. We also have to take the variation of the action with respect to the components of the  independent connection. The variation of the Riemann tensor with respect to the connection is:

\begin{IEEEeqnarray}{rCl}
 \delta R^\alpha{}_{\beta \mu \nu} &=& \partial_\mu \delta \Gamma^\alpha_{\nu \beta} - \partial_\nu \delta \Gamma^\alpha_{\nu \beta} + \delta \Gamma^\alpha_{\mu \lambda} \Gamma^\lambda_{\nu \beta}+  \Gamma^\alpha_{\mu \lambda} \delta \Gamma^\lambda_{\nu \beta} - \delta \Gamma^\alpha_{\nu \lambda} \Gamma^\lambda_{\mu \beta}-  \Gamma^\alpha_{\nu \lambda} \delta \Gamma^\lambda_{\mu \beta} \IEEEeqnarraynumspace \\
 &=& \nabla_\mu \delta \Gamma^\alpha_{\nu \beta} - \nabla_\nu \delta \Gamma^\alpha_{\mu \beta} - \underbrace{(\Gamma^\lambda_{\mu \nu} - \Gamma^\lambda_{\nu \mu})}_{2 S_{\mu \nu}{}^\lambda} \delta \Gamma^\alpha_{\lambda \beta}
\end{IEEEeqnarray}

With this result let us proceed to take the variation of the Lagrangian:

\begin{IEEEeqnarray}{rCl}
 \delta (R \sqrt{-g} ) &=& g^{\nu \beta} \left ( \nabla_\alpha \delta \Gamma^\alpha_{\nu \beta} - \nabla_\nu \delta \Gamma^\alpha_{\alpha \beta} - 2 S^\lambda{}_{\alpha \nu} \delta \Gamma^\alpha_{\lambda \beta} \right ) \sqrt{-g}  \\
 &=& \nabla_\alpha \left \{ ( g^{\nu \beta} \delta \Gamma^\alpha_{\nu \beta}  - g^{\alpha \beta} \delta \Gamma^\sigma_{\sigma \beta} ) \sqrt{-g}  \right \} \\
 && - \delta \Gamma^\alpha_{\nu \beta} \left \{ \nabla_\alpha (g^{\nu \beta} \sqrt{-g} ) - \delta^\nu{}_\alpha \nabla_\sigma ( g^{\sigma \beta} \sqrt{-g}) - 2S_{\alpha \lambda}{}^\nu g^{\lambda \beta} \sqrt{-g} \right \} \nonumber
\end{IEEEeqnarray}

At this point we are interested in rewriting some of the terms, so that we can separate the surface terms from the rest. The first term is of the form $\nabla_\alpha (J^\alpha \sqrt{-g})$, but in general we have that $\nabla_\mu \sqrt{-g} = \partial_\mu \sqrt{-g} - \Gamma^\lambda_{\mu \lambda} \sqrt{-g}$, so $\nabla_\alpha (J^\alpha \sqrt{-g}) = \partial_\alpha (  J^\alpha \sqrt{-g}) - J^\lambda S_{\lambda \alpha}{}^\alpha{}$. With this, the variation of the Lagrangian is:

\begin{IEEEeqnarray}{rCl}
 \delta (R \sqrt{-g} ) &=& \partial_\alpha \left \{ ( g^{\nu \beta} \delta \Gamma^\alpha_{\nu \beta}  - g^{\alpha \beta} \delta \Gamma^\sigma_{\sigma \beta} ) \sqrt{-g}  \right \} - ( g^{\nu \beta} \delta \Gamma^\alpha_{\nu \beta}  - g^{\alpha \beta} \delta \Gamma^\sigma_{\sigma \beta} ) S_{\alpha \lambda}{}^\lambda \IEEEeqnarraynumspace  \\
 && - \delta \Gamma^\alpha_{\nu \beta} \left \{ \nabla_\alpha (g^{\nu \beta} \sqrt{-g} ) - \delta^\nu{}_\alpha \nabla_\sigma ( g^{\sigma \beta} \sqrt{-g}) - 2S_{\alpha \lambda}{}^\nu g^{\lambda \beta} \sqrt{-g} \right \} \nonumber
\end{IEEEeqnarray}

The first term is a total derivative that under the integral gives a surface term whose variation is $0$. The variation of the action is:

\begin{IEEEeqnarray}{rCl}
 \delta S &=& \int \left \{ \nabla_\alpha (g^{\nu \beta} \sqrt{-g} ) - \delta^\nu{}_\alpha \nabla_\sigma ( g^{\sigma \beta} \sqrt{-g}) - 2S_{\alpha \lambda}{}^\nu g^{\lambda \beta} \sqrt{-g}  \right. \\
 && \left. + 2 g^{\nu \beta} S_\alpha{}^\lambda{}_\lambda \sqrt{-g} - 2 \delta^\nu{}_\alpha S^{\beta \lambda}{}_\lambda \sqrt{-g} \right \} \delta \Gamma^\alpha_{\nu \beta} \df^4 x  \nonumber
\end{IEEEeqnarray}

In this thesis, I will work with the \emph{a priori} assumption that the torsion vanishes, the role of torsion will be discussed in section \ref{3sec:roletor}. At this point we have to set the torsion to $0$ and take the variation just to the symmetric part of the connection, so we need to symmetrize the equation multiplying $\delta \Gamma^\alpha_{\nu \beta}$. With this assumption, the variation of the action reads:

\begin{equation}
 \delta S = \int \left \{ \nabla_\alpha (g^{\nu \beta} \sqrt{-g} ) - \frac{1}{2}\delta^\nu{}_\alpha \nabla_\sigma ( g^{\sigma \beta} \sqrt{-g}) - \frac{1}{2}\delta^\beta{}_\alpha \nabla_\sigma ( g^{\sigma \nu} \sqrt{-g})  \right \} \delta \Gamma^\alpha_{(\nu \beta)} \df^4 x
\end{equation}

From which we obtain the second set of equations of motion:

\begin{equation}
 \frac{1}{\sqrt{-g}} \nabla_\alpha (g^{\nu \beta} \sqrt{-g} ) - \frac{1}{2}\delta^\nu{}_\alpha \frac{1}{\sqrt{-g}} \nabla_\sigma ( g^{\sigma \beta} \sqrt{-g}) - \frac{1}{2}\delta^\beta{}_\alpha \frac{1}{\sqrt{-g}} \nabla_\sigma ( g^{\sigma \nu} \sqrt{-g}) = 16 \pi l_p^2 H_\alpha{}^{\beta \nu}
\end{equation}

Where $H_\alpha{}^{\beta \nu} \equiv \frac{1}{\sqrt{-g}} \frac{\delta \Lagr_M \sqrt{-g}}{\delta \Gamma^\alpha_{\nu \beta}}$ corresponds to the coupling of matter to the connection and is called the hypermomentum tensor. Let us consider that matter does not couple to the connection\footnote{This will be the case if the matter action consists of the usual Lagrangians.}. In this case the equation can be simplified if we contract indices $\alpha$ and $\beta$, from which we obtain $-\frac{3}{2} \nabla_\sigma ( g^{\sigma \nu} \sqrt{-g}) =0$. The second set of equations of motion simplifies to:

%
%

\begin{equation}
 \nabla_\alpha (\sqrt{-g} g^{\nu \beta}) = 0 \label{3eq:connectioneq}
\end{equation}

This is an algebraic system of equations that tells us that the independent connection must be the Levi-Civita connection of the metric $g$. This fact, together with equation \ref{3eq:metriceq} gives the same equations as General Relativity in Riemannian formalism.

For the $\Lagr_G=R$ Lagrangian, the metric-affine formalism does not offer new solutions with respect to GR, although it is always welcome to impose less artificial restrictions in our theory. If we want to obtain new solutions we will need a different Lagrangian so eq. \ref{3eq:connectioneq} changes, which will generate a difference between the independent connection and the Levi-Civita connection of $g$. 



\subsection{General Lagrangian}

Now we can consider a Lagrangian that depends in a general way of the Riemann tensor of the curvature and the metric, $\Lagr_G = f(R^\alpha{}_{\beta \mu \nu}, g_{\mu \nu})$. The variation of the action with respect to the metric gives us one set of equations of motion:


\begin{equation}
 \frac{\partial f}{\partial g^{(\mu \nu)}} - \frac{f}{2} g_{\mu \nu} =8 \pi l_P^2 T_{\mu \nu} 
\end{equation}

Unlike the $\Lagr_G=R$ case, these equations in the Metric-Affine formalism are different from those of the Riemannian formalism for the same Lagrangian. In particular, the equations in the Riemannian formalism include second derivatives of the Ricci tensor, or in other words, fourth-order derivatives of the metric. The equations in the Metric-Affine formalism are of lower order, and therefore there will not be ghost instabilities in these theories.

Now let us focus in the second set of equations of motion. Let us define $P_\alpha{}^{\beta \mu \nu} \equiv \frac{\partial f}{R^\alpha{}_{\beta \mu \nu}}$ and let us perform the variation of the gravity Lagrangian with respect to the connection:

\begin{IEEEeqnarray}{rCl}
 \delta (\Lagr_G \sqrt{-g}) &=& P_\alpha{}^{\beta \mu \nu} \delta R^\alpha{}_{\beta \mu \nu} \sqrt{-g} \\
 &=& P_\alpha{}^{\beta \mu \nu} \left ( \nabla_\mu \delta \Gamma^\alpha_{\nu \beta} - \nabla_\nu \delta \Gamma^\alpha_{\mu \beta} - 2 S_{\mu \nu}{}^\lambda \delta \Gamma^\alpha_{\lambda \beta} \right ) \sqrt{-g} \\
 &=& \nabla_\mu \left [ \left ( P_\alpha{}^{\beta \mu \nu} \delta \Gamma^\alpha_{\nu \beta} -  P_\alpha{}^{\beta \nu \mu} \delta \Gamma^\alpha_{\nu \beta} \right ) \sqrt{-g} \right ]\\
 && - \nabla_\mu \left [ \left ( P_\alpha{}^{\beta \mu \nu} -  P_\alpha{}^{\beta \nu \mu}  \right ) \sqrt{-g} \right ] \delta \Gamma^\alpha_{\nu \beta}  - 2 S_{\mu \nu}{}^\lambda \delta \Gamma^\alpha_{\lambda \beta} \sqrt{-g}\nonumber \\
 &=& \partial_\mu \left [ \left ( P_\alpha{}^{\beta \mu \nu} \delta \Gamma^\alpha_{\nu \beta} -  P_\alpha{}^{\beta \nu \mu} \delta \Gamma^\alpha_{\nu \beta} \right ) \sqrt{-g} \right ] \label{3eq:VariationTorsion}\\
&& -  \left [ \left ( P_\alpha{}^{\beta \mu \nu} \delta \Gamma^\alpha_{\nu \beta} -  P_\alpha{}^{\beta \nu \mu} \delta \Gamma^\alpha_{\nu \beta} \right ) \sqrt{-g} \right ] S_{\mu \lambda}{}^\lambda \nonumber\\
 && - \nabla_\mu \left [ \left ( P_\alpha{}^{\beta \mu \nu} -  P_\alpha{}^{\beta \nu \mu}  \right ) \sqrt{-g} \right ] \delta \Gamma^\alpha_{\nu \beta}  - 2 S_{\mu \nu}{}^\lambda \delta \Gamma^\alpha_{\lambda \beta} \sqrt{-g}\nonumber
\end{IEEEeqnarray}

The total derivatives will become surface terms under integration and will not contribute to the variation of the action. Again, we will consider no torsion $S_{\alpha \beta}{}^\lambda=0$, so we will have to symmetrize the indices that multiply the symmetric part of the connection:

\begin{IEEEeqnarray}{rCl}
 \delta (\Lagr_G \sqrt{-g}) &=& \frac{1}{2} \nabla_\mu \left [ \left ( P_\alpha{}^{\beta \mu \nu} + P_\alpha{}^{\nu \mu \beta}-  P_\alpha{}^{\beta \nu \mu} -  P_\alpha{}^{\nu \beta \mu} \right ) \sqrt{-g} \right ] \delta \Gamma^\alpha_{(\nu \beta)} \IEEEeqnarraynumspace 
\end{IEEEeqnarray}

The variation of the matter Lagrangian will give the hypermomentum tensor as before. The second set of equations of motion are:

\begin{equation}
 \frac{1}{2} \nabla_\mu \left [ \left ( P_\alpha{}^{\beta \mu \nu} + P_\alpha{}^{\nu \mu \beta}-  P_\alpha{}^{\beta \nu \mu} -  P_\alpha{}^{\nu \beta \mu} \right ) \sqrt{-g} \right ] = 16 \pi l_p^2 H_\alpha{}^{\nu \beta} \label{3eq:EqConnectionGeneral}
\end{equation}

These equations will generate differences between the independent connection and the Levi-Civita connection of the metric.

\subsubsection{f(R,Q) Lagrangian}

A particularly interesting subset of Lagrangians are those of the type $f(R,Q)$, which depend on the curvature scalars $R$ and $Q\equiv R^{\mu \nu} R_{\mu \nu}$ with $R^{\mu \nu} \equiv g^{\mu \alpha} g^{\nu \alpha} R_{\alpha \beta} (\Gamma)$. In this case we will have that the variation of $f(R,Q)$ with respect to the Riemann tensor is:

\begin{IEEEeqnarray}{rCl}
 P_\alpha{}^{\beta \mu \nu} &=& \partial_R f \delta_\alpha{}^\mu g^{\beta \nu} + 2 \partial_Q f \delta_\alpha{}^\mu R^{\beta \nu}
\end{IEEEeqnarray}

Let us remember that $R^{\beta \nu}$ does not need to be symmetric, so $Q$ can be written as $Q=R^{(\mu \nu)}R_{(\mu \nu)} + R^{[\mu \nu]}R_{[\mu \nu]}$. We are going to consider that the torsion vanishes, and that matter does not couple to the connection. The two sets of equations of motion are:

\begin{IEEEeqnarray}{rCl}
 8\pi l_p^2 T_{\mu \nu} &=& \partial_R f R_{(\mu \nu)} + 2 \partial_Q f (R_{(\alpha \mu)} R_{(\beta \nu)} + R_{[\alpha \mu]} R_{[\beta \nu]}) g^{\alpha \beta}- \frac{f}{2} g_{\mu \nu} \label{3eq:metriceqfRQ}\\
 0 &=& \nabla_\alpha [(\partial_R f g^{\beta \nu} + 2 \partial_Q f R^{(\beta \nu)}) \sqrt{-g} ]\\
 &&-\frac{1}{2} \delta_\alpha{}^\nu \nabla_\mu [(\partial_R f g^{\beta \mu} + 2 \partial_Q f R^{\beta \mu}) \sqrt{-g} ]\nonumber\\
 &&-\frac{1}{2} \delta_\alpha{}^\beta \nabla_\mu [(\partial_R f g^{\nu \mu} + 2 \partial_Q f R^{\nu \mu}) \sqrt{-g} ]\nonumber
\end{IEEEeqnarray}


In order to solve these equations is useful to construct an auxiliary metric $h$ such that:

\begin{equation}
 (h^{-1})^{\beta \nu} \sqrt{-h} = (\partial_R f g^{\beta \nu} + 2 \partial_Q f R^{(\beta \nu)}) \sqrt{-g} \label{3eq:hg1}
\end{equation}

With this auxiliary metric it is possible to write the independent connection in terms of the Levi-Civita connection of $h$ and the non-metricity (with respect to $h$):

\begin{equation}
 Q_{\mu \alpha \beta}(h) \equiv - \nabla_\mu h_{\alpha \beta} \qquad \Gamma^\sigma_{\alpha \beta} = {\sigma \brace \alpha \beta}(h) + \underbrace{\frac{1}{2} (h^{-1})^{\sigma \mu}\left ( Q_{\alpha \beta \mu} + Q_{\beta \alpha \mu} - Q_{\mu \alpha \beta} \right )}_{L_{\alpha \beta}{}^\sigma}
\end{equation}

With this change, the second set of equations of motion can be written as:

\begin{IEEEeqnarray}{rCl}
 0&=&\nabla_\alpha [(h^{-1})^{\beta \nu} \sqrt{-h}]-\frac{1}{2} \delta_{\alpha}{}^\nu \nabla_\mu [(h^{-1})^{\beta \mu} \sqrt{-h}]-\frac{1}{2} \delta_{\alpha}{}^\beta \nabla_\mu [(h^{-1})^{\nu \mu} \sqrt{-h}] \nonumber \\
 &&-  \delta_\alpha{}^\nu \nabla_\mu (\partial_Q f R^{[\beta \mu]} \sqrt{-g}) -  \delta_\alpha{}^\beta \nabla_\mu (\partial_Q f R^{[\nu \mu]} \sqrt{-g}) \label{3eq:fQRconnect}
\end{IEEEeqnarray}

It is possible to simplify this last expression if we realize that contracting $\alpha$ and $\beta$ we obtain:

\begin{equation}
 5 \nabla_\mu (\partial_Q f R^{[\nu \mu]} \sqrt{-g}) = -\frac{3}{2}\nabla_\mu [(h^{-1})^{\nu \mu} \sqrt{-h}] \label{3eq:eqPhi1}
\end{equation}

With this relation it is possible to simplify eq. \ref{3eq:fQRconnect} (note that this way we are losing 4 of the 40 equations that determine the symmetric connection, and we will need to go back to this last equation to solve completely the connection):

\begin{equation}
 0=\nabla_\alpha [(h^{-1})^{\beta \nu} \sqrt{-h}]-\frac{1}{5} \delta_{\alpha}{}^\nu \nabla_\mu [(h^{-1})^{\beta \mu} \sqrt{-h}]-\frac{1}{5} \delta_{\alpha}{}^\beta \nabla_\mu [(h^{-1})^{\nu \mu} \sqrt{-h}]
\end{equation}

Using the decomposition of the connection as the Levi-Civita connection of $h$ plus the non-metricity, the last equation reads:

\begin{equation}
 0 = (h^{-1})^{\beta \lambda} L_{\alpha \lambda}{}^\nu + (h^{-1})^{\nu \lambda} L_{\alpha \lambda}{}^\beta - (h^{-1})^{\beta \nu} L_{\alpha \lambda}{}^\lambda - \frac{1}{5} \delta_\alpha{}^\beta (h^{-1})^{\lambda \sigma} L_{\lambda \sigma}{}^\nu - \frac{1}{5} \delta_\alpha{}^\nu (h^{-1})^{\lambda \sigma} L_{\lambda \sigma}{}^\beta
\end{equation}

In this equation, it appears both the full non-metric part of the connection $L_{\alpha \beta}{}^\nu$ and different traces of it. It is possible to use this expression to write $L_{\alpha \beta}{}^\nu$ in terms of its own traces and the metric. To do that we have to cycle the free indices and add or subtract the new expressions accordingly and multiply by $h$ to obtain: 

\begin{equation}
 L_{\alpha \beta}{}^\nu = \frac{1}{2} \left \{ \delta_\beta{}^\nu L_{\alpha \lambda}{}^\lambda + \delta_\alpha{}^\nu L_{\beta \lambda}{}^\lambda - h_{\alpha \beta} (h^{-1})^{\nu \sigma} L_{\sigma \lambda}{}^\lambda + \frac{2}{5} h_{\alpha \beta} (h^{-1})^{\sigma \lambda} L_{\sigma \lambda}{}^\nu \right \}
\end{equation}

Multiplying this equation by $(h^{-1})^{\alpha \beta}$ we obtain that $\frac{1}{5} (h^{-1})^{\sigma \lambda} L_{\sigma \lambda}{}^\nu = -(h^{-1})^{\nu \sigma} L_{\sigma \lambda}{}^\lambda$. Using this relation and defining $\Phi_\alpha \equiv L_{\alpha \lambda}{}^\lambda$ we finally have:

\begin{equation}
 L_{\alpha \beta}{}^\nu = \frac{1}{2} \left \{ \delta_\alpha{}^\nu \Phi_\beta + \delta_\beta{}^\nu \Phi_\alpha - 3 h_{\alpha \beta} (h^{-1})^{\nu \lambda} \Phi_\lambda \right \}
\end{equation}

The non-metric part of the connection can be written just as a contribution of the vector $\Phi$. The full independent connection is the Levi-Civita connection of $h$ plus this contribution. To completely determine the connection, we still need to determine the vector $\Phi$ through equation \ref{3eq:eqPhi1}, which will be a dynamical equation. But before doing that, let us write the Ricci tensor of the connection as the Ricci tensor of the metric $h$ plus terms that depend on $\Phi$ using eq. \ref{3eq:RicciW}:

\begin{IEEEeqnarray}{rCl}
 R_{(\beta \nu)} (\Gamma)&=&R_{\beta \nu}(h) - \frac{3}{4}(\Phi_\beta \Phi_\nu + h_{\beta \nu} (h^{-1})^{\lambda \sigma} \Phi_\lambda \Phi_\sigma ) \\
 R_{[\beta \nu]} (\Gamma)&=& \frac{1}{2}(\partial_\beta \Phi_\nu - \partial_\nu \Phi_\beta)
\end{IEEEeqnarray}

$R_{[\beta \nu]}$ works like an ``Electromagnetic tensor'' of the vector potential $\Phi$. Now, equation \ref{3eq:eqPhi1} reads:

\begin{equation}
 \nabla_\mu(\partial_Q f R^{[\nu \mu]} \sqrt{-g}) = \frac{3}{2} (h^{-1})^{\nu \lambda} \Phi_\lambda
\end{equation}

Which is a dynamical equation for the vector $\Phi$. In equation \ref{3eq:metriceqfRQ}, it is possible to move the curvature terms that depend on $\Phi$ to the other side of the equation, and think of them as a contribution to the energy-momentum tensor. For a Lagrangian $f=R+R_{[\alpha \mu]} R_{[\beta \nu]} g^{\alpha \beta} g^{\mu \nu}$, the theory would be equivalent to the Einstein-Proca system, with $\Phi$ the Proca field (\cite{Olmo:2013lta},\cite{Buchdahl:1979ut}). Other Lagrangians would be non-linear generalizations of this system.

If we are not interested in describing a Proca field, we can set $\Phi=0$, $R_{[\mu \nu]}=0$. Let us note that disregarding this Proca field, the connection is completely determined by the auxiliary metric $h$, which in turn is related to $g$ through eq. \ref{3eq:hg1}. If we define a matrix $\Sigma$ as:
\begin{equation}
 \Sigma_\alpha{}^\nu \equiv (\partial_R f) \delta_\alpha{}^\nu + 2 (\partial_Q f)g^{\nu \beta} R_{\beta \alpha}
\end{equation}

Then the relation between $h$ and $g$ can be written as:

\begin{equation}
 (h^{-1})^{\mu\nu}= \frac{g^{\mu\alpha} \Sigma_\alpha{}^\nu}{\sqrt{\det \Sigma}} \qquad h_{\mu \nu}=  \sqrt{\det \Sigma} (\Sigma^{-1})^\alpha{}_\nu g_{\mu\alpha} \label{3eq:hg2}
\end{equation}

This matrix $\Sigma$ can be obtained from the stress-energy tensor alone: the matter content of the space-time is the one who dictates the relation between the space-time metric and the auxiliary metric. For a reasonable theory in which we expect to recover GR at low curvatures ($f(R,Q) \sim R$), this $\Sigma$ matrix will become the identity for vacuum, and will only introduce changes when there is matter present. In the next chapter I will show how to solve these equations of motion, and in particular, electrovacuum solutions for a family of quadratic Lagrangians in a static and spherically symmetric geometry.


\subsection{Role of Torsion in Metric-Affine Formalism\label{3sec:roletor}}

Setting the torsion to zero is a choice we have made during the work of this thesis. The resulting theories are simpler than if we had considered torsion, yet they will still provide new and exciting features. The opposite scenario, where non-metricity is considered to be zero but torsion allowed to be free, is called Einstein-Cartan theory. Torsion couples to fermionic fields and would be of special importance if we are studying solutions in which they are present, as there would be new interactions depending if torsion vanishes or not.

I also want to highlight the differences of considering vanishing torsion \emph{a priori} instead of \emph{a posteriori} (see \cite{Olmo:2013lta} for further details). We could have worked in the case where torsion is allowed to be free but then treat only solutions in which the torsion is zero. These solutions would be different from the ones obtained with the a priori approach, and it can be shown that the equations of motion are manifestly different. If we take the variation of the action in eq. \ref{3eq:VariationTorsion}, obtain the equation of motion with torsion, and then set the torsion to 0, we would obtain:

\begin{equation}
  \nabla_\mu \left [ \left (P_\alpha{}^{\beta \nu \mu} - P_\alpha{}^{\beta \mu \nu}    \right ) \sqrt{-g} \right ]
\end{equation}

which is manifestly different from eq. \ref{3eq:EqConnectionGeneral}, unless $P_\alpha{}^{\beta [\mu \nu]}=P_\alpha{}^{\nu[\mu \beta]}$. In general, solutions for the equations of motion with a connection with torsion, will result in the torsion being different from 0, even if matter does not couple to it. Although in general, torsion will not vanish, we can look for solutions in which it vanishes (which will be a particular subset of all the solutions). If torsion vanishes, the Ricci tensor will always be symmetric (which would remove the dynamical vector degree of freedom that happens removing torsion \emph{a priori}).
The equations for a $f(R,Q)$ theory with vanishing torsion \emph{a posteriori} are equivalent to the equations with vanishing torsion \emph{a priori} and considering a symmetric Ricci tensor, which are:

\begin{eqnarray}
 (\partial_R f)  R_{\mu \nu} - \frac{f}{2} g_{\mu \nu} + 2 (\partial_Q f) R_{\mu \alpha} R_{\beta \nu} g^{\alpha \beta}&=& 8 \pi l_P^2 T_{\mu \nu}\\
 \nabla_\lambda \left [ \sqrt{-g}((\partial_R f) g^{\mu \nu} + 2 (\partial_Q f) R_{\alpha \beta}g^{\mu \alpha}g^{\nu \beta} ) \right ] &=& 0 
\end{eqnarray}

The solutions obtained in both approaches are the same, although the full theory is different. In particular, if we consider perturbations of these solutions, in the \emph{a posteriori} approach we should consider perturbations that give rise to torsion.

\section{Motivation}


As we have seen, GR can be obtained in both the Riemannian and the Metric-Affine formalisms from the Lagrangian $\Lagr_G=R$. One might think that the Riemannian formalism is simpler because it depends on less variables; but one could also argue that forcing the connection to be the Levi-Civita connection of the metric is an artificial restriction that greatly increases the complexity of the problem. 

In the Metric-Affine formalism there is a greater number of equations, but those are of lesser degree than in the Riemannian formalism. For the GR Lagrangian $\Lagr_G=R$, the equations in Riemannian formalism are second-order differential equations on the metric tensor, meanwhile in the Metric-Affine formalism, they are first-order differential equations on the connection (although both are equivalent). For a more general Lagrangian $\Lagr_G=f(R^\alpha{}_{\beta \mu \nu}, g_{\mu \nu})$, the equations in Riemannian formalism will contain up to fourth-order derivatives of the metric\footnote{There are some gravity theories --known as Lovelock gravity-- in which the Lagrangian is constructed in such a way that the fourth-order derivatives in the equations of motion cancel out \cite{Lovelock:1971yv}. For these theories, the field equations in the torsionless case are the same in both the Riemannian formalism and the Metric-Affine formalism (\cite{Exirifard:2007da}, \cite{Borunda:2008kf}). }, meanwhile in the Metric-Affine formalism, they will be second-order in the connection.

As pointed out in section \ref{2subsec:ghosts}, the fourth-order derivatives that appear in the Riemannian formalism lead to new degrees of freedom in the theory, some of them with negative energy. From a classical point of view, this can lead to violations of causality. The Metric-Affine formalism is free from this problem. This makes the Metric-Affine formalism a compelling way to introduce modifications to General Relativity. Also, second-order equations fit better to the requirements of quantization of physical theories, which is an additional positive feature of the metric-affine approach versus the metric one. 

The Metric-Affine approach has already been used to study the cosmic speed-up problem \cite{Guendelman:2013sca}, and also to remove the big bang singularity through a bouncing cosmology \cite{Avelino:2012ue}. In the next chapters we will see how the Metric-Affine formalism can be used to treat the problem of singularities in black holes.

\subsection{Analogy with Bravais Crystals}

A particularly nice way to understand Metric-Affine theory is through a totally different system that can also be described in these terms: the Bravais crystals (\cite{Lobo:2014nwa}, \cite{Olmo:2015bha}). Crystalline structures are discrete systems that can be described in the continuum limit using the language of differential geometry. An ideal Bravais crystal can be constructed translating a point (atom) in three crystallographic directions repeatedly. A perfect crystal is a deformation of an ideal crystal. Both ideal and perfect crystals can be described using Riemannian geometry. However, real crystals contain defects in its structure and must be described using a metric-affine approach (\cite{Falk1981}, \cite{1986GMMWJ..66..284K}, \cite{Kroner1990}, \cite{Kupferman2015361}, \cite{2015arXiv150802003K}). The type of defects we are going to consider are vacancies/interstitials (point-like defects) and dislocations (one-dimensional defects). These defects are usually dynamical and can move through the crystal, or recombine with themselves and with other types of defects, perhaps upon the effect of heat or external forces.

\begin{figure}[h!]
\begin{tabular}{lr}
  \includegraphics[width=.4\linewidth]{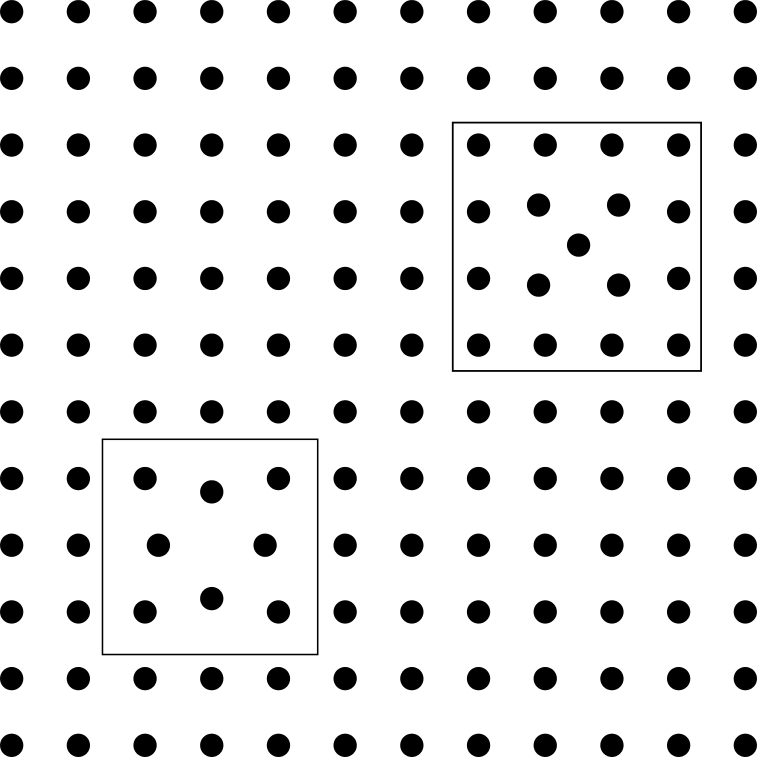}\ \ \
 &
 \ \ \ \includegraphics[width=.4\linewidth]{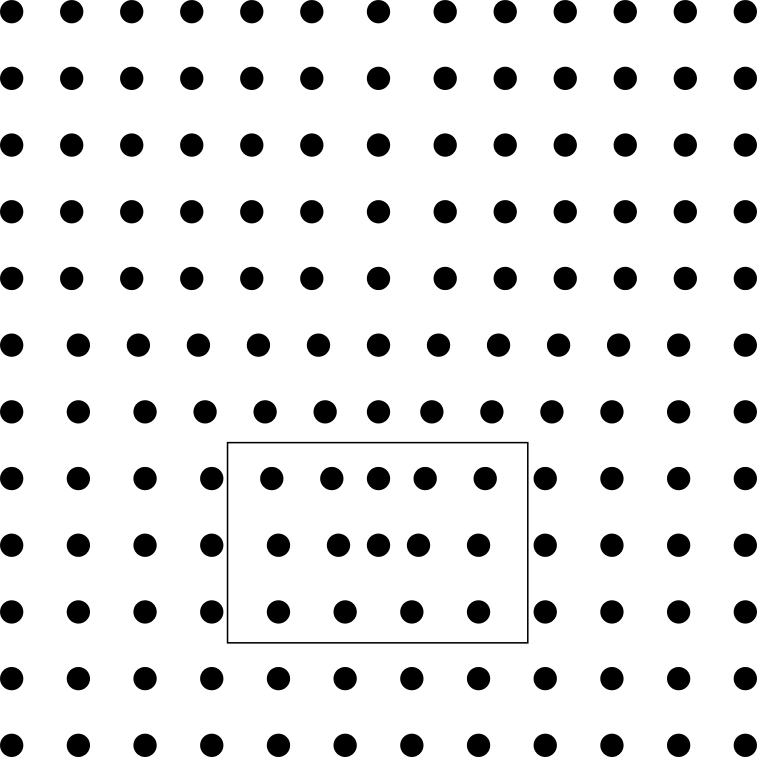}
\end{tabular}
 \caption{Example of defects in a crystal are interstitials/vacancies (on the left) and dislocations (on the right)}\label{5fig:defects}
\end{figure}

The continuum limit is obtained as we take the lattice spacing to 0 while keeping the matter density constant and also the density of defects constant. In a crystal, at each point (atom) there is a vector space defined by the three crystallographic directions. For a cubic ideal crystal, the three crystallographic directions $\{ e_a \}$ match the directions defined by a Cartesian coordinate system $\{\frac{\partial}{\partial x^i} \}$. The distance between two points in the same direction is measured by lattice step counting. If we consider two arbitrary points in a neighbourhood separated by $\df x$, the distance is given by the metric $\df s^2 = \delta_{ij} \df x^i \df x^j$. The parallel transport of a vector $v$ is introduced considering that $v$ is parallel transported if $\partial_i v^j =0$.

A perfect crystal is a deformation of an ideal crystal, and so, the crystallographic directions will be related to the Cartesian basis vectors through a transformation $e_a = A^i{}_a \frac{\partial}{\partial x^i}$. Since the perfect crystal has no defects, the matrix $A$ can be written in terms of a coordinate transformation $A^i{}_a=\frac{\partial x^i}{\partial \tilde{x}^a}$. The metric that describes our crystal now can be written as $g_{ab}=A^i{}_a A^j{}_b \delta_{ij}$. The condition of parallelism now becomes $\partial_{c} v^a = \Gamma^a_{cb} v^b$ where $\Gamma^a_{cb} = (A^{-1})^a{}_l \partial_c A^l{}_b$ is the lattice connection. Two vectors are parallel according to the lattice connection if they were parallel in the ideal crystal, before deformation. 


The description of a perfect crystal is just taking the euclidean metric and writing it in the basis of the crystallographic directions. This is because the step counting procedure is equivalent to the euclidean distance. However, in the case of a crystal with defects, interstitials and vacancies break the step counting procedure, and it is not possible to work out a metric in the same way as before. Besides, dislocations makes the matrix $A^i{}_a$, that relates the crystallographic directions to the Cartesian directions, no longer derivable from a coordinate transformation. Therefore, the lattice connection will present torsion $S_{cb}{}^a = (A^{-1})^a{}_l \partial_{[c} A^l{}_{b]}$, if there is a non-vanishing density of dislocations.

To understand the effect of point-like defects we can construct an auxiliary metric $h$, which is compatible with the connection $\Gamma$, if we use the step counting procedure as if no point defects were present (removing the interstitials and filling the vacancies). The metric that describes the crystal will be related to the auxiliary one through some transformation. If $N^+$ and $N^-$ are the densities of interstitials and vacancies, and these defects are isotropic, both metrics are related as $g=(1-N^-+N^+)^2 h$. Since $h$ is parallel transported with $\Gamma$, ${}^\Gamma \nabla h=0$,  then  $g$ is not, ${}^\Gamma \nabla g \neq 0$. Point-like defects generate non-metricity.

We can see the similarities with the metric-affine approach in gravity. The crystallographic defects play the role of the matter stress-energy density in the space-time. When there are no defects, the geometry is effectively Riemannian; this would also be the case of the space-time when the matter content of the universe is vacuum. If there are defects in the crystal or matter in the space-time, the geometry becomes non-metric, but in both cases the parallel transport is defined by an auxiliary metric. This auxiliary metric is related to the physical one through a matrix that depends on the defects/matter (see eq. \ref{3eq:hg2}), becoming the identity for an ideal crystal/vacuum. The theory of defects plays a fundamental role in the physics of crystalline solids, and although defects seem to be objects that play a role in the micro-scale, they give raise to global properties such as plasticity, viscosity and viscoelasticity. In the same spirit, Metric-Affine theories might be fundamental not only in the description of the high curvature regions of space-time, but also in the cosmological properties of our universe. 

%
%

\cleardoublepage
\chapter{Geonic Wormhole}\label{SolChap}

%

In this chapter, we will see how to get black hole solutions of the space-time in the Metric-Affine framework. As we discussed in the last chapter, to obtain solutions different from GR we need two things: an energy-momentum tensor different from vacuum, and a gravity Lagrangian different from the GR Lagrangian, $R$. For the energy-momentum tensor, we will study the case of a spherically symmetric electrovacuum field. In this way we will obtain charged black hole solutions that can be compared to charged black holes in GR, which are given by the Reissner-Nordström metric. For the gravity Lagrangian, we will consider a general Lagrangian that depends on the curvature scalars $R$, $Q\equiv R_{\alpha \beta} R^{\alpha \beta}$. Then we will choose a particular quadratic Lagrangian, $\Lagr_G = R+l_P^2(a R^2 +Q)$, and study the solutions obtained from it.

We will see that this new charged black hole solutions are really similar to the GR solution, but instead of a central singularity, they have a wormhole structure near the centre. Besides, they have no sources that generate the charges that characterize the solutions. The charge arises as a topological magnitude associated to the electric flux that passes through the geometry. This makes these solutions to be in agreement with Wheeler's definition of \emph{geons}. 

\section{General Method for Solving a Space-time with a $f(R,Q)$ Action and Spherical Symmetry}

We start from the action:

\begin{equation}
 S=\frac{1}{16 \pi l_P^2} \int f(R,Q) \sqrt{-g} \df^4 x + \int \Lagr_m \sqrt{-g} \df^4x
\end{equation}
where $\Lagr_m$ is the matter Lagrangian. Imposing that the variation of the action with respect to the metric and the connection is zero, we obtain the equations of motion:

\begin{eqnarray}
 (\partial_R f)  R_{\mu \nu} - \frac{f}{2} g_{\mu \nu} + 2 (\partial_Q f) R_{\mu \alpha} R_{\beta \nu} g^{\alpha \beta}&=& 8 \pi l_P^2 T_{\mu \nu} \label{4eq:metric}\\
 \nabla_\lambda \left [ \sqrt{-g}((\partial_R f) g^{\mu \nu} + 2 (\partial_Q f) R_{\alpha \beta}g^{\mu \alpha}g^{\nu \beta} ) \right ] &=& 0 \label{4eq:connection}
\end{eqnarray}
where $T_{\mu \nu} = \frac{2}{\sqrt{-g}} \frac{\delta \Lagr\sqrt{-g}}{\delta g^{\mu \nu}}$ is the energy-momentum tensor. It is assumed that the matter Lagrangian does not depend on the independent connection, and that torsion vanishes. We will also consider the Ricci tensor $R_{\mu \nu}$ to be symmetric\footnote{As we saw in the last chapter, there are two approaches to vanishing torsion, we can consider vanishing torsion \emph{a priori}, before performing the variation of the action, in which the antisymmetric part of $R_{\mu \nu}$ describes a vectorial degree of freedom, akin to a Proca field, in which we are not interested and can disregard; or we can consider vanishing torsion \emph{a posteriori}, in which we are studying the particular solutions of the theory in which torsion vanishes, which makes the Ricci tensor naturally symmetric.}.

This system of equations is apparently very complicated. We have to solve for the metric $g_{\mu \nu}$ and the independent connection $\Gamma^\alpha_{\beta \gamma}$, but they appear mixed in the equations. The procedure to solve this equations will be the following \cite{Olmo:2012nx}:

\begin{itemize}
 \item First, we will define a mixed object $P^\mu{}_\nu \equiv g^{\mu \alpha} R_{\alpha \nu}$ that depends both on the metric and the connection. The trace of $P^\mu{}_\nu$, and the trace of $P^\mu{}_\alpha P^\alpha{}_\nu$ gives us $R$ and $Q$ respectively. With this object, eqs. \ref{4eq:metric} can be written as an algebraic matrix equation on $P^\mu{}_\nu$, which can be solved, and from it, obtain $R$ and $Q$.
 \item Second, we will use eqs. \ref{4eq:connection} to define an auxiliary metric $h_{\mu \nu}$ so that the independent connection is the Levi-Civita connection of this metric. The Ricci tensor $R_{\mu \nu}$ can be written either in terms of first derivatives of $\Gamma^\alpha_{\beta \gamma}$, or second derivatives of $h_{\mu \nu}$.
 \item The metric $g_{\mu \nu}$ is related to $h_{\mu\nu}$ through a transformation matrix $\Sigma^\mu{}_\nu$. This transformation depends on $P^\mu{}_\nu$, $R$ and $Q$, but those are already known. With this transformation, we can write eqs. \ref{4eq:metric} only in terms of $h$ and its derivatives, instead of $(g,\Gamma)$.
 \item Now it is possible to integrate $h$ from those equations, and from there, get the metric $g$ and the connection $\Gamma$.
\end{itemize}

Let us follow this procedure as much as we can without specifying a particular Lagrangian and energy-momentum tensor. The first step is to rewrite eq. \ref{4eq:metric} using the mixed object $P^\mu{}_\nu \equiv g^{\mu \alpha} R_{\alpha \nu}$:

\begin{equation}
 2(\partial_Q f)P^\mu{}_\alpha P^\alpha{}_\nu +(\partial_R f)P^\mu{}_\nu - \frac{f}{2} \delta^\mu{}_\nu = 8 \pi l_P^2 T^\mu{}_\nu
\end{equation}

This is an algebraic equation on $P^\mu{}_\nu$ of second degree. To obtain $P^\mu{}_\nu$, first we have to complete the square:

\begin{eqnarray}
 \left ( P^\mu{}_\alpha + \frac{\partial_R f}{ 4 \partial_Q f} \delta^\mu{}_\alpha \right )\left ( P^\alpha{}_\nu + \frac{\partial_R f}{ 4 \partial_Q f} \delta^\alpha{}_\nu \right )=\frac{1}{2 \partial_Q f} \left ( \left ( \frac{f}{2}+\frac{(\partial_R f)^2}{8 \partial_Q f}\right ) \delta^\mu{}_\nu + 8 \pi l_P^2 T^\mu{}_\nu \right ) \label{4eq:metricsquares}
\end{eqnarray}

From this equation, we would like to say that $\left ( P^\mu{}_\alpha + \frac{\partial_R f}{ 4 \partial_Q f} \delta^\mu{}_\alpha \right )$ is the ``square root'' of the right hand side. But this is a matrix equation, and it has multiple square roots. Agreement in the low curvature regime should select the correct one. Then we will obtain an equation that gives us $P^\mu{}_\nu$ in terms of $T^\mu{}_\nu$, $R$, and $Q$. The trace of this new equation and the trace of its square gives a system of equations that depend only in $R$, $Q$ and $T^\mu{}_\nu$. Known $T^\mu{}_\nu$ gives us the value of the curvature scalars, and hence, also of $P^\mu{}_\nu$.
%
%

The next step is to construct an auxiliary metric $h$ so that the independent connection is the Levi-Civita connection of this metric. This implies that $h$ must fulfil:

\begin{equation}
 \nabla_\lambda \left [ \sqrt{-h} h^{\mu \nu} \right ] = 0
\end{equation}

Comparing this equation to eq. \ref{4eq:connection}, we get:
\begin{equation}
 \sqrt{-h} h^{\mu \nu} =  \sqrt{-g}((\partial_R f) g^{\mu \nu} + 2 (\partial_Q f) R_{\alpha \beta}g^{\mu \alpha}g^{\nu \beta} ) = \sqrt{-g} g^{\mu \alpha} \Sigma_\alpha{}^\nu
\end{equation}

Where we have defined a matrix $\Sigma_\alpha{}^\nu$ that depends on already known quantities $P^\mu{}_\nu$, $R$ and $Q$:

\begin{equation}
 \Sigma_\alpha{}^\nu \equiv (\partial_R f) \delta_\alpha^\nu + 2 (\partial_Q f)P^\nu{}_\alpha
\end{equation}

This matrix gives us the relation between $h$ and $g$:

\begin{equation}
 (h^{-1})^{\mu \nu} \equiv h^{\mu\nu}= \frac{g^{\mu\alpha} \Sigma_\alpha{}^\nu}{\sqrt{\det \Sigma}} \qquad h_{\mu \nu}=  \sqrt{\det \Sigma} (\Sigma^{-1})^\alpha{}_\nu g_{\mu\alpha}\label{4eq:htog}
\end{equation}

Using this auxiliary metric, eq. \ref{4eq:metric} can be rewritten:

\begin{equation}
 h^{\mu \alpha} R_{\alpha \nu} = \frac{1}{\sqrt{\det \Sigma}} \left ( \frac{f}{2} \delta^\mu{}_\nu + 8 \pi l_P^2 T^\mu{}_\nu \right ) \label{4eq:metric3}
\end{equation}

We have disentangled the original equations. In this equation, the left hand side contains only the auxiliary metric $h$ and its derivatives. The right hand side contains only the functions $T^\mu{}_\nu$, $R$, $Q$ and $P^\mu{}_\nu$, which we already solved algebraically. It is possible to integrate this equation to obtain $h$ and then use $\Sigma$ to get the metric $g$.

\subsection{Spherically Symmetric Electrovacuum Field}

In order to obtain charged solutions in the spirit of the Reissner-Nordström black hole, we must introduce the electromagnetic sector in the matter action, and consider no sources\footnote{Except perhaps in the central region of the geometry, like in the singularity of the Reissner-Nordström black hole.}. The action for the electromagnetic sector and its energy-momentum tensor are:

\begin{equation}
 S_m = -\frac{1}{16 \pi l_P^2} \int F_{\alpha \beta} F^{\alpha \beta} \sqrt{-g} \df^4 x  \label{4eq:EM-action}
\end{equation}
\begin{equation}
 T_\mu{}^\nu = -\frac{1}{4\pi} \left ( F_\mu{}^\alpha F_\alpha{}^\nu-\frac{F_\alpha{}^\beta F_\beta{}^\alpha}{4}\delta_\mu{}^\nu \right )
\end{equation}

where $F_{\mu \nu}=(\df A)_{\mu \nu} = \partial_\mu A_\nu - \partial_\nu A_\mu$ is the electromagnetic field and $A$ is the potential. The sourceless equations of motion are:

\begin{eqnarray}
 \df F&=&0 \label{4eq:EMmov1}\\
 \df (*F)&=&0 \qquad \Rightarrow \qquad \nabla_\mu F^{\mu \nu}=0 \label{4eq:EMmov2}
\end{eqnarray}

Since we are considering a static and spherically symmetric solution, the metric can be written as $\df s^2 = g_{tt}(r) \df t^2 + g_{rr}(r) \df r^2 + r^2 \df \Omega^2$, and the components of the field $F_{\mu \nu}$ depend only on the coordinate $r$. With this information, eq. \ref{4eq:EMmov2} can be integrated and the only non-zero component of the field strength tensor is $F^{tr}$: 

\begin{equation}
 F^{tr} = \frac{q}{r^2} \frac{1}{\sqrt{-g_{tt} g_{rr}}} = -F^{rt} \label{4eq:EMflux}
\end{equation}

Where $q$ is an integration constant that corresponds to the charge measured by computing the electric flux that passes through a surface that encloses the centre of the geometry $\int_S *F = 4 \pi q$. Now, the corresponding energy-momentum tensor in spherical coordinates $(t,r,\theta,\phi)$ has this simple form:

\begin{equation}
 T_\mu{}^\nu = \frac{q^2}{8\pi r^4}\left ( \begin{array}{cccc} -1 & 0 & 0 & 0 \\ 0 & -1 & 0 & 0 \\ 0 & 0 & 1 & 0 \\ 0 & 0 & 0& 1 \end{array} \right ) \label{4eq:EMVacuumT}
\end{equation}

\subsection{Charged Black Hole for a Generic $\Lagr_G=f(R,Q)$}

Now that we have the energy-momentum tensor for a spherically symmetric electrovacuum field, we can continue constructing the solution. Substituting $T^\mu{}_\nu$ into eq. \ref{4eq:metricsquares} we obtain:

\begin{equation}
\left ( P^\alpha{}_\nu + \frac{\partial_R f}{ 4 \partial_Q f} \delta^\alpha{}_\nu \right )^2 =\frac{1}{2 \partial_Q f} \left ( \begin{array}{cccc} \lambda_-^2 & 0 & 0 & 0 \\ 0 & \lambda_-^2 & 0 & 0 \\ 0 & 0 & \lambda_+^2 & 0 \\ 0 & 0 & 0& \lambda_+^2 \end{array} \right ) 
\end{equation}

Where we have defined:

\begin{equation}
 \lambda_\pm^2\equiv\frac{f}{2}+\frac{(\partial_R f)^2}{8 \partial_Q f}\pm\frac{l_P^2 q^2}{r^4} \label{4eq:lambdadefinition}
\end{equation}

We want to solve this equation for $P^\alpha{}_\nu$. To do so, we need to take the square root of the right hand side, but it has multiple solutions of the form:

\begin{equation}
 \left ( P^\alpha{}_\nu + \frac{\partial_R f}{ 4 \partial_Q f} \delta^\alpha{}_\nu \right ) = \frac{1}{\sqrt{2 \partial_Q f}} \left ( \begin{array}{cc} \pm \lambda_- \hat{U} & 0 \\ 0 & \pm \lambda_+ \hat{U} \end{array} \right )   \label{4eq:MSqrt}
\end{equation} 
where $\hat{U}$ is a 2x2 matrix with value:
\begin{equation}
 \hat{U}= \left ( \begin{array}{cc} 1 & 0 \\ 0 & 1 \end{array} \right ) \text{ or } \left ( \begin{array}{cc} \cos{\alpha} & \sin{\alpha} \\ \sin{\alpha} & -\cos{\alpha}\end{array} \right )
\end{equation}

In order to see which one of these solutions corresponds to $\left ( P^\alpha{}_\nu + \frac{\partial_R f}{ 4 \partial_Q f} \delta^\alpha{}_\nu \right )$ we have to look at the low curvature regime. We expect that far away from the central region, as the radius tends to infinity, the curvature will tend to $0$ and the gravity Lagrangian will tend to the GR one. These conditions can be summed up in:
\begin{equation}
 r\rightarrow \infty \qquad R_{\mu \nu} \rightarrow 0 \qquad f \approx R \qquad \partial_R f \approx 1
\end{equation}
In this regime, the left hand side of eq. \ref{4eq:MSqrt} and $\lambda_\pm$ are\footnote{We are considering $f(R,Q)$ with second-order or higher corrections to the GR action; therefore $\frac{1}{\partial_Q f}$ dominates over $R$.}:
\begin{equation}
 \left ( P^\alpha{}_\nu + \frac{\partial_R f}{ 4 \partial_Q f} \delta^\alpha{}_\nu \right )\approx \frac{1}{ 4 \partial_Q f} \delta^\alpha{}_\nu \qquad \lambda_\pm \approx \frac{1}{\sqrt{8 \partial_Q f}}
\end{equation}
which leads to:
\begin{equation}
 \frac{1}{ 4 \partial_Q f} \delta^\alpha{}_\nu = \frac{1}{ 4 \partial_Q f} \left ( \begin{array}{cc} \pm \hat{U} & 0 \\ 0 & \pm  \hat{U} \end{array} \right )
\end{equation}


For this equation to hold, we have to take both signs positive and $\hat{U}$ must be the $2\times2$ identity. This selects the correct way to take the square root, which is:

\begin{equation}
 \left ( P^\alpha{}_\nu + \frac{\partial_R f}{ 4 \partial_Q f} \delta^\alpha{}_\nu \right ) =\frac{1}{\sqrt{2\partial_Q f}} \left ( \begin{array}{cccc} \lambda_- & 0 & 0 & 0 \\ 0 & \lambda_- & 0 & 0 \\ 0 & 0 & \lambda_+ & 0 \\ 0 & 0 & 0& \lambda_+ \end{array} \right ) 
\end{equation}

Obtaining the tensor $P^\alpha{}_\nu$ is straightforward:

\begin{equation}
P^\alpha{}_\nu =\left ( \begin{array}{cccc}  
\frac{\lambda_-}{\sqrt{2\partial_Qf}}-\frac{\partial_R f}{4\partial_Q f} & 0 & 0 & 0 \\ 
0 & \frac{\lambda_-}{\sqrt{2\partial_Qf}}-\frac{\partial_R f}{4\partial_Q f} & 0 & 0 \\ 
0 & 0 & \frac{\lambda_+}{\sqrt{2\partial_Qf}}-\frac{\partial_R f}{4\partial_Q f} & 0 \\ 
0 & 0 & 0& \frac{\lambda_+}{\sqrt{2\partial_Qf}}-\frac{\partial_R f}{4\partial_Q f} \end{array} \right ) 
\end{equation}

Taking the trace of this object and its square, we obtain the following system of two equations for $R$ and $Q$:

\begin{IEEEeqnarray}{rCl}
R&=&\frac{2}{\sqrt{2\partial_Qf}}(\lambda_-+ \lambda_+)-\frac{\partial_R f}{\partial_Q f} \label{4eq:RQsystem}\\
Q &=& \frac{1}{\partial_Q f} \left ( \lambda_-^2+\lambda_+^2\right )+ \frac{(\partial_R f)^2}{4 (\partial_Q f)^2} -(\lambda_++\lambda_-)\frac{\partial_R f}{\sqrt{2}(\partial_Q f)^\frac{3}{2}} 
\end{IEEEeqnarray}

From which $R$ and $Q$ can be obtained. Now we can write the matrix $\Sigma$ that gives us the relation between the auxiliary metric $h$ and $g$:

\begin{equation}
 \Sigma_\alpha{}^\nu = (\partial_R f) \delta_\alpha^\nu + 2 (\partial_Q f)  P^\nu{}_\alpha =\left ( \begin{array}{cccc}  
 \sigma_- & 0 & 0 & 0 \\ 
 0 & \sigma_- & 0 & 0 \\ 
 0 & 0 & \sigma_+ & 0 \\ 
 0 & 0 & 0& \sigma_+ \end{array} \right )
\end{equation}

where we have defined:

\begin{equation}
 \sigma_\pm\equiv\lambda_\pm \sqrt{2\partial_Q f} + \frac{\partial_R f}{2}
\end{equation}

The determinant of $\Sigma$ takes a simple form:

\begin{equation}
 \det{\Sigma}=\sigma_+^2\sigma_-^2
\end{equation}


\subsection{Solutions for Quadratic Gravity}

We have reached the point where we have to specify a particular $f(R,Q)$ Lagrangian to advance. In chapter \ref{Int2} we introduced quadratic gravity, as a way to have a renormalizable quantum theory of gravity, although it was ultimately insufficient because the resulting theory suffered from ghost instabilities (in the Riemannian formalism). In any case, that theory could be seen as an effective theory at low energies, and it would be natural that the lowest order corrections of most extensions of GR have quadratic terms in the curvature. We can try this Lagrangian, but for the Metric-Affine formalism. The Lagrangian would be the GR Lagrangian plus quadratic terms that depend linearly on $R^2$, $Q=R_{\alpha \beta}R^{\alpha \beta}$ and $K=R_{\alpha \beta \mu \nu} R^{\alpha \beta \mu \nu}$. It is possible to remove one of the curvature scalars by noticing that the integral of the quantity $K-4Q+R^2$, called Gauss-Bonnet term, is a topological invariant of the space-time, and does not affect the equations of motion. We choose to remove the dependency on $K$, and have our Lagrangian to depend only on $R^2$ and $Q$. We can also realize that, because the electromagnetic energy-momentum tensor is traceless, the scalar curvature vanishes, and therefore the black hole solutions will be the same no matter which coefficient goes with the $R^2$ term. With these considerations in mind, we propose the following Lagrangian $\Lagr_G=R+l_P^2(aR^2+Q)$, where $a$ has not been specified. If we substitute it into eqs. \ref{4eq:lambdadefinition} and \ref{4eq:RQsystem}, we get

\begin{equation}
\lambda_\pm = \frac{\sqrt{2}}{l_P}\left ( \frac{1}{4} \pm \frac{l_P^4 q^2}{r^4}  \right ) \qquad R=0 \qquad Q=\frac{4 l_P^4 q^4}{r^8} \label{4eq:RQinvariants}
\end{equation}

%
%

The components of the matrix $\Sigma$ become:

\begin{equation}
 \sigma_\pm = 1 \pm \frac{2 l_P^4 q^2}{r^4}
\end{equation}


which take a very convenient form, since we can express now Eq. \ref{4eq:metric3} as:

\begin{equation}
 h^{\mu \alpha} R_{\alpha \nu} = \frac{q^2 l_P^2}{r^4} \left ( \begin{array}{cccc}  
 -\frac{1}{\sigma_+} & 0 & 0 & 0 \\ 
 0 & -\frac{1}{\sigma_+} & 0 & 0 \\ 
 0 & 0 & \frac{1}{\sigma_-}& 0 \\ 
 0 & 0 & 0& \frac{1}{\sigma_-} \end{array} \right ) \label{4eq:hdiffeq}
\end{equation}

where on the left hand side, there is only dependence on $h$ and its derivatives, and the right hand side is completely known. It is time to choose coordinates and solve these equations. In spherical coordinates, we can write the metric $g$ as:

\begin{equation}
 g = g_{tt} \df t^2 + g_{rr} \df r^2 + r^2 \df \Omega^2 
\end{equation}

We recall from eq. \ref{4eq:htog} that $h^{-1}=\frac{\Sigma g^{-1}}{\sqrt{\det{\Sigma}}}$. So, in these coordinates, $h$ is expressed as:

\begin{equation}
 h = h_{tt} \df t^2 + h_{rr} \df r^2 + \tilde{r}^2(r) \df \Omega^2
\end{equation}

with
\begin{equation}
 g_{tt}=\frac{h_{tt}}{\sigma_+} \qquad g_{rr} = \frac{h_{rr}}{\sigma_+} \qquad r^2=\frac{\tilde{r}^2}{\sigma_-} \label{4eq:htogcomponents}
\end{equation}
%
%
%

However, it is more convenient to choose a set of coordinates in which the metric $h$ is written as:

\begin{equation}
 h = -A(x) \df t^2 + \frac{1}{A(x)} \df x^2 + \tilde{r}^2(x) \df \Omega^2 
\end{equation}
and then transform to the usual $(t,r,\theta,\phi)$ coordinates. First thing to note is that $x$ can be expressed just as a function of $r$. Second, it is that the expression of $T^\mu{}_\nu$ using the coordinate $x$ is the same as using the coordinate $r$ (eq. \ref{4eq:EMVacuumT}), the same happens for eq. \ref{4eq:hdiffeq}. In this set of coordinates, the components of the tensor $h^{\mu \alpha} R_{\alpha \nu}$ are:

\begin{IEEEeqnarray}{rClCl}
 h^{t \alpha} R_{\alpha t} &=& - \frac{\tilde{r} (\partial_x \partial_x A) + 2 (\partial_x \tilde{r} )(\partial_x A)}{2\tilde{r}}&=&-\frac{q^2 l_P^2}{r^4} \frac{1}{\sigma_+} \IEEEeqnarraynumspace \label{4eq:Rtt}\\
 h^{r \alpha} R_{\alpha r} &=& - \frac{\tilde{r} (\partial_x \partial_x A) + 2 (\partial_x \tilde{r} )(\partial_x A) + 4 A (\partial_x \partial_x \tilde{r})}{2\tilde{r}} &=&-\frac{q^2 l_P^2}{r^4} \frac{1}{\sigma_+}\IEEEeqnarraynumspace  \label{4eq:Rrr}\\
 h^{\theta \alpha} R_{\alpha \theta} = h^{\phi \alpha} R_{\alpha \phi} &=& - \frac{\tilde{r} (\partial_x \tilde{r} ) (\partial_x A) + \tilde{r} A (\partial_x \partial_x \tilde{r}) + A (\partial_x \tilde{r})^2-1}{\tilde{r}^2}&=&\frac{q^2 l_P^2}{r^4} \frac{1}{\sigma_-}\IEEEeqnarraynumspace  \label{4eq:Rphiphi}
\end{IEEEeqnarray}

Subtracting equation \ref{4eq:Rtt} from \ref{4eq:Rrr}, we get $\partial_x \partial_x\tilde{r} =0$, which implies $\tilde{r}=k x$, where $k$ is a constant. It is possible to absorb this constant into a coordinate redefinition $x' = \sqrt{k}x$, $t'= t/\sqrt{k}$, $A' = kA$, so that $\tilde{r}=x$. With this and eq. \ref{4eq:htogcomponents}, the relation between the coordinate $r$ and $x$ is known:

\begin{equation}
x=r \sqrt{\sigma_-} \qquad \df x= \frac{\sigma_+}{\sqrt{\sigma_-}} \df r \label{4eq:dxtodr}
\end{equation}

The function $A$ is the only piece of information left to know of $h$. Now that $\tilde{r}$ is known, eq. \ref{4eq:Rphiphi} is a differential equation for $A$:

\begin{equation}
  1 - A - x (\partial_x A)=\frac{q^2 l_P^2}{r^2}
\end{equation}

We make the ansatz $A=1-\frac{2M(x)}{x}$, which results in:

\begin{equation}
   \partial_x M = \frac{q^2 l_P^2}{2r^2} 
\end{equation}

In terms of the coordinate $r$:

\begin{equation}
 \partial_r M = \frac{q^2 l_P^2}{2r^2} \frac{\partial x}{\partial r}  = \frac{q^2 l_P^2}{2r^2} \frac{\sigma_+}{\sqrt{\sigma_-}}
\end{equation}

This expression can be directly integrated. Before doing so, it is useful to define the charge radius $r_q$ and a critical radius $r_c$:

\begin{equation}
 r_q \equiv l_P q \qquad r_c \equiv \sqrt{\sqrt{2} l_P r_q} \qquad \sigma_\pm = 1 \pm \frac{r_c^4}{r^4}\label{4eq:sigmadef}
\end{equation}

The integral gives\footnote{It is possible to do this integral in several ways, and the result can be written in different but equivalent expressions. The result presented in the text has been chosen to be easy to compare to the Reissner-Nordström solution of GR. In order to do the integral, it is better to use a normalized coordinate $z=r/r_c$, so that:
\begin{IEEEeqnarray}{rCl}
\partial_z M &=& \frac{r_q^2}{2 r_c} \frac{z^4+1}{z^4\sqrt{z^4-1}}\\
M(z)&=& M_0 + \frac{r_q^2}{2r_c} \int^{\infty}_{z} \frac{z^{\prime 4}+1}{z^{\prime 4}\sqrt{z^{\prime 4}-1}} \df z^\prime \label{4eq:IntegralG}
\end{IEEEeqnarray}
The limits of integrations are chosen so that $\lim_{z\rightarrow \infty} M = M_0$. Making the indefinite integral with \emph{Mathematica} and then selecting the appropriate integration constant gives:
\begin{IEEEeqnarray}{rCl}
M&=& M_0 + \frac{r_q^2}{2 r_c}  \frac{4 z^6 \, _2F_1\left(-\frac{3}{4},\frac{1}{2};\frac{1}{4};\frac{1}{z^4}\right)+\sqrt{z^4-1} \left(1-4 z^4\right)}{3 z^3} \\ 
&=& M_0+\frac{r_q^2}{2 r_c} \left \{-\frac{\sqrt{z^4-1}}{z^3}+ \frac{4 z^6 \, _2F_1\left(-\frac{3}{4},\frac{1}{2};\frac{1}{4};\frac{1}{z^4}\right)+\sqrt{z^4-1} \left(4-4 z^4\right)}{3 z^3} \right \} \\ 
&=& M_0-\frac{r_q^2 \sqrt{\sigma_-}}{2r}+ \frac{ 2 r_q^2 z^3}{3 r_c}\left \{ \, _2F_1\left(-\frac{3}{4},\frac{1}{2};\frac{1}{4};\frac{1}{z^4}\right)-(\sigma_-)^\frac{3}{2} \right \} 
\end{IEEEeqnarray}
Which is the result presented in the text. However, it is possible to use the binomial expansion of the integrand in eq. \ref{4eq:IntegralG} and integrate each of the terms separately to write the integral as:
\begin{equation}
 M=M_0 + \frac{r_c^3}{4 l_P^2}\left ( \frac{1}{2} \sqrt{1-\frac{1}{z^4}} z^2 \left(\, _2F_1\left(\frac{1}{2},\frac{3}{4};\frac{3}{2};1-z^4\right)+\, _2F_1\left(\frac{1}{2},\frac{7}{4};\frac{3}{2};1-z^4\right)\right)-\frac{\sqrt{2} \pi ^{3/2}}{3 \Gamma \left(\frac{3}{4}\right)^2} \right )
\end{equation}
which is the method found in \cite{Olmo:2012nx}.
}:

\begin{equation}
 M = M_0 - \frac{r_q^2 \sqrt{\sigma_-}}{2r}+\frac{r_q^2}{r_c}G \left (\frac{r}{r_c} \right )
\end{equation}

with

\begin{equation}
 G \left (\frac{r}{r_c} \right ) = \frac{2 r^3}{3r_c^3}\left [ {}_2F_1 \left ( -\frac{3}{4},\frac{1}{2};\frac{1}{4};\frac{r_c^4}{r^4} \right ) - (\sigma_-)^\frac{3}{2} \right ]
\end{equation}
where $M_0$ is an integration constant\footnote{That has dimensions of [length]. The mass of the geometry would be $M_0/l_P^2$ in mass units.} and ${}_2F_1 \left ( -\frac{3}{4},\frac{1}{2};\frac{1}{4};\frac{r_c^4}{r^4} \right )$ is an hypergeometric function. $G(r/r_c)$ goes to $0$ as $1/r^5$ for $r$ going to infinity,  and its value when $r= r_c$ is $\frac{2\sqrt{\pi }}{3}\frac{ \Gamma \left(\frac{1}{4}\right)}{\Gamma \left(-\frac{1}{4}\right)} \ \simeq -0.874019 $. If we call $r_S\equiv2M_0$, we can write function $A$ as:

\begin{equation}
 A=1-\frac{r_s}{x}+\frac{r_q^2}{r^2}+\frac{2 r_q^2}{ r_c}\frac{G(\frac{r}{r_c})}{x} \label{4eq:A}
\end{equation}

The components of the metric $h$ are in the $(t,r,\theta,\phi)$ coordinates are:

\begin{IEEEeqnarray}{rCl}
 h &=& -A \df t^2 + \frac{1}{A} \df x^2 + x^2 \df \Omega^2 \\
 &=& -A \df t^2 + \frac{1}{A}\frac{(\sigma_+)^2}{\sigma_-} \df r^2 + r^2 \sigma_- \df \Omega^2
\end{IEEEeqnarray}

And thus, we finally obtain the components of the metric $g$:

\begin{equation}
 g=-\frac{A}{\sigma_+} \df t^2 + \frac{1}{A}\frac{\sigma_+}{\sigma_-} \df r^2 + r^2 \df \Omega^2 \label{4eq:metricg}
\end{equation}

\section{Geometry of Solutions for Quadratic Gravity}

The geometry has been constructed adding quadratic corrections to the GR action, and letting the connection to be independent of the metric. In the low curvature regime, the quadratic corrections become negligible, and the independent connection becomes the Levi-Civita connection of the metric. In this regime the quadratic lagrangian tends to the GR gravity Lagrangian, and we expect our solution to recover the GR solution. However, these quadratic corrections become dominant for high curvatures, and as we approach the singularity in the GR solution, we expect to find a totally different picture in our geometry.

\subsection{Large $r$ limit}

The Reissner-Nordström metric (\cite{ANDP:ANDP19163550905}, \cite{Nordstrom:1918}), that describes a charged black hole in GR, takes the form:

\begin{equation}
 \df s^2=-F \df t^2 + \frac{1}{F} \df r^2 + r^2\df \Omega^2 \qquad \text{with} \qquad F=1-\frac{r_S}{r}+\frac{r_q^2}{r^2}
\end{equation}

If we compare $F$ to $g_{tt}=-\frac{A}{\sigma_+}$, we can see from eq. \ref{4eq:A} that the factor $A$ already takes a very similar form to $F$. However, it has two differences, as the factor that goes with the mass, $\frac{r_S}{x}$, is inversely proportional to $x$ instead of $r$, and there is an additional factor $\frac{2 r_q^2}{ r_c}\frac{G(\frac{r}{r_c})}{x}$. The conversion from the coordinate $x$ to $r$ is explicitly:

\begin{equation}
 x=r\sqrt{\sigma_-}= r \sqrt{1-\frac{r_c^4}{r^4}} \qquad r^2=\frac{x^2+\sqrt{x^4+4r_c^4}}{2} \label{4eq:xtor}
\end{equation}

We can see that for large radius, $x$ is almost equal to $r$ up to corrections of order $1/r^4$. Let us see what kind of corrections the metric has: The factor $\frac{2 r_q^2}{ r_c}\frac{G(\frac{r}{r_c})}{x}$ is a correction of order $1/r^6$, and from the definition, $\sigma_\pm = 1 + O\left ( \frac{1}{r^4} \right )$. The metric components in the $r\rightarrow\infty$ limit are:

\begin{equation}
 g_{tt} = -\left (1-\frac{r_s}{r}+\frac{r_q^2}{r^2}-\frac{r_c^4}{r^4}+O\left ( \frac{r_c^5}{r^5} \right ) \right )
\end{equation}
\begin{equation}
 g_{rr} =  \left (1-\frac{r_s}{r}+\frac{r_q^2}{r^2}-\frac{2 r_c^4}{r^4}+O\left ( \frac{r_c^5}{r^5} \right ) \right )^{-1}
\end{equation}

Which are corrections of order $1/r^4$ with respect to GR, which will be negligible for radius $r \gg r_c$. Given that $r_c$ is the geometric mean of the charge radius and the Planck length, it will be extremely small respect to the Schwarzschild radius in a typical astronomical black hole. Let us see what kind of correction the curvature invariants get. For the Reissner-Nordström solution metric, the curvature scalars are:

\begin{equation}
 R_{GR}=0 \qquad (R_{\mu \nu} R^{\mu \nu})_{GR}=\frac{4 r_q^4}{r^8} \qquad (R^\alpha{}_{\beta \mu \nu} R_\alpha{}^{\beta \mu \nu})_{GR} = \frac{12 r_S^2}{r^6} - \frac{48 r_S r_q^2}{r^7} + \frac{56 r_q^4}{r^8}
\end{equation}

If we compute the curvature invariants of the metric of our charged black hole\footnote{Note that these are different from the curvature invariants of the independent connection, that appear in eq. \ref{4eq:RQinvariants}}, we find:

\begin{IEEEeqnarray}{rCl}
 R(g) &\approx& -\frac{48 r_c^8}{r^{10}}+... \\
 (R_{\mu \nu} R^{\mu \nu})(g) &\approx& \frac{4 r_q^4}{r^8}-64\frac{r_q^4 l_P^2}{r^{10}}+...\\
 (R^\alpha{}_{\beta \mu \nu} R_\alpha{}^{\beta \mu \nu})(g) &\approx& \frac{12 r_S^2}{r^6} - \frac{48 r_S r_q^2}{r^7} + \frac{14 r_q^4}{r^8} + \frac{144 r_S r_c^4}{r^9} + ...
\end{IEEEeqnarray}

As we can see, the curvature scalars only get small corrections respect to the GR solution and our charged black hole solution quickly converges to the Reissner-Nordström solution for $r\gg r_c$.

\subsection{$r\rightarrow r_c$ limit}

If we look at the geometry for small values of the radius, we start seeing differences with respect to GR. The radius $r=r_c$ is a critical value at which $\sigma_-=0$, $x=0$, $r_S/x \rightarrow \infty$. It seems that there will be a divergence in the components of the metric, unless there is a cancellation of some kind. In this section we will study the behaviour of the metric around $r_c$. Before expanding the metric around $r_c$, it will be useful to rewrite the parameters $r_q$, $r_S$, in terms of dimensionless ones, like the number of elementary charges $N_q$, and a dimensionless parameter $\delta_1$ related to the mass-charge ratio. It will be also useful to use a dimensionless coordinate $z=r/r_c$. With these changes we have:

\begin{equation}
 \delta_1 \equiv \frac{r_q^2}{r_S r_c} \qquad q = N_q e = N_q \sqrt{\alpha_{_{EM}}}\qquad \sigma_\pm = 1 \pm \frac{1}{z^4}
\end{equation}

These two dimensionless parameters have associated two critical values:

\begin{equation}
 \delta_c \equiv -\frac{1}{2 G(1)} =  -\frac{3}{4 \sqrt{\pi }}\frac{\Gamma \left(-\frac{1}{4}\right)}{ \Gamma \left(\frac{1}{4}\right)}  \simeq 0.57207 \qquad N_c \equiv \sqrt{\frac{2}{\alpha_{_{EM}}}} \simeq 16.55
\end{equation}
where $\alpha_{_{EM}}$ is the fine structure constant. With these definitions, we can expand the $g_{tt}$ component of the metric around $r_c$ ($z=1$):

\begin{IEEEeqnarray}{rCl}
 g_{tt} &\approx& \frac{N_q}{4N_c} \frac{\left (1-\frac{\delta_1}{\delta_c} \right )}{\delta_1} \left ( \frac{1}{\sqrt{z-1}} + \frac{9}{4}\sqrt{z-1} + \frac{29}{32} (z-1)^\frac{3}{2} + ... \right ) \\
 &&- \frac{1}{2} \left ( 1 - \frac{N_q}{N_c} \right ) - \left ( 1-\frac{2 N_q}{3N_c} \right ) (z-1)+...\nonumber
\end{IEEEeqnarray}

In general, the component $g_{tt}$ does indeed diverge as $r\rightarrow r_c$, but in a smoother way than in GR: $(r-r_c)^{-\frac{1}{2}}$ instead of $r^{-2}$. The sign of the divergence depends on $\delta_1$, whether it is greater or lesser than $\delta_c$. More surprisingly, if $\delta_1 = \delta_c$, there is no divergence at all! It is remarkable that a small variation in the charge-to-mass ratio of the black hole could change the nature of the black hole so much. Let us keep looking into the metric components, and then we will analyse in detail this behaviour. The expansion of $g_{rr}$ around $r_c$ is:

\begin{equation}
 g_{rr} \approx \frac{N_c}{N_q}\frac{\delta_1}{1-\frac{\delta_1}{\delta_c}}\frac{1}{\sqrt{z-1}}-2\frac{N_c}{N_q}\left ( 1- \frac{N_c}{N_q} \right ) \left (\frac{\delta_1}{1-\frac{\delta_1}{\delta_c}} \right )^2+ ... 
\end{equation}

This series will only converge for $(z-1) \ll \frac{(1-\frac{\delta_1}{\delta_c})}{\delta_1}$, and therefore it is unsuitable to study the case $\delta_1=\delta_c$. It is better to look at the expansion of $g_{rr}^{-1}$:

\begin{equation}
 g_{rr}^{-1} \approx -\frac{N_q}{N_c} \frac{1-\frac{\delta_1}{\delta_c}}{\delta_1}\sqrt{z-1}+2\left (1- \frac{N_q}{N_c} \right ) (z-1) + ...\label{4eq:grr-1expansion}
\end{equation}

We can see that $g_{rr}^{-1}=0$ at $r_c$ no matter the value of $\delta_1$. So $g_{rr}$ always diverges, even in the case the component $g_{tt}$ is completely regular. This is a hint that $r$ might not be a good coordinate to describe the central region. Actually, it would be possible to absorb the divergence of $g_{rr}$ into the definition of a new radial coordinate. A more rigorous approach would be to check if the curvature invariants diverge or not for $\delta_1=\delta_c$. If they do not, there should be a better coordinate to describe the metric $g$. The curvature invariants around $r=r_c$ can be expanded as:

\begin{IEEEeqnarray}{rCl}
 r_c^2 R(g) &=& \left ( \frac{16 N_q}{3 N_c}-4 \right ) + O(z-1) \\
 &&-\frac{N_q}{2N_c} \frac{ 1-\frac{\delta_1}{\delta_c}}{\delta_1} \left ( \frac{1}{(z-1)^\frac{3}{2}}+O\left (\frac{1}{(z-1)^\frac{1}{2}}\right )\right ) \nonumber
\end{IEEEeqnarray}



\begin{IEEEeqnarray}{rCl}
 r_c^4 (R_{\mu \nu} R^{\mu \nu})(g) &=& \left ( 10 + \frac{86 N_q^2}{9 N_c^2}-\frac{52N_q}{3N_c} \right ) + O(z-1) \\
 &&+\frac{N_q}{N_c}\left (\frac{1-\frac{\delta_1}{\delta_c}}{\delta_1} \right ) \left ( \frac{6 N_c - 5 N_q}{3N_c (z-1)^\frac{3}{2}} + O \left ( \frac{1}{\sqrt{z-1}} \right ) \right ) \nonumber \\
 &&+\frac{N_q^2}{N_c^2}\left (\frac{1-\frac{\delta_1}{\delta_c}}{\delta_1} \right )^2 \left ( \frac{1}{8(z-1)^3}+O\left ( \frac{1}{(z-1)^2} \right ) \right ) \nonumber
\end{IEEEeqnarray}


\begin{IEEEeqnarray}{rCl}
 r_c^4 (R^\alpha{}_{\beta \mu \nu} R_\alpha{}^{\beta \mu \nu})(g) &=& \left ( 16 + \frac{88 N_q^2}{9 N_c^2} -\frac{64N_q}{3N_c}\right ) + O(z-1) \\
 &&+\frac{N_q}{N_c}\left (\frac{1-\frac{\delta_1}{\delta_c}}{\delta_1} \right ) \left ( \frac{4 N_q - 6 N_c}{3N_c (z-1)^\frac{3}{2}} + O \left ( \frac{1}{\sqrt{z-1}} \right ) \right ) \nonumber \\
 &&+\frac{N_q^2}{N_c^2}\left (\frac{1-\frac{\delta_1}{\delta_c} }{\delta_1} \right)^2 \left ( \frac{1}{4(z-1)^3}+O\left ( \frac{1}{(z-1)^2} \right ) \right ) \nonumber
\end{IEEEeqnarray}

In general, the curvature invariants diverge at $r_c$, in a smoother way than GR. But when $\delta_1=\delta_c$, the curvature is finite, and the geometry is completely regular. This confirms that the divergence of $g_{rr}$ in the $\delta_1=\delta_c$ case is a consequence of an unsuitable choice of coordinates. In fact, the divergence of the component $g_{rr}$ can be traced back to the factor $\sigma_-^{-1}$ that appears in eq. \ref{4eq:metricg}. If we absorb this factor into a new coordinate, and we are in the case $\delta_1=\delta_c$, then the expression of $g$ would be completely regular.

\subsection{Coordinate Choices}

There are multiple ways to absorb the $\sigma_-^{-1}$ factor in the $g_{rr}$ component through a coordinate change. The most straightforward way would be to find a coordinate $y$ such that $\df y = \frac{1}{\sqrt{\sigma_-}} \df r$. Although this approach is perfectly valid, it leads to a complicated expression for $y(r)$. A more friendly way is to use again the coordinate $x$ that we introduce to calculate the auxiliary metric $h$. This coordinate has a very direct expression for $x(r)$. Let us recall:
\begin{equation}
x=r \sqrt{\sigma_-} = r \sqrt{1-\frac{r_c^4}{r^4}} \qquad \df x= \frac{\sigma_+}{\sqrt{\sigma_-}}  \df r \qquad r^2=\frac{x^2+\sqrt{x^4+4r_c^4}}{2}
\end{equation}

In this coordinates the metric $g$ is written:

\begin{equation}
 g=-\frac{A}{\sigma_+} \df t^2 + \frac{1}{A \sigma_+} \df x^2 + x^2 \sigma_- \df \Omega^2 \label{4eq:metricgx}
\end{equation}

The expansion of the component $g_{xx}$ is:

\begin{IEEEeqnarray}{rCl}
 g_{xx} &=&  -\frac{N_c}{N_q}\frac{\delta_1}{2\left ( 1-\frac{\delta_1}{\delta_c}\right )} \frac{|x|}{r_c}-\frac{N_c}{N_q}\frac{1-\frac{N_c}{2N_q}}{ \left ( \frac{1-\frac{\delta_1}{\delta_c}}{\delta_1} \right )^2}\frac{x^2}{r_c^2}+O(x^3)\\
 g_{xx}^{-1} &=& -2\frac{N_q}{N_c}\left ( \frac{1-\frac{\delta_1}{\delta_c}}{\delta_1} \right ) \frac{r_c}{|x|}+ 2\left ( 1 -\frac{N_q}{N_c} \right )+ \frac{N_q}{N_c}\left ( \frac{1-\frac{\delta_1}{\delta_c}}{\delta_1}\right )\frac{|x|}{r_c} + O(x^2)\IEEEeqnarraynumspace  \label{4eq:gxxexpansion}
\end{IEEEeqnarray}

Which is perfectly regular at $r=r_c$ if $\delta_1 = \delta_c$. It is easy to see why $r$ is not a good coordinate at $r_c$ if $x$ is a good one: $\left (\frac{\partial r}{\partial x}\right )=0$ at $r=r_c$, and the Jacobian of the transformation is degenerated. In these coordinates, the component $g_{tt}$ is expanded as:

\begin{equation}
 g_{tt} = \frac{1}{2}\frac{N_q}{N_c}\frac{\left ( 1-\frac{\delta_1}{\delta_c}\right )}{\delta_1} \frac{r_c}{x} - \frac{1}{2}\left ( 1-\frac{N_q}{N_c} \right ) + \frac{1}{4}\frac{N_q}{N_c}\frac{\left ( 1-\frac{\delta_1}{\delta_c}\right )}{\delta_1} \frac{x}{r_c}-\frac{1}{4}\left ( 1- \frac{2}{3}\frac{N_q}{N_c} \right ) \frac{x^2}{r_c^2} + O(x^3)
\end{equation}

\begin{figure}[h!]
 \includegraphics[width=.8\linewidth]{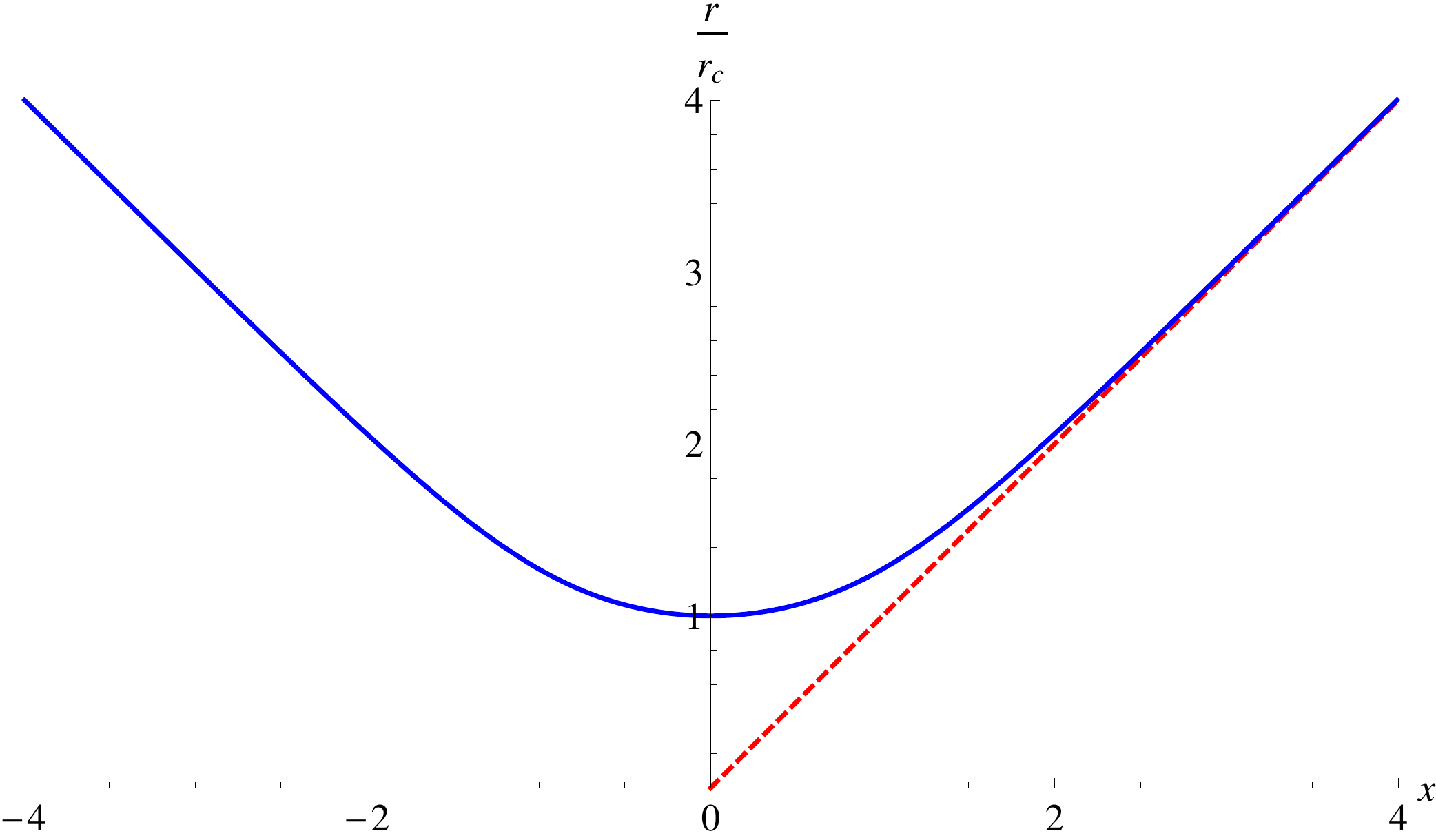}
 \caption{$r$ as a function of $x$ in units of $r_c$. The dotted line represents $r=x$.}\label{4fig:x-r}
\end{figure}

In fig. \ref{4fig:x-r} we have plotted $x(r)$. We can see that the coordinate $x$ is almost equal to $r$, except a few units of $r_c$ near the central region $x=0$  ($r=r_c$). At that point, the derivative of $r$ respect to $x$ is $0$, and it has a minimum value. $r$ is not a good coordinate any more, but $r^2(x)$ is still meaningful as the metric component of the angular sector of the metric, and as such, $4\pi r^2(x)$ is the area of the 2-spheres of constant $x$.

\FloatBarrier

\subsection{Beyond $r_c$, Geonic Wormhole}

Fig. \ref{4fig:x-r} gives us a hint about the topology of the solutions: The $r>r_c$ region corresponds to the $x>0$ region, but what about $x<0$? This region is not mapped by the $(t,r,\theta,\phi)$ coordinates. What happens to an inertial observer when it reaches the $x=0$ surface?

Let us consider the $\delta_1=\delta_c$ case. As we have seen in the previous section, in this case there is no divergence of curvature at $r=r_c$ ($x=0$). If we also have $N_q\neq N_c$, all the components of the metric are finite and non-zero at that point:

\begin{IEEEeqnarray}{rCl}
 g &\approx& -\left( \frac{1}{2} \left ( 1- \frac{N_q}{N_c}\right ) + \left ( 1- \frac{2}{3}\frac{N_q}{N_c} \right ) \frac{x^2}{4 r_c^2} + O(x^3) \right ) \df t^2  \\
 &&+ \left ( 2 \left ( 1- \frac{N_q}{N_c}\right ) +\left ( 1- \frac{2}{3}\frac{N_q}{N_c} \right ) \frac{x^2}{r_c^2}  + O(x^3) \right )^{-1} \df x^2 \nonumber\\
 &&+ \left ( r_c+ \frac{x^2}{4 r_c}  + O(x^3) \right )^2 \df \Omega \nonumber
\end{IEEEeqnarray}

It is natural that the path of the observer can be extended beyond $x=0$ to the realm of negative $x$. The observer would see how the radius of the 2-spheres reaches to a minimum, but beyond that point it starts growing again. As $x$ becomes more negative, the effective radius $r$ of the 2-sphere grows like $|x|$. For $x\rightarrow -\infty$, the geometry has an asymptotically flat region, a mirror image of the $x\rightarrow \infty$ region, which are connected at $x=0$ through a \emph{wormhole} of area $4 \pi r_c^2$. 


As the observer moves towards $x=0$, it feels an electric flux that comes from the centre. But what happens when it crosses to the other side? Where is the charge located in this geometry?. In this work, the charge $q$ has been defined as an integration constant in eq. \ref{4eq:EMflux}. This definition corresponds to the locally measured charge that can be obtained computing the electric flux $\Phi=\int_S *F = 4 \pi q$ that passes through a bidimensional surface $S$ that encloses the centre of the geometry. In GR, if $S$ is taken as 2-spheres of constant $r$ and we take $r\rightarrow 0$, the charge can be traced back to the central singularity. However, for the wormhole, if $S$ is taken as 2-spheres of constant $x$ and take $x\rightarrow -\infty$ we see that the charge is nowhere: the lines of force of the electric field enter from one side of the wormhole, and exit through the other side, creating the illusion of a positively charged object in one side and negatively charged one on the other\footnote{The change of sign of the locally measured charge is due to the fact that a positive orientation for the surface $S$ seen from one of the asymptotic flat regions looks like a negative orientation from the other side.}. The observer would feel the electric flux coming from the same direction all the time; only that it can be interpreted as coming from a positive (negative) charge as it moves towards the wormhole, or coming from a negative (positive) charge left behind after it crosses the wormhole.


The flux density crossing the wormhole throat is an universal quantity:
\begin{equation}
 \frac{\Phi}{4\pi r_c^2}=\frac{1}{2 l_P^2}
\end{equation}
this is true regardless of the mass, charge, or if the geometry is has a curvature divergence ($\delta_1 \neq \delta_c$). This suggests that the wormhole structure may not be a property only of the $\delta_1 = \delta_c$ case, but a general property of the black holes in this model. From the point of view of the equations, Maxwell's equations are solved everywhere, no matter the value of $\delta_1$, including the wormhole throat, which gives a finite energy-momentum tensor for all space-time. This is in contrast to the Reissner-Nordström solution of GR, in which the energy-momentum tensor is ill-defined at $r=0$, which also makes Einstein's equations meaningless at that region. In the metric-affine case, the equations for the metric and the connection show no problems at the wormhole throat. We are only left with studying the fate of observers crossing the wormhole throat when $\delta_1\neq \delta_c$, which contains curvature divergences. We will study the divergent case in the next chapter.

This geometry possesses a genuine wormhole structure supported by the electric field that passes through it. Unlike other known wormhole solutions that can be found in the literature (\cite{Morris:1988tu}, \cite{Lobo:2009ip}, \cite{Visser:1995cc}), this solution comes out naturally from the field equations and does not need exotic matter sources to generate a pre-designed geometry. We want to emphasize that the gravitational equations and the electromagnetic equations are satisfied everywhere, even in the cases with divergences of curvature. This wormhole is a sourceless gravitational-electromagnetic entity that is consistent with Wheeler's definition of \emph{geons} (\cite{Wheeler:1955zz}, \cite{Misner:1957mt}). These objects are particle-like in the classical sense, without the need of introducing singularities into the space-time. Moreover, they should be stable for topological reasons, and in case the geometry does not have horizons, they would not evaporate via Hawking radiation.

\subsection{Horizons and Conformal Diagrams of the Geonic Wormhole}

To get a global idea of the geometry it is useful to analyse if the geometry has horizons and see what is the nature of the hypersurface $x=0$. Horizons are located where the component $g_{tt}$ vanishes. For an astronomical black hole, the horizons in the GR solution are located much further than $r_c$, so no differences are expected in this geometry. That is not the case for a microscopic black hole. Depending on the value of $\delta_1$ and $N_q$ the cases are:

\begin{itemize}
 \item $\delta_1 < \delta_c$: In this case, there is a single event horizon on each side of the wormhole for any value of $N_q$. The hypersurface $x=0$ is space-like. We will call this case \emph{Schwarzschild-like} for its similarities with the GR solution.
 \begin{figure}[h!]
  \centering
  \includegraphics[width=.7\linewidth]{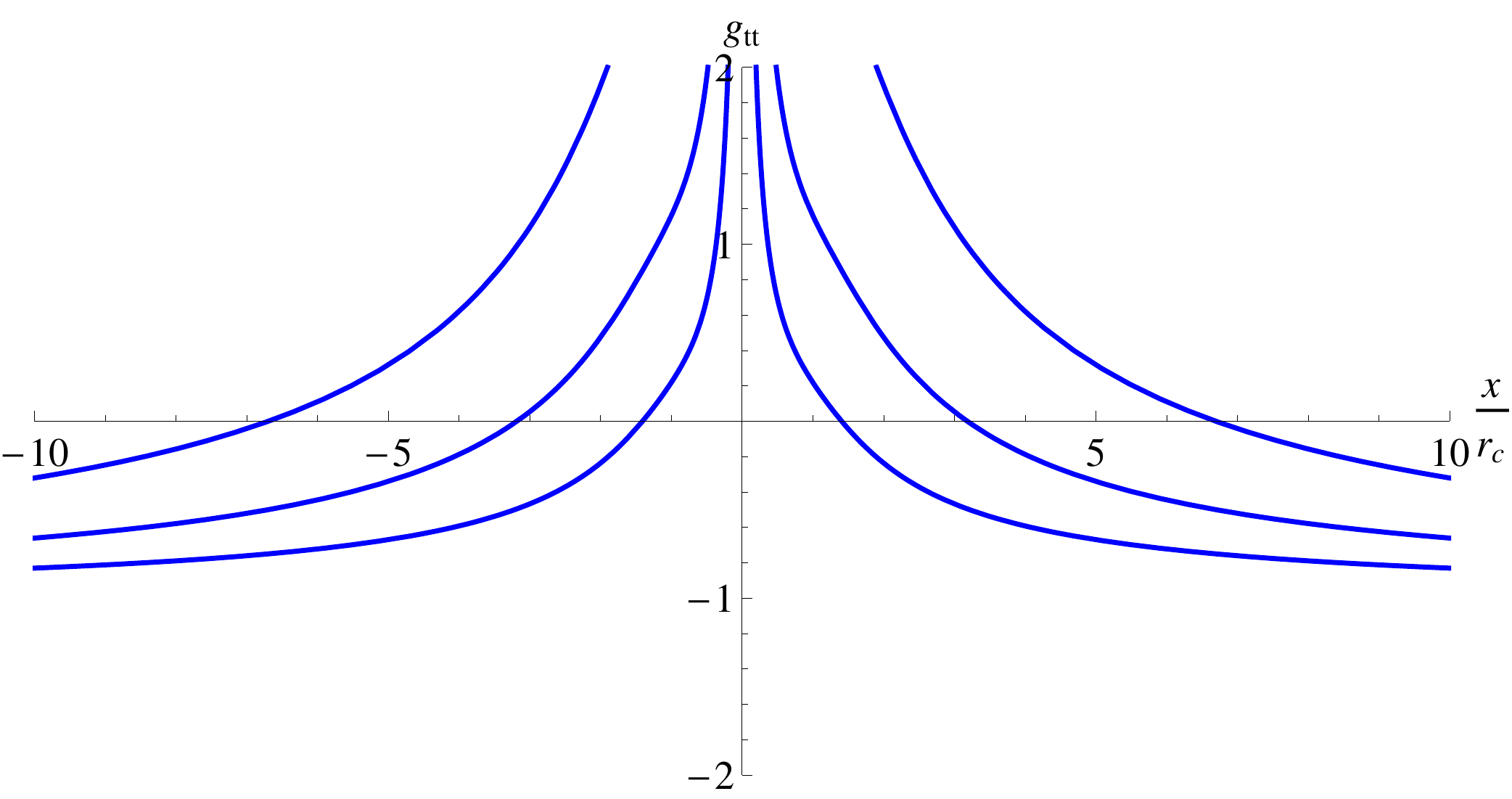}
  \caption{Function $g_{tt}$ in terms of the coordinate $x$, for a wormhole with $\delta_1 < \delta_c$ and different values of $N_q$}
 \end{figure}
 \item $\delta_1 > \delta_c$: In this case it is possible to find two horizons (if $N_q$ is high), no horizon (if $N_q$ is low), or one degenerated horizon (if $N_q$ is a critical value that depends on $\delta_1$) on each side of the wormhole. The hypersurface $x=0$ is always time-like in this case. We will call this type of geometry \emph{Reissner-Nordström-like}. Let us note that since we are working with a fixed value of $\delta_1$, high values of the charge imply even higher values of the mass, so high $N_q$ is equivalent to the $M>Q$ of Reissner-Nordström, meanwhile low $N_q$ is equivalent to the naked singularity in Reissner-Nordström, $Q>M$.
 \begin{figure}[h!]
  \centering
  \includegraphics[width=.7\linewidth]{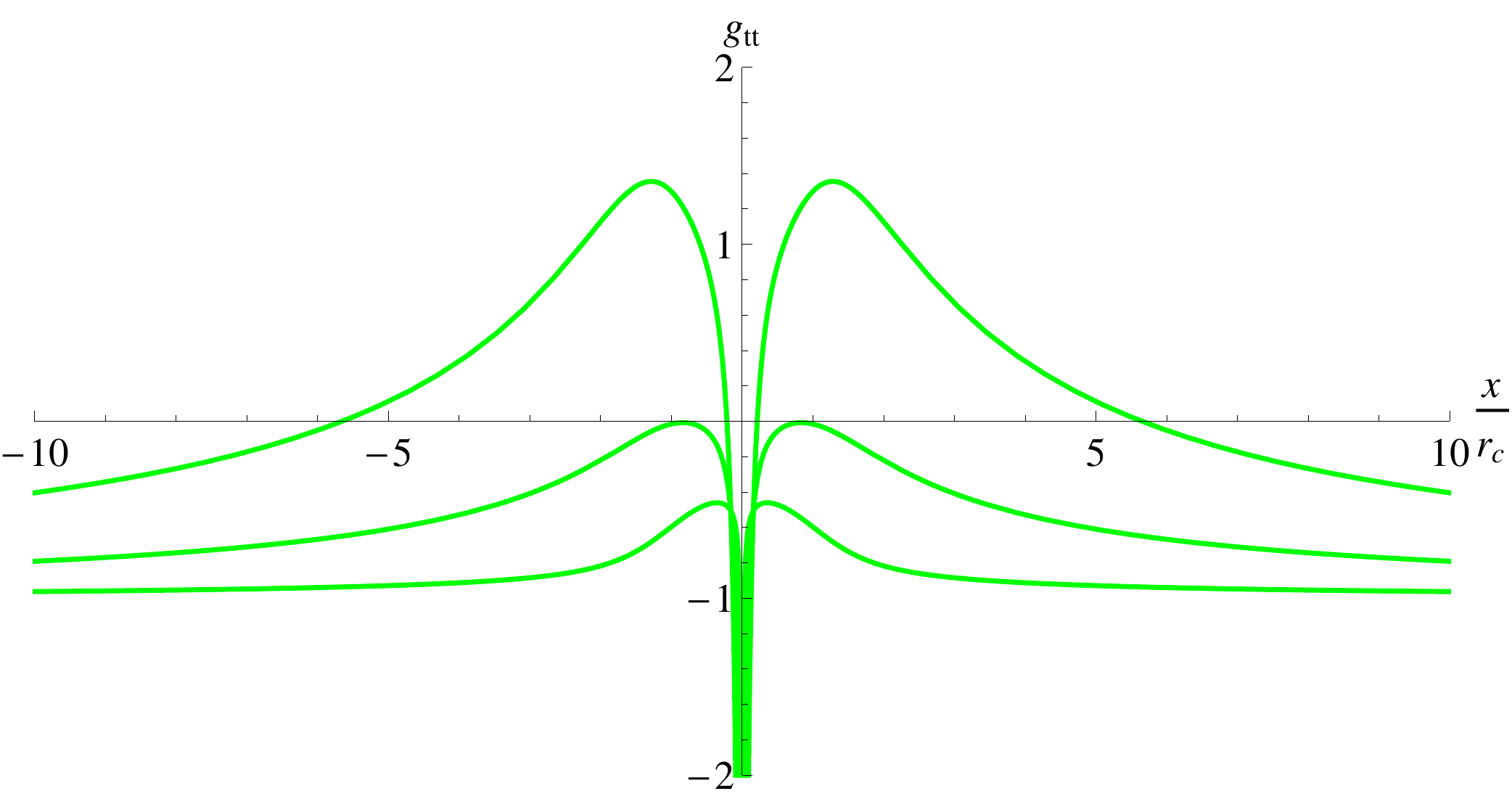}
  \caption{Function $g_{tt}$ in terms of the coordinate $x$, for a wormhole with $\delta_1 > \delta_c$ and different values of $N_q$}
 \end{figure}
 \item $\delta_1 = \delta_c$: If $N_q>N_c$, there is one horizon on each side of the wormhole and $x=0$ is a space-like hypersurface. If $N_q=N_c$, the horizons meet at the wormhole throat, creating a degenerated horizon there, and $x=0$ is a null hypersurface. If $N_q<N_c$ there is no horizon and $x=0$ is time-like hypersurface; in this case it is possible to cross the wormhole and come back to the same asymptotic region. We will refer to the $\delta_1=\delta_c$ case as the \emph{smooth} case, because the metric and curvature are finite everywhere.
  \begin{figure}[h!]
  \centering
  \includegraphics[width=.7\linewidth]{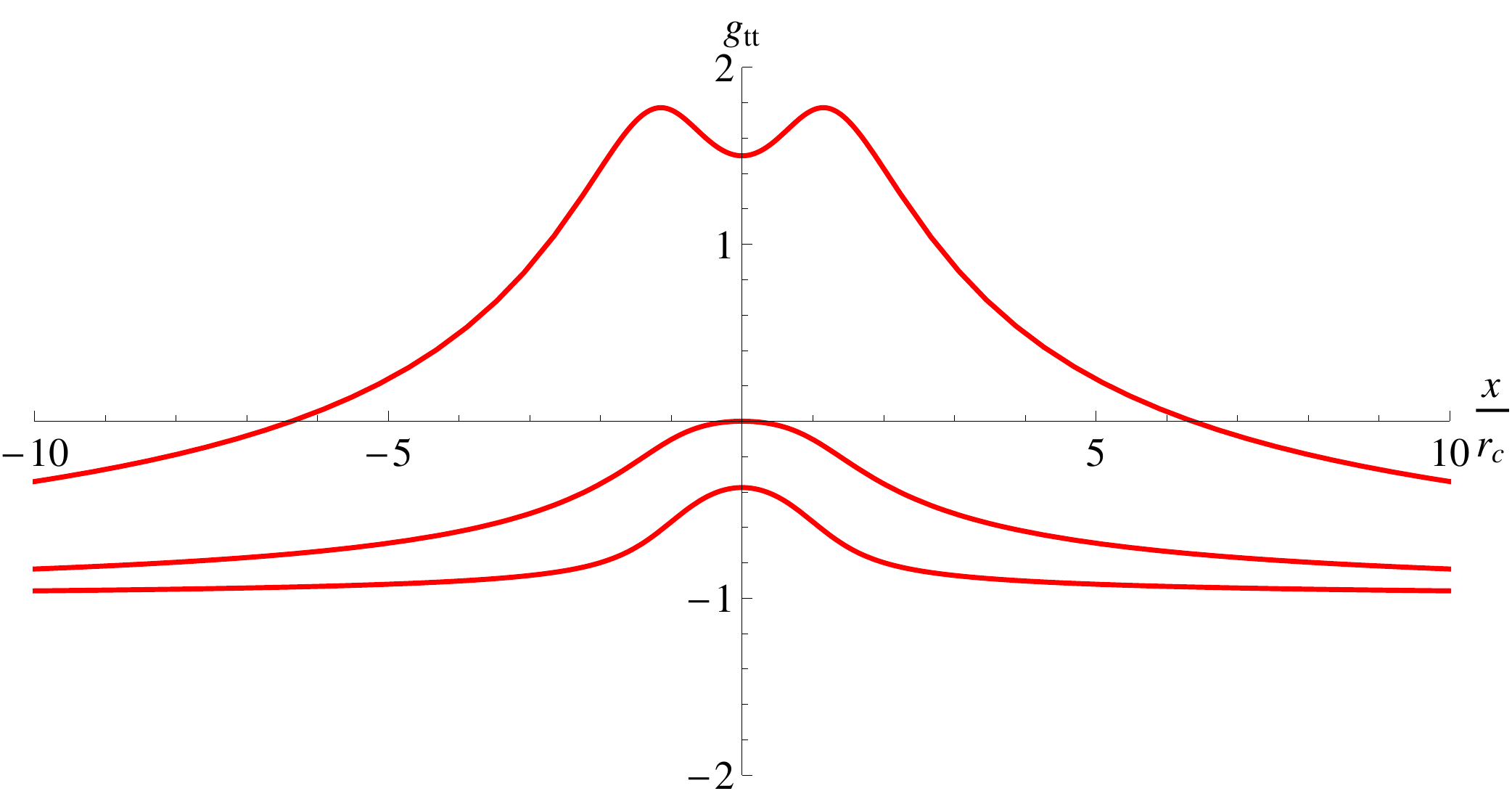}
  \caption{Function $g_{tt}$ in terms of the coordinate $x$, for a wormhole with $\delta_1 = \delta_c$ and $N_q=4 N_c, N_c, 0.25 N_c$}
 \end{figure}
\end{itemize}

\begin{figure}[p!]
\begin{tabular}{ccc}
   \includegraphics[width=.25\linewidth]{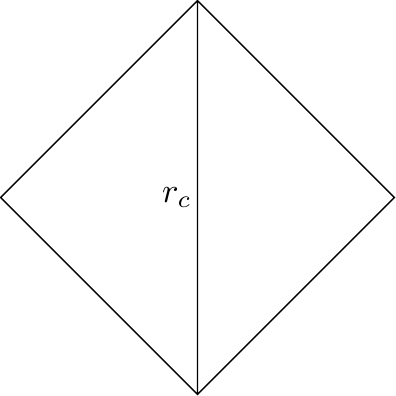}
  &  
   \includegraphics[width=.3\linewidth]{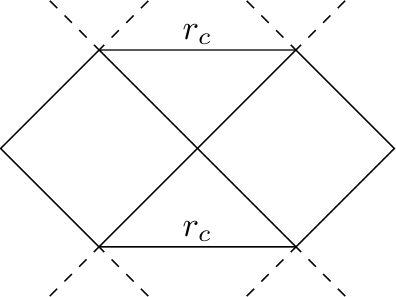}
  &
   \includegraphics[width=.2\linewidth]{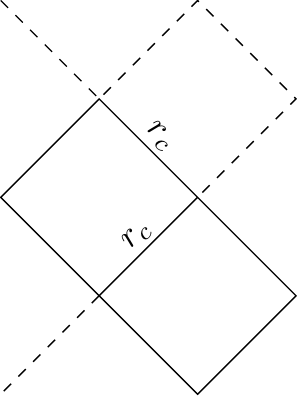}
  \\
  (a) & (b) & (c)\\
  \vspace{.5cm} \\
   \includegraphics[width=.25\linewidth]{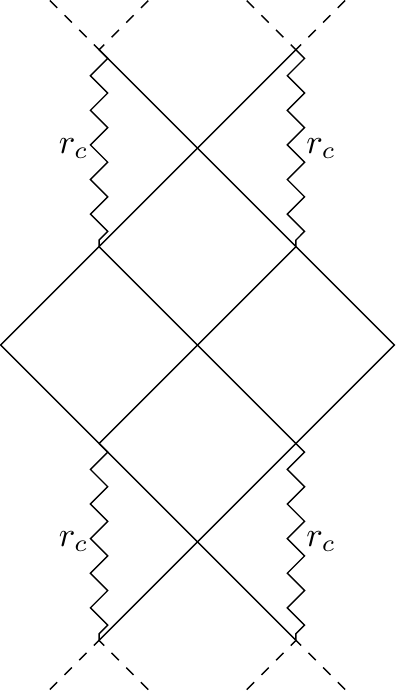}
  &
   \includegraphics[width=.175\linewidth]{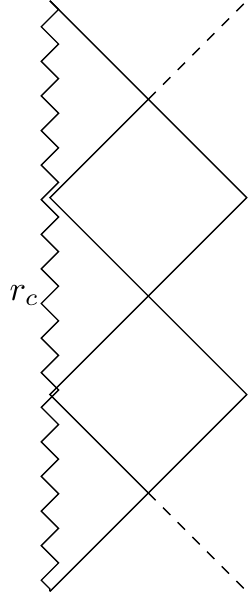}
  &
   \includegraphics[width=.15\linewidth]{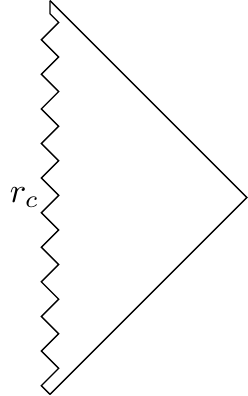}
     \\
  (d) & (e) &(f) \\
  \vspace{.5cm} \\
   &
    \includegraphics[width=.3\linewidth]{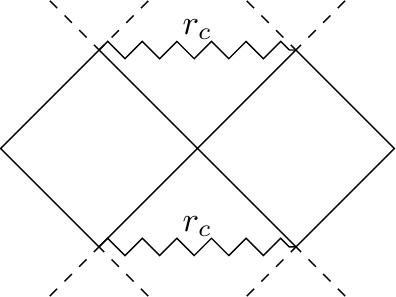}
  &\\
  &(g)&
 \end{tabular}
 \caption{Penrose diagrams for: (a)(b)(c) Smooth case with $N_q < N_c$, $N_q > N_c$, $N_q = N_c$, (d)(e)(f) Reissner-Nordström-like case and different values of $N_q$, (g) Schwarzschild-like case.}\label{4fig:Penrose}
\end{figure}

This information can be summed up in a conformal diagram. These are shown in figure \ref{4fig:Penrose}.

\clearpage

\subsection{Euclidean Embeddings}

The properties of this type of wormholes clashes with our natural intuition about curvature divergences. Physical properties such as the electric flux density and the integrated electromagnetic energy:
\begin{equation}
 \mathcal{E}=-\int  F^{\alpha \beta} F_{\alpha \beta} / (16 \pi) \sqrt{-g} \df^4x
\end{equation}
are finite and as well-behaved in the smooth case as in the case with $\delta_1 \neq \delta_c$ \cite{Lobo:2013prg}. In order to get a better understanding of the differences and similarities between the smooth and the divergent case, it is useful to embed the spatial equatorial section of the geometry in Euclidean 3D space, and visualize the geometry directly. Let us consider the surface $t=\text{constant}$, $\theta=\pi/2$. The induced metric in this surface is:

\begin{equation}
 \df l^2=g_{rr} \df r^2 + r^2 \df \phi^2
\end{equation}

On the other hand, the euclidean metric written in cylindrical coordinates is:

\begin{equation}
 \df l^2 = \df \xi^2 + \df r^2 + r^2 \df \phi^2
\end{equation}

To embed the equatorial slicing of the wormhole into euclidean space we must find a surface that satisfies an equation $\xi=\xi(r)$ so that $1+\left ( \frac{\partial \xi}{\partial r} \right )^2=g_{rr}$. Since the main interest is to illustrate the curvature divergence region and compare it against the smooth case, we can take the $g_{rr}$ expansion in the $r\approx r_c$ region, shown in eq. \ref{4eq:grr-1expansion}, for the most simple cases: the horizonless smooth case ($\delta_1 = \delta_c$, $N_q < N_c$) and the Reissner-Nordström-like ``naked singularity'' ($\delta_1 > \delta_c$, $N_q$ low) to write:

\begin{IEEEeqnarray}{rClCl}
 \df l^2 &=& \frac{N_c}{N_q}\frac{\delta_1}{1-\frac{\delta_1}{\delta_c}} \frac{1}{\sqrt{\frac{r}{r_c}-1}} \df r^2 + r^2 \df \phi^2 &\qquad& \text{if } \delta_1\neq\delta_c \\
 \df l^2 &=& \frac{N_c}{2(N_q-N_c)} \frac{1}{\frac{r}{r_c}-1} \df r^2 + r^2 \df \phi^2 &\qquad& \text{if } \delta_1=\delta_c 
\end{IEEEeqnarray}

Since both cases diverge as $r\rightarrow r_c$, it is possible to approximate $\frac{\partial \xi}{\partial r} \approx \sqrt{g_{rr}}$. Integrating this equation we get:

\begin{IEEEeqnarray}{rClCl}
 \xi (r) &=& \pm \frac{4}{3}  \sqrt{\frac{N_c}{N_q} \frac{\delta_1}{1-\frac{\delta_1}{\delta_c}}} \left ( \frac{r}{r_c}-1 \right )^{\frac{3}{4}}r_c&\qquad& \text{if } \delta_1\neq\delta_c \\
 \xi (r) &=& \pm 2 \sqrt{\frac{N_c - N_q}{2N_c}}\left ( \frac{r}{r_c}-1 \right )^{\frac{1}{2}} r_c &\qquad& \text{if } \delta_1=\delta_c 
\end{IEEEeqnarray}

\begin{figure}[h!]
 \begin{tabular}{cc}
  \includegraphics[width=.5\linewidth]{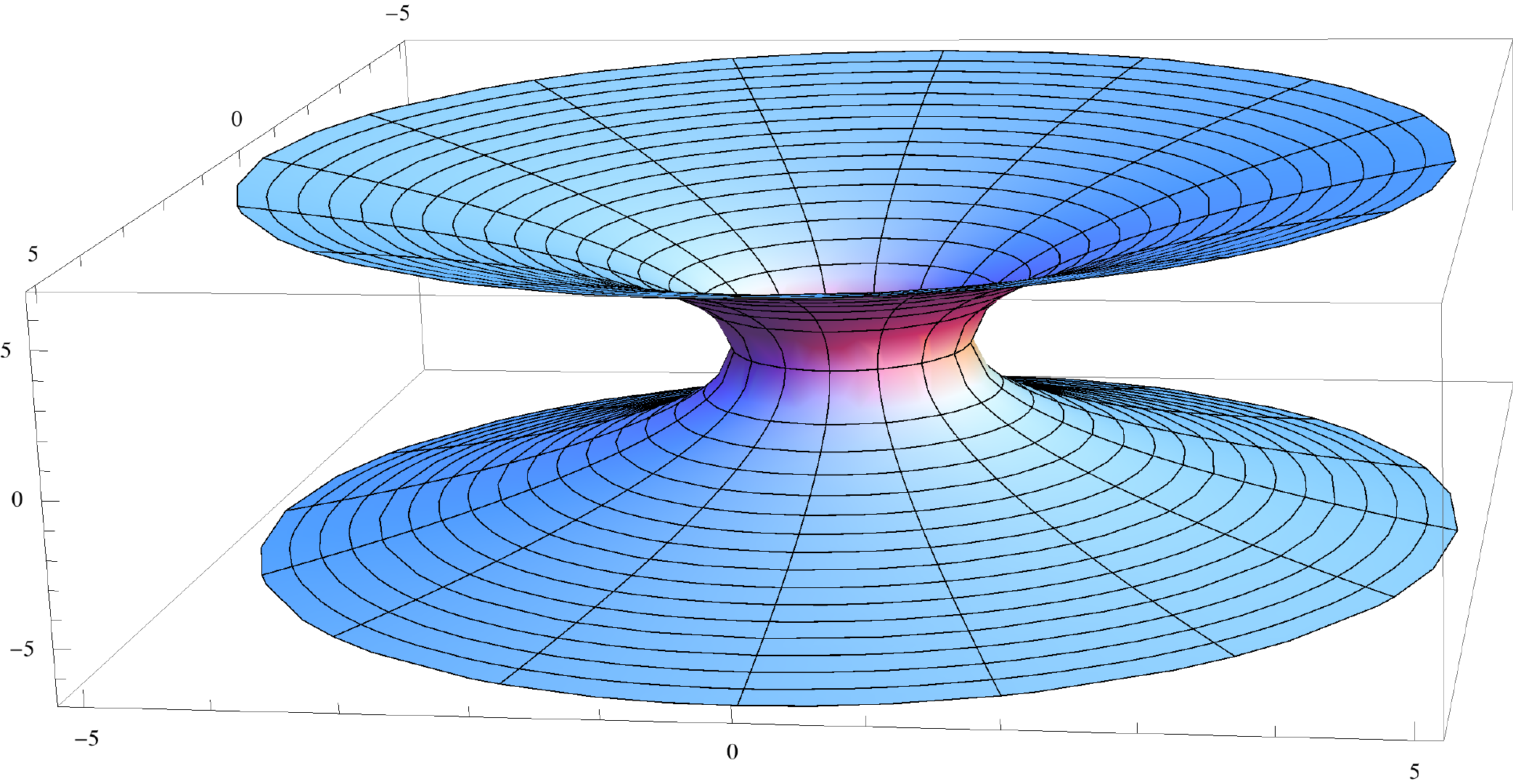}
  &
  \includegraphics[width=.5\linewidth]{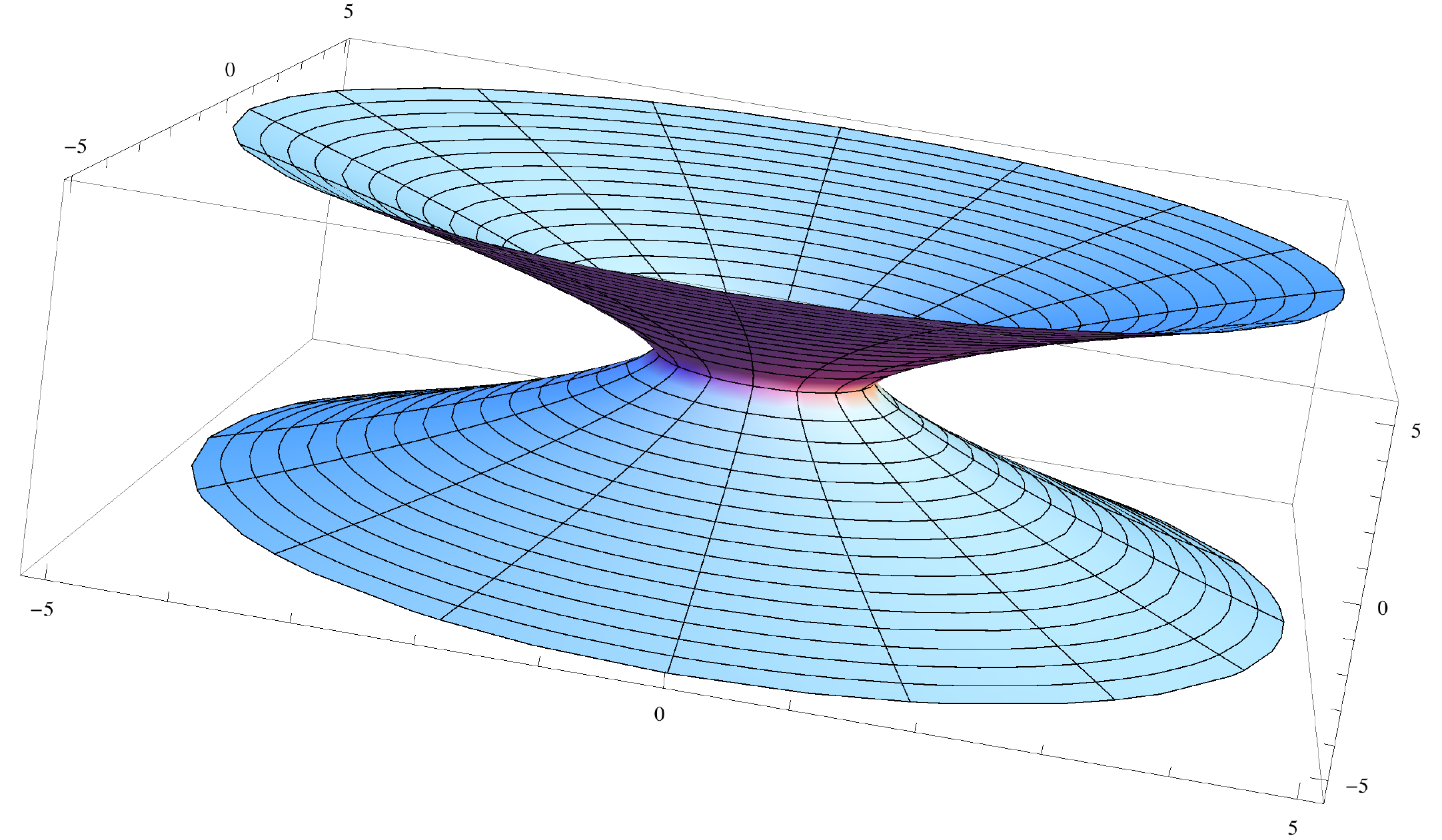}
  \\
  \includegraphics[width=.5\linewidth]{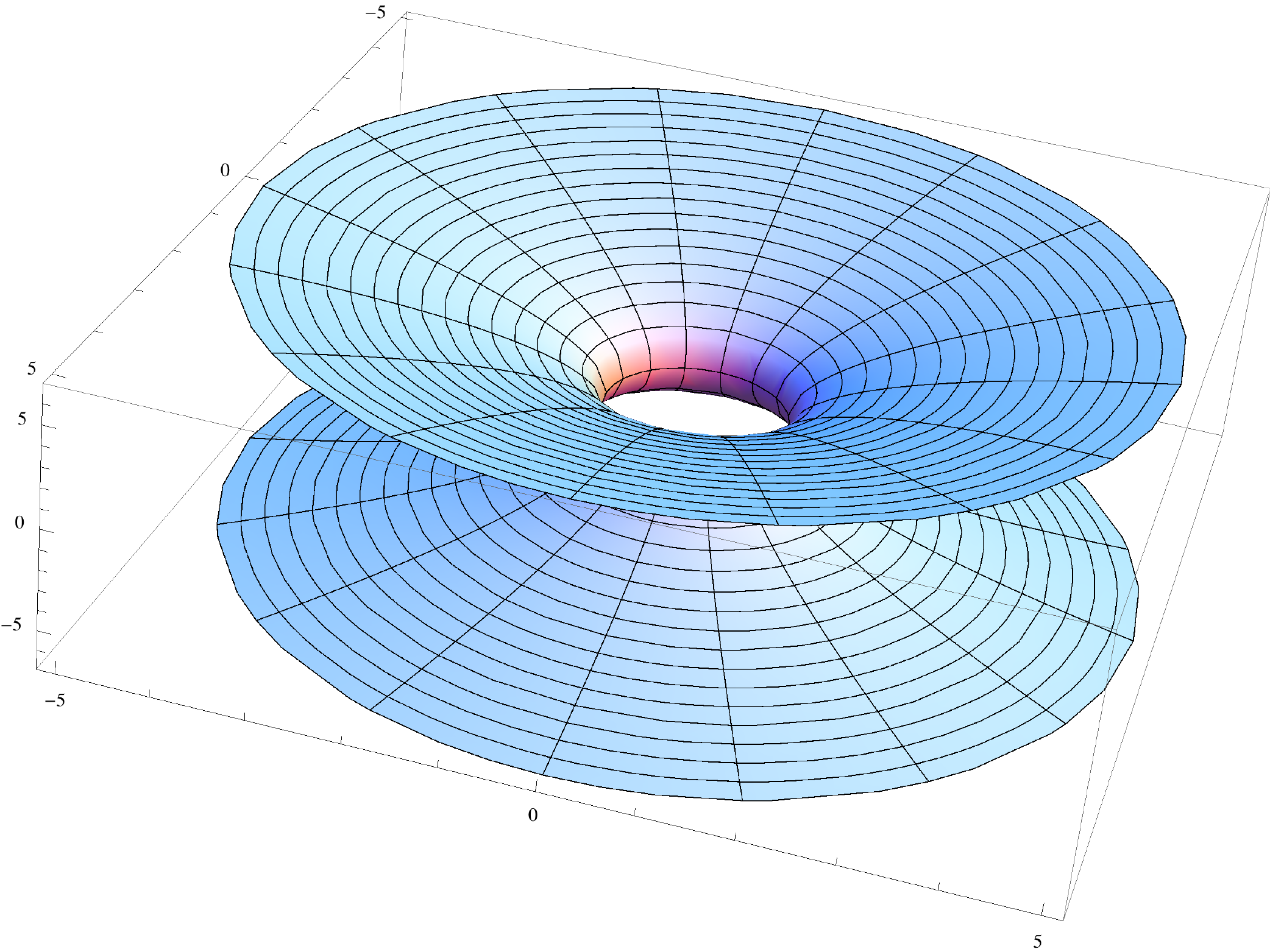}
  &
  \includegraphics[width=.65\linewidth]{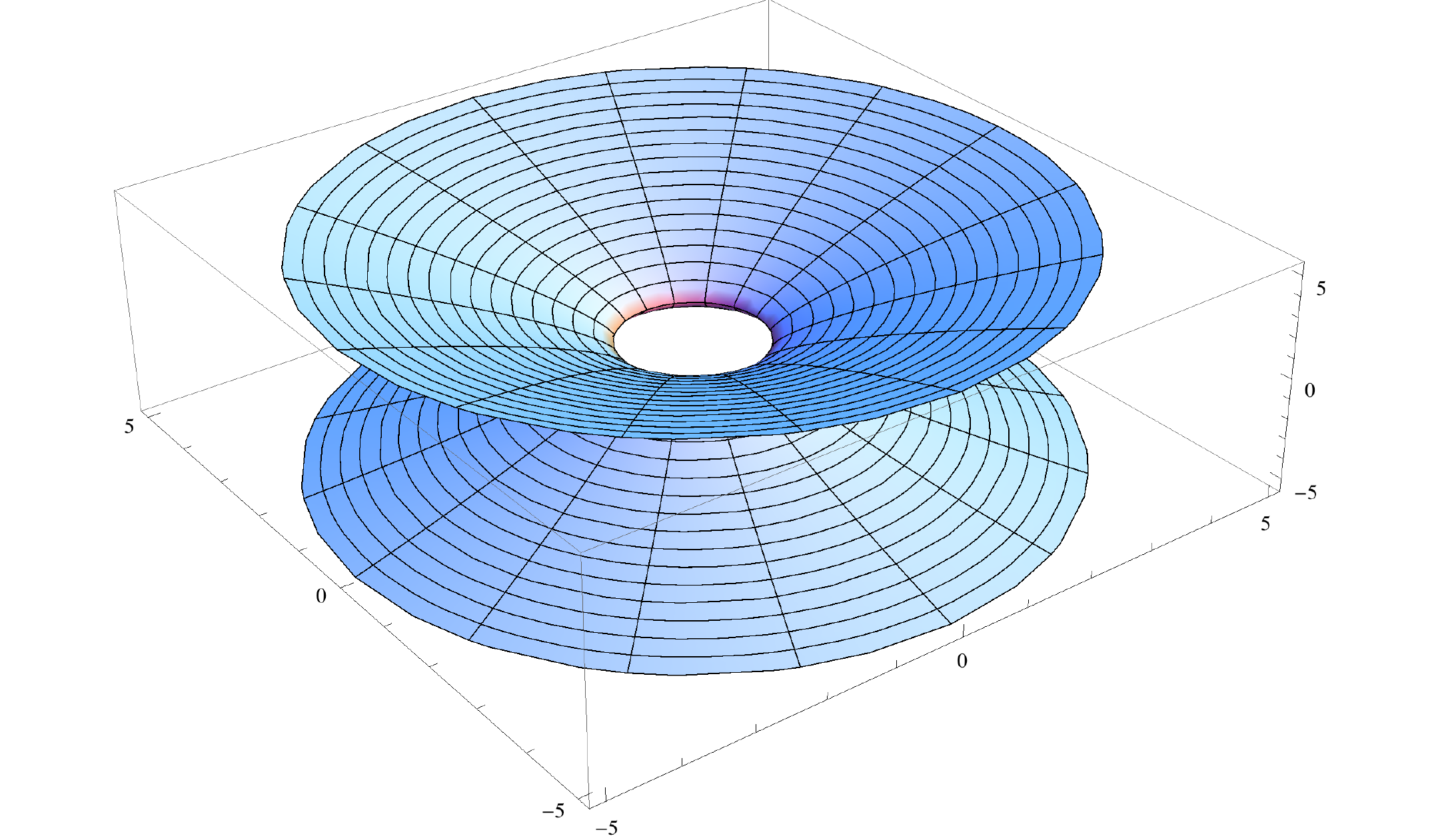}
 \end{tabular}
 \caption{Euclidean embedding of the $\theta=\frac{\pi}{2}$ spatial section for a smooth wormhole ($\delta_1=\delta_c$) on the left, and a wormhole curvature divergences on its throat  ($\delta_1 \neq \delta_c$) on the right.}\label{4fig:WH3D}
\end{figure}

\begin{figure}
 \centering
 \includegraphics[width=.6\linewidth]{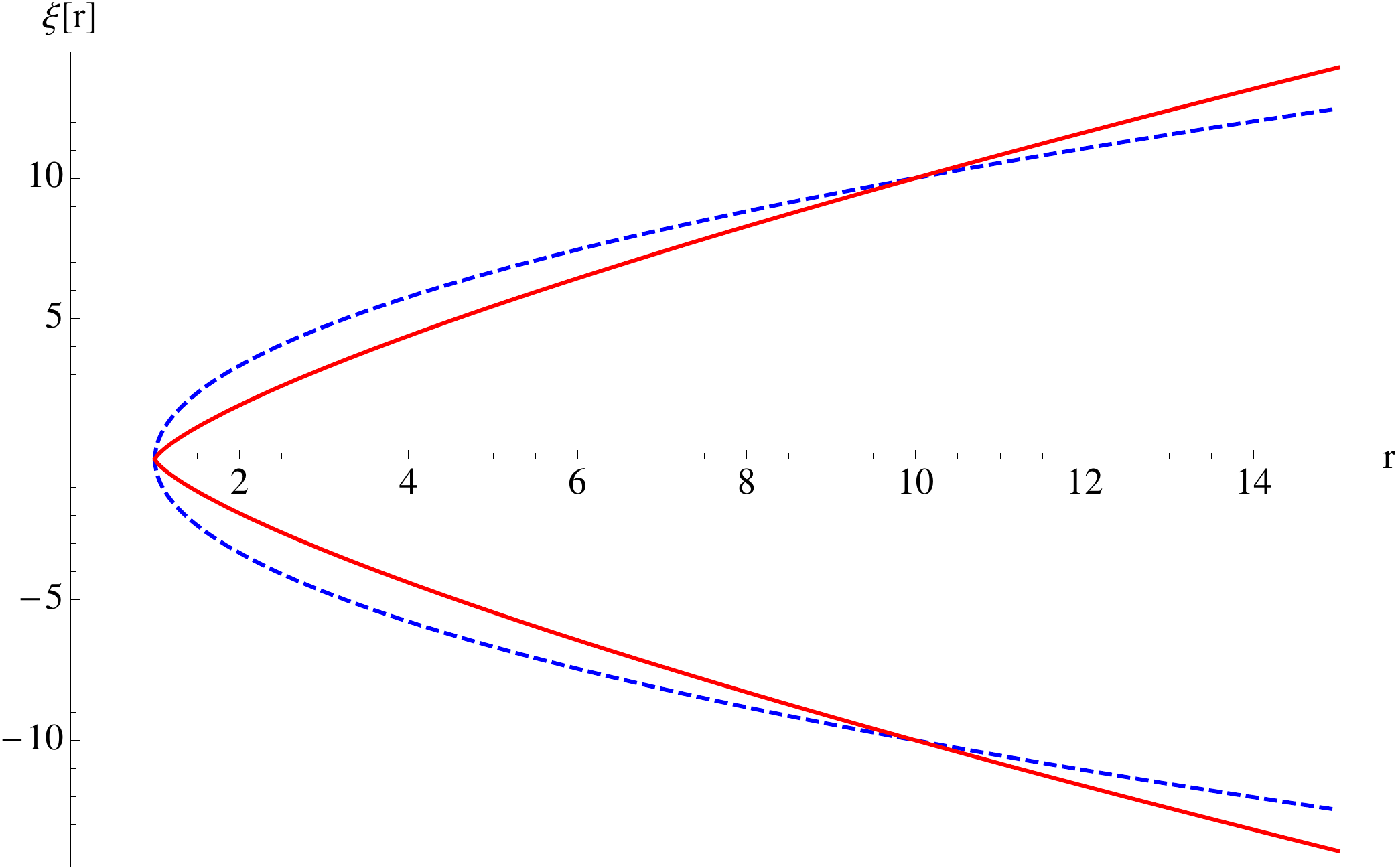}
 \caption{$\xi(r)$ as a function of $r$. The blue dashed curve represents the smooth wormhole configuration, while the red continuous curve represents the wormhole with curvature divergences at its throat. The curves have been normalized to make them coincide at $\frac{r}{r_c}=10$.}\label{4fig:WH2D}
\end{figure}

In Fig. \ref{4fig:WH3D} we can see the wormhole structure on both the smooth and the divergent case. In the divergent case, the wormhole throat presents a vertex, which is responsible of the divergent curvature. In Fig. \ref{4fig:WH2D}, we can see the radial section of the embedded surfaces. It is worth noting that this vertex \emph{is not} like the vertex of a polyhedron, where two faces meet at an angle. Indeed, the inverse function $r(\xi) \approx r_c+\kappa |\xi|^\frac{4}{3}$ is both continuous and differentiable at $r_c$ ($\xi=0$), but its second derivative diverges, which in turn causes the curvature scalars to diverge.

\cleardoublepage
\chapter{Geodesics}\label{GeoChap}

In the last chapter we saw how, for a quadratic Lagrangian in the metric-affine formalism, there are charged black hole solutions where curvature scalars are bounded and the metric components are continuous and differentiable everywhere, with a wormhole structure instead of a singularity. However, in general, the solutions presented a divergence of curvature on a spherical surface of area $\pi r_c^2$, but still showed the wormhole structure. Besides, in any case the charge and the mass are a consequence of the topology of the solution and do not come from any sources, and the equations form the electromagnetic field, the metric and the connection are well-defined and solved everywhere.

All the nice properties of this geometry would be meaningless if physical observers (represented by geodesics and congruences of geodesics) do not have a well defined evolution when crossing the wormhole throat, in particular in the cases where there is a curvature divergence. If we could not extend the geodesics through this wormhole throat would imply the existence of a singularity in our space-time, and the theory would face the same problems as GR. In this chapter we will study the geodesics and congruences of geodesics, and will stablish the regularity of these solutions.



\section{Geodesics of the Geonic Wormhole}



In the introduction, we defined geodesics as the paths ``as straight as possible'', which are followed by unaccelerated observers. We recall from section \ref{1sec:Geo}, that a geodesic curve described in the affine parametrization satisfies the following equation:

\begin{equation}
 u^\mu \nabla_\mu u^\nu = \frac{\df^2 \gamma^\nu}{\df \lambda^2} + \Gamma^\nu_{\alpha \beta} \frac{\df \gamma^\alpha}{\df \lambda}\frac{\df \gamma^\beta}{\df \lambda} = 0 \label{5eq:Geo}
\end{equation}

In the introduction we considered the connection $\Gamma^\nu_{\alpha \beta}$ to be the Levi-Civita connection of the metric; after all, GR does not have any other affine structure, and this way the geodesics are also the paths that extremize the length between two points. However, in metric-affine theories we have the Levi-Civita connection of the metric and the independent connection. This gives rise to two different covariant derivatives, that give two different ways of measuring acceleration, and two different sets of geodesics. We are only interested in the paths of unaccelerated observers, such as  test particles. The question is which of these two ways of measuring acceleration is the physical one. To answer that, we should look into the equations of motion of our particles.

%

If the matter Lagrangian is constructed in such a way that the matter fields do not couple to the connection, then the equations of motion will know nothing about this structure, and the paths of test particles will follow geodesics derived from the Levi-Civita connection associated to the metric. Until now, we have only written the Lagrangian of the electromagnetic sector $\Lagr_\text{EM} = -\frac{1}{16 \pi l_P^2} F_{\alpha \beta} F^{\alpha \beta}$ with $F_{\mu \nu} = (\df A)_{\mu \nu}$, which we used to construct the black hole solutions in the previous chapter. This Lagrangian is insensitive to the details of the connection. Therefore, light rays will follow the null geodesics of the metric. To check this, we can write the equations of motion (eq. \ref{4eq:EMmov2}) as $\partial_\mu (\sqrt{-g} F^{\mu \nu})=0$, and see that they do not depend on the independent connection.

It would be possible for other matter fields to couple to the connection, and then the geodesics of the independent connection could be important. However, this would lead to violations of the EEP which have not been yet observed. In our approach, the independent connection modifies the equations that give raise to the metric in terms of the matter distribution, but does not couple directly to the matter. For that reason, we will restrict ourselves to study geodesics of the Levi-Civita connection of the metric.

In order to find the geodesic paths, we will work in the $(t,x,\phi,\theta)$ coordinates. Let us recall from eq. \ref{4eq:metricgx} that the metric in these coordinates can be written as:

\begin{equation}
 g=-\frac{A}{\sigma_+} \df t^2 + \frac{1}{A \sigma_+} \df x^2 + \underbrace{x^2 \sigma_-}_{r^2} \df \Omega^2 \label{5eq:metricgx}
\end{equation}

The definitions of $\sigma_\pm$, $A$ and $x$ can be found in the previous chapter in eqs. \ref{4eq:sigmadef}, \ref{4eq:A} and \ref{4eq:dxtodr}. As we did in the introduction, in section \ref{1sec:Geo}, we can make use of the symmetries of the geometry to obtain conserved quantities that simplify the analysis. First of all, because of spherical symmetry the geodesics lie on a plane, and we can rotate our coordinate system so that plane is $\theta= \frac{\pi}{2}$ without loss of generality. Second, if the geodesics are time-like, we can normalize its tangent vector to $-1$; if they are null, the norm of the tangent vector is $0$. Third, the symmetries under rotations and temporal translations gives us two conserved quantities: $E=\frac{A}{\sigma_+}\frac{\df \gamma^t}{\df \lambda}$, $L=r^2 \frac{\df \gamma^\phi}{\df \lambda}$. For time-like geodesics, $E$ can be interpreted as the total energy per unit mass, and $L$ as angular momentum per unit mass. In the case of light rays, it is not possible to normalize the tangent vector and consequently, $E$ and $L$ lack meaning by themselves; but the quotient $L/E$ can be interpreted as the apparent impact parameter as seen from the asymptotically flat infinity. The condition that the tangent vector to the geodesics has to be normalized to $0$ or, $-1$ gives us another equation:

\begin{equation}
 -\kappa = -\frac{A}{\sigma_+} \left ( \frac{\df \gamma^t}{\df \lambda} \right )^2 + \frac{1}{A \sigma_+} \left ( \frac{\df \gamma^x}{\df \lambda} \right )^2 + r^2(x) \left ( \frac{\df \gamma^\phi}{\df \lambda} \right )^2
\end{equation}

Substituting the value of the conserved quantities, this equation gives us the radial component of the tangent vector:

\begin{equation}
 \frac{1}{\sigma_+} \left ( \frac{\df \gamma^x}{\df \lambda} \right ) = \pm \sqrt{E^2 - \frac{A}{\sigma_+} \left ( \kappa +\frac{L^2}{r^2(x)} \right )} \label{5eq:xcomp}
\end{equation}

where $\kappa=0$ or $1$ depending if the geodesic is null or time-like. In principle, these equations can be integrated, as we did in the introduction. However, it is possible to obtain lots of information from the form of this equation alone. This equation, that describes the movement in the radial direction, is analogue to the movement of a particle with energy $E^2$ in a one-dimensional potential:
%
\begin{equation}
 V(x)=\frac{A}{\sigma_+} \left ( \kappa + \frac{L^2}{r^2(x)} \right )
\end{equation}

This potential has two parts: a repulsive centrifugal part, $\frac{A}{\sigma_+}\frac{L^2}{r^2(x)}$ and an attractive part $\frac{A}{\sigma_+} \kappa$. The centrifugal part is like the typical centrifugal term for big radius. However, unlike the typical centrifugal term it does not diverge to infinity at any point because the radius has a minimum value. This means that this repulsive part of the potential will not be able to keep geodesics off the centre, if the geodesic has enough energy.


The zeroes of the potential correspond to the zeroes of $A/\sigma_+$, the $g_{tt}$ component of the metric, and signal the horizons of the geometry. The regions where the potential is negative correspond to the regions in which $x$ is time-like and where the geodesic cannot remain stationary: Since $E^2$ is always positive, then $\frac{\df x}{\df \lambda}$ must be different from $0$, which is what was expected from the time-like nature of the coordinate $x$.

At the extrema of the potential there will be stationary circular orbits, i.e. $\left ( \frac{\df \gamma^x}{\df \lambda} \right )=0$, $\left ( \frac{\df^2 \gamma^x}{\df \lambda^2} \right )=0$. This happens to geodesics with energy equal to the value of the potential ($E^2=V(x_0)$) and lying at a extremum ($\left. \frac{\partial V}{\partial x} \right |_{x=x_0}=0$). If the extremum is a minimum, the orbit is stable and perturbations would make the orbit oscillate around the minimum. If it is a maximum, the orbit is unstable and any perturbation would ``knock'' the geodesic out of the orbit. In the regions where the potential is negatively valued it is impossible to fulfil the condition $E^2=V(x_0)$. In that region we will have $\frac{\df x}{\df \lambda} \neq 0$, and a geodesic would orbit away, no matter if there is an extremum of the potential. However, if we have a negatively valued region lying between two horizons, a geodesic with low $E^2$ could oscillate around this region, crossing through each horizon out and back again in each oscillation.

\subsection{Radial Null Geodesics}

Radial null geodesics are characterized by $\kappa = 0$, $L=0$. The potential is zero in this case and insensitive to the details of $A(x)$. The only difference with GR will be a shift in the affine parameter as the geodesic gets near the wormhole throat, caused by the factor $\sigma_+$. Eq. \ref{5eq:xcomp} transforms to:

\begin{equation}
 \frac{1}{\sigma_+} \left ( \frac{\df \gamma^x}{\df \lambda} \right ) = E 
\end{equation}

This equation admits exact solutions in terms of hypergeometric functions of the form:

\begin{equation}
 \pm E \ \lambda(x) = \left \{ \begin{array}{ll}
                              {}_2F_1[-\frac{1}{4}, \frac{1}{2}, \frac{3}{4}; \frac{r_c^2}{r^4} ]r \qquad & \text{if } x \geq 0 \\
                              2\lambda_0-{}_2F_1[-\frac{1}{4}, \frac{1}{2}, \frac{3}{4}; \frac{r_c^2}{r^4}] r \qquad & \text{if } x \leq 0
                             \end{array} \right.
\end{equation}

\begin{figure}[h!]
 \centering
 \includegraphics[width=.5\linewidth]{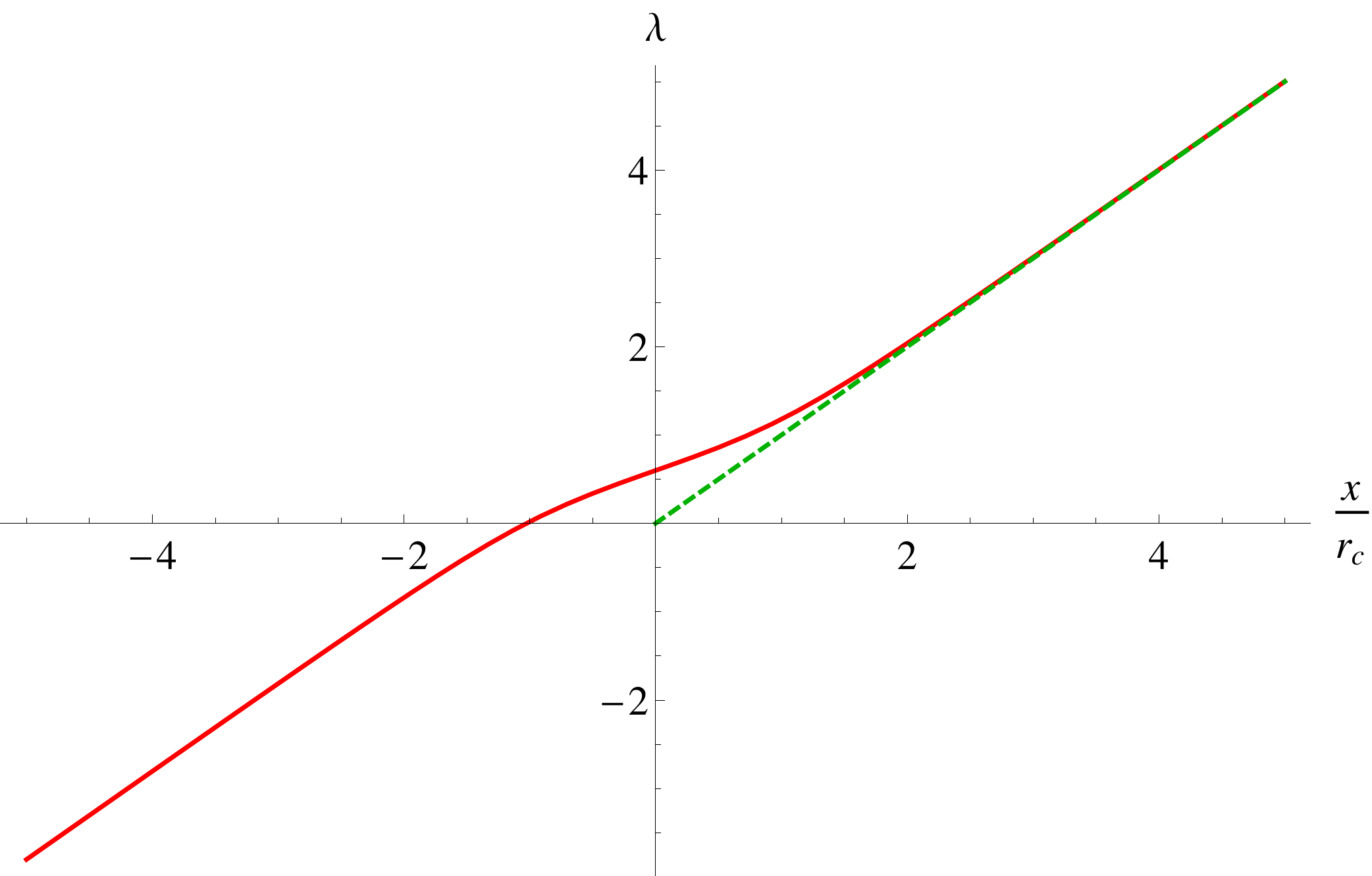}
 \caption{Affine parameter $\lambda(x)$ as a function of the radial coordinate $x$ (in units of $r_c$) for radial null geodesics. In solid red for the wormhole space time, in dotted green for GR. $E=1$ for this plot. The affine parameter experiences a shift with respect to the GR case passing through the wormhole.}\label{5fig:RadNull}
\end{figure}

Where $\lambda_0={}_2F_1[-\frac{1}{4}, \frac{1}{2}, \frac{3}{4};1] r_c \approx 0.59907 r_c$.
The integration of this equation has been plotted in fig. \ref{5fig:RadNull}; note that this result is independent of the value of $\delta_1$, i.e, independent on the the presence or not of curvature divergences at the wormhole throat.

\subsection{Null Geodesics with $L\neq 0$}
In the case $\kappa = 0$, $L\neq0$, the potential shows the typical centrifugal barrier $V\approx\frac{L}{r^2}$ at large distances ($r\rightarrow \infty$). As the geodesic approaches the centre of the geometry, the centrifugal barrier starts to be modified by the $A$ factor. Near the wormhole throat, when $x\rightarrow 0$, the centrifugal term tends to a constant value $\frac{L}{r_c^2}$ and the behaviour of the potential depends on the behaviour of $A$. Consequently, it changes radically depending on the value of $\delta_1$ and $N_q$. 

Near the wormhole throat the potential can be approximated as:

\begin{equation}
 V(x) \approx -\frac{1}{2} \frac{N_q}{N_c} \frac{1-\frac{\delta_1}{\delta_c}}{\delta_1} \frac{L^2}{r_c  x} + \frac{1}{2} \frac{L^2}{r_c^2} \left ( 1 - \frac{N_q}{N_c} \right ) + \frac{1}{12} \frac{N_q}{N_c} \frac{L^2}{r_c^4} x^2+O(x^4) \label{5eq:Vapprox}
\end{equation}

\begin{figure}[p!]
\centering
\begin{tabular}{rl}
 \includegraphics[width=0.5\textwidth]{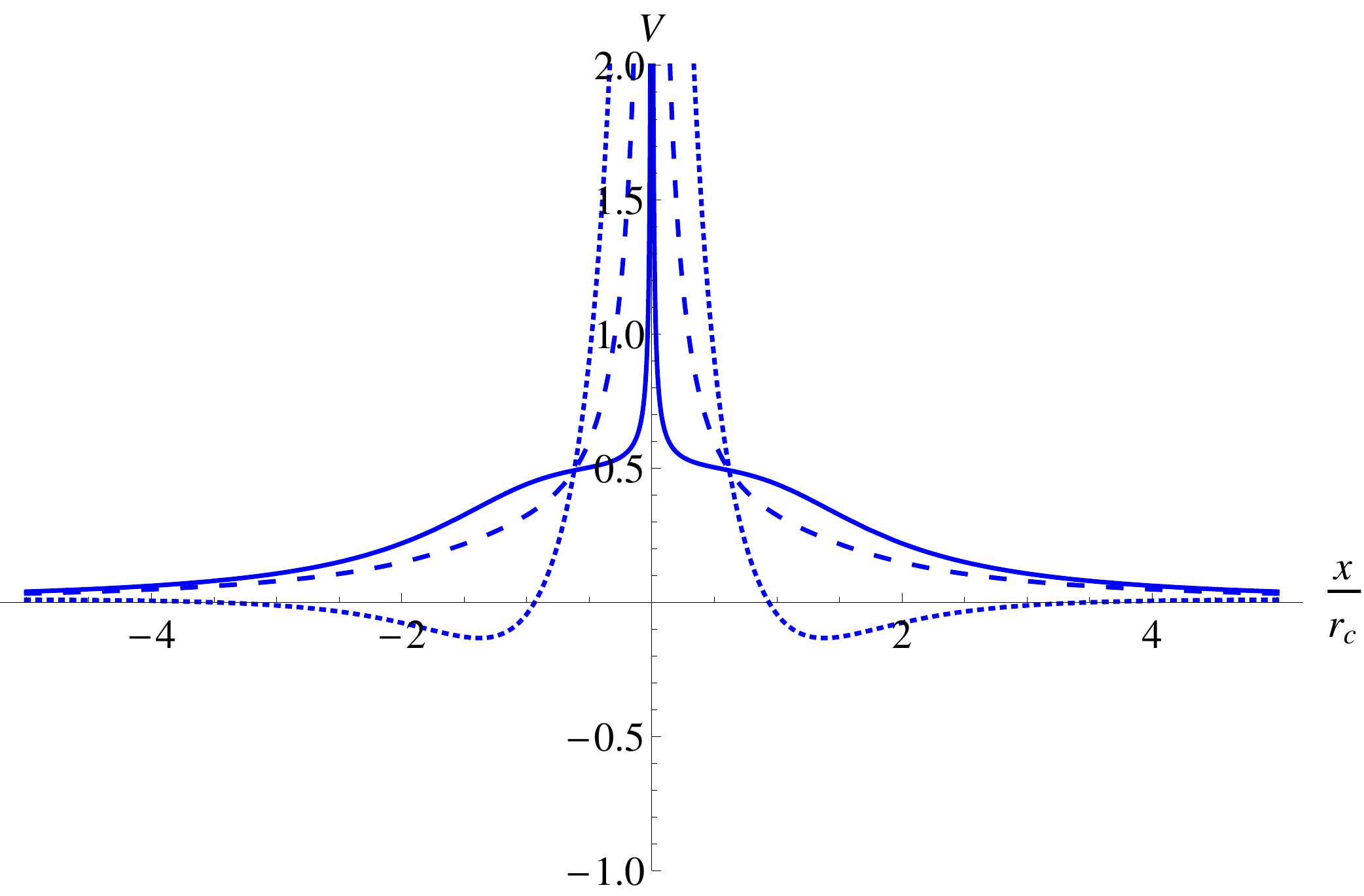} &
 \includegraphics[width=0.5\textwidth]{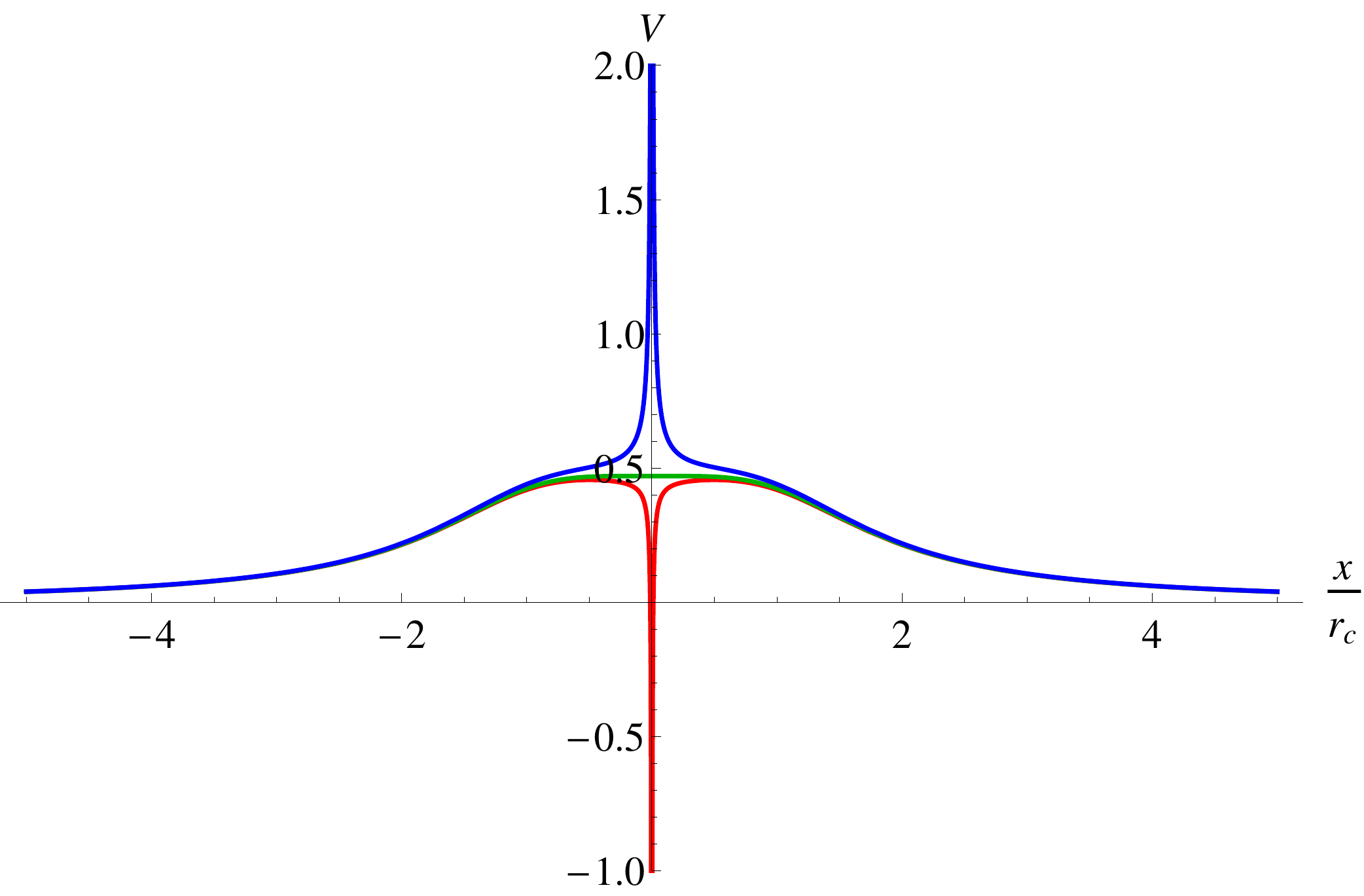} \\
  \includegraphics[width=0.5\textwidth]{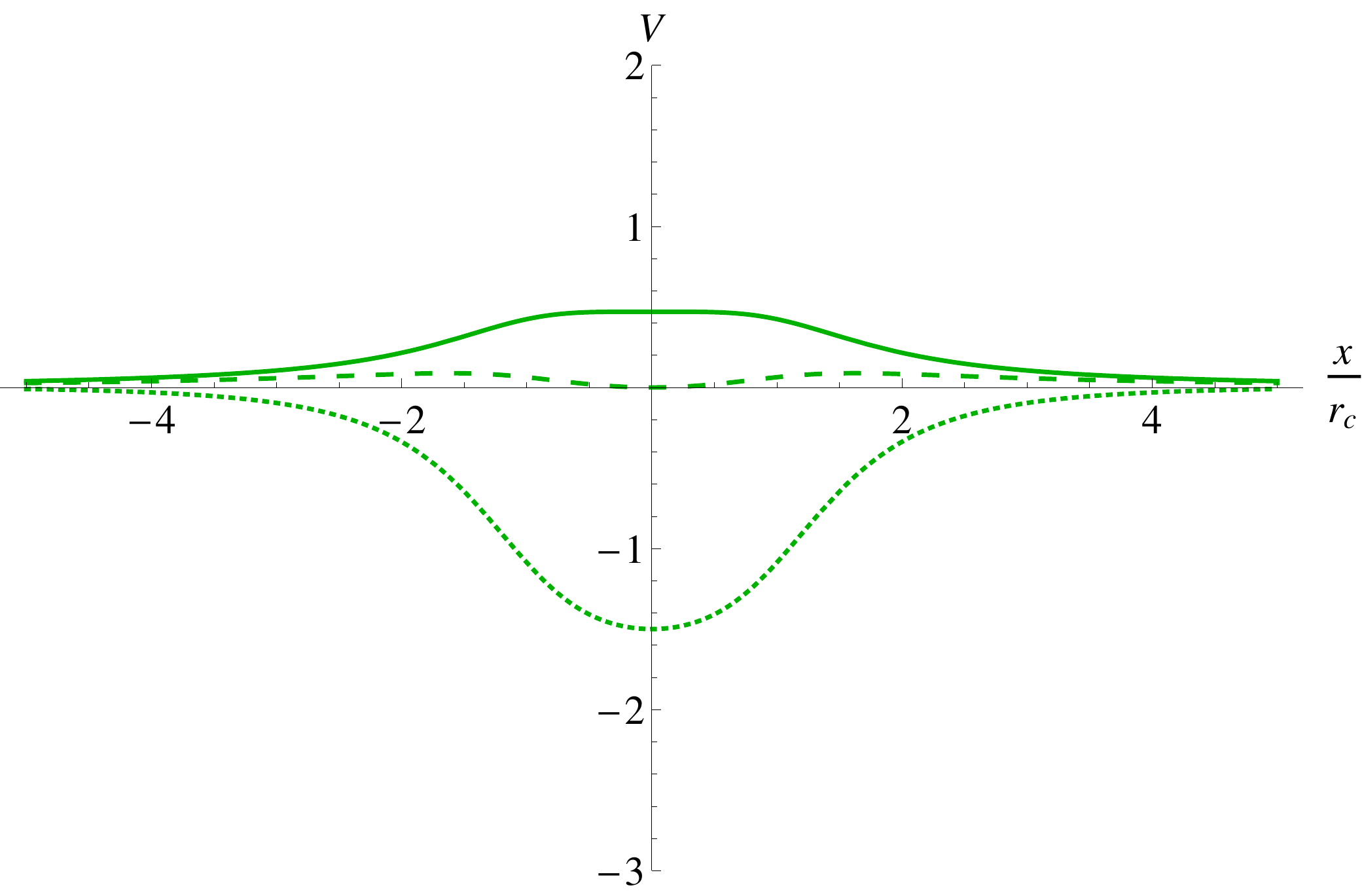} &
 \includegraphics[width=0.5\textwidth]{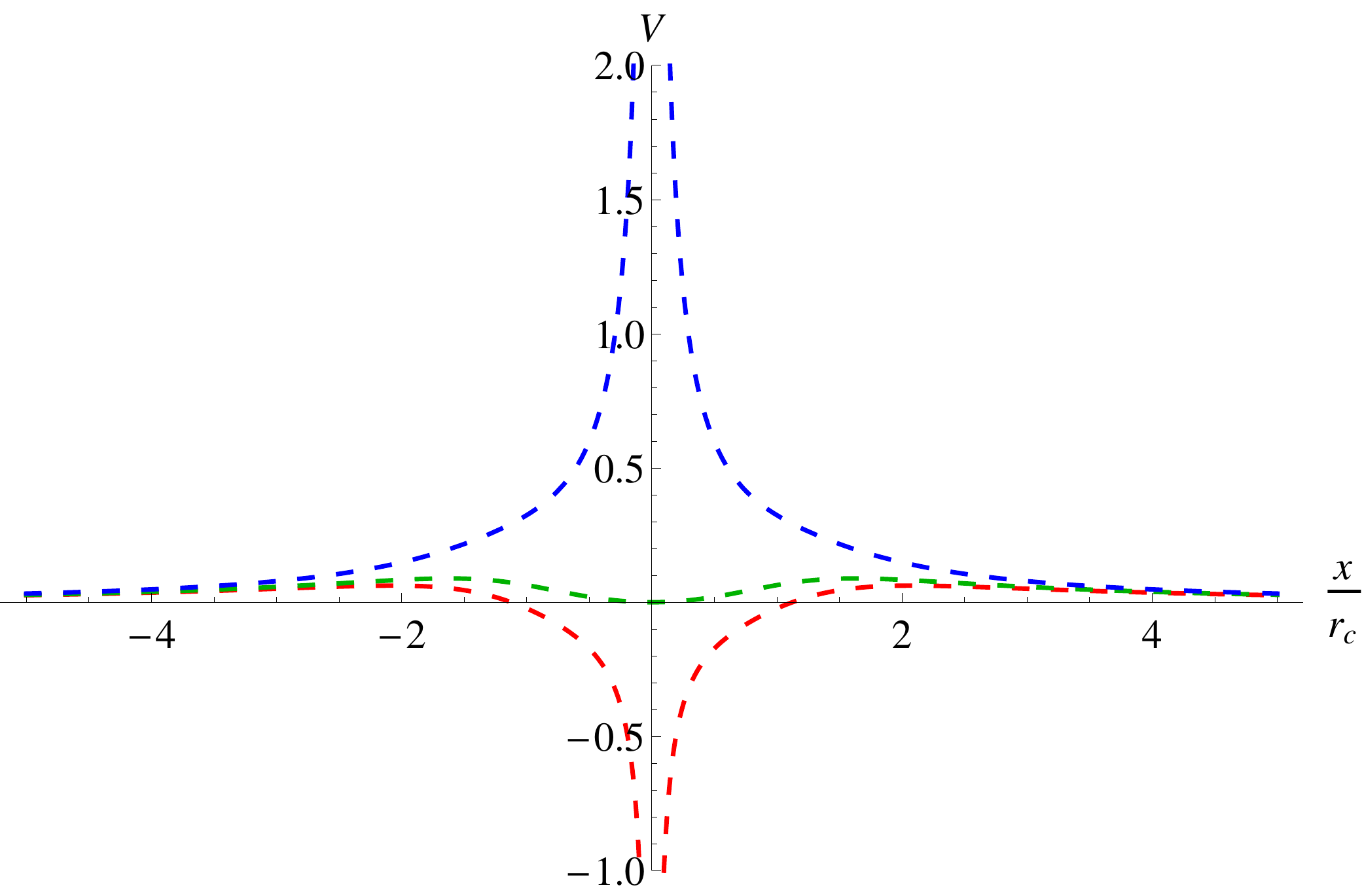} \\
  \includegraphics[width=0.5\textwidth]{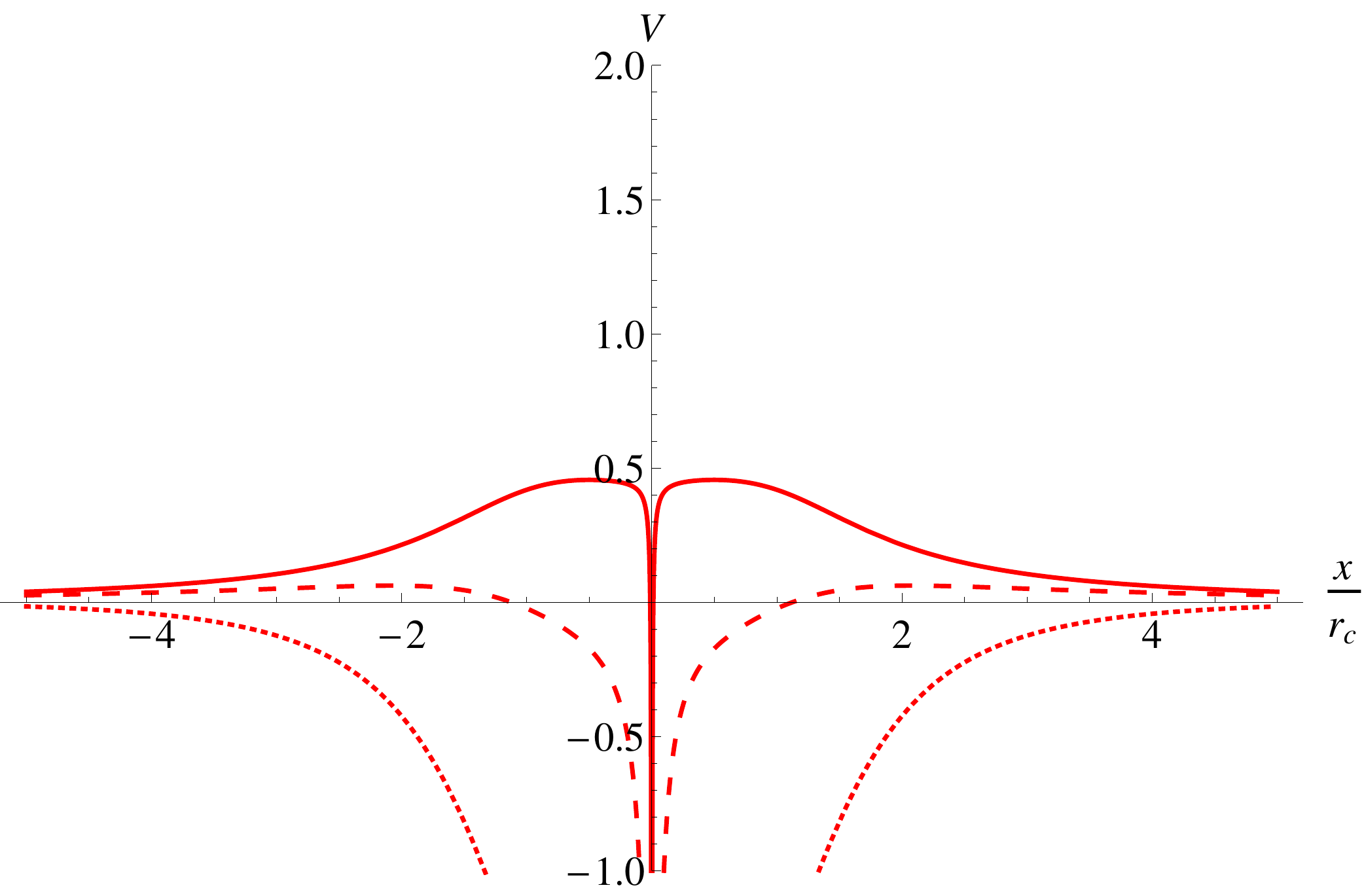} &
 \includegraphics[width=0.5\textwidth]{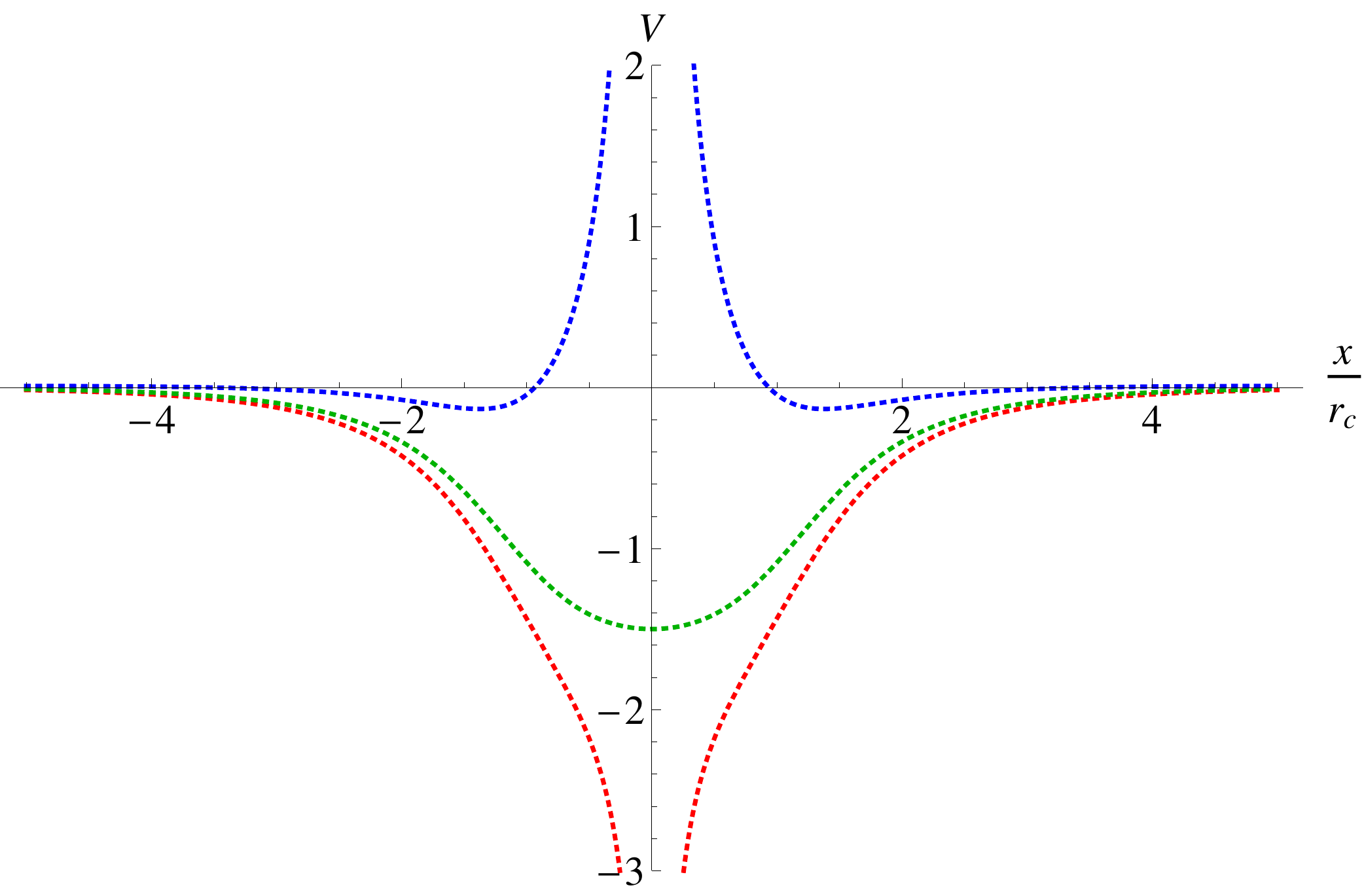} 
\end{tabular}
 \caption{Representation of the effective potential $V(x)$ for null geodesics with $L=1$ for different scenarios. Each plot on the left represents the potential for three different number of charges but the same value of $\delta_1$. Each plot on the right represents three different values of $\delta_1$ but the same number of charges. Blue, green and red lines correspond with geometries with $\delta = 1.5 \delta_c, \delta_c, 0.9 \delta_c$ respectively. Continuous, dashed and dotted lines correspond to $N_q=1, N_c, 4N_c$.}\label{5fig:l1k0}
\end{figure}

If $\delta_1$ is different from $\delta_c$, the potential will diverge, and whether $\delta_1$ is greater or lesser than $\delta_c$ will make the potential be an infinite barrier and the wormhole throat a time-like region, or will make the potential be an infinite well and time-like region and the wormhole throat a space-like region. If $\delta_1=\delta_c$, the potential will be regular and the behaviour will change depending on the number of charges:

The different cases are summed up here:

\begin{itemize}
 \item $\delta_1 < \delta_c$ (Schwarzschild-like case): In this case, the potential becomes infinitely attractive at the wormhole throat. Between the asymptotic region ($x\rightarrow\infty$ where a free-falling observer feels a centrifugal barrier and the potential is repulsive) and the black hole horizon (where the potential is attractive) there must be a point where the potential reaches a maximum. That point is a critical point where there are unstable photon orbits around the black hole. All geodesics with ``energy'' $E^2$ greater than that value will reach the wormhole throat. We recall that the ``energy'' of a null-like geodesic does not have meaning on itself, and that is the apparent impact parameter $L/E$ the meaningful parameter. This means that all geodesics that at infinity have an apparent impact parameter less than certain value will fall into the wormhole.
 \item $\delta_1 > \delta_c$ (Reissner-Nordström-like case): In this case, there is an infinite repulsive barrier at $x=0$ which makes all the geodesics bounce at some $r>r_c$. This prevents the geodesics from reaching the wormhole as in the Reissner-Nordström solution of GR, where $L\neq0$ geodesics cannot reach the central singularity. Between the infinite barrier and the $r\rightarrow\infty$ region where the centrifugal barrier dominates, the potential may have a local maximum and a minimum (see fig. \ref{5fig:l1k0}).
 \item $\delta_1 = \delta_c$ (smooth case): In this case, the potential is finite through the wormhole throat. At the wormhole throat, the potential always has a minimum. The value of this minimum will be positive, negative or zero, depending if the number of charges $N_q$ is less, more, or exactly the critical number $N_c$, respectively. As in the Schwarzschild-like case, the potential will always have a maximum between the asymptotic infinity and the wormhole throat.
\end{itemize}

\subsubsection{Stationary Null Orbits}

We have seen that in the geonic wormhole there may be stationary null orbits, where a photon could remain spinning around the black hole. This is also the case in GR and it would be interesting to compare both scenarios.

In the Schwarzschild solution of GR\footnote{$\df s^2 = -F \df t^2 + 1/F \df r^2 +r^2 \df \Omega^2$ with $F=1-r_S/r$}, there are unstable photon circular orbits at radius $r=3/2 r_S$. This radius is a critical point, any observer that comes closer to the black hole than this radius must spend energy not to fall to it\footnote{This applies to observers that come from the asymptotic infinity. Observers that are inside $r=3/2 r_S$ but come from the white hole part of the black hole may already have enough radial velocity to scape.}.

In the Reissner-Nordström solution of GR\footnote{$\df s^2 = -F \df t^2 + 1/F \df r^2 +r^2 \df \Omega^2$ with $F=1-r_S/r+r_q^2/r^2$}, the situation is similar but a bit different, depending on the charge to mass ratio. The potential is:

\begin{equation}
 V(r) = \left ( 1 - \frac{r_s}{r} + \frac{r_q^2}{r^2} \right )\frac{L^2}{r^2}
\end{equation}

If the charge is much bigger than the mass, the solution is of the naked singularity type. In this case, the potential is dominated by the term $V \sim \frac{L^2 r_q^2}{r^4}$ and there are no stationary null orbits. For greater values of the mass, the term that goes with the mass in the potential becomes more important and at the point $r_S^2 = \frac{2}{9} r_q^2$ there is a stationary null orbit lying on the inflection point of the potential. From that point on, there are two branches of null orbits: An unstable one at the maximum of the potential, and a stable one, at the minimum of the potential and closer to the singularity. When $r_S^2 = r_q^2$ a horizon appears at the point of the stable null orbits. For $r_S^2 > 4r_q^2$, the stable branch of the null orbits lies in a region where the potential is negative, and consequently, photons can not remain stationary at that point. Also, the unstable branch becomes the critical point beyond that any observer in free fall must spend energy not to fall into the horizon.

In the following plots we compare the photon stable orbits for the Reissner-Nordström solution of GR with the ones from the Palatini wormhole, with respect to the number of charges and given a fixed value of $\delta_1$. The mass of the black hole can be obtained from these two parameters as $M l_P^2=\frac{l_P N_q^\frac{3}{2} \alpha_{_{EM}}^\frac{3}{2}}{2^\frac{5}{4}  \delta_1}$. This means that for a fixed value of $\delta_1$, low values of $N_q$ means even lower values of $M$, which is the regime of the naked singularity of Reissner-Nordström; and higher values of $N_q$ correspond to the solution with two horizons of GR.

For the metric-affine wormhole, the null stationary orbits will depend on the value of $\delta_1$. If the geometry is smooth ($\delta_1 = \delta_c$), there will be a stable branch on the wormhole throat as long as it is not behind a horizon ($N_q < N_c$). Also, there will always be an unstable branch at the maximum of the potential outside the horizons of the geometry. For the Schwarzschild-like wormhole, there is only a branch of unstable stationary null orbits. For the Reissner-Nordström-like wormhole, there will be two branches, as in GR, but displaced.

\begin{figure}[h!]
\centering
 \includegraphics[width=0.75\textwidth]{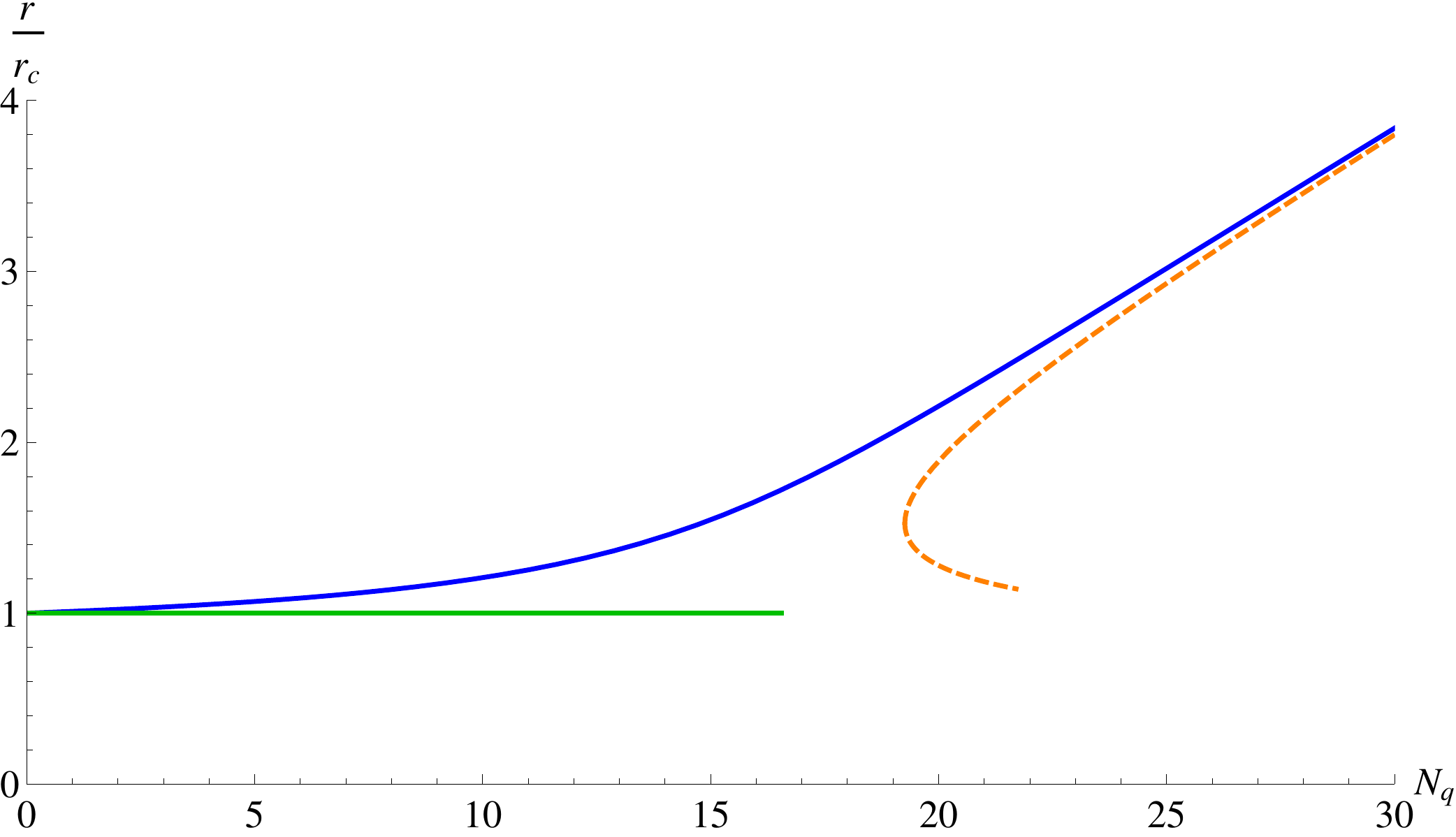}
 \caption{Radius of the stationary null orbits in units of $r_c$ (vertical axis) against the number of charges (horizontal axis) for a black hole with $\delta_1=\delta_c$. The blue (upper) and green (flat) solid lines are for the wormhole configuration, the dashed (orange) line is for the RN black hole of GR. Notice that the stable (flat) branch of stationary orbits ends at $N_q=N_c$. The upper solid (blue) curve smoothly tends to the GR prediction for large values of $N_q$. \label{fig:NullOrb1}}
\end{figure}

\begin{figure}[h!]
\centering
 \includegraphics[width=0.75\textwidth]{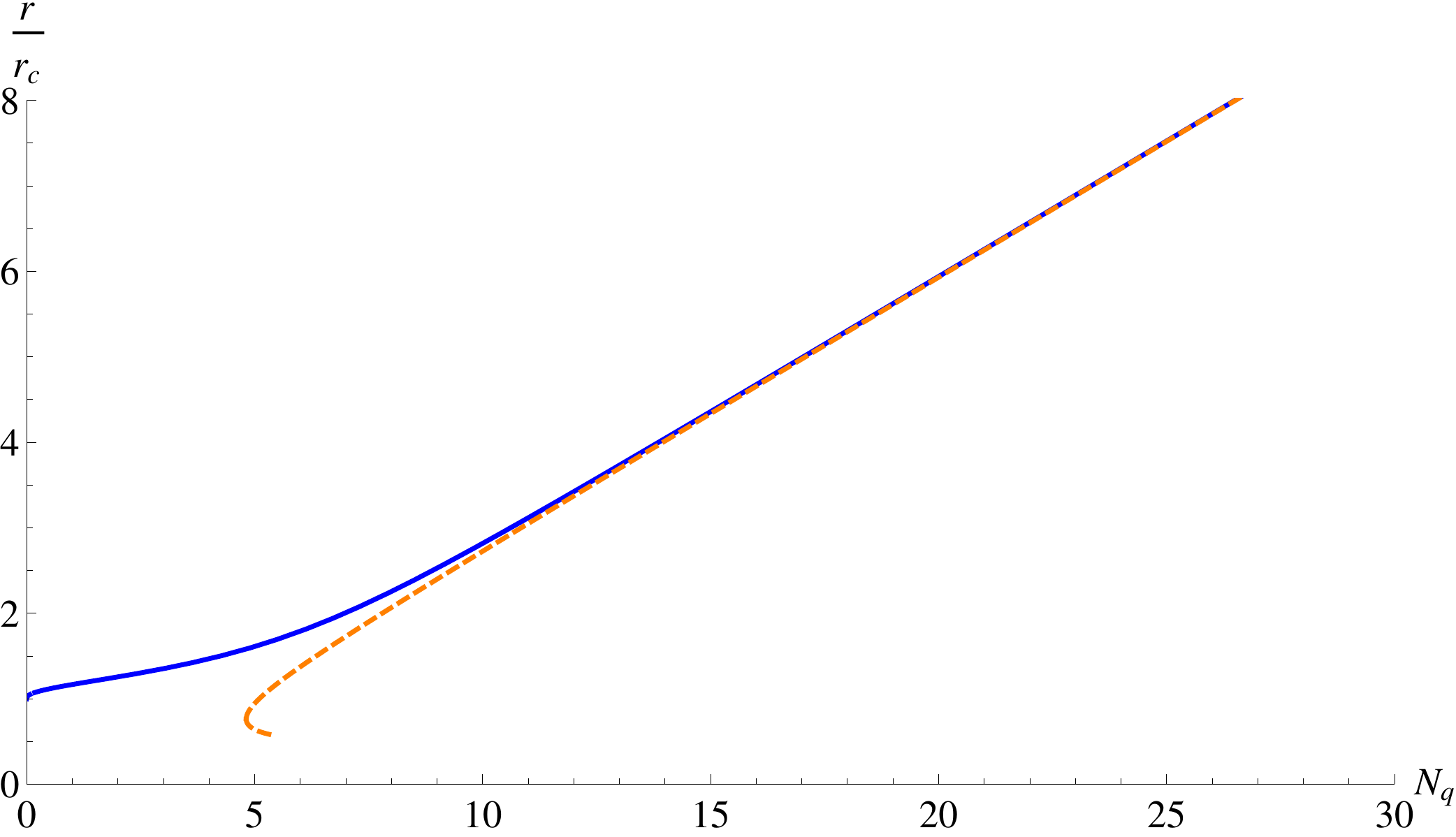}
 \caption{Radius of the stationary null orbits in units of $r_c$ (vertical axis) against the number of charges (horizontal axis) for a black hole with $\delta_1=0.5*\delta_c$. The solid (blue) line is for the wormhole, the dashed (orange) line is for the RN black hole of GR. \label{fig:NullOrb2}}
\end{figure}

\begin{figure}[h!]
\centering
 \includegraphics[width=0.75\textwidth]{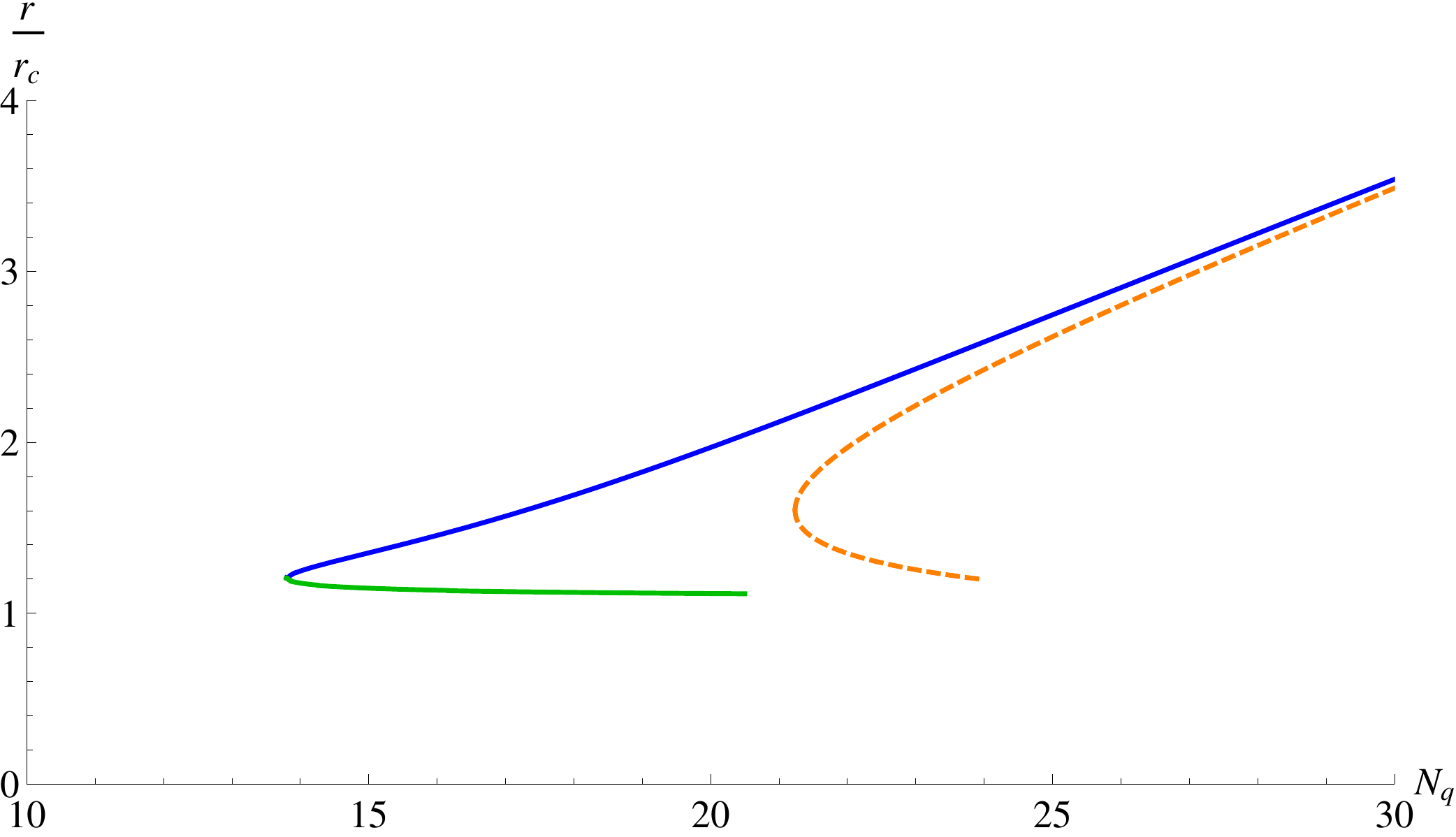}
 \caption{Radius of the stationary null orbits in units of $r_c$ (vertical axis) against the number of charges (horizontal axis) for a black hole with $\delta_1=1.05*\delta_c$. The solid blue  and green lines are for the Palatini black hole, the dashed (orange) line is for the RN black hole of GR. \label{fig:NullOrb3}}
\end{figure}

It is worth pointing out that the position of unstable branch of null orbits tends to $r \simeq \frac{3 \sqrt{\alpha_{_{EM}}}}{2 \sqrt{2}} \frac{N_q}{\delta_1} r_c = \frac{3}{2} r_S$ as the number of charges grow. As we have fixed the value of the parameter $\delta_1$, then the mass of the solution will depend on the number of charges through the relation: $r_S= r_q^{3/2}/(\delta_1 l_P^{1/2})$, so increasing the number of charges implies increasing even more the mass of the black hole, and the solution becomes more similar to the Schwarzschild geometry, which has unstable photon orbits at $r=3/2 r_S$. 

\FloatBarrier

\subsection{Radial Time-like Geodesics}

\begin{figure}[p!]
\centering
\begin{tabular}{rl}
 \includegraphics[width=0.5\textwidth]{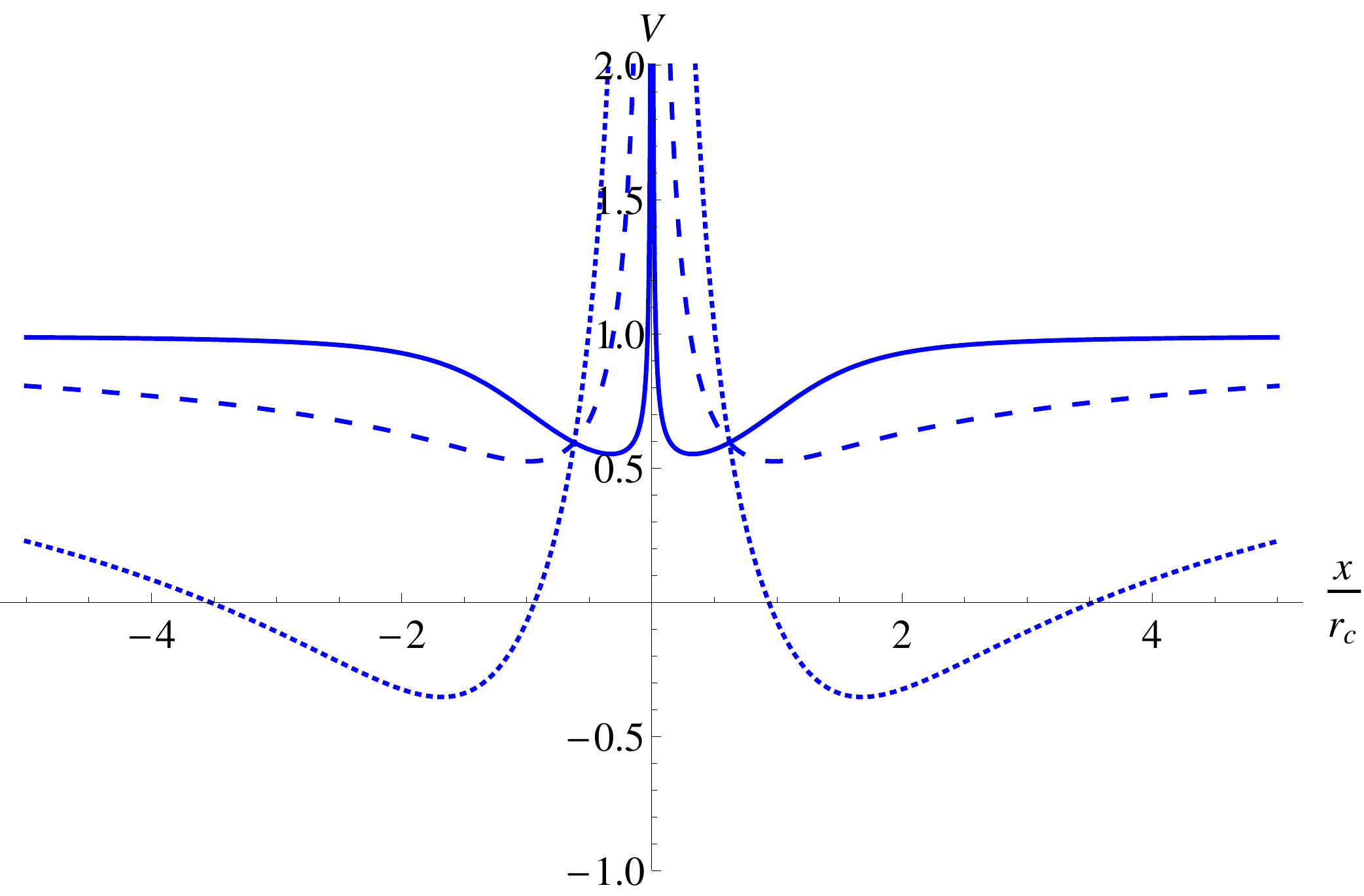} &
 \includegraphics[width=0.5\textwidth]{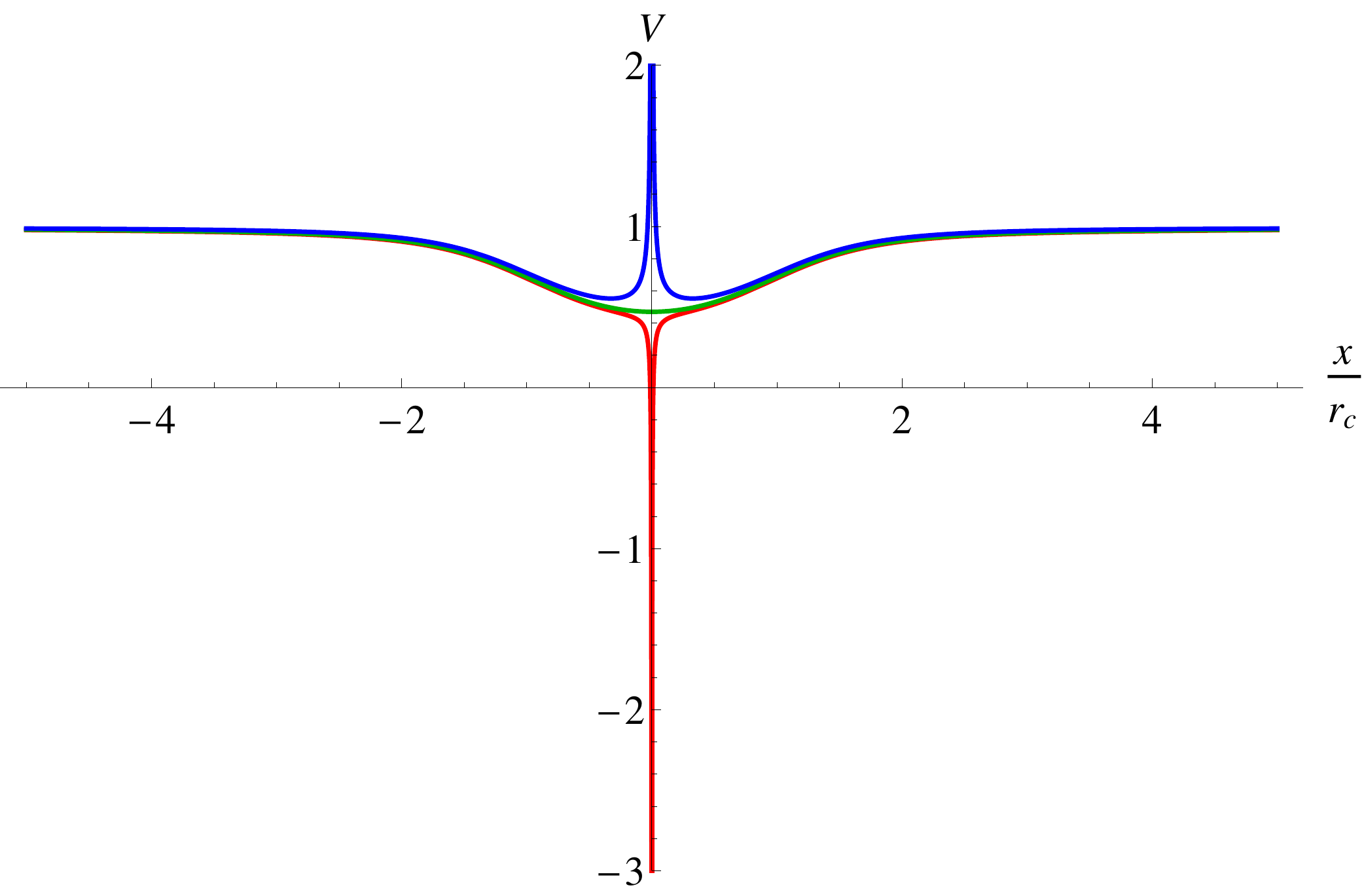} \\
  \includegraphics[width=0.5\textwidth]{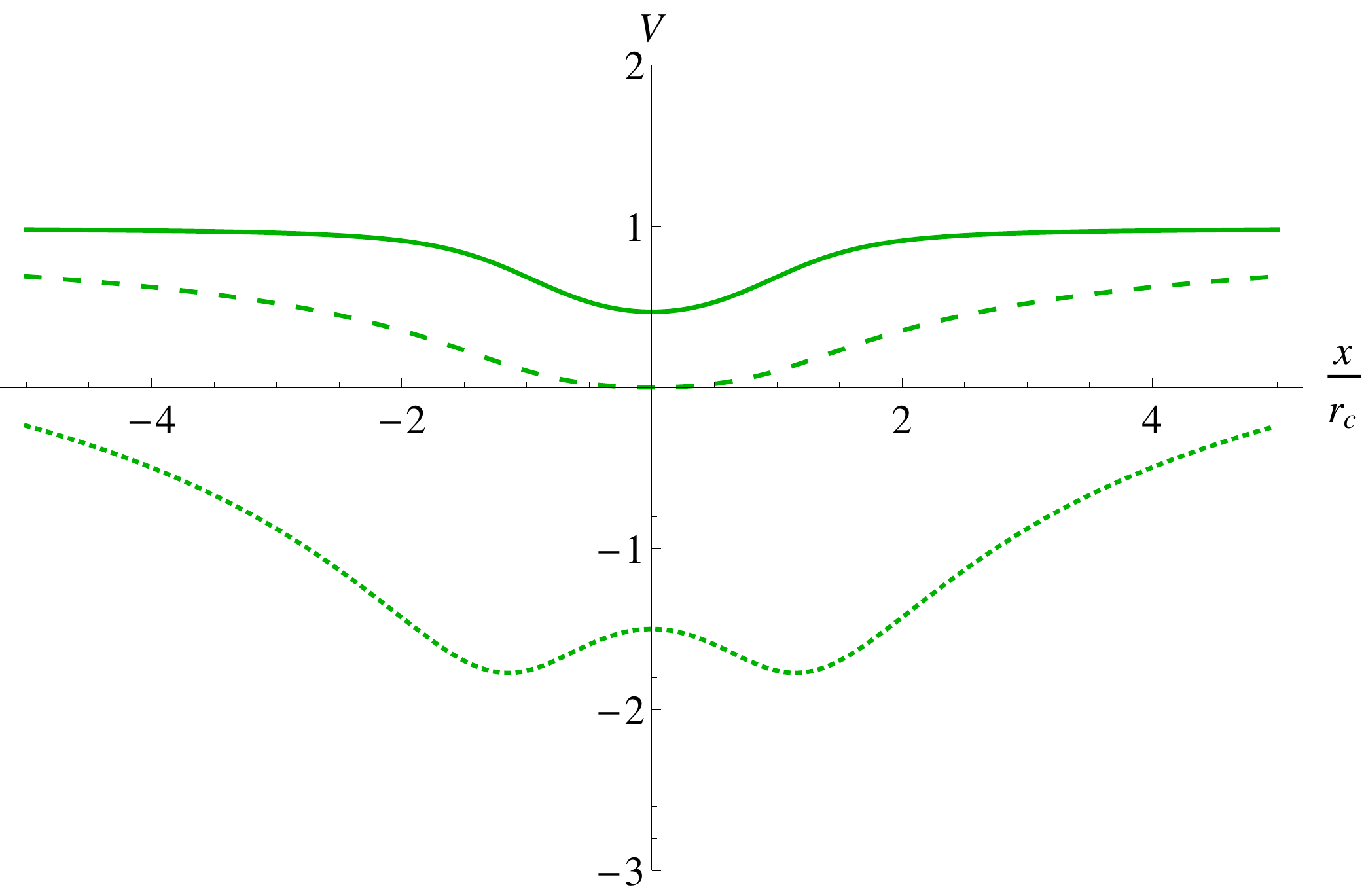} &
 \includegraphics[width=0.5\textwidth]{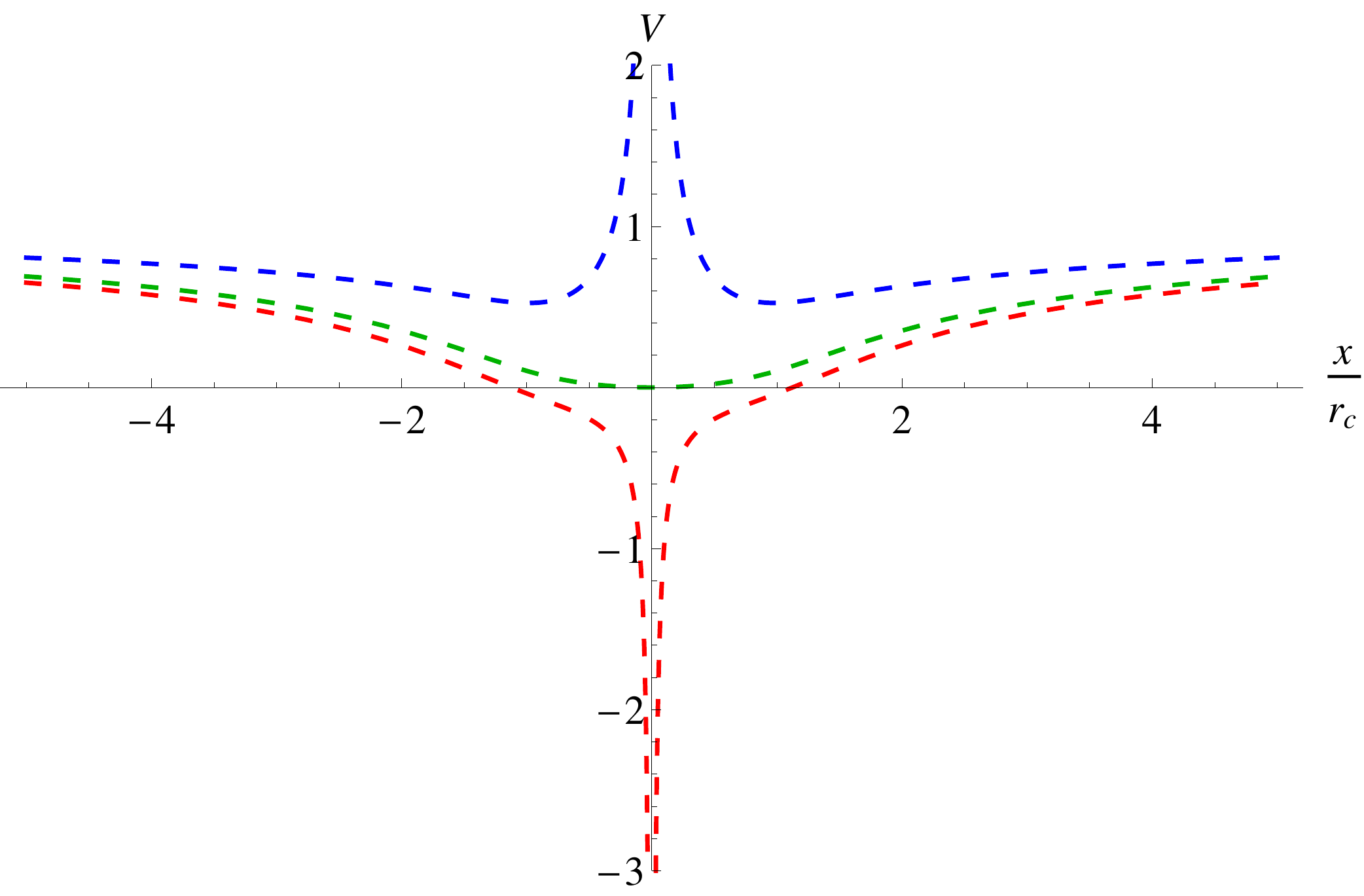} \\
  \includegraphics[width=0.5\textwidth]{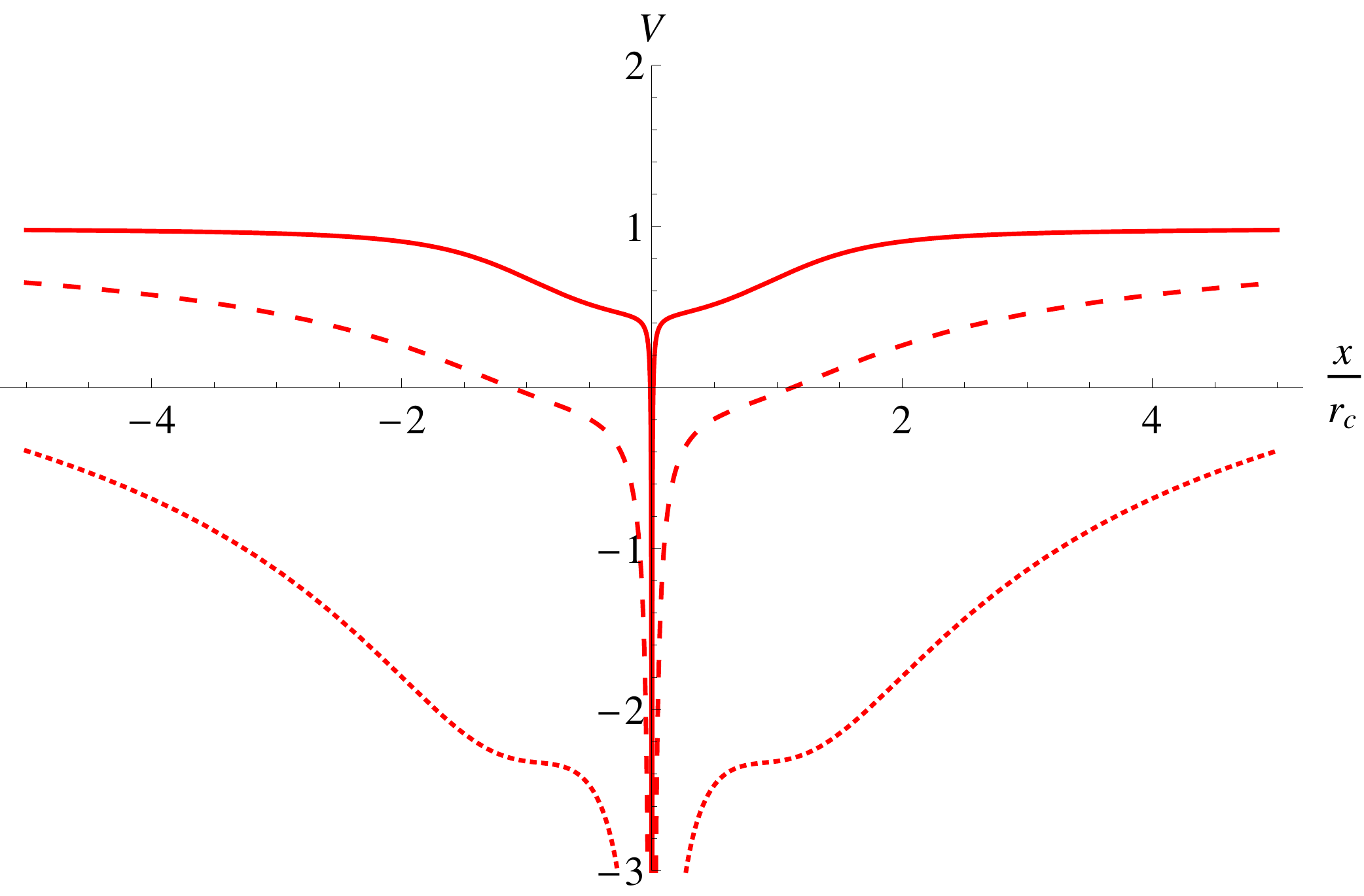} &
 \includegraphics[width=0.5\textwidth]{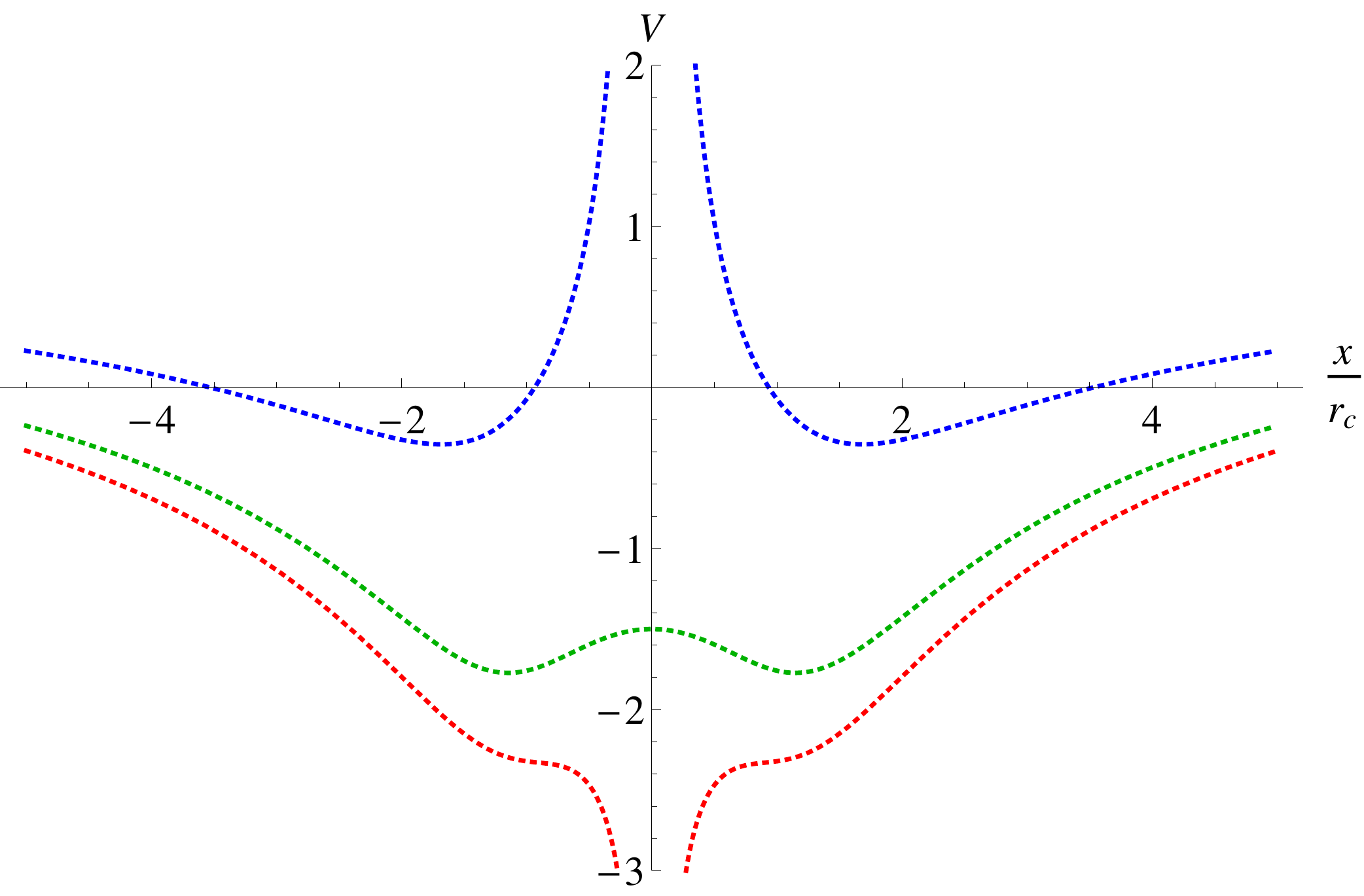} 
\end{tabular}
 \caption{Representation of the effective potential $V(x)$ for time-like geodesics with $L=0$ for different scenarios. Each plot on the left represents the potential for three different number of charges but the same value of $\delta_1$. Each plot on the right represents three different values of $\delta_1$ but the same number of charges. Blue, Green and Red lines correspond with geometries with $\delta = 1.5 \delta_c, \delta_c, 0.9 \delta_c$ respectively. Continuous, dashed and dotted lines correspond to $N_q=1, N_c, 4N_c$.}\label{5fig:l0k1}
\end{figure}

For radial time-like geodesics, we have $\kappa=1$, $L=0$. In this case, the potential far from the wormhole is attractive and behaves as in GR: $V\simeq 1 - \frac{r_S}{x}$. Near the centre of the geometry, the potential behaves very much like in the case before (see eq. \ref{5eq:Vapprox}), but without the factor $\frac{L^2}{r_c^2}$. The potential has been plotted for the different cases in fig. \ref{5fig:l0k1}.

\begin{itemize}
 \item $\delta_1 < \delta_c$ (Schwarzschild-like case): The potential is just a well, with no local maximum or minimum between infinity and the wormhole throat. Any geodesic directed towards the centre will reach the wormhole throat.
 \item $\delta_1 > \delta_c$ (Reissner-Nordström-like case): The potential has an infinite barrier at the wormhole throat, and so, all time-like geodesics will be repelled and cannot reach $r=r_c$. The potential presents a minimum between the attractive part at infinity and the infinite barrier at the wormhole throat. If the black hole is of the naked divergence type, the potential will be positive at the minimum, and a geodesic can remain stationary there, not orbiting around it, but standing at a constant value of the angle $\phi$. If it is not, the minimum will have a negative value and lie between the internal and external horizon of the geometry; and a geodesic will at most oscillate around the minimum, entering and exiting each horizon in each oscillation. We should note that after each oscillation, the observer reaches a different region than the one it started, unless we identify horizons in such a way that it comes back to the same region (in that case, there would be closed time-like curves, which can be problematic).
 \item $\delta_1 = \delta_c$ (Smooth case): In this case, the potential presents a minimum at the wormhole throat. As in the RN case, if there are no horizons, a massive particle can remain stationary at the wormhole throat. On the other hand, if the geometry has horizons, the minimum of the potential will be negative, and a geodesic will oscillate crossing the horizons to the other side of the geometry and entering back. As in the previous case, after each oscillation, the observer would reach a region of the space-time different from the one it started.
\end{itemize}

\subsection{Time-like Geodesics with $L \neq 0$}

\begin{figure}[p!]
\centering
\begin{tabular}{rl}
 \includegraphics[width=0.5\textwidth]{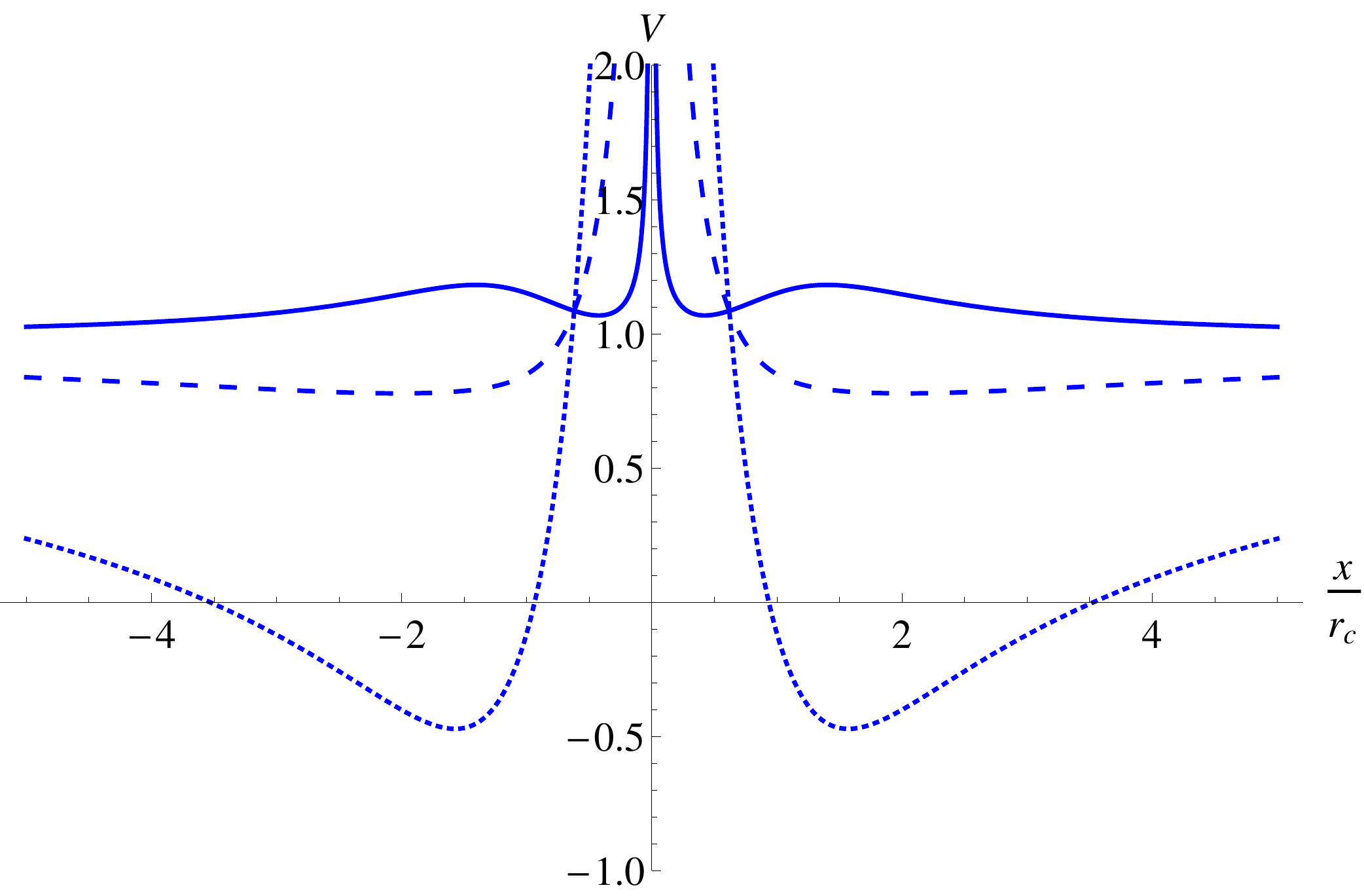} &
 \includegraphics[width=0.5\textwidth]{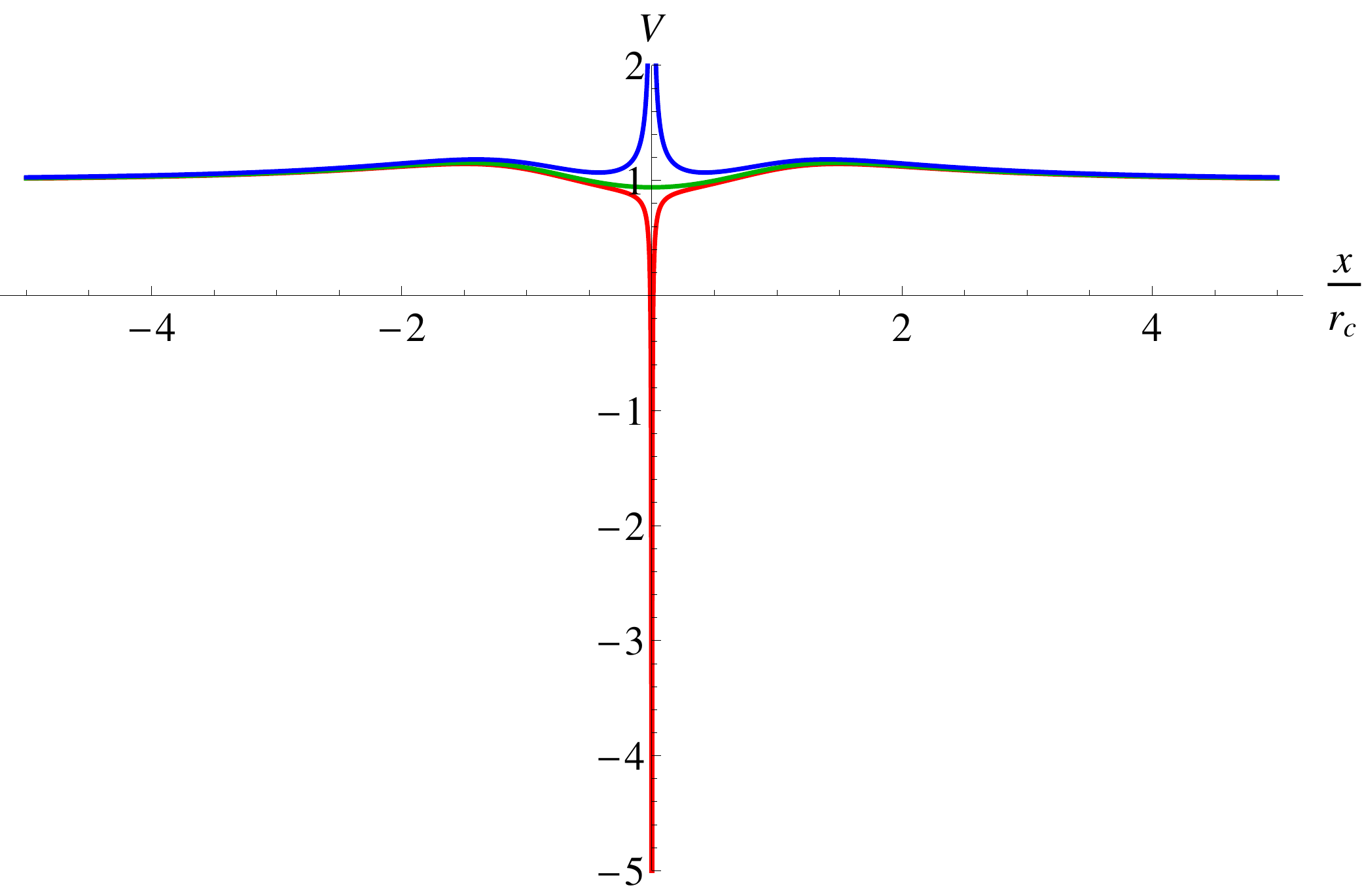} \\
  \includegraphics[width=0.5\textwidth]{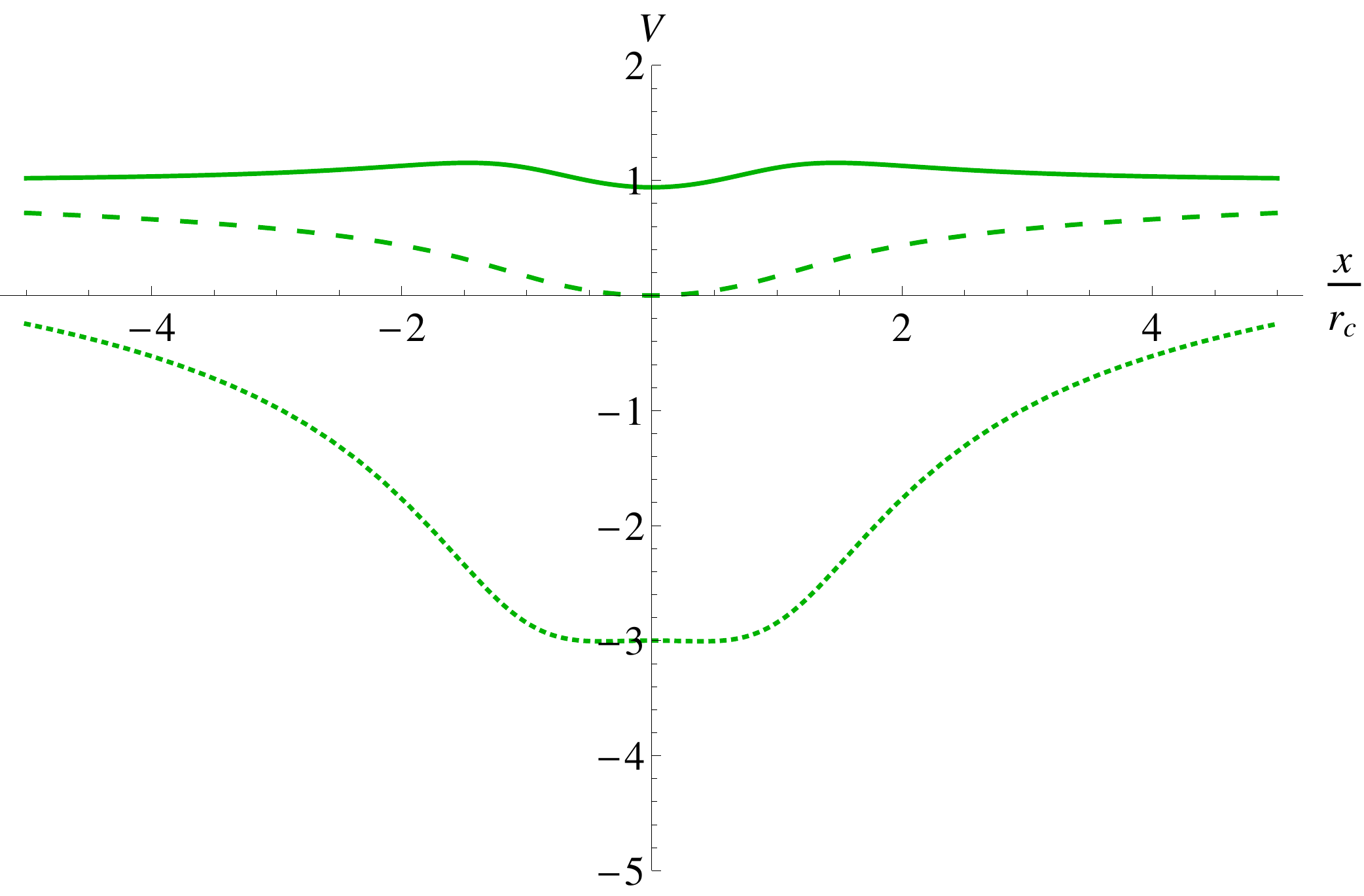} &
 \includegraphics[width=0.5\textwidth]{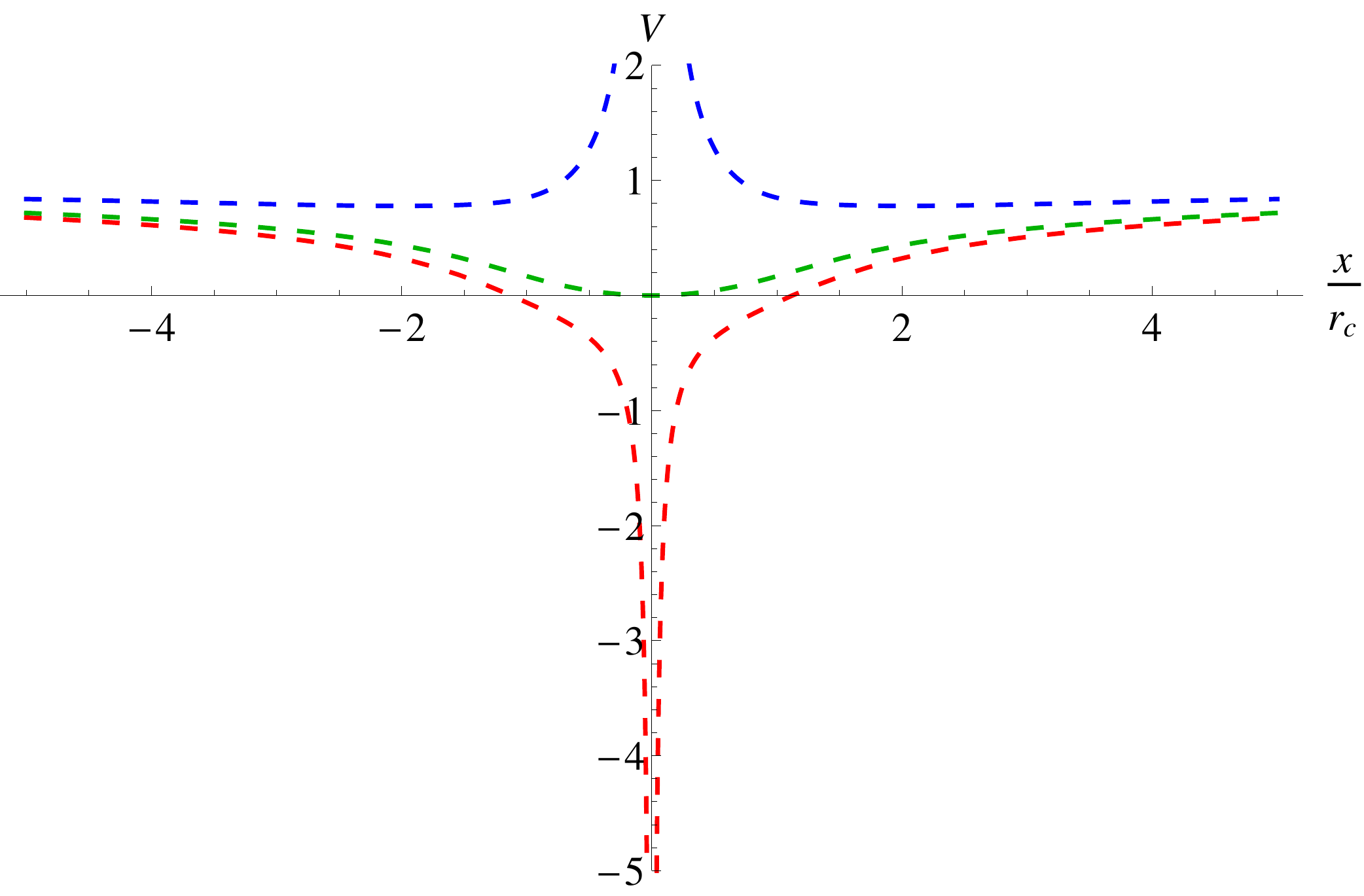} \\
  \includegraphics[width=0.5\textwidth]{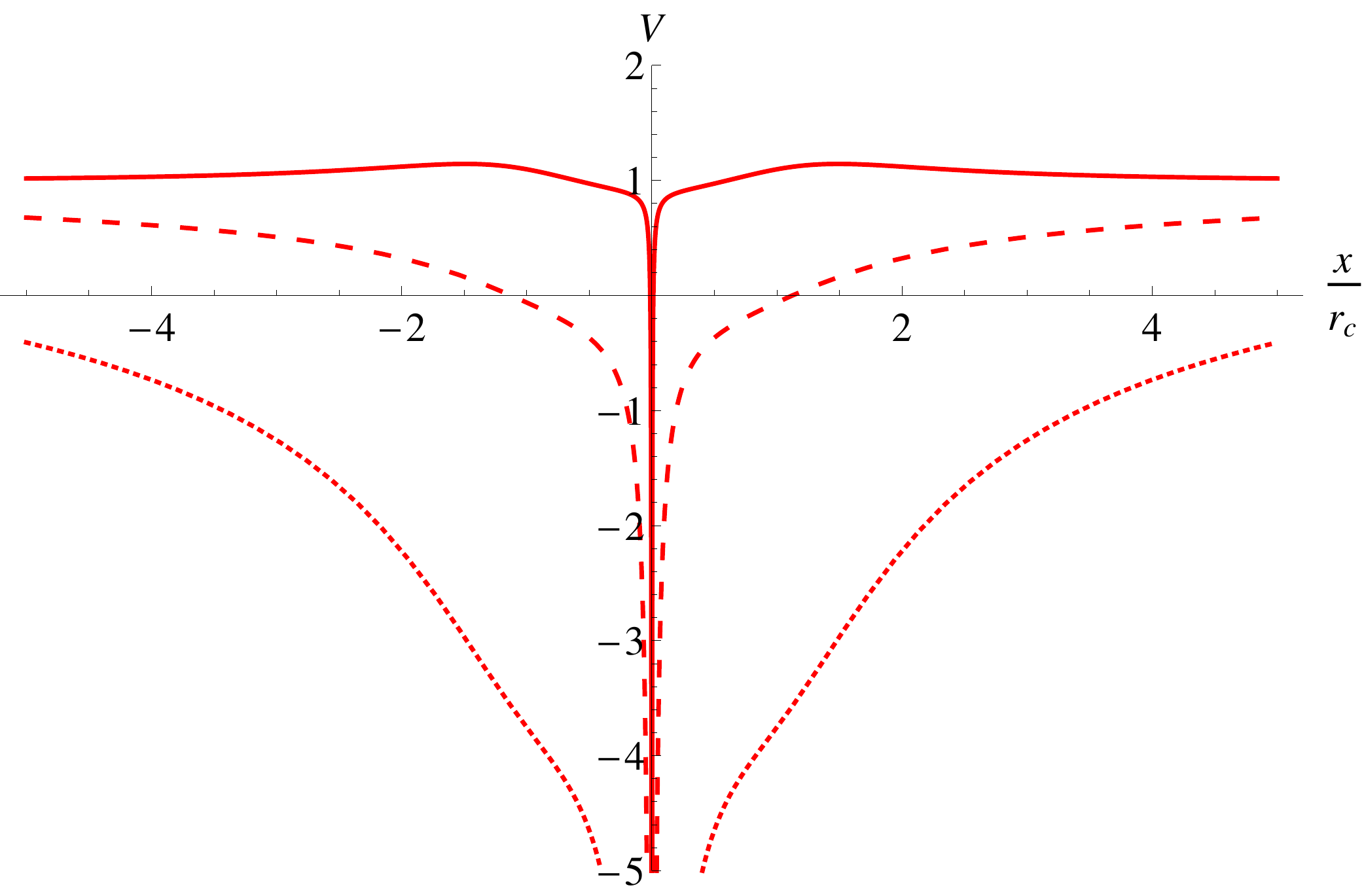} &
 \includegraphics[width=0.5\textwidth]{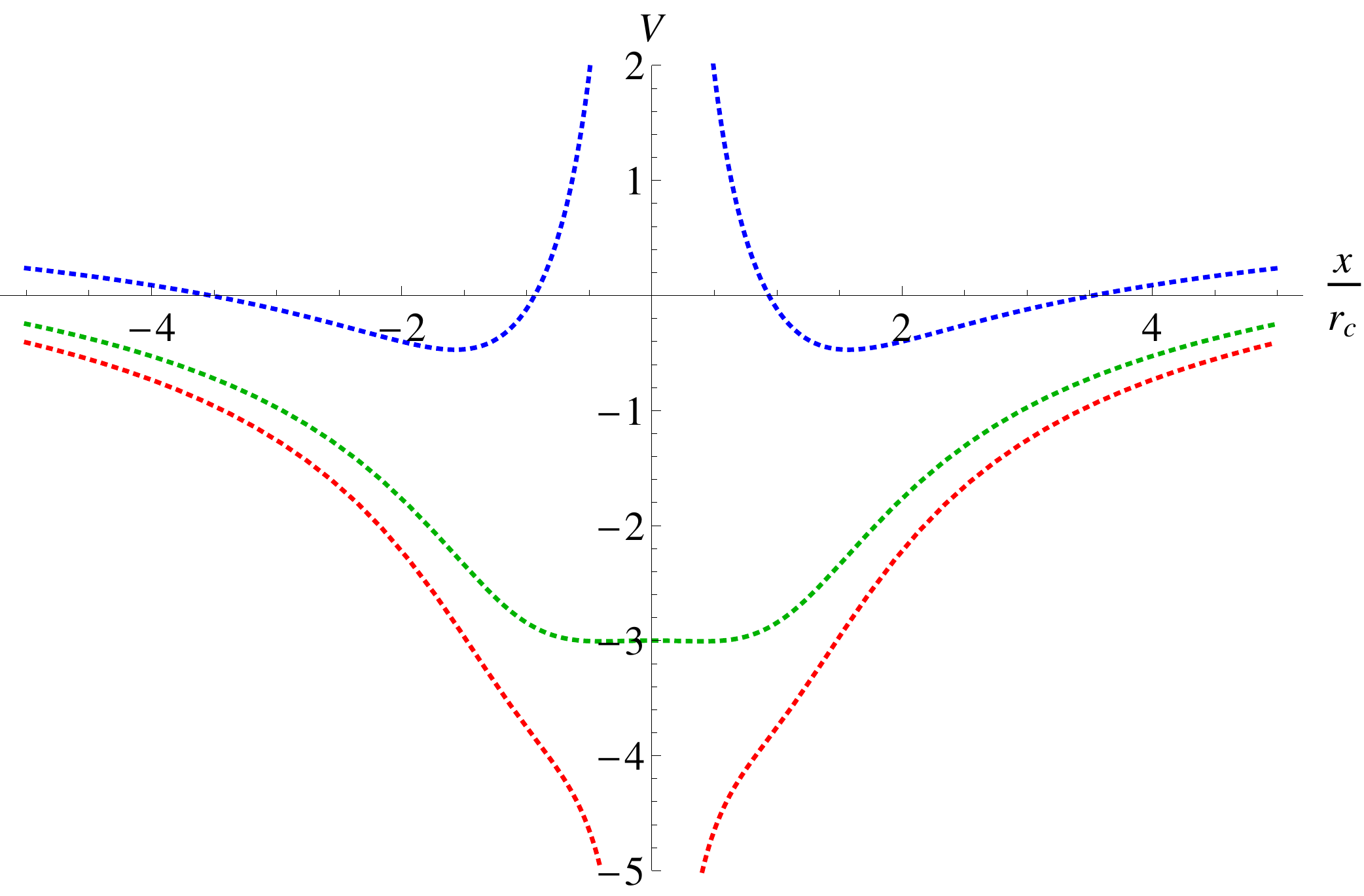} 
\end{tabular}
 \caption{Representation of the effective potential $V(x)$ for time-like geodesics with $L=1$ for different scenarios. Each plot on the left represents the potential for three different number of charges but the same value of $\delta_1$. Each plot on the right represents three different values of $\delta_1$ but the same number of charges. Blue, Green and Red lines correspond with geometries with $\delta = 1.5 \delta_c, \delta_c, 0.9 \delta_c$ respectively. Continuous, dashed and dotted lines correspond to $N_q=1, N_c, 4N_c$.}\label{5fig:l1k1}
\end{figure}

The potential for time-like geodesics with non-zero angular momentum ($\kappa=1$, $L\neq0$) is the sum of the two previous cases. At infinity, the geodesics feel the attractive potential, $V\simeq 1 - \frac{r_S}{r}$. Going closer to the centre of the geometry, they will feel the centrifugal term $\frac{L^2}{r^2}$. Depending on the value of $L^2$, there will be a minimum of the potential between the region the attractive potential dominates and the region the centrifugal term dominates. Those minima correspond to the orbits of a particle with mass (or a planet) would have around the black hole. If $L^2$ is low and we are in a wormhole geometry with horizons, the centrifugal term might not dominate at any point. This would be the case of a planet whose trajectory gets too close to the black hole and reaches the event horizon because the centrifugal barrier is not repulsive enough. The potential has been plotted for the different cases in fig. \ref{5fig:l1k1}.

\FloatBarrier

\section{Extension of Geodesics}

As we have seen, geodesics do reach the $x=0$ wormhole throat. Up to that point, there is no problem integrating the geodesic paths, because the potential is finite. For the smooth case, there is no problem integrating the geodesics through the wormhole, because the potential is also finite at that point. However, the potential becomes infinite at the wormhole throat for the Schwarzschild-like ($\delta_1<\delta_c$) and Reissner-Nordström-like ($\delta_1>\delta_c$) cases. In the geodesic equation, the coefficients of the Levi-Civita connection would diverge.

Therefore, we should try to check if there is a unique way of extending the geodesics using a Pfaff equation as we did in section \ref{1sec:ExtDisc}. But we already did that, as the geometries (i) and (iii) we studied in that section correspond to the $(t,x)$ and $(x,\phi)$ parts of the Schwarzschild-like geonic wormhole. For the Reissner-Nordström-like case, both time-like radial geodesics and null angular geodesics are repelled before reaching the wormhole throat; only null radial geodesics can reach it. But radial null geodesics can be integrated without problems through the wormhole throat.

\section{Congruences}\label{5sec:congruences}

In the previous section we have seen that some of the geodesics of the geometry can reach the wormhole throat. In the case the geometry is smooth, the geodesics reach the other side without problems. However, in general there will be a curvature divergence at the wormhole throat, and the geodesics still cross the wormhole in a continuous and unique way. What happens to an object with non-vanishing physical size that reaches the divergence?

In the introduction we studied that the mathematical tool to describe the forces that act upon a rigid object are described by the evolution of a congruence of geodesics. In sections \ref{1sec:CongSph} and \ref{1sec:CongSchw} we studied the evolution of the geodesics for a space-time with spherical symmetry and for Schwarzschild in particular. In the Schwarzschild geometry, every congruence that reached the curvature divergence was crushed to zero volume. In the geonic wormhole, the curvature divergence is milder --- $R^\alpha{}_{\beta \mu \nu} R_\alpha{}^{\beta \mu \nu} \propto 1/(r-r_c)^3$ instead of $\propto 1/r^6$ ---, and the divergent region is located at a sphere instead of at a point. We will follow what we did in the introduction an see if there are any differences with GR. 

In this section, we will work in the coordinates $(t,y,\theta,\phi)$, with $y$ such that $\df y = \frac{1}{\sigma_+} \df x$ (or equivalently $\df y = \frac{1}{\sqrt{\sigma_-}} \df r$), so that the metric of the geonic wormhole in these coordinates is written in analogue way as in eq. \ref{1eq:sphmetric}. In these coordinates the metric looks like:

\begin{equation}
 g=-\frac{A}{\sigma_+} \df t^2 + \frac{\sigma_+}{A } \df y^2 + r^2(y) \df \Omega^2 \label{5eq:metricgy}
\end{equation}

The components of the the tangent vector to a time-like radial geodesic are:

\begin{equation}
 u^\mu = \left  (\left ( \frac{\sigma_+}{A} \right ) E, \sqrt{E^2-\left ( \frac{A}{\sigma_+} \right )},0,0 \right ) \label{5eq:tangentv}
\end{equation}
and the components of the deviation vectors that describe the congruence can be written as:

\begin{IEEEeqnarray}{rCl}
 Z_{(1)}&=&B(\lambda) \left (\left ( \frac{\sigma_+}{A} \right ) \sqrt{E^2-\left ( \frac{A}{\sigma_+} \right )},E,0,0 \right ) \label{5eq:Z1}\\
 Z_{(2)}&=&P(\lambda) (0,0,1,0) \\
 Z_{(3)}&=&Q(\lambda) (0,0,0,\frac{1}{\sin \theta})
\end{IEEEeqnarray}

The equations for the modulus of the deviation vectors (eqs. \ref{1eq:PQeq}, \ref{1eq:Beq}) now look like:

\begin{equation}
 P(\lambda) = P_0 + C \int \frac{\df \lambda}{r^2(\lambda)}\label{5eq:PQeq}
\end{equation}

\begin{equation}
 \ddot{B}(\lambda)+\frac{\partial_y \partial_y \left ( \frac{A}{\sigma_+} \right )}{2}B(\lambda)=0\label{5eq:Beq}
\end{equation}

Near the wormhole throat we can make the following approximations:

\begin{equation}
r(\lambda) \approx r_c \qquad \sigma_+ \approx 2 \qquad A \approx-\frac{a}{|y|} \qquad \text{with } a \equiv \frac{N_q}{2 N_c}\left ( \frac{1-\frac{\delta_1}{\delta_c}}{\delta_1} \right ) r_c
\end{equation}

In this approximation, it is possible to integrate the component $u^y = \frac{\df y}{\df \lambda}$ to obtain $y \approx (\frac{9}{8} a \lambda^2)^\frac{1}{3}$ (We have chosen the parameter $\lambda$ such that at $\lambda=0$ the geodesic reaches the curvature divergence). Then last equation just turns into:

\begin{equation}
 \ddot{B}(\lambda)-\frac{4}{9 \lambda^2}B(\lambda)=0\label{5eq:Beq2}
\end{equation}

Now the value of the functions $P$, $Q$, $B$, can be obtained directly:

\begin{IEEEeqnarray}{rCl}
 P(\lambda) &\simeq& K_P (\lambda - \lambda_i)\\
 Q(\lambda) &\simeq& K_Q (\lambda - \lambda_i)\\
 B(\lambda) &\simeq& K_B \left (\frac{1}{|\lambda|^\frac{1}{3}} - \frac{|\lambda|^\frac{4}{3}}{|\lambda_i|^\frac{5}{3}} \right )
\end{IEEEeqnarray}
where $K_P$, $K_Q$, $K_B$ are integration constants. The volume transported by the congruence is given by:

\begin{equation}
 V(\lambda) = |B(\lambda) P(\lambda) Q(\lambda)| r^2(\lambda)
\end{equation}
which behaves as $V(\lambda) \propto 1/\lambda^\frac{1}{3} \propto 1/y^\frac{1}{2}$. Rather than vanishing as in GR, this volume diverges to infinity. This is due to the behaviour of the radial deviation vector $Z_{(1)}$, whose modulus grows without bound as $1/r^\frac{1}{2}$ when $\lambda\rightarrow 0$. The behaviour of $Z_{(1)}$ is the same as the one found in the standard Schwarzschild black hole (see eq. \ref{1eq:Z1}). In both cases, the behaviour of $g_{rr}$ ($g_{yy}$) goes like $1/r$ ($1/y$). The angular part, however, is different, because near the Schwarzschild singularity, the modulus of $Z_{(2)}$, $Z_{(3)}$ vanish like $r^\frac{1}{2}$, but in the wormhole have a finite value $K_{(P/Q)} \lambda_i$.

This behaviour is significantly different from that observed for geodesic congruences in the Schwarzschild geometry. The volume does not go to 0, and consequently, this would not be considered a strong singularity (never mind that we already established that the geodesics have an unique extension and therefore, the geometry is non-singular). The volume does go to $\infty$ and some authors regard this as a \emph{deformationally strong} singularity (\cite{Nolan:2000rn}), however, we will argue that this will not cause the unavoidable disruption of physical body crossing such region, unlike strong singularities. Also, the modules of the deviation vectors of a congruence do not vanish as we approach the wormhole throat, in general; this means that there are no conjugate points between the origin of the geodesics and the wormhole throat, which is a fundamental piece of the singularity theorems to stablish that geodesics are not extendible in GR for certain initial conditions (see Section \ref{2sec:SingTh}).

Let us study more carefully what it means that the volume transported by the congruence diverges. After all, we are considering an infinitesimal volume. What does it mean that something infinitesimal becomes infinite? To proceed, we will consider a family of freely falling observers, all with a reference energy E and zero angular momentum, their tangent vector given by eq. \ref{5eq:tangentv}. We will find it useful to write the line element of our space-time in coordinates adapted to this family of observers. We can thus define a new time coordinate that corresponds to the proper time of each observer, its corresponding basis vector is $\partial_\lambda = u^t \partial_t + u^y \partial_y$. We could also try to define a coordinate $\bar{\xi}$ such that its corresponding basis vector would be $\partial_{\bar{\xi}} = u^y \sigma_+/A \partial_t + A u^t / \sigma_+ \partial_y$. This vector has unit norm and points in the same direction as the Jacobi field $Z_{(1)}$ given in eq. \ref{5eq:Z1}. Unfortunately, this choice leads to $[\partial_\lambda, \partial_{\bar{\xi}}] \neq 0$, and therefore, does not define coordinates. However, one can verify that $\partial_\xi = u^y (u^y \sigma_+/A \partial_t + A u^t / \sigma_+ \partial_y)$ does define a coordinate basis. The new coordinates have the following expression in terms of the old ones:

\begin{equation}
 \lambda(y,t)=-E t + \int_0^y \frac{u^y(y^\prime)}{A(y^\prime)} \df y^\prime \qquad \xi(y,t) = -t + \int_0^y \frac{u^t(y^\prime)}{u^y(y^\prime)} \df y^\prime
\end{equation}

It is also possible to get the old coordinate $y$ in terms of ($\lambda$, $\xi$) inverting the following relation:

\begin{equation}
 \lambda - E \xi = \int_0^y \frac{1}{u^y (y^\prime)} \df y^\prime
\end{equation}

From this last relation it is easy to see that the wormhole throat ($y=0$) is found at $\lambda - E \xi = 0$. We can write the metric in these new coordinates:

\begin{equation}
 \df s^2 = -\df \lambda^2 + (u^y)^2 \df \xi^2 + r^2(\lambda, \xi) \df \Omega^2 \label{5eq:metriclx}
\end{equation}

With this coordinates, the tangent vector of the family of observers with energy $E$ is simply $\partial_\lambda$, and the congruence consists of the curves of constant $\xi$. Near the wormhole throat, the metric can be approximated as: 

\begin{equation}
 \df s^2 \approx - \df \lambda^2 + \left ( \frac{3}{a} |\lambda - E \xi | \right )^{-\frac{2}{3}}  \df \xi^2 \label{5eq:metriclxapp}
\end{equation}

This expression confirms what we know: the spatial distance between two infinitesimally close geodesics diverges as we approach the wormhole throat as $\left ( \frac{3}{a} |\lambda - E \xi | \right )^{-\frac{2}{3}}$. However, if we consider two geodesics separated by a finite \emph{comoving} length $l_\xi \equiv \xi_1-\xi_0$, the physical spatial distance $l_\text{Phys}\equiv \int_{\xi_0}^{\xi_1} |g_{\xi \xi}| \df \xi$ is finite and given by:

\begin{equation}
 l_\text{Phys} \approx \left ( \frac{a}{3} \right )^\frac{1}{3} \frac{1}{E} \left | |\lambda - E \xi_0 | - |\lambda - E \xi_1 | \right |
\end{equation}

This result is very important and puts forward that infinitesimal quantities that are divergent should be treated carefully. Given that the physical length for any comoving separation remains finite, and that the angular sector is well-behaved at the wormhole throat, then any object with finite (not infinitesimal) volume, will remain finite as it crosses the wormhole.

A possible gedankenexperiment to better understand the geometry would be the following: Let us consider two geodesics located at $\xi_0$ and $\xi_1$. An observer following the geodesic at $\xi_0$ sends continually a laser pulse to another observer at $\xi_1$, and when $\xi_1$ receives it, he/she sends it back to $\xi_0$. The observer at $\xi_0$ annotates the time that the laser pulse takes to travel to $\xi_1$ and back. This would be a way to measure the distance between $\xi_0$ and $\xi_1$ as they approach the wormhole throat. What would the results be?

From the metric (eq. \ref{5eq:metriclxapp}) we can obtain the paths of null geodesics, which satisfy:

\begin{equation}
 \frac{\df \xi}{\df \lambda} = \pm \left | \frac{3}{a} ( \lambda - E \xi ) \right |^\frac{1}{3}
\end{equation}

\begin{figure}
 \centering
 \includegraphics[width=.5\linewidth]{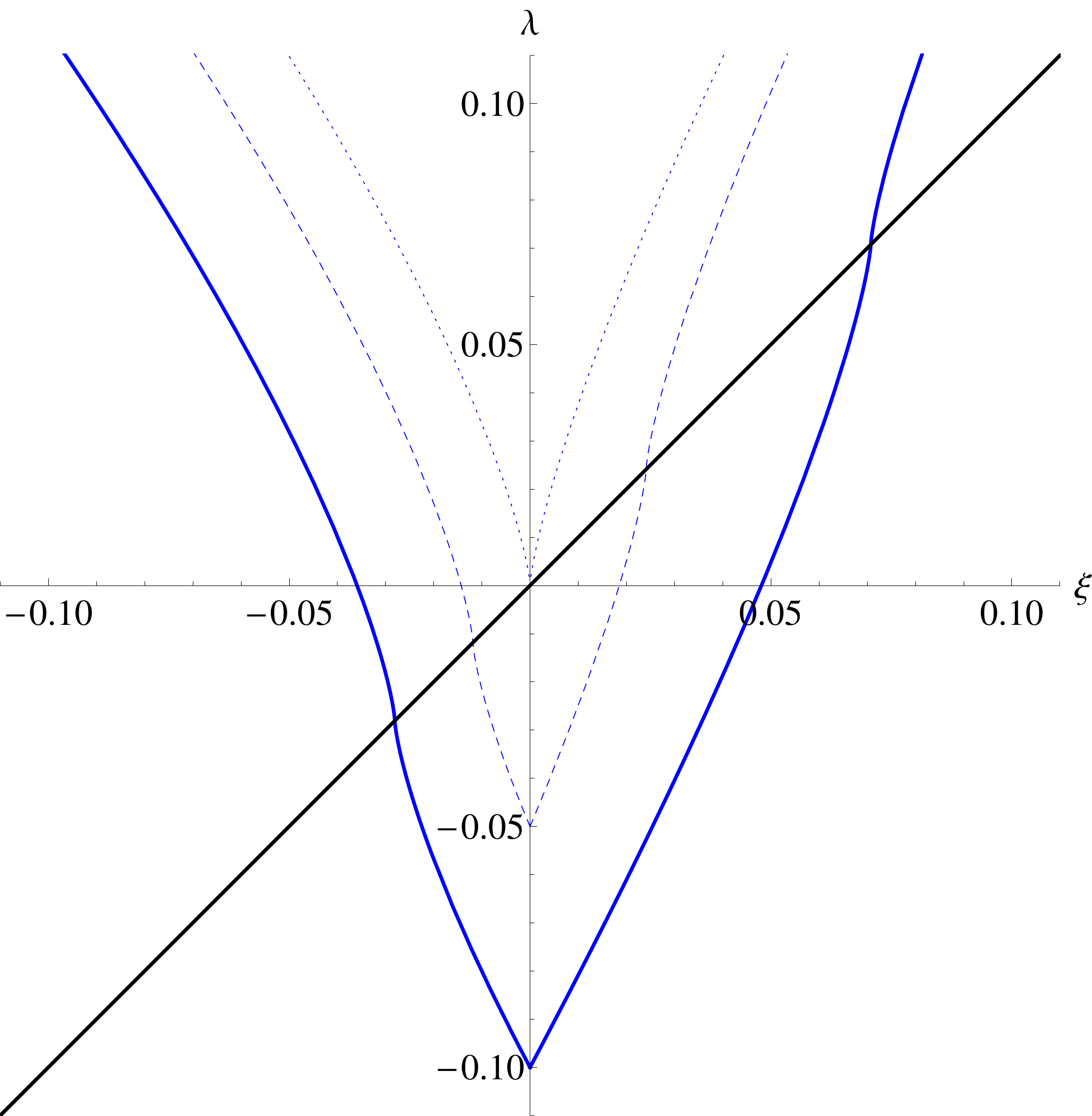}
 \caption{Trajectories of light rays emitted by a freely falling observer, ($\xi = 0$) at different times shortly before reaching the wormhole throat, for a Schwarzschild-like configuration ($\delta_1 < \delta_c$). The wormhole throat is located at the oblique (solid black) line $\lambda - E\xi = 0$. The rays going to the left/right represent ingoing/outgoing geodesics. Given that the observer is inside an event horizon, both ingoing and outgoing light rays end up crossing the wormhole throat.}\label{5fig:LightRayWH}
\end{figure}

\begin{figure}
 \centering
 \includegraphics[width=.6\linewidth]{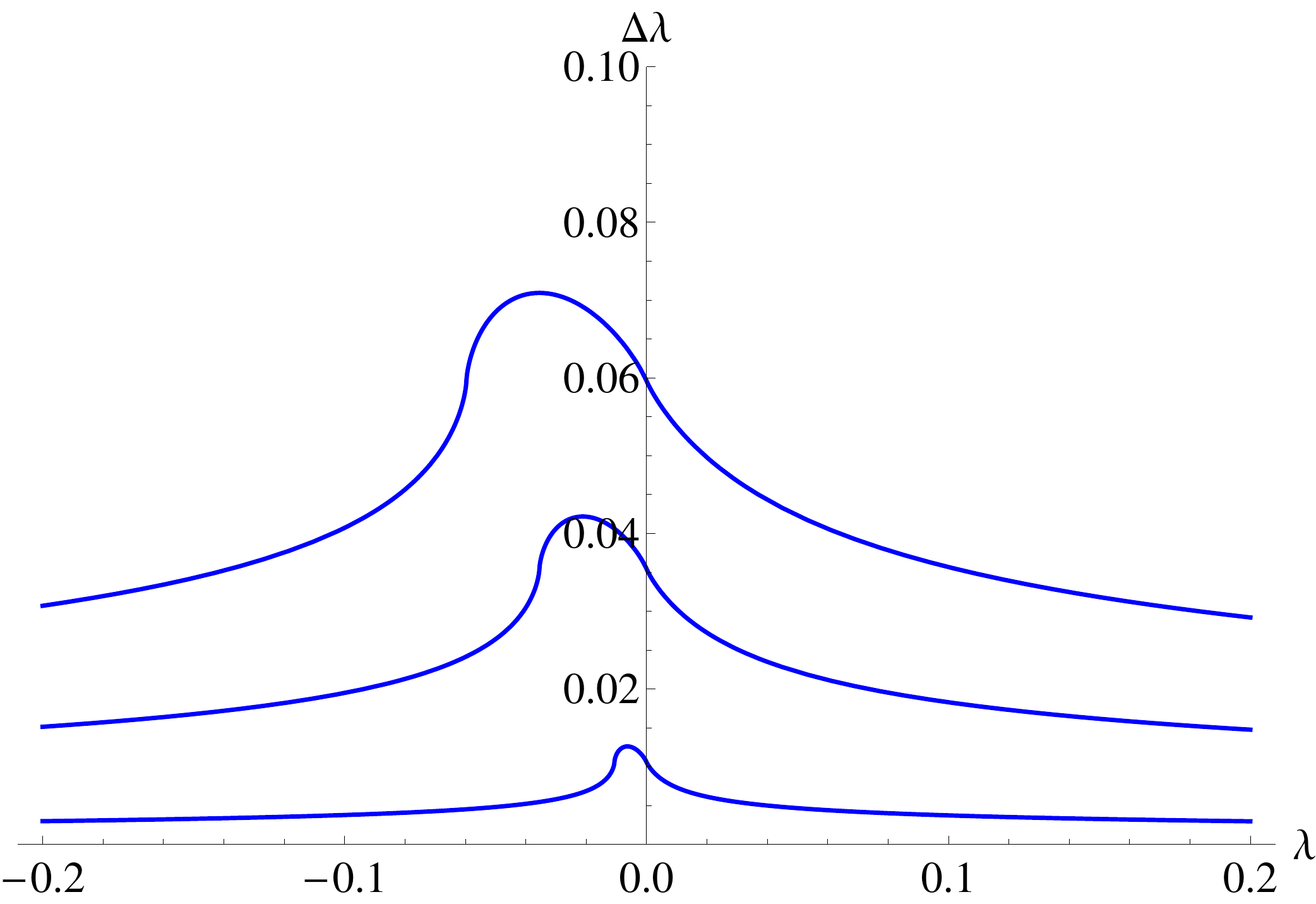}
 \caption{Representation of the proper time $\Delta \lambda$ that a light ray takes in a round trip from a fiducial free-falling observer located at $\xi=0$ with respect to another separated radially by a comoving distance $\xi = 0.01r_c, 0.005r_c, 0.001r_c$, versus the value of the proper time $\lambda$ at which the light ray was sent. At $\lambda=0$ the observer encounters the curvature divergence. Light rays sent soon before reaching the wormhole will encounter the divergence on their way, causing an additional delay (the ``bumps'' in the plot) in travelling time. This graph confirms that the travelling time is finite at all moments and tends to 0 as the comoving distance tends to 0.}\label{5fig:LightRayTimeWH}
\end{figure}

It is possible to integrate this equation numerically to obtain the paths of light rays. In figure \ref{5fig:LightRayWH} there are represented the light cones of an infalling observer. In figure \ref{5fig:LightRayTimeWH}, we can see the time the light ray takes to go from $\xi_0$ to $\xi_1$ and back, as measured by $\xi_0$ with respect to the proper time in which the light ray was originally sent (in the graph, the proper time has been set such that at $\lambda=0$, the observer at $\xi_0$ crosses the wormhole). We have represented this time for three different comoving separations between $\xi_0$ and $\xi_1$. We can see that the distance measured in this way is always finite, and vanishes as the comoving distance tends to $0$. There is a ``bump'' in travelling time as the light ray encounters the curvature divergence in its path. However, this is not problematic from a physical point of view, as both geodesics are causally connected at all times.

\FloatBarrier

\cleardoublepage
\chapter{Waves}\label{WaveChap}


During this thesis, we have characterized singularities in terms of classical aspects, studying the geodesics that physical observers follow in the geometry. However, fundamental particles are quantum in nature, and are more aptly described in terms of the propagation of a wave. In this chapter we will consider how a curvature divergence affects the propagation of a scalar wave, and will study the scattering of waves through it. Then, in a naked divergence scenario (for it is the simplest one), we will calculate the transmission coefficient for different partial waves and the transmission cross section for a plane wave.

\section{Scalar Waves and Regularity}

In this chapter, we will follow the work of Giveon et al. \cite{Giveon:2004yg} did for the wave propagation in Reissner-Nordström black hole in GR. Let us consider that matter is described by a scalar field. The wave equation for a scalar field is:

\begin{equation}
 \nabla^\mu \nabla_\mu \Psi = m^2 \Psi
\end{equation}
or equivalently:

\begin{equation}
 \frac{1}{\sqrt{|g|}} \partial_\mu (g^{\mu \nu} \sqrt{|g|} \partial_\nu \Psi) = m^2 \Psi \label{6eq:SW2}
\end{equation}

Let us study the well-posedness of this equation in the wormhole geometry we derived in Chapter \ref{SolChap}. In this section, it is convenient to use coordinates $(t,y,\theta,\phi)$ with $y$ such that $\df y = \frac{1}{\sigma_+} \df x$ (or equivalently $\df y = \frac{1}{\sqrt{\sigma_-}} \df r$). $y$ is approximately $\sim \frac{x}{2}$ near the wormhole, and $\sim x$ in the asymptotic infinity. This coordinate has a difficult expression in terms of $x$ (or $r$), but the determinant of the metric takes a simpler form:

\begin{equation}
 \df s^2 = - \frac{A}{\sigma_+} \df t^2 + \frac{\sigma_+}{A} \df y^2 +r^2(y) \df \Omega^2 \qquad \sqrt{|g|} = r^2(y) \sin(\theta) 
\end{equation}

Because the metric is static and has spherical symmetry, we can decompose the scalar field $\Psi$ into a sum of partial waves $\Psi_{\omega l m}$ of given angular momentum (denoted by the quantum numbers $l$ and $m$) and frequency $\omega$, and each of them satisfies the wave equation. Moreover, each partial wave can be separated into a spatial, a temporal and an angular part:

\begin{equation}
\Psi_{\omega l m}=\frac{\psi_{\omega l}(y)}{r} Y_{l m}(\theta,\phi) e^{i \omega t}
\end{equation}
where $Y_{l m}$ are the spherical harmonics. Eq. \ref{6eq:SW2} turns into:

\begin{equation}
 \frac{\sigma_+}{A}\omega^2 \Psi_{\omega l m} + \left [ \frac{1}{r\psi_{\omega l}} \partial_y \left ( \frac{A}{\sigma_+} r^2 \partial_y \frac{\psi_{\omega l}}{r} \right ) \right ] \Psi_{\omega l m} - \frac{l(l+1)}{r^2} \Psi_{\omega l m} - m^2 \Psi_{\omega l m}= 0
\end{equation}
for each partial wave. Our main interest is understanding the behaviour of the wave equation near the curvature divergence. To make the analysis simpler, we will study the case of a naked wormhole\footnote{Besides being a simpler case, the naked wormhole case does not have the interior horizon that in GR creates the phenomenon of mass inflation \cite{Poisson:1989zz}. If we look at the causal structure of Reissner-Nordström in fig. \ref{1fig:PenRN}a, the inner horizon $r_-$ sees all the information of the asymptotic region $r\rightarrow \infty$ (in a blue-shifted way). All the energy of this information and its backreaction generates curvature divergences at the horizon. The Metric-Affine approach has some peculiarities which would make the analysis different in this case (see \cite{MartinezAsencio:2012xn}, \cite{Lobo:2014zla}, for a discussion on the subject).}. In this case $A>0$ everywhere, and the coordinate $y$ is always space-like. Because of this, it is possible to construct a tortoise coordinate $\df y^* = \frac{\sigma_+}{A} \df y$ which will be well-defined everywhere. Using this coordinate, the differential equation for the radial function of the partial waves can be written as:

\begin{equation}
 \partial_{y^*} \partial_{y^*} \psi_{\omega l} + \left [ \omega^2 - \underbrace{\left ( \frac{1}{r} \partial_{y^*} \partial_{y^*} r + \left ( m^2 + \frac{l(l+1)}{r^2} \right ) \frac{A}{\sigma_+} \right )}_{V_{\text{eff}}} \right ] \psi_{\omega l}=0 \label{6eq:waveeq}
\end{equation}

This is a Schrödinger-like equation with effective potential $V_{\text{eff}}$. In the limit of $r\gg r_c$, $\sigma_\pm \rightarrow 1$, $A\rightarrow F$, where $F$ is the metric function $-g_{tt}$ of the Reissner-Nordström geometry, this effective potential tends to the usual one in GR $V_\text{eff}=\left ( \frac{1}{r} \partial_r F+\left (m^2+ \frac{l(l+1)}{r^2} \right ) \right ) F$; the scattering and absorption of waves by black holes in GR has already been studied (\cite{Giveon:2004yg}, \cite{Sanchez:1976}, \cite{Crispino:2009xt}). Using this equation, it should be possible to obtain solutions for a given partial wave (two independent solutions given $\omega$ and $l$). However, we are concerned about the behaviour of the effective potential at the wormhole throat, whether it poses a problem to the propagation of scalar waves or not. For this purpose we can approximate the effective potential near the wormhole throat as:

\begin{equation}
 V_{\text{eff}} \approx \frac{\kappa}{|y^*|^\frac{1}{2}} + O(|y^*|^\frac{1}{2}) \label{6eq:VeffApprox}
\end{equation}
with:

\begin{equation}
 \kappa \equiv \sqrt{\frac{(\delta_1-\delta_c)N_q}{\delta_1 \delta_c N_c}}\frac{N_c [m^2 r_c^2 +1 + l(l+1)]-N_q)}{N_c (2 r_c)^\frac{3}{2}} \label{6eq:kappadef}
\end{equation}

The effective potential is divergent at $y=0$. Will this cause a singular behaviour of the wave function at this point? The answer is no: A linear second-order differential equation of the type $[a(x) \partial_{xx}+ b(x) \partial_x + c(x)]\phi =0$ is \emph{regular-singular} at a point $x_0$ if $b(x)$ has at most a first-order pole and $c(x)$ at most a second-order pole at $x_0$ with $a(x_0)$ normalized to $1$ (see \cite{Giveon:2004yg}). This guarantees that the solution will not have an essential singularity at $x_0$, although it can have poles or branch cuts. The leading behaviour of the two independent solutions at the regular-singular point is given by $\phi_{1,2} \sim (x-x_0)^{\rho_{1,2}}$, with the characteristic exponents $\rho_{1,2}$ being the solutions of the quadratic equation\footnote{If $\rho_1=\rho_2=\rho$, then one of the solutions has leading behaviour as $(x-x_0)^\rho \log(x-x_0)$.}:

\begin{equation}
 a(x_0) \rho (\rho-1)+b(x_0) (x-x_0) \rho+c(x_0) (x-x_0)^2=0
\end{equation}

In our case, $c(y) \sim 1/y^\frac{1}{2}$ and the leading behaviour of the solution at $y=0$ is simply one solution that is constant, and other that is linear. If we consider potentials of the type $V=-v_0/y^\alpha$, the solutions for the potentials with $0<\alpha < 2$ will be well-behaved, the ones with $\alpha >2$ will be singular, and the ones with $\alpha=2$ will depend on the value of $v_0$, and be well-behaved if $v_0<1/4$.


Eq. \ref{6eq:waveeq} has two independent solutions $\psi_{\omega l}^1$, $\psi_{\omega l}^2$. When $r\rightarrow \infty$, the effective potential vanishes and each of the independent solutions is a combination of complex exponentials:

\begin{IEEEeqnarray}{rCl}
 \lim_{y^*\rightarrow + \infty} \psi_{\omega l}^{1,2}(y^*) &=& A_{\omega l,\text{in}}^{1,2} e^{-i k y^*} + A_{ \omega l,\text{out}}^{1,2} e^{i k y^*} \\
 \lim_{y^*\rightarrow- \infty}\psi_{\omega l}^{1,2}(y^*) &=& B_{ \omega l,\text{in}}^{1,2} e^{i k y^*} + B_{ \omega l, \text{out}}^{1,2} e^{-i k y^*} 
\end{IEEEeqnarray}


With $\omega = \sqrt{k^2+m^2}$. Any other linear combination of these two solutions will be a solution too. In particular we could look for a combination $\psi_{\omega l, \text{in}}^+$ that corresponds to having an incoming wave from $y^*\rightarrow \infty$ that interacts with the geometry and then is outgoing in both sides of the wormhole. This combination then satisfies $\lim_{y^*\rightarrow -\infty} \psi_{\omega l, \text{in}}^+ = e^{-iky^*}$. In an analogue way we can define $\psi_{\omega l, \text{out}}^+$ , $\psi_{\omega l, \text{in}}^-$ and $\psi_{\omega l, \text{out}}^-$, which are combinations such that $\lim_{y^*\rightarrow -\infty} \psi_{\omega l, \text{out}}^+ = e^{iky^*}$, $\lim_{y^*\rightarrow +\infty} \psi_{\omega l, \text{in}}^- = e^{iky^*}$, $\lim_{y^*\rightarrow +\infty} \psi_{\omega l, \text{out}}^- = e^{iky^*}$, respectively. Each of these solutions will give a corresponding partial wave, that we will label the same way, i.e. $\Psi_{\omega l m, \text{in}}^+ = \psi_{\omega l, \text{in}}^+(r)/r \ Y_{l m}(\theta,\phi) e^{i \omega t}$. A generic wave can be decomposed into a sum of partial waves of different frequency, angular momentum and any combination of two independent solutions for a given $\omega$ and $l$. For example, we can decompose a generic wave using $\Psi_{\omega l m, \text{in}}^+$ and  $\Psi_{\omega l m, \text{out}}^+$:

\begin{equation}
 \Psi = \sum_{l=0}^\infty \sum_{m=-l}^l \int_{0}^\infty \left \{ f_{l m, \text{in}}^+(k) \Psi_{\omega l m, \text{in}}^+ +f_{l m, \text{out}}^+(k) \Psi_{\omega l m, \text{out}}^+ \right \} \df k
\end{equation}
where, $\{ f_{l m, \text{in}}^+(k)$, $f_{l m, \text{out}}^+(k)\}$ specifies the particular combination of partial waves that make $\Psi$. But we can work with any combination we want, so it is also possible to decompose it using $\Psi_{\omega l m, \text{in}}^+$ and $\Psi_{\omega l m, \text{in}}^-$:

\begin{equation}
 \Psi = \sum_{l=0}^\infty \sum_{m=-l}^l \int_{0}^\infty \left \{ g_{l m, \text{in}}^+(k) \Psi_{\omega l m, \text{in}}^+ +g_{l m, \text{in}}^-(k) \Psi_{\omega l m, \text{in}}^- \right \} \df k \label{6eq:Psigg}
\end{equation}
or with $\Psi_{\omega l m, \text{in}}^-$ and  $\Psi_{\omega l m, \text{out}}^-$:

\begin{equation}
 \Psi = \sum_{l=0}^\infty \sum_{m=-l}^l \int_{0}^\infty \left \{ h_{l m, \text{in}}^-(k) \Psi_{\omega l m, \text{in}}^- +h_{l m, \text{out}}^-(k) \Psi_{\omega l m, \text{out}}^- \right \} \df k \label{6eq:hdecomp}
\end{equation}

Where $\{g_{l m, \text{in}}^+(k)$, $g_{l m, \text{in}}^-(k)\}$ and $\{ h_{l m, \text{in}}^-(k)$, $h_{l m, \text{out}}^-(k)\}$ will have some relation between themselves\footnote{Let us note that in general $g_{l m, \text{in}}^+(k) \neq f_{l m, \text{in}}^+(k)$} and to $\{ f_{l m, \text{in}}^+(k)$, $f_{l m, \text{out}}^+(k)\}$.

Now we would like to determine if we can solve a generic initial value problem in the naked wormhole setting. It is possible to formulate this problem in two ways: in the Cauchy formulation, the field and its time derivative are specified on a space-like hypersurface; in the characteristic formulation, the field is specified on two null hypersurfaces. We can see that there is a problem with the Cauchy formulation, as (for the naked wormhole) any space-like hypersurface will cross the wormhole throat, and we do not yet know what initial conditions are physically acceptable there. However, in the characteristic formulation, we can choose as our null hypersurfaces the null past infinity and null future infinity in the positive $y^*$ side of the wormhole, usually denoted by $\mathcal{I}^-$ and $\mathcal{I}^+$. In Eddington-Finkelstein coordinates $u=t+y^*$, $v=t-y^*$, these two hypersurfaces are the hypersurface of constant $v$ in the limit $v\rightarrow -\infty$, and of constant $u$ in the limit $u\rightarrow +\infty$. If the scalar field is given in those hypersurfaces, it should be possible to obtain the particular composition of partial waves, and with them obtain the value of the scalar field everywhere else. For example, for a massless scalar field, in the limit $y^*\rightarrow +\infty$ we have $\Psi_{\omega l m, \text{in}}^- \rightarrow \frac{e^{i \omega u}}{r} Y_{lm}(\theta, \phi)$, and $\Psi_{\omega l m, \text{out}}^- \rightarrow \frac{ e^{i \omega v}}{r} Y_{lm}(\theta, \phi)$. So, if we know $\Psi$ in these hypersurfaces, we can Fourier decompose them and obtain $h_{l m, \text{in}}^-(\omega)$ and $h_{l m, \text{out}}^-(\omega)$:

\begin{IEEEeqnarray}{rCl}
 h_{l m, \text{in}}^-(\omega) &=& \lim_{v\rightarrow -\infty} \frac{1}{2 \pi} \int \Psi(u,v,\theta,\phi) Y_{lm}(\theta, \phi) r(u,v) e^{-i \omega u} \sin (\theta) \df u \df \theta \df \phi \IEEEeqnarraynumspace \\ 
 h_{l m, \text{out}}^-(\omega) &=& \lim_{u\rightarrow +\infty} \frac{1}{2 \pi} \int \Psi(u,v,\theta,\phi) Y_{lm}(\theta, \phi) r(u,v) e^{-i \omega v} \sin (\theta) \df v \df \theta \df \phi  \IEEEeqnarraynumspace 
\end{IEEEeqnarray}

With these two functions we know the evolution of $\Psi$ in the full space-time through eq. \ref{6eq:hdecomp}. It is more natural to take as null hypersurfaces for our initial value problem the null past infinity of both sides of the wormhole. In this case, in the limit $y^* \rightarrow -\infty$, we have $\Psi_{\omega l m, \text{in}}^- \rightarrow \left ( A_{\omega l} \frac{e^{i \omega u}}{r} + B_{\omega l} \frac{e^{i \omega v}}{r} \right ) Y_{lm}(\theta, \phi)$, and $\Psi_{\omega l m, \text{out}}^- \rightarrow \left ( C_{\omega l} \frac{e^{i \omega u}}{r} + D_{\omega l} \frac{e^{i \omega v}}{r} \right )  Y_{lm}(\theta, \phi)$. In this case, $h_{l m, \text{in}}^-(\omega)$ and $ h_{l m, \text{out}}^-(\omega)$ can be obtained from the following equations:

\begin{IEEEeqnarray}{l}
 h_{l m, \text{in}}^-(\omega) = \lim_{v\rightarrow -\infty} \frac{1}{2 \pi} \int \Psi(u,v,\theta,\phi) Y_{lm}(\theta, \phi) r(u,v) e^{-i \omega u} \sin (\theta) \df u \df \theta \df \phi \IEEEeqnarraynumspace \\ 
 B_{\omega l} h_{l m, \text{in}}^-(\omega) + D_{\omega l} h_{l m, \text{out}}^-(\omega) = \nonumber \\
 \qquad = \lim_{u\rightarrow -\infty} \frac{1}{2 \pi} \int \Psi(u,v,\theta,\phi) Y_{lm}(\theta, \phi) r(u,v) e^{-i \omega v} \sin (\theta) \df v \df \theta \df \phi  
\end{IEEEeqnarray}

\section{Transmission Through The Wormhole Throat}

Now that we have seen that the propagation of a wave is smooth through the curvature divergence, we are interested in understanding what happens to a plane wave sent towards the wormhole throat. At the throat, the potential will be an infinite barrier or an infinite well depending on the sign of $\kappa$, which in turn depends on $l$, $m$, $\delta_1$ and $N_q$. For the potential well, there will be bounded states if $\omega^2 < m^2$. In the case of the infinite barrier, there can be propagation through the wormhole throat due to tunnel effect.

If we consider a high frequency wave sent towards the wormhole, all the features of the potential will be irrelevant except for the behaviour at the throat. The centrifugal part of the effective potential will not deter the wave to interact with the curvature divergence. Then, we can calculate the transmission and the cross section just by looking at the interaction of the wave with the approximated potential at the wormhole throat.

\subsubsection{Classical Scattering}

Let us first review the scattering of a scalar field in flat space-time. In the classical scattering experiment, a monochromatic plane wave is sent along the direction $z$ towards a static and spherically symmetric potential. Part of the wave will pass unscattered through the potential, part of the wave will interact with it and produce a spherically outgoing wave, and maybe part of the wave will be absorbed. Far from the potential, the wave is essentially free $(\nabla^\mu \nabla_\mu - m^2) \Psi=0$ and can be approximated by the following form:

\begin{equation}
 \Psi \underset{r\rightarrow \infty}{\simeq} \left ( e^{ikz} + f_\omega(\theta) \frac{e^{ikr}}{r} \right ) e^{i \omega t} \label{6eq:Scatteq}
\end{equation}
where we can see the plane wave and the spherically outgoing wave, the intensity of which depends on the angle respect to the direction of the plane wave. As the potential is spherically symmetric and static, we can write the wave as a sum of partial waves, and each of them can be written as:

\begin{equation}
 \Psi_{\omega l m}=\frac{\psi_{\omega l}(y)}{r} Y_{l m}(\theta,\phi) \exp(-i \omega t)
\end{equation}

For big radii, the waves are free, and the radial part of the partial wave behaves like\footnote{The equation for the radial part of the wave is of second-order, and consequently, has two independent solutions. However, either imposing well behavedness of the solution at the origin, or boundary conditions (for example, for the scattering of waves by a hard ball), will select just one of them.} $\psi_{\omega l}(r) \underset{r\rightarrow \infty}{\simeq} A_{{\text{in,} \omega l}} \exp(-i \omega r) + A_{{\text{out,} \omega l}} \exp(i \omega r)$. Now we would like to write the equation \ref{6eq:Scatteq} in terms of partial waves. For this, we have to look first how a plane wave $e^{ikz}$ decomposes into waves of known angular momentum:

\begin{IEEEeqnarray}{rCl}
e^{i kz} &=&  \sum_{l=0}^\infty i^l \sqrt{4 \pi (2l +1)} j_l(k r) Y_{l 0}(\theta, \phi)\\
&\underset{r\rightarrow \infty}{\simeq}&  \sum_{l=0}^\infty i^l \sqrt{4 \pi (2l +1)} \frac{1}{2kr}(e^{i k r}+e^{-i k r}) Y_{l 0}(\theta, \phi)
\end{IEEEeqnarray}
where $j_l(kr)$ is the spherical Bessel function of the first kind. The wave function that corresponds to the scattering experiment described in eq. \ref{6eq:Scatteq} must be made of a combination of partial waves $\Psi_{\omega l m}$ with quantum number $m=0$, such that the incoming part at infinity is the same as the incoming part of the plane wave $e^{ikz}$. This combination then must be:

\begin{equation}
 \Psi = \sum_{l=0}^\infty i^l \sqrt{4 \pi (2l +1)}  \frac{1}{2 k } \frac{1}{A_{\text{in,} \omega l}} \frac{ \psi_{\omega l}  }{r} e^{i \omega t} Y_{l 0}(\theta, \phi)
\end{equation}

Then, if we sum and subtract the outgoing part of the plane wave, we can separate the wave function into the incoming plane wave and the scattered spherically outgoing wave.

\begin{IEEEeqnarray}{rCl}
 \Psi &\underset{r\rightarrow +\infty}{\simeq}& \sum_{l=0}^\infty i^l \sqrt{4 \pi (2l +1)}  \frac{1}{2 k } \frac{1}{A_{\text{in,} \omega l}} \frac{ A_{\text{in,} \omega l} e^{-i k r} + A_{{\text{out,} \omega l}} e^{i k r} }{r} e^{i \omega t} Y_{l 0}(\theta, \phi)\\ 
 &=&\sum_{l=0}^\infty i^l \sqrt{4 \pi (2l +1)}  \frac{1}{2 k } \frac{1}{r} \left ( e^{-ikr} + e^{ikr} - e^{ikr} +  \frac{A_{{\text{out,} \omega l}}}{A_{\text{in,} \omega l}} e^{i k r} \right ) e^{i \omega t} Y_{l 0}(\theta, \phi) \IEEEeqnarraynumspace \\ 
 &=& e^{i(kz +\omega t)} + \underbrace{\sum_{l=0}^\infty i^l \sqrt{4 \pi (2l +1)} \frac{1}{2 k} \left ( \frac{A_{\text{out,} \omega l}}{A_{\text{in,} \omega l}}-1 \right ) Y_{l 0}(\theta, \phi) }_{=f_\omega(\theta)}\frac{e^{i k r}}{r} e^{i \omega t} 
\end{IEEEeqnarray}


In this way, we have reduced a complex problem (calculating $f_\omega(\theta)$) to several simpler ones, i.e., calculate the value of the relation $A_{\text{out,} \omega l}/A_{\text{in,} \omega l}$ for partial waves with different values of the angular momentum $l$. Obviously, there is an infinite number of partial waves, but beyond certain value of the angular momentum $l$ they will not contribute to the scattering process ($A_{\text{out,} \omega l}/A_{\text{in,} \omega l} \approx 1$), because the centrifugal barrier will repel them before they interact with the scatterer.

The free scalar wave equation has a conserved current associated to charge conservation $j^\mu = i(\Psi \nabla^\mu \Psi^* - \Psi^* \nabla^\mu \Psi)$, such that $\nabla_\mu j^\mu=0$. Since the scattering experiment is stationary, we have that time derivative of the charge density vanishes $\partial_t j^t=0$, and so, the charge current density is conserved in the spatial dimensions: $\vec{\nabla} \cdot \vec{j}=0$. We can then interpret the integral of spherically outgoing current through a sphere at infinity as coming from the current density of the plane wave (which is constant) interacting with a scatterer with an effective area. This area is known as the elastic scattering cross section:

\begin{IEEEeqnarray}{rCl}
 \sigma_\text{el} &=& \frac{\int_{S_\infty} i(\Psi_\text{sph.} \vec{\nabla} \Psi_\text{sph.}^* - \Psi_\text{sph.}^* \vec{\nabla} \Psi_\text{sph.}) \cdot \df \vec{S}}{|i(\Psi_\text{plane} \vec{\nabla} \Psi_\text{plane}^* - \Psi_\text{plane}^* \vec{\nabla} \Psi_\text{plane})|}\\
 &=& \int |f_\omega(\theta)|^2 \df \Omega \\
 &=& \frac{\pi}{k^2} \sum_{l=0}^\infty (2l +1) \left | \frac{A_{\text{out,} \omega l}}{A_{\text{in,} \omega l}}-1 \right |^2
\end{IEEEeqnarray}

The current density might not be conserved by the full wave equation with the potential. If it is conserved, we will have $|A_{\text{in,} \omega l}|^2=|A_{\text{out,} \omega l}|^2$ for every partial wave; if it is not, the outgoing current of each partial wave will be less than the incoming current. Again, we can think of this lost current as the interaction of the plane wave with a target that absorbs the wave with an effective area called the inelastic cross section\footnote{Having a look at eq. \ref{6eq:Scatteq} it seems there is no lost current. Moreover, it seems that there is current being generated from the origin. This is an artefact of the approximation we have taken when $r\rightarrow \infty$, because the total current of the plane wave is infinite, and taking out (or creating) a finite amount of current will not change the form of the wave solution in the asymptotic limit.}. The sum of the lost current for all partial waves divided by the the current density of the plane wave gives us the inelastic cross section:

\begin{equation}
  \sigma_\text{in} = \frac{\pi}{k^2} \sum_{l=0}^\infty (2l +1) \left ( 1- \left | \frac{A_{\text{out,} \omega l}}{A_{\text{in,} \omega l}} \right |^2 \right )
\end{equation}

\subsubsection{Scattering Experiment in a Wormhole Geometry}

The propagation of plane wave in the geonic wormhole geometry looks very similar to the classical scattering. In this case, there are two asymptotic infinite regions where the effective potential is negligible and the scalar field is essentially free. Then, the solutions of the radial part of the partial wave will behave in the asymptotic regions like:
\begin{IEEEeqnarray}{rCl}
 \psi_{\omega l}(x) &\approx& A_{{\text{in,} \omega l}} e^{-i \omega x} + A_{{\text{out,} \omega l}} e^{i \omega x} \qquad \text{if }x\rightarrow \infty \label{6eq:psiinf} \\
 \psi_{\omega l}(x) &\approx& B_{{\text{in,} \omega l}} e^{i \omega x} + B_{{\text{out,} \omega l}} e^{-i \omega x} \qquad \text{if }x\rightarrow -\infty
\end{IEEEeqnarray}

We want to consider the case in which we are sending plane waves from the $x >0$ side, and there are no incoming waves from the $x <0$ side. This condition implies that, of the two independent solutions of the wave equation, we have to work only with the combination such that $B_{{\text{in,} \omega l}}=0$. With these partial waves we can construct a combination that behaves like a plane wave plus a spherical outgoing wave on the $x>0$ side, and other spherical outgoing wave on the $x<0$ side. The elastic cross section takes the same form as in the classical scattering:

\begin{IEEEeqnarray}{rCl}
 \sigma_\text{el} = \frac{\pi}{k^2} \sum_{l=0}^\infty (2l +1) \left | \frac{A_{\text{out,} \omega l}}{A_{\text{in,} \omega l}}-1 \right |^2
\end{IEEEeqnarray}

But now there is a new phenomenon, where the current might reach the $x<0$ side of the wormhole. In an analogue way, we define the transmission cross section:

\begin{IEEEeqnarray}{rCl}
 \sigma_\text{tr} = \frac{\pi}{k^2} \sum_{l=0}^\infty (2l +1) \left | \frac{B_{\text{out,} \omega l}}{A_{\text{in,}\omega l}} \right |^2
\end{IEEEeqnarray}

Where $| B_{\text{out,} \omega l}/A_{\text{in,}\omega l}|^2$ is $T_l$, the \emph{transmission coefficient} of the partial wave $l$. The incoming current will go to either side of the wormhole, but can also be lost. The inelastic scattering is defined as:

\begin{equation}
  \sigma_\text{in} = \frac{\pi}{k^2} \sum_{l=0}^\infty (2l +1) \left ( 1- \left | \frac{A_{\text{out,} \omega l}}{A_{\text{in,} \omega l}} \right |^2 - \left | \frac{B_{\text{out,} \omega l}}{A_{\text{in,}\omega l}} \right |^2\right )
\end{equation}

\subsubsection{Transmission Coefficients for the Geonic Wormhole}

In order to simplify the problem, we are going to consider the scattering of high frequency waves. In this regime, all the features of the potential are transparent to the wave, except for the leading behaviour at the wormhole throat (eq. \ref{6eq:VeffApprox}). It is possible to study the problem considering only one parameter that has the information from the frequency and the potential. We can redefine our coordinates and constant $\kappa$:

\begin{equation}
 \tilde{y} = |\kappa|^\frac{2}{3} y^* \qquad \alpha = |\kappa|^{-\frac{2}{3}} \omega \label{6eq:alphayprime}
\end{equation}
so that the wave equation takes a really simple form near the wormhole:

\begin{equation}
 \partial_{\tilde{y}}\partial_{\tilde{y}} \psi + (\alpha^2\pm \frac{1}{\sqrt{\tilde{y}}}) \psi = 0
\end{equation}

with the $\pm$ sign being the sign of $\kappa$ (infinite well or barrier). It is possible to compute numerically the transmission from this equation, which will depend only on the parameter $\alpha$. We can set initial conditions $\psi$, $\partial_{\tilde{y}}\psi$ at some point, and the computer can obtain the solution everywhere else.

To obtain the transmission amplitude for a mode of frequency $\alpha$, we can do the following: Choose two points $\tilde{y}_{\text{in}}$, $\tilde{y}_{\text{out}}$, each one at one side of the wormhole and far from it, where the potential has no effect and the wave equation takes the form of eq. \ref{6eq:psiinf}. Set initial conditions $\psi$, $\partial_{\tilde{y}} \psi$ at $\tilde{y}_\text{out}$ taking into account that $B_{\text{in,} \omega l}=0$. Then we run the computation and obtain the value of $\psi$, $\partial_{\tilde{y}} \psi$ everywhere else, in particular at $\tilde{y}_{\text{in}}$. Then, the value of $A_{\text{in,} \omega l}$, $A_{\text{out,} \omega l}$ can be easily recovered: 

\begin{IEEEeqnarray}{rCl}
 A_{\text{in,} \omega l}&=&\frac{1}{2}(\psi(\tilde{y}_{\text{in}})+ \frac{1}{\alpha} \partial_{\tilde{y}} \psi(\tilde{y}_{\text{in}}))\\
 A_{\text{out,} \omega l}&=&\frac{1}{2}(\psi(\tilde{y}_{\text{in}})- \frac{1}{\alpha} \partial_{\tilde{y}} \psi(\tilde{y}_{\text{in}}))\\
 B_{\text{out,} \omega l}&=&\frac{1}{2}(\psi(\tilde{y}_{\text{out}})+ \frac{1}{\alpha} \partial_{\tilde{y}} \psi(\tilde{y}_{\text{out}}))
\end{IEEEeqnarray}

The transmission coefficient is simply $B_{\text{out,} \omega l}^2/A_{\text{in,} \omega l}^2$. There is a problem, however, as the potential $1/\sqrt{\tilde{y}}$ we have obtained for the behaviour near the wormhole throat is long-range, and therefore, the solution is never well approximated far from the wormhole by eq. \ref{6eq:psiinf}. The full potential (without approximations) decays appropriately, but if we use the full potential in the computations, we cannot do the rescaling of the coordinates and the simplification that leads all the parameters of the problem being described by $\alpha$, which seemed convenient. A possible way around this issue, is to describe the wave solution far from the wormhole using the WKB approximation:
\begin{IEEEeqnarray}{rCl}
 \psi_{\omega l}(\tilde{y}) &\approx& \tilde{A}_{{\text{in,} \omega l}} \frac{1}{(\alpha^2-1/\sqrt{\tilde{y}})^\frac{1}{4}} e^{-i \int \sqrt{\alpha^2-1/\sqrt{\tilde{y}}} \df \tilde{y}} \nonumber \\
 &+& \tilde{A}_{{\text{out,} \omega l}} \frac{1}{(\alpha^2-1/\sqrt{\tilde{y}})^\frac{1}{4}} e^{i \int \sqrt{\alpha^2-1/\sqrt{\tilde{y}}} \df \tilde{y}}  \qquad \text{if }x\rightarrow \infty  \\
 \psi_{\omega l}(\tilde{y}) &\approx& \tilde{B}_{{\text{in,} \omega l}} \frac{1}{(\alpha^2-1/\sqrt{\tilde{y}})^\frac{1}{4}} e^{i \int \sqrt{\alpha^2-1/\sqrt{\tilde{y}}} \df \tilde{y}}  \nonumber \\
 &+& \tilde{B}_{{\text{out,} \omega l}} \frac{1}{(\alpha^2-1/\sqrt{\tilde{y}})^\frac{1}{4}} e^{-i \int \sqrt{\alpha^2-1/\sqrt{\tilde{y}}} \df \tilde{y}}  \qquad \text{if }x\rightarrow -\infty
\end{IEEEeqnarray}

\begin{mdframed}
\textbf{Quick aside: The WKB approximation}

The WKB method is an approximation in which we assume that the rate of change of the modulus of the wave function to be slower than the rate of change of phase. A scalar field can be expressed in the form $\psi(x)=A(x) e^{iB(x)}$ where $A$ and $B$ are real functions. A wave equation that has the form $\partial_x\partial_x \psi = Q(x) \psi$, with $Q>0$, gives raise to the following equation ($Q(x)=\alpha^2-\frac{1}{\sqrt{x}}$ in our case):
\begin{equation} 
\underbrace{\partial_x \partial_x A}_{\sim 0}+2i (\partial_x A)(\partial_x B) + i A \partial_x \partial_x B - A (\partial_x B)^2=A Q
\end{equation}
If we neglect the $\partial_x \partial_x A$ term, the real and imaginary part of the equation gives solution for $A$ and $B$: 
\begin{equation}
 B(x)=\pm \int \sqrt{Q(x)} \df x \qquad A(x)=k\ Q(x)^{-\frac{1}{4}}
\end{equation}
Where $k$ in an integration constant. There are two independent solutions, one incoming and other outgoing. It is worth noting that under the WKB approximation, the solutions are incoming everywhere or outgoing everywhere, so there is no reflection at any point. It is possible to get transmission and reflection coefficients by matching WKB solutions in different regions of space where the approximation is valid (for example, one region where $Q>0$, and other where $Q<0$), although one must be careful during the matching procedure so that probability is conserved. It is easy to check that the WKB approximation as presented here conserves the current:
\begin{equation}
 J_x(x) =i(\psi \partial_x \psi^* - \psi^* \partial_x \psi) = i A^2(x) \partial_x B(x) = \pm k^2
 \end{equation}
 This current corresponds to the 1-dimensional problem, but works the same for a three dimensional problem expressed in the radial tortoise coordinate.
\end{mdframed}

\bigskip

Taking these functions as a good approximation of the wave function far from the wormhole, we can obtain $\tilde{A}_{\text{in,} \omega l}$, $\tilde{A}_{\text{out,} \omega l}$ and $\tilde{B}_{\text{out,} \omega l}$ from the values of $\psi(\tilde{y})$ and $\partial_x(\tilde{y})$ in a similar fashion as when the wave function could be approximated by exponential functions. However, as the potential is long-range, it will introduce phases respect to the amplitudes in the full potential. Nevertheless, the transmission coefficient will not change due to these phases.

\begin{figure}[h!]
 \centering
 \includegraphics[width=.5\linewidth]{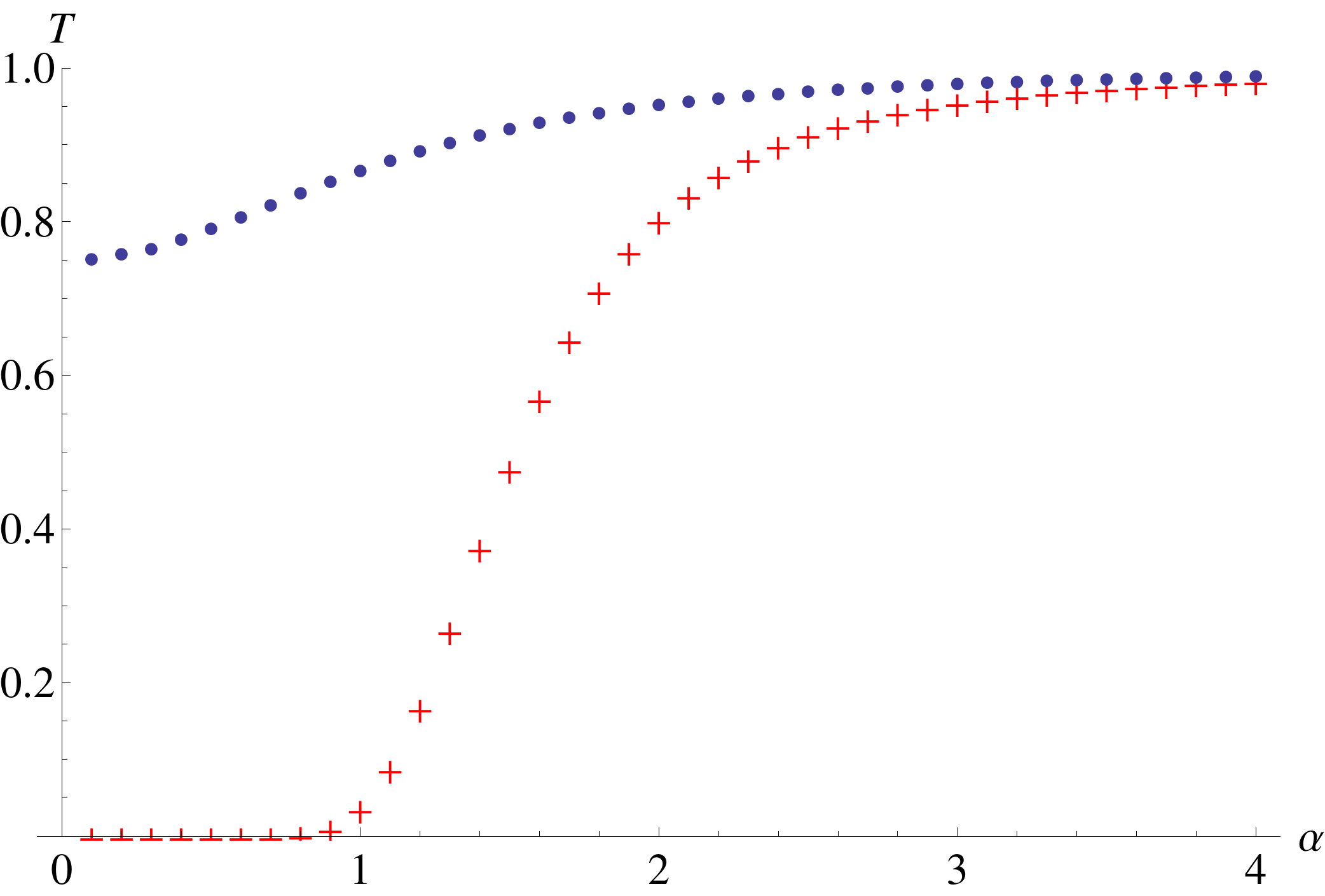}
\caption{Transmission coefficient for the wave equation $\partial_{\tilde{y}}\partial_{\tilde{y}} \psi + (\alpha^2\pm \frac{1}{\sqrt{\tilde{y}}}) \psi = 0$, versus the coefficient alpha ($\alpha = |\kappa|^{-\frac{2}{3}} \omega$). The blue dots are for the plus sign in the equation ($\kappa$ negative), while the red crosses are for the minus sign.}\label{6fig:TransCoef}
\end{figure}

In figure \ref{6fig:TransCoef}, the transmission coefficient as a function of $\alpha$ is represented. When the potential is an infinite well ($\kappa<0$), the transmission is mostly transmitted, and the transmission tends to $1$ as $\alpha$ increases. When the potential is an infinite barrier ($\kappa>0$), the plot shows the typical sigmoid profile of barrier experiments.

Let us consider a given geometry ($\delta_1$, $N_q$ given) with $\kappa>0$ and fixed angular momentum $l$. For waves with low values of $\omega$, the transmission coefficient will vanish, meanwhile for high values of $\omega$, the transmission will be $1$. If we have $\kappa<0$ instead, the transmission will be $1$ for high values of $\omega$, and will be $T\sim 0.75$ for low values.

Consider now a case of a given geometry ($\delta_1$, $N_q$ given) and fixed $\omega$, and let us change the value of the angular momentum $l$. Let us consider that $\kappa$ is positive for $l=0$. When $l$ increases, $\kappa$ also increases (as $l^2$, see eq. \ref{6eq:kappadef}), and therefore $\alpha$ decreases. As we can see from fig. \ref{6fig:TransCoef}, for low values of $\alpha$, the wave is reflected almost completely. Now let us consider that $\kappa$ is negative for $l=0$. Meanwhile $\kappa$ is negative, all waves are transmitted almost completely. But since $\kappa$ increases with $l$, there will be certain value of the angular momentum for which $\kappa$ flips signs. At that point $\alpha \gg 1$ (because $\kappa \sim 0$) and the partial waves is still transmitted almost completely. But as we keep increasing $l$, $\kappa$ will be more and more positive, $\alpha$ will decrease, and we will reach to a point where the partial wave is almost completely reflected.
%
%

A very simple approximation would be to consider that every partial wave with $\alpha>1.5$ or $\kappa$ negative is transmitted, and all the rest are reflected. In this approximation, for every frequency $\omega$ there is a critical value of the angular momentum $l_\text{max}(\omega)$, such that every partial wave with angular momentum greater than that is reflected. As large values of $\kappa$ are dominated by the $l^2$ contribution, and looking at eq. \ref{6eq:alphayprime}, we can see that for large value of $\omega$ we will have $l_\text{max} \propto \omega^\frac{3}{4}$. 

We can use this approximation to reach a very simple formula for the transmission cross-section. 

\begin{equation}
 \sigma_\text{tr} \simeq \frac{\pi}{\omega^2} \sum_{l=0}^{l_\text{max}} (2l+1)= \frac{\pi (l_\text{max}+1)^2}{\omega^2} \propto \omega^{-\frac{1}{2}} 
\end{equation}

\begin{figure}[h!]
 \centering
 \includegraphics[width=.5\linewidth]{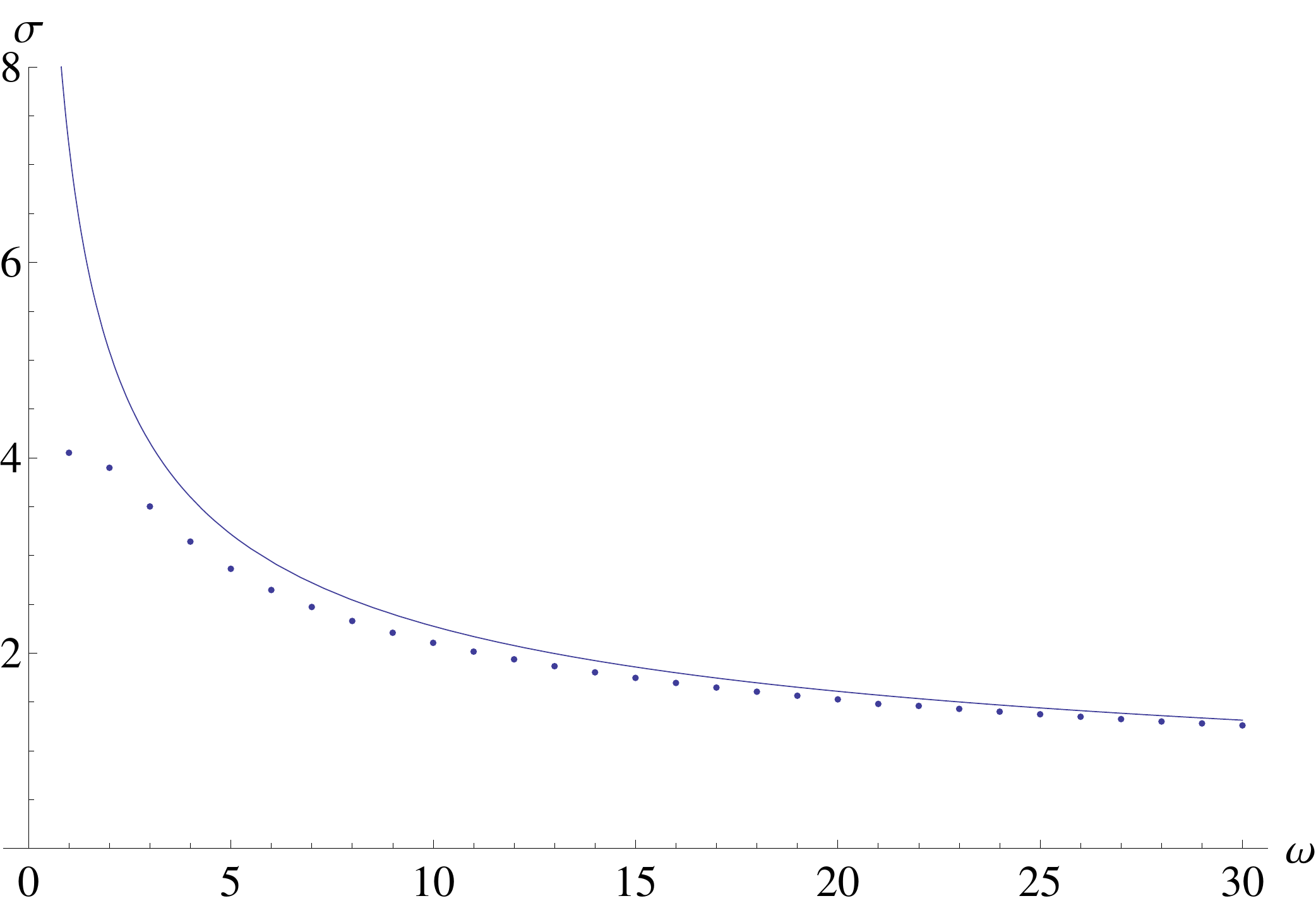}
\caption{Transmission cross section (in $r_c^2$ units) versus frequency (in $c \ r_c$ units) for a naked divergence with $\delta_1=1$, $N_q=10$, calculated numerically (dots) for the exact $V_\text{eff}$. The continuous line shows the approximation $\sigma \propto \omega^\frac{1}{2}$.}\label{6fig:CrossSect}
\end{figure}

In figure \ref{6fig:CrossSect} we can check that this approximation works well for high frequencies.

\cleardoublepage
\chapter{Wormholes in d-Dimensions}\label{DDim}

In the last chapters we have studied how charged black holes for certain family of $f(R,Q)$ theories in the metric-affine formalism avoid the central singularity that happens in GR. It would be interesting to check whether this is a property of the particular theory we have studied or if it is a more general property of the metric-affine formalism. An easy first step is to consider a different family of theories, and more than 4 dimensions. 

In this chapter, I will present an extension of GR called Born-Infeld inspired gravity, study its properties and obtain black hole solutions in an arbitrary number of dimensions. Then I will study how these new solutions are non-singular too.


\section{Born-Infeld Theory}
\subsection{Born-Infeld Model for Electromagnetism}

In 1934, Born and Infeld proposed a non-linear theory for electromagnetism \cite{Born:1934gh}. The goal of this new theory was to be able to describe an electron in such a way that its electric potential (and the energy density associated to a point-like charge) remained finite at the origin. The proposed theory is described by the following action.

\begin{equation}
 S= - b^4 \int \left \{ \sqrt{-\det \left ( \eta_{\mu \nu} + \frac{F_{\mu \nu}}{b^2} \right )}  -1 \right \} \df^4 x
\end{equation}
where $b$ is an energy scale. In the limit of $b \gg F_{\mu \nu}$, this theory recovers Maxwell's equations. In other words, for radius greater than $r_0\sim \sqrt{\alpha_\text{EM}}b^{-1}$ we recover the Maxwell's description of the electron, but inside that radius, the electromagnetic field will change and remains finite. This Lagrangian might seem unnatural, but is analogous to the change of the action of a non-relativistic free particle to a relativistic one, where $\frac{1}{2} m v^2 \approx m c^2 \left ( \sqrt{1- \frac{mv^2}{mc^2}} -1 \right )$ when $mc^2 \gg mv^2$.

Although with the advent of Dirac's equation and QFT, a new classical model for the electron was no longer needed, this theory remained interesting for many of its properties. The Born-Infeld model has causal propagation and does not present birefringence or shock waves (\cite{Boillat:1970gw}, \cite{Aiello:2006cz}), which are a common occurrence in non-linear theories. It also preserves the electric-magnetic duality of Maxwell's equations (\cite{Gibbons:1995cv}). If we expand the lagrangian, the quartic terms reproduce the effective action of one-loop SUSY QED (\cite{Deser:1998rj}). It also appears naturally in the context of strings, in the low energy limit of the electromagnetic action in D-branes (\cite{Tseytlin:1997csa}, \cite{Gauntlett:1997ss}).

\subsection{Born-Infeld inspired Gravity}

As the Born-Infeld model is successful regularizing the electric potential for an electron, it would be natural to see if it is possible to do something similar with gravity to regularize the curvature that a point-like mass generates around itself. In 1998 Deser and Gibbons \cite{Deser:1998rj} proposed the following action:

\begin{equation}
 S=b^4 \int \df^4 x \sqrt{-\det(a g_{\mu \nu} + b R_{\mu \nu} + c X_{\mu \nu})}
\end{equation}
where $X_{\mu \nu}$ is a tensor that contains terms of quadratic and higher order in $R_{\mu \nu}$,  which must be chosen so the theory is ghost free. The ghost problem arises in the Riemannian approach, but the metric-affine formulation is free from this problem and allows us to propose a simpler action for gravity \cite{Banados:2010ix} (in $d$-dimensions):

%

\begin{equation}
S= \frac{1}{8 \pi l_P^{d-2} \epsilon} \int  \left \{ \sqrt{-\det(g_{\mu \nu} + \epsilon R_{\mu \nu} (\Gamma))} - \lambda \sqrt{-\det(g_{\mu \nu})} \right \} \df^d x + S_m
\end{equation}
which will not contain ghost due to the generic second-order field equations, similar as other metric-affine theories. Sometimes this formulation is called 'Eddington-inspired Born-Infeld gravity'. It has drawn a notable amount of interest with applications in astrophysics (\cite{Pani:2011mg},\cite{Pani:2012qd}), cosmology (\cite{Banados:2008rm},\cite{Scargill:2012kg},\cite{Harko:2013xma}, \cite{Odintsov:2014yaa}, \cite{Avelino:2012ue}) and black hole physics (\cite{Olmo:2013gqa}, \cite{Bazeia:2016rlg}). A recent review on Born-Infeld gravity can be found in \cite{BeltranJimenez:2017doy}.

As we did for the $f(R,Q)$ Lagrangian, we are going to consider a theory with a symmetric Ricci tensor $R_{\mu \nu}$. $\epsilon$ is a parameter with dimensions of $[\text{length}]^2$ which plays the same role as $b$ in the Born-Infeld model for electromagnetism. In the limit case of $\epsilon \rightarrow 0$, the expansion of the Lagrangian\footnote{We note that $\det(\mathbb{I}+\epsilon R^\mu{}_\nu)\simeq 1 + \epsilon R + \epsilon^2 (R^2- R^{\alpha \beta} R_{\alpha \beta})/2 + O(\epsilon^3)$} gives the GR Lagrangian plus a cosmological constant $\Lambda_\text{eff}=\frac{\lambda-1}{\epsilon}$.

\section{Charged Black Holes in an Arbitrary Number of Dimensions}

First step we have to take to obtain charged black hole solutions is to derive the equations of motion of the theory. To simplify this derivation, we can define the following auxiliary metric:

\begin{equation}
 q_{\mu \nu} \equiv g_{\mu \nu} + \epsilon R_{\mu \nu} \label{7eq:defq}
\end{equation}

With this, the action can be written as:

\begin{IEEEeqnarray}{rCl}
S&=& \frac{1}{8 \pi l_P^{d-2} \epsilon} \int  \left \{ \sqrt{|q| }- \lambda \sqrt{|g|} \right \}  \df^d x + S_m \\
&=& \frac{1}{8 \pi l_P^{d-2} \epsilon} \int  \left \{ \frac{\sqrt{|q| }}{\sqrt{|g|}} - \lambda  \right \}  \sqrt{|g|}\df^d x + S_m
\end{IEEEeqnarray}

So it can be seen as a $f(R^\alpha{}_{\beta \mu \nu}, g_{\mu \nu})$ theory. The variation of the auxiliary metric in terms of the variations of the metric and the Ricci curvature is:

\begin{IEEEeqnarray}{rCl}
 \delta \sqrt{|q|} &=& \frac{1}{2} \sqrt{|q|} (q^{-1})^{\alpha \beta} \delta q_{\alpha \beta} \\
 &=&\frac{1}{2} \frac{\sqrt{|q|}}{\sqrt{|g|}} \sqrt{|g|}(q^{-1})^{\alpha \beta} (\delta g_{\alpha \beta} + \epsilon \delta R_{\alpha \beta})
\end{IEEEeqnarray}

Obtaining the equations of motion for the variation of the metric is straightforward. The equations of motion for the variation of the connection can be obtained using eq.\ref{3eq:EqConnectionGeneral} with $P_\alpha{}^{\beta \mu \nu}= \partial f/ \partial R^\alpha{}_{\beta \mu \nu} = \sqrt{|q|/|g|} (q^{-1})^{\beta \nu} \delta_\alpha{}^\mu$ and then use the relation obtained contracting indices $\alpha$ and $\nu$ to simplify the expression. The two sets of equations of motion are:

\begin{IEEEeqnarray}{rCl}
 \frac{\sqrt{-q}}{\sqrt{-g}} (q^{-1})^{\mu \nu} - \lambda g^{\mu \nu} &=& -8 \pi l_P^{d-2} \epsilon T^{\mu \nu} \label{7eq:BImetriceq}\\
 \nabla_\lambda (\sqrt{-q} q^{\mu \nu} ) &=& 0 \label{7eq:BIconnectioneq}
\end{IEEEeqnarray}

From the last equation we see that the connection is the Levi-Civita connection of the auxiliary metric $q_{\mu \nu}$. To solve these equations we can define a mixed object $\Omega^\mu{}_\nu=g^{\mu \alpha} q_{\alpha \nu}$, $|\Omega|=|q|/|g|$. With this object we can write eq. \ref{7eq:BImetriceq} as:

\begin{equation}
 |\Omega|^\frac{1}{2} (\Omega^{-1})^\mu{}_\nu = \lambda \delta^\mu{}_\nu - \epsilon 8 \pi l_P^{d-2} T^\mu{}_\nu
\end{equation}

And now it is possible to obtain the value of $\Omega$ if we know the energy-momentum tensor $T^\mu{}_\nu$:

\begin{equation}
 (\Omega^{-1})^\mu{}_\nu = \frac{\lambda \delta^\mu{}_\nu - \epsilon 8 \pi l_P^{d-2} T^\mu{}_\nu}{|\lambda \delta^\mu{}_\nu - \epsilon 8 \pi l_P^{d-2} T^\mu{}_\nu|^\frac{1}{d-2}} \label{7eq:OmegaSol}
\end{equation}

Now we can rewrite eq. \ref{7eq:defq} as $\epsilon R_{\mu \nu} = q_{\mu \nu} - g_{\mu \nu}$, and multiply it by $q^{-1}$:

\begin{equation}
 \epsilon (q^{-1})^{\mu \alpha} R_{\alpha \nu} = \delta^\mu{}_\nu - (\Omega^{-1})^\mu{}_\nu \label{7eq:RqSol}
\end{equation}

The left hand side only depends on $q$, and the right hand side is completely known if we know $\Omega$. Therefore, if we know $T^\mu{}_\nu$, we can obtain $\Omega$, then $q$, and finally we can obtain the space-time metric using that $g_{\mu \nu} = (\Omega^{-1})_\mu{}^\alpha q_{\alpha \nu}$.

%
%
%

\subsection{Electrovacuum Stress-Energy Tensor in d-Dimensions}

The matter content of a charged black hole solution is a spherically symmetric electrovacuum field. As in the $f(R,Q)$ theories in metric-affine formalism, it is important that the energy-momentum tensor is not vacuum, because in that case the independent connection becomes the Levi-Civita connection of the space-time metric $g$. This is also the case in Born-Infeld gravity, as we can see that if $T_\mu{}^\nu$ vanishes, then from equation \ref{7eq:OmegaSol} we can see that $\Omega$ is the identity, and the equations would be equivalent to the ones obtained in the Riemannian formalism.



To describe a spherically symmetric electrovacuum field we have to start from the pure Electromagnetic action without sources:

\begin{equation}
 S_m = -\frac{l_P^{d-4}}{16 \pi} \int F_{\alpha \beta} F^{\alpha \beta} \sqrt{-g} \df^d x 
\end{equation}
where $F=\df A$ comes from the electromagnetic potential. The dimensional constant has been chosen so that the charge $q=\int_S *F$ is dimensionless. The Euler-Lagrange equations without sources are:

\begin{eqnarray}
 \df F&=&0\\
 \df (*F)&=&0 \qquad \Rightarrow \qquad \nabla_\mu F^{\mu \nu}=0
\end{eqnarray}

To solve this equations we are going to consider a space-time with an $n$-dimensional ($n=d-2$) maximally symmetric subspace, which could be spherical, flat or hyperbolic. The metric of such space can be written as:

\begin{equation}
 \df s^2 = g_{tt} \df t^2 +g_{xx} \df x^2+r(x)^2 \df \Omega^2_n \label{7eq:gproposed}
\end{equation}
with $\df \Omega^2_n$ being the line element of the maximally symmetric subspace. The solution for the equations is:

\begin{equation}
 F^{tx} = \frac{q}{r(x)^{d-2} \sqrt{-g_{tt} g_{xx}}}
\end{equation}
where $q$ is a dimensionless integration constant that corresponds with the charge. The energy-momentum tensor is generically:

\begin{equation}
 T_\mu{}^\nu = -\frac{l_P^{d-4}}{4\pi} \left ( F_\mu{}^\alpha F_\alpha{}^\nu-\frac{F_\alpha{}^\beta F_\beta{}^\alpha}{4}\delta_\mu{}^\nu \right )
\end{equation}

And substituting the value of the electromagnetic field we have found, we have a energy-momentum tensor separated into two blocks:

\begin{equation}
 T_\mu{}^\nu = \frac{Xl_P^{d-4}}{8\pi}\left ( \begin{array}{cc} -\mathbb{I}_{2\times 2} & 0 \\ 0 & \mathbb{I}_{n\times n} \end{array} \right ) \label{7eq:EMVacuumTd}
\end{equation}
with $X=q^2/r(x)^{2(d-2)}$, and $\mathbb{I}_{2\times 2},\ \mathbb{I}_{n\times n}$ are the identity in $2$ and $n$ dimensions respectively.

\subsection{Solution for Spherically Symmetric and Static Electrovacuum Field}

Now that we know the value of $T_\mu{}^\nu$ in our space-time, we can proceed to solve equation \ref{7eq:OmegaSol}:

\begin{equation}
 (\Omega^{-1})_\mu{}^\nu = \left ( \begin{array}{cc} \frac{(\lambda+X \epsilon l_P^{2d-6})^\frac{d-4}{d-2}}{\lambda-X \epsilon l_P^{2d-6}}\mathbb{I}_{2\times 2} & 0 \\ 0 & \frac{1}{(\lambda+X \epsilon l_P^{2d-6})^\frac{2}{d-2}}\mathbb{I}_{n\times n} \end{array} \right )\equiv \left ( \begin{array}{cc} \frac{1}{\Omega_{(2)}}\mathbb{I}_{2\times 2} & 0 \\ 0 & \frac{1}{\Omega_{(n)}} \mathbb{I}_{n\times n}\end{array} \right )
\end{equation}

Now we have to substitute the value of $\Omega^{-1}$ into eq. \ref{7eq:RqSol}:

\begin{equation}
 \epsilon q^{\mu \alpha} R_{\alpha \nu} = \left ( \begin{array}{cc} \frac{\Omega_{(2)}-1}{\Omega_{(2)}}\mathbb{I}_{2\times 2} & 0 \\ 0 & \frac{\Omega_{(n)}-1}{\Omega_{(n)}} \mathbb{I}_{n\times n}\end{array} \right ) \label{7eq:RqMatrix}
\end{equation}

To solve this equation we can propose a functional form for the auxiliary metric $q_{\mu \nu}$ with the symmetries of the geometry:

\begin{equation}
 q_{\mu \nu} = -A(x) \df t^2 + \frac{1}{A(x)} \df x^2 + \tilde{r}(x)^2 \df \Omega_n^2
\end{equation}

Comparing with the space-time metric we proposed in eq. \ref{7eq:gproposed}, the components of both metrics are related as:

\begin{equation}
 g_{tt} = \frac{q_{tt}}{\Omega_{(2)}} \qquad g_{xx} = \frac{q_{xx}}{\Omega_{(2)}} \qquad r^2(x) = \frac{\tilde{r}^2(x)}{\Omega_{(n)}}\label{7eq:gtoq}
\end{equation}

As we want to solve for a space-time with an $n$-dimensional maximally symmetric subspace (in this case, with spherical symmetry), it is interesting to separate the contributions to the curvature of this subspace from the ones of the $(t,x)$ coordinates. Let us consider a metric that can generically be separated as $q=q^{(2)}_{ab}+\tilde{r}^2 q^{(n)}_{ij}$ where the indices $(a,b,c)$ take values in $\{ t, x\}$, the indices $(i,j,l)$ take value in $\{ \theta_1,..., \theta_{n} \}$, and the function $\tilde{r}$ depends on the $(t,x)$ coordinates. Then if $\Gamma^{(2)}$ and $\Gamma^{(n)}$ are the Levi-Civita connections of $q^{(2)}$ and $q^{(n)}$ respectively, the components of the Levi-Civita connection of $q$ are:

\begin{IEEEeqnarray}{lCl}
 \Gamma^a_{bc} = (\Gamma^{(2)})^a_{bc} & \qquad &  \Gamma^i_{jl} = (\Gamma^{(n)})^i_{jl} \label{7eq:GammaBlock} \\
 \Gamma^i_{aj} = \delta^i{}_j \frac{1}{\tilde{r}} \partial_a \tilde{r} &\qquad & \Gamma^a_{i j} = -q^{ab}q^{(n)}_{ij} \tilde{r} \partial_b \tilde{r} \nonumber \\
 \Gamma^a_{bi} =0 &\qquad &\Gamma^i_{ab} =0 \nonumber
\end{IEEEeqnarray}

To obtain the Ricci tensor of $q$, we have to substitute these values into:

\begin{equation}
 R_{\mu\nu} = \partial_\alpha \Gamma^\alpha_{\mu \nu} - \partial_\nu \Gamma^\alpha_{\alpha \mu} + \Gamma^\alpha_{\alpha\beta} \Gamma^\beta_{\mu \nu} - \Gamma^\alpha_{\nu \beta} \Gamma^\beta_{\mu \alpha}
\end{equation}

The $(x,t)$ part gives:

\begin{IEEEeqnarray}{rCl}
 R_{ab}&=&R^{(2)}_{ab} - \partial_a \Gamma^i_{ib} + \Gamma^i_{ic} \Gamma^c_{ab}-\Gamma^i_{bj} \Gamma^j_{ia}\\
 &=&R^{(2)}_{ab} -(d-2) \frac{1}{\tilde{r}} \nabla_a \nabla_b\tilde{r}
\end{IEEEeqnarray}

Meanwhile the $( \theta_1,..., \theta_{n} )$ part gives:

\begin{IEEEeqnarray}{rCl}
 R_{ij}&=&R^{(n)}_{ij}+\partial_a \Gamma^a_{ij}+\Gamma^l_{la} \Gamma^a_{ij} -\Gamma^a_{il} \Gamma^l_{ja}-\Gamma^l_{ia}\Gamma^a_{jl} \\
 &=& R^{(n)}_{ij} - (d-3)  q^{(n)}_{ij} q^{ab} (\partial_a\tilde{r}) (\partial_b\tilde{r}) - q^{(n)}_{ij}\tilde{r} \partial_a \partial^a\tilde{r} 
\end{IEEEeqnarray}

For a maximally symmetric $n$-dimensional space the Riemann tensor is equal to $(R^{(n)})^i{}_{jlm} = k (\delta^i{}_l q^{(n)}_{jm} - \delta^i{}_m q^{(n)}_{jl})$ with $k=1,0,-1$ depending if it is spherical, flat or hyperbolic; and the Ricci tensor is equal to $(R^{(n)})_{ij} = (n-1) k q^{(n)}_{ij}$. Therefore:

\begin{equation}
 R_{ij} = q^{(n)}_{ij} \left [ (d-3)(k-q^{ab} (\partial_a \tilde{r}) (\partial_b \tilde{r})) - \tilde{r} \partial_a \partial^a \tilde{r} \right ]
\end{equation}

With this, it is possible to substitute into eq. \ref{7eq:RqMatrix} and solve for $q$. The Ricci tensor has only three independent components, because there are only diagonal terms and all the components of the maximally symmetric subspace are all the same:

\begin{IEEEeqnarray}{lClCl}
 R^t{}_t &=& -\frac{ \partial_x \partial_x A}{2} - (d-2)\frac{\partial_x A}{2} \frac{\partial_x \tilde{r}}{\tilde{r}} &=&  \frac{\Omega_{(2)}-1}{\epsilon \Omega_{(2)}}\IEEEeqnarraynumspace  \\
 R^x{}_x &=& -\frac{ \partial_x \partial_x A}{2}-(d-2) \frac{\partial_x A}{2} \frac{\partial_x \tilde{r}}{\tilde{r}} - (d-2) A \frac{1}{\tilde{r}} \partial_x \partial_x \tilde{r} &=&\frac{\Omega_{(2)}-1}{\epsilon \Omega_{(2)}} \IEEEeqnarraynumspace \\
 R^i{}_i &=& \frac{1}{\tilde{r}^2(x)} \left \{ (d-3)(k-A (\partial_x \tilde{r})^2) - A \partial_x \partial_x \tilde{r} - \tilde{r}(\partial_x \tilde{r})( \partial_x A) \right \} &=&\frac{\Omega_{(n)}-1}{\epsilon \Omega_{(n)}}\IEEEeqnarraynumspace  \label{7eq:Rii}
\end{IEEEeqnarray}

Since we have $R^t{}_t -R^x{}_x =0$, this implies that $\tilde{r}(x)=x$. Then equation \ref{7eq:Rii} reads:

\begin{equation}
 \frac{1}{x^2} \left \{ (d-3)(k-A) -x \partial_x A \right \}= \frac{\Omega_{(n)}-1}{\epsilon \Omega_{(n)}}
\end{equation}

Now we take the ansatz:

\begin{equation}
 A(x) = k- \frac{2 M(x)}{(d-3) x^{d-3}} \label{7eq:defA}
\end{equation}

So that we have a differential equation for $M$:

\begin{equation}
 \frac{2 \epsilon \partial_x M}{d-3}= x^{d-2}\frac{\Omega_{(n)}-1}{\Omega_{(n)}}
\end{equation}

Since we know $\Omega_{(n)}$ in terms of the function $r(x)$, we should try to rewrite this differential equation using $r$ as a variable. Let us recall the relation between $r$ and $x$ from eq. \ref{7eq:gtoq}:

\begin{equation}
 x^2=\Omega_{(n)} r^2 \qquad  \frac{\partial x}{\partial r} = \frac{\Omega_{(2)}}{\sqrt{\Omega_{(n)}}}
\end{equation}

Using $r$ as a coordinate we can rewrite the differential equation for $M$ so that we can solve directly:

\begin{equation}
 \frac{2 \epsilon \partial_r M}{d-3}= r^{d-2} (\Omega_{(n)})^\frac{d-5}{2} (\Omega_{(n)}-1) \Omega_{(2)}
\end{equation}

But before solving it, we are going to define certain constants and study the relation between $x$ and $r$. In particular $\epsilon$ has dimensions of length squared and I will assume that it has negative value\footnote{The case with $\epsilon>0$ leads to a different relation between $x$ and $r$ which requires a new study.}. We will impose now that the maximally symmetric subspace is spherical and that the constant $\lambda$ is equal to $1$, so there is no effective cosmological constant\footnote{In \cite{Guendelman:2013sca} the case $\lambda \neq 1$ was treated for $d=4$.}, for simplicity:

\begin{equation}
 \epsilon \equiv - l_\epsilon^2 \qquad r_c^{2d-4} \equiv q^2 l_P^{2d-6} l_\epsilon^2 \qquad z\equiv\frac{r}{r_c} \qquad X \epsilon l_P^{2d-6} = -\frac{1}{z^{2d-4}}
\end{equation}

With these definitions, the elements of the matrix $\Omega$ have a simpler expression:

\begin{equation}
 \Omega_{(2)} = \frac{1+\frac{1}{z^{2d-4}}}{\left (1-\frac{1}{z^{2d-4}} \right )^{\frac{d-4}{d-2}}} \qquad \Omega_{(n)} = \left (1-\frac{1}{z^{2d-4}} \right )^{\frac{2}{d-2}} 
\end{equation}

The relation between the radial coordinate of the space-time $r$, and $x$ is:

\begin{equation}
 r^{d-2}=\frac{|x|^{d-2}+\sqrt{x^{2(d-2)}+4r_c^{2(d-2)}}}{2}
\end{equation}

\begin{figure}[h!]
\centering
 \includegraphics[width=.7\linewidth]{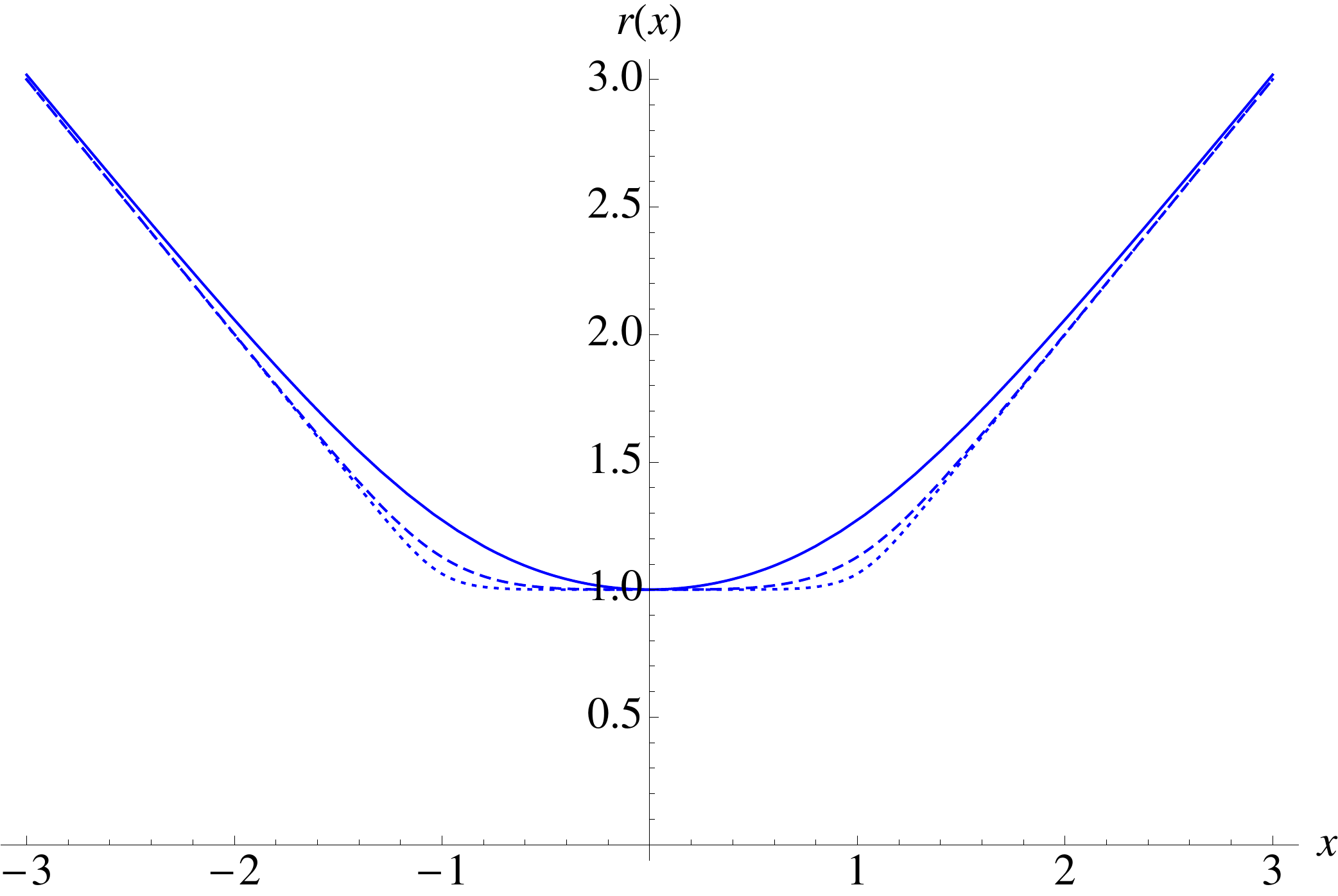}
 \caption{Representation of the radial coordinate $r$ in terms of the coordinate $x$ for different number of dimensions (solid d=4, dashed d=6, dotted d=10)}\label{7fig:x-r}
\end{figure}

As we can see in fig. \ref{7fig:x-r}, $r$ reaches a minimum at $x=0$ ($r=r_c$) and then grows again for negative values of $x$. This is the same wormhole structure that we had in the 4-dimensional case. However, we still need to study the completeness of the geodesics that reach the sphere of minimum radius to determine if the geometry is indeed extended for physical observers. Now that we know how to write the equations in terms of $x$, we can integrate and obtain the following solution for $M$, $A$ and the metric:

\begin{IEEEeqnarray}{rCl}
 M(r)&=&M_0 \left (1 + \delta_1 G_d(\frac{r}{r_c}) \right ) \\
 A(r)&=&1-\frac{2M_0 \left (1 + \delta_1 G_d(\frac{r}{r_c}) \right )}{(d-3)x^{d-3}} = 1-\frac{1-\delta_1 G_d(z)}{\delta_2 \Omega_{(n)}^\frac{d-3}{2} z^{d-3}}\\
 g &=& -\frac{A}{\Omega_{(2)}} \df t^2 + \frac{1}{A \Omega_{(2)}} \df x^2 + r^2(x) \df \Omega_n^2
\end{IEEEeqnarray}
where $M_0$ is an integration constant\footnote{Which has dimensions of [length]${}^{d-3}$.} and $\delta_1,\delta_2$ are constants equal to:

\begin{equation}
 \delta_1 \equiv \frac{(d-3) r_c^{d-1}}{2 M_0 l_\epsilon^2} \qquad \delta_2 = \frac{(d-3) r_c^{d-3}}{2M_0}=\delta_1 \frac{l_\epsilon^2}{r_c^2}
\end{equation}
and $G$ is a function whose value is:

\begin{IEEEeqnarray}{rCl}
G_d(z)&\equiv& -\frac{z^{(3-d)}}{(d-3)} {_2F_1}\left[\frac{1}{d-2},\frac{d-3}{2d-4},\frac{3d-7}{2d-4};\frac{1}{z^{2d-4}}\right]  \\
&&+\frac{z^{(d-1)}}{d-1}{_2F_1}\left[\frac{1}{d-2},-\frac{(d-1)}{2d-4},\frac{d-3}{2d-4};\frac{1}{z^{2d-4}}\right]\nonumber \\
&&-\frac{z^{(d-1)}}{d-1}(1-z^{4-2d})^\frac{(d-1)}{(d-2)} \nonumber
\end{IEEEeqnarray}


The particular functional form of the solution shown in the text is not as straightforward to compare to the GR solution as the one I gave in the 4-dimensional case, where we could clearly see that the solution was very similar to the Reissner-Nördstrom solution, but with $x$ replacing the radial coordinate $r$ in some terms, and an additional term that vanished at infinity and had a constant value at $r=r_c$. The $d$-dimensional case is actually very similar, but to see it clearly we will have to work out the series expansion of the metric function at infinity and at $z=1$. At infinity $A$ behaves like:

\begin{IEEEeqnarray}{rCl}
 \lim_{r\rightarrow \infty} A &=&1-\frac{2 M_0}{(d-3)r^{d-3}} + \frac{4 M_0 r_c^{d-3} \delta_1}{(d-3)^2(d-2)r^{2d-6}}+ \frac{2 M_0 r_c^{2d-4}}{(d-2)r^{3d-7}} +  O(\frac{1}{r^{3d-6}}) \IEEEeqnarraynumspace  \\
 &=& 1-\frac{2 M_0}{(d-3)r^{d-3}} + \frac{q^2 l_P^{2d-8}}{(d-3)(d-2)r^{2d-6}} + \frac{2 M_0 r_c^{2d-4}}{(d-2)r^{3d-7}} + O(\frac{1}{r^{3d-6}}) \IEEEeqnarraynumspace 
\end{IEEEeqnarray}

Which is the behaviour of the GR solution in $d$-dimensions ($-g_{tt}=1-\frac{2 M_0}{(d-3)r^{d-3}} + \frac{q^2 l_P^{2d-8}}{(d-3)(d-2)r^{2d-6}}$) plus corrections that become very small far from the centre. Although we are looking at the components of the auxiliary metric, the value of the space-time metric is very similar because the two are related via $\Omega_{(2)/(n)}$, which at large radius tends to $\Omega_{(2)/(n)}\rightarrow 1 + O(1/r^{2d-4})$.

Now we are interested in the behaviour of $A$ near $r=r_c$, which is the limit of $z\rightarrow 1$. Let us look first the behaviour of the function $G_d$:

\begin{equation}
\lim_{z\to 1} G_d(z)\approx -\frac{1}{\delta_d}+a_d(z-1)^{\frac{d-3}{d-2}}+b_d(z-1)^{\frac{d-1}{d-2}} + O((z-1)^\frac{d}{d-2})
\end{equation}

$a_d$, $b_d$, $\delta_d$ are constants that just depend on the dimension of the space-time. Its particular value is:

\begin{IEEEeqnarray}{rCl}
\delta_d^{-1}&\equiv&\frac{-\pi\csc\left(\frac{\pi}{d-2}\right)}{(d-1)\Gamma\left[\frac{1}{d-2}\right]}\left(\frac{\Gamma\left[\frac{d-3}{2d-4}\right]}{\Gamma\left[\frac{d-5}{2d-4}\right]}-\frac{(d-1)}{(d-3)}\frac{\Gamma\left[\frac{3d-7}{2d-4}\right]}{\Gamma\left[\frac{3(d-3)}{2d-4}\right]}\right) \\
a_d&\equiv& \frac{\pi\csc\left[\frac{\pi}{d-2}\right]2^{\frac{d-3}{d-2}}(d-2)^{-\frac{1}{d-2}}}{\Gamma\left[\frac{1}{d-2}\right]\Gamma\left[\frac{d-5}{d-2}\right]}\\
 b_d&\equiv& -\frac{(2(d-2))^{\frac{d-1}{d-2}}}{d-1}
\end{IEEEeqnarray}

Looking at the definition of $A$ and the expansion of $G_d$ around $z=1$, we can see that $A$ will diverge at $z=1$ in general. There is an exceptional case, $\delta_1=\delta_d$, in which the auxiliary metric is finite and smooth at $z=1$. The expansion of $A$ around $z=1$ is\footnote{We just need to use the definition of $A$ and the expansion of $1/(\Omega_{(n)}^\frac{d-3}{2} z^{d-3}) \approx ((2d-4)(z-1))^{-\frac{d-3}{d-2}} \left ( 1+ (z-1) (d-3)/(2d-4)  \right)$.}:

\begin{IEEEeqnarray}{rCl}
A(z)&\approx & \frac{(\delta_1-\delta_d)}{\delta_2\delta_d}\frac{(2(d-2))^{-\frac{d-3}{d-2}}}{(z-1)^{\frac{d-3}{d-2}}}\left(1+\frac{(d-3)}{2(d-2)}(z-1)\right)\\
&+&1-\frac{\delta_1(2(d-2))^{-\frac{d-3}{d-2}}}{\delta_2} \left(a_d-\frac{2b_d(d-2)}{(d-1)}(z-1)^{\frac{2}{d-2}}\right.\nonumber\\
&+&\left.\frac{2a_d(d-3)}{(2d-5)}(z-1)+\frac{2b_d}{(d-1)}(z-1)^{\frac{d}{d-2}}\right) \nonumber
\end{IEEEeqnarray}



However, in this case the auxiliary metric and the space-time metric differ greatly, because $\Omega_{(2)} \propto 1/(z-1)^\frac{d-4}{d-2}$. So except in the 4 dimensional case, both metrics will have different behaviour at $z=1$. Let us look at the expansion of $g_{tt}$:


\begin{IEEEeqnarray}{rCl}
g_{tt}&\approx & -\frac{(2(d-2))^{\frac{d-4}{d-2}}}{2}\left[\frac{(\delta_1-\delta_d)}{\delta_2\delta_d}\frac{(2(d-2))^{-\frac{d-3}{d-2}}}{(z-1)^{\frac{1}{d-2}}} \right. \\
&+& \left(1-\frac{a_d\delta_1(2(d-2))^{-\frac{d-3}{d-2}}}{\delta_2} \right)(z-1)^{\frac{d-4}{d-2}} \nonumber\\
&+& \frac{(\delta_1-\delta_d)}{\delta_2\delta_d}\frac{(4d-7)}{(2(d-2))^{\frac{2d-5}{d-2}}}(z-1)^{\frac{d-3}{d-2}}\nonumber\\
&+& \left.\frac{\delta_1b_d}{\delta_2} \frac{(2(d-2))^{\frac{1}{d-2}}}{(d-1)}(z-1)\right] \nonumber
\end{IEEEeqnarray}

If $\delta_1\neq \delta_d$, the first term will dominate over the rest. In terms of the coordinate $x$, near $x\to 0$ it turns into:

\begin{equation}
g_{tt}\approx -\frac{(\delta_1-\delta_d)}{2\delta_2\delta_d}\frac{r_c}{|x|}
\end{equation}

For $\delta_1= \delta_d$, we find instead:

\begin{IEEEeqnarray}{rCl}
g_{tt}&\approx&-\frac{1}{2}\left(1-\frac{a_d\delta_d(2d-4)^{-\frac{d-3}{d-2}}}{\delta_2} \right)\left(\frac{|x|}{r_c}\right)^{d-4}\\&-&\frac{b_d\delta_d}{\delta_2(d-1)(2(d-2))^{\frac{1}{d-2}}}\left(\frac{|x|}{r_c}\right)^{d-2} \nonumber
\end{IEEEeqnarray}

The component $g_{tt}$ no longer diverges as we approach $x=0$. Instead, depending on the value of $\delta_2$ it goes to $0$ as $|x|^{d-4}$ (and be positive or negative), or goes to $0$ as $|x|^{d-2}$ (positive). In all cases, the inverse of the metric will diverge and therefore the curvature scalars will be also divergent at $x=0$.

 

\subsubsection{Kretschmann Curvature Scalar}

To check that this geometry is indeed divergent at $x=0$ and that it is not a by-product of choosing an inadequate set of coordinates, we can compute the Kretschmann curvature scalar. For this purpose we will first separate the space-time metric in two blocks $g=g^{(2)}_{ab}+r^2g^{(n)}_{ij}$, where the indices $(a,b,c)$ take values in the coordinates $\{t,x\}$, whereas the indices $(i,j,l,m,n)$ take values in the coordinates $\{ \theta_1,..., \theta_{n} \}$, and $r$ is a function of $(t,x)$. Each metric $g^{(2)}$ and $g^{(n)}$  will have its own connection $\Gamma^{(2)}$ and $\Gamma^{(n)}$ (see eqs. \ref{7eq:GammaBlock}), and Riemann tensor $R^{(2)}$ and $R^{(n)}$. Then the non-zero components of the Riemann tensor of $g$ can be expressed in terms of $R^{(2)}$, $R^{(n)}$, and derivatives of the function $r$:

\begin{IEEEeqnarray}{rCl}
 R^a{}_{bcd} &=& (R^{(2)})^a{}_{bcd} \\
 R^a{}_{ibj} = - R^a{}_{ijb} &=& -g^{(n)}_{ij} r \nabla_b \nabla^a r \\
 R^i{}_{jmn} &=& (R^{(n)})^i{}_{jmn} + (\delta^i{}_n g^{(n)}_{mj}-\delta^i{}_m g^{(n)}_{nj} (\nabla_a r)(\nabla^a r)\\
 R^i{}_{ajb} = -R^i{}_{abj} &=& - \delta^i{}_j \frac{\nabla_a \nabla_b r}{r}
\end{IEEEeqnarray}

The covariant derivative symbol here refers to the derivative according to Levi-Civita connection of $g$ (The independent connection plays no role here). In addition to this, we can make use that in our geometry, $g^{(n)}$ corresponds to a maximally symmetric subspace, whose Riemann tensor is: 

\begin{equation}
 (R^{(n)})^i{}_{jlm} = k (\delta^i{}_l g^{(n)}_{jm} - \delta^i{}_m g^{(n)}_{jl})
\end{equation}
with $k=1,0\text{ or }-1$ depending if the geometry is spherical, flat or hyperbolic. With this, we can express the Kretschmann scalar as:

\begin{IEEEeqnarray}{rCl}
 K&=& K^{(2)} + \frac{4(d-2)}{r^2 g_{xx}^2} \left [ \frac{ (\partial_x g_{tt})^2}{4 g_{tt}^2} (\partial_x r)^2+ \left ( \partial_x \partial_x r - \frac{\partial_x g_{xx}}{2 g_{xx}} (\partial_x r) \right )^2 \right ] \\
&& + \frac{2(d-2)(d-3)}{r^4} \left ( k - \frac{(\partial_x r)^2}{g_{xx}} \right )^2 \nonumber
\end{IEEEeqnarray}

And the Kretschmann scalar $K^{(2)}$ of the $(t,x)$ subspace is:

\begin{equation}
 K^{(2)}= \frac{1}{4 g_{xx}^2} \left [ \frac{\partial_x g_{tt}}{g_{tt}} \frac{\partial_x g_{xx}}{g_{xx}} + \left ( \frac{\partial_x g_{tt}}{g_{tt}} \right )^2-2\frac{\partial_x \partial_x g_{tt}}{g_{tt}} \right ]^2
\end{equation}

Now we can study the behaviour of the Kretschmann depending on the different cases:

\begin{itemize}
 \item $\delta_1 \neq \delta_d$: In this case we have $A \propto 1/|x|^{d-3}$, which in turn implies $g_{tt} \propto 1/|x|$ and $g_{xx} \propto 1/|x|^{2d-7}$. The Kretschmann diverges like $1/|x|^{4d-10}$
 \item $\delta_1 = \delta_d$, $\delta_2 \neq a_d\delta_d(2d-4)^{-\frac{d-3}{d-2}}$: In this case, $g_{tt} \propto |x|^{d-4}$, $g_{xx} \propto{|x|^{d-4}}$ and the Kretschmann diverges like $1/|x|^{2d-4}$
 \item $\delta_1 = \delta_d$, $\delta_2 = a_d\delta_d(2d-4)^{-\frac{d-3}{d-2}}$: In this case, $g_{tt} \propto |x|^{d-2}$, $g_{xx} \propto{|x|^{d-2}}$ and the Kretschmann diverges like $1/|x|^{2d}$
\end{itemize}

As we can see, there will be a curvature divergence at $r=r_c$ no matter the value of the parameters of the black hole (as long as $d>4$).

\section{Geodesics}

In $d=4$, these charged black holes are equivalent to the one we studied in chapter \ref{SolChap} (equal if $l_\epsilon = l_P$). In this case $\Omega_{(2)}$ is equal to $\sigma_+$, which has a finite value at $r=r_c$. Then, if $\delta_1=\delta_d$, the geometry has no curvature divergences. In 4 dimensions, even if $\delta_1\neq\delta_d$ (which causes curvature divergences to appear), the space-time is not singular because the geodesics can be extended in a unique way to the other side of the wormhole.

We are interested if this is also the case for dimensions greater than 4, and the geometry is indeed regular. This case is different because no matter the value of $\delta_1$, the factor $\Omega_{(2)}$ always diverges at $r=r_c$, and that causes the curvature to be divergent, too. However, there are still many similarities with the 4 dimensional case. In particular, the radial function $r(x)$ that gives us the size of the symmetric $(d-2)$-dimensional subspace has a minimum value, $r_c$. Also, the behaviour of the metric near the wormhole throat is of the type $g_{tt} \sim \frac{1}{|x|}$ when $\delta_1 \neq \delta_c$, which is the same behaviour as in the 4 dimensional case.


%


As we did in the introduction and chapter \ref{GeoChap}, we can use the symmetries of the geometry to simplify the problem. First, because of spherical symmetry, the movement of geodesics will lie on a plane, and we can rotate our coordinate system so that the movement can be described with only one angle we will call $\phi$. Second, we can normalize the tangent vector of time-like geodesics to $-1$; if the geodesics are null, the norm of the tangent vector will be $0$. Third, because of the symmetries of the geometry there are two conserved quantities: $E=\frac{A}{\Omega_{(2)}} \frac{\df \gamma^t}{\df \lambda}$, $L=r^2 \frac{\df \gamma^\phi}{\df \lambda}$, which as usual can be interpreted as the energy and angular momentum per unit mass if the geodesics are time-like. If the geodesics are null, then $L/E$ can be interpreted as the apparent impact parameter as seen from the asymptotically flat infinity. This conditions give raise to the following for the tangent vector of the geodesic:

\begin{equation}
 - \kappa = -\frac{A}{\Omega_{(2)}} \left ( \frac{\df \gamma^t}{\df \lambda} \right )^2 + \frac{1}{A \Omega_{(2)}} \left ( \frac{\df \gamma^x}{\df \lambda} \right )^2 + r^2(x) \left ( \frac{\df \gamma^\phi}{\df \lambda} \right )^2
\end{equation}

Substituting the conserved quantities in this equation and solving for the radial component of the tangent vector:

\begin{equation}
 \frac{1}{\Omega_{(2)}} \left ( \frac{\df \gamma^x}{\df \lambda} \right ) = \pm \sqrt{E^2-\frac{A}{\Omega_{(2)}} \left ( \kappa + \frac{L^2}{r^2(x)} \right )}\label{7eq:diffx}
\end{equation}

The motion of geodesics in the radial direction is analogue to the movement of a particle of energy $E^2$ in a one dimensional potential:

\begin{equation}
 V_\text{eff}= \frac{A}{\Omega_{(2)}} \left ( \kappa + \frac{L^2}{r^2(x)} \right ) =g_{tt} \left ( \kappa + \frac{L^2}{r^2(x)} \right )
\end{equation}

As we are interested in the behaviour near the wormhole throat, we can use the expansion of $g_{tt}$ around $x=0$. A difference respect to the 4-dimensional case is that this time we have to take into account that $\Omega_{(2)}$ is no longer constant and might modify substantially the way the equation is integrated. In terms of the value of $\delta_1$ we will find a different behaviour of the geodesics:

\begin{itemize}
 \item $\delta_1>\delta_d$: In this case $g_{tt}$ is negative at $x=0$, and the wormhole throat is a time-like hypersurface. The potential is an infinite barrier at the throat $V_\text{eff} \propto 1/|x|$. The geodesics in the direction of the wormhole throat will find this barrier and will be repelled.
 
 \item $\delta_1<\delta_d$: In this case $g_{tt}$ is positive at $x=0$, and the wormhole throat is a space-like hypersurface. Every geodesic that crosses the horizon has the wormhole throat in its future and will reach it for a finite value of its affine parameter. The potential in this case is an infinite well, $V_\text{eff} \propto -1/|x|$. We also have $\Omega_{(2)} \propto x^{4-d}$. Integrating eq. \ref{7eq:diffx} we find that the affine parameter goes as $(\lambda - \lambda_0) \propto \pm x |x|^{d-7/2}$.
 
 \item $\delta_1 = \delta_d$: In this case $g_{tt}$ vanishes at $x=0$ (in $d>4$), and the wormhole throat is a null hypersurface. The behaviour of $g_{tt}$ in this case goes as either $|x|^{d-4}$ or $|x|^{d-2}$. In any case, the effective potential can be disregarded respect to $E^2$ and the affine parameter behaves as $(\lambda-\lambda_0) \propto x |x|^{d-4}$
\end{itemize}

In all the cases, the affine parameter can be extended smoothly through $r=r_c$ and the geometry is that of a wormhole. 

As the metric is not well defined at that region we should be more careful and look at the extensions of geodesics as we did in section \ref{1sec:ExtDisc}. In the case of $\delta_1>\delta_d$, geodesics do not reach the wormhole throat and we do not have to worry about them. In the case $\delta_1 < \delta_d$, the  $(t,x)$ and $(x,\theta_i)$ part of the metric are equivalent to the geometries (i) and (iii) we studied. The case $\delta_1=\delta_d$ corresponds to geometry (iv). In all these cases we found a well-defined Pfaff system whose integral manifolds are the geodesics.

\cleardoublepage
\chapter{Conclusions}

One of the most important open questions in gravitation theory is the existence of space-time singularities. Singular geometries are characterized by incomplete geodesics, which physically would correspond to an observer disappearing from the space-time. Several modified gravity theories try to avoid space-time singularities in some way.

For this purpose, we have considered modifications to the GR Lagrangian, such as quadratic gravity and Born-Infeld gravity, in the Metric-Affine formalism. In this formalism, the connection (from which the Riemann curvature tensor is constructed) is considered to be independent from the metric, and we let the variational principle to dictate its value. We find that, in these theories, the connection is the Levi-Civita connection of the metric in vacuum, as in the Riemannian formalism; but we start seeing differences when the amplitude of the energy-momentum tensor is high, such as near the central region of a charged black hole. We note that this formalism is used in solid state physics, as it is needed to describe the geometry of continuum systems such as Bravais crystals. The density of defects in the crystal is a source of non-metricity, in an analogous way as the matter density in gravity in the Metric-Affine formalism. These are very enticing models, because they add higher curvature corrections to GR (which are expected to come from the full quantum description of gravity) in a way that the theory remains ghost-free. 

We have studied charged black holes for quadratic gravity in this formalism, and found that the solution of the metric has the form:

\begin{equation}
 g=-\frac{A}{\sigma_+} \df t^2 + \frac{1}{A \sigma_+} \df x^2 + r^2(x) \df \Omega^2
\end{equation}
where $\sigma_+\equiv 1 + r_c^4/r^4$, and the radius of the 2-spheres is a function of the coordinate $x$:

\begin{equation}
 r^2=\frac{x^2+\sqrt{x^4+4r_c^4}}{2} 
\end{equation}
which has a minimum value $r_c\equiv 2^\frac{1}{4} \sqrt{l_P r_q}$, which depends on the charge of the black hole, giving the structure of a wormhole to the space-time. For big values of the radius compared to $r_c$, the Reissner-Nordström solution of GR is recovered: $A$ tends to $1-r_S/r+r_q^2/r^2$ and $r\simeq x$. Near $x=0$, $A$ behaves like $+1/|x|$, $-1/|x|$ or a constant value, depending if this relation between the mass and charge $\delta_1 \equiv r_q^2/(r_S r_c)$ is greater, lesser or equal to a critical value $\delta_c \simeq 0.57207$. We find that the curvature scalars for the $\delta_1=\delta_c$ case are finite everywhere, but diverge for $\delta_1 \neq \delta_c$.

We have established that this geometry is non-singular in three different ways:
\begin{itemize}
 \item First, we have studied the geodesic completeness of the geometry, which is the standard criterion for a space-time to be singular. In many cases geodesics do reach the wormhole throat, but are always complete --crossing to the other side of the geometry-- no matter the value of the mass and charge of the black hole, even in the case where the curvature scalars diverge. In the literature, the concepts of curvature divergence and singularity are often identified; here we have put forward an example in which these two concepts are clearly different. We have provided a new avenue to avoid singularities, different from other strategies such as putting bounds to the curvature, at a level more closely related to the geometric meaning of singularity.
 \item Since geodesics are an idealization of an observer when its size is much smaller than the change in curvature, we have also studied congruences of geodesics to represent objects with finite size. We have established that the individual components of an object with finite size do not lose causal contact passing through the wormhole throat: Even though two infinitesimally close time-like geodesics seem to suffer an infinite stretching in the radial direction, the spatial distance between non-infinitesimally separated geodesics is always finite (see figs. \ref{5fig:LightRayWH} and \ref{5fig:LightRayTimeWH}). This guarantees the effective transmission of interactions among the constituents of the body.
 \item Since ultimately matter is quantum in nature, and more aptly described by a wave, we have studied propagation of waves through the wormhole throat. For that, we have separated each wave into partial waves that satisfy the following equation (expressed in the tortoise coordinate $y^*$):

\begin{equation}
 \partial_{y^*} \partial_{y^*} \psi_{\omega l} + \left [ \omega^2 - \underbrace{\left ( \frac{1}{r} \partial_{y^*} \partial_{y^*} r + \left ( m^2 + \frac{l(l+1)}{r^2} \right ) \frac{A}{\sigma_+} \right )}_{V_{\text{eff}}} \right ] \psi_{\omega l}=0
\end{equation}
where the leading behaviour of the potential is $ V_{\text{eff}} \approx \kappa/|y^*|^\frac{1}{2}$. The solutions of this differential equation have \emph{regular} behaviour at the wormhole throat ($y^*=0$), which can be either linear or constant. After this, we  were able to make computations such as the transmission cross-section for one of these wormholes (without horizons).
\item A fourth case of physical interest for future study is the fate of observers with bounded accelaration (\cite{Scarr:2016dyg}, \cite{Friedman:2016tye}, \cite{Olmo:2017}), and the study of congruences of such observers.
\end{itemize}


These wormhole solutions are constructed without the need for exotic matter, unlike other known wormholes found in the literature. The electromagnetic and gravitational equations are solved everywhere and show that the space-time does not contain a source of matter or charge, just an electric field passing through the wormhole throat. These solutions are sourceless gravitational-electromagnetic entities known as geons. These objects are particle-like in the classical sense, without the need for introducing singularities into the space-time.

Finally, we considered the study of charged black holes in higher dimensions, and in Born-Infeld gravity. We see that the results we obtained in 4 dimensions also apply in this case, although there is no longer a particular mass-to-charge ratio in which the curvature scalars are regular everywhere. Therefore we conclude that quadratic gravity in 4 dimensions is not a particular case, and that the Metric-Affine formalism may avoid singularities for a wide range of theories. This has received further confirmation with several interesting examples within the family of $f(R)$ theories (\cite{Olmo:2015axa}, \cite{Bambi:2015zch}, \cite{Olmo:2016tra}, \cite{Olmo:2011ja}, \cite{Bejarano:2017fgz}). 

In the light of the work in this thesis, we should reconsider the usual approach dealing with singularities. Perhaps it is not the job of quantum gravity to solve this problem, but maybe it is a necessary step to take in the classical description of gravity, before we can attempt a quantization. After all, we have introduced new geometrical tools, that are artificially restricted in GR. In condensed matter physics, these tools are necessary to describe crystals with defects, as opposed to perfect crystals in which Riemannian geometry is enough; and this defects are essential to understand the global properties of the crystal as a whole. This makes us wonder which is the true underlying geometry of our universe. In this thesis, we have considered that the matter Lagrangian only couples to the metric, but not to the independent connection; but it is an open question whether this is true, and we should also look for possible violations of EEP, specially near the most curved regions of the space-time. This work also changes the picture of naked singularities. The cosmic censorship conjecture is no longer needed in these kind of theories, and the naked wormholes are a feature that could be searched for, perhaps as remnants from primordial black holes (\cite{Lobo:2013prg}, \cite{Olmo:2013gqa}, \cite{Chen:2014jwq}).

A fine point to discuss would be whether it is a bit contradictory to highlight the fact that curvature divergences are harmless in Metric-Affine theories, when a quantum theory of gravity would possibly regularize these infinities. That would be the expected result of a perturbative quantization, however, it has been argued in the literature that such a scheme cannot capture the features a quantum theory of gravity should have (such as the regularization of the self-energy of a point particle, see introduction of \cite{ashtekar1991lectures}). Also, let us note that a perturbative quantization cannot reproduce the non-trivial topology of the solutions we have found. The work in this thesis is closer to the spirit of non-perturbative quantizations of gravity, such as LQG, and the solutions presented here could be a suitable effective description in certain limit (see \cite{Gambini:2013ooa} for an example of non-trivial topology in LQG). Still, it is typically expected that a non-perturbative quantization scheme might also regularize the curvature divergences. Following the condensed matter analogy, we can take a look at ``wormholes'' constructed gluing sheets of graphene together with carbon nanotubes. The realization in nature of these wormholes does not have anything divergent about it, but the mathematical description in the continuum limit does contain a curvature divergence at the throat, because the structure of rings used to join the sheets with the nanotubes is overlooked in this limit (see \cite{Gonzalez:2009je}). Thus, we consider that these theories can be the classical limit of a non-perturbative quantization of gravity, which retain key features such as non-trivial topologies, and whose curvature divergences are not a problem for the description of the physics of the space-time.

\cleardoublepage
\pagestyle{empty}
\chapter*{Agradecimientos}
\addcontentsline{toc}{chapter}{Agradecimientos}

En primer lugar, me gustaría mostrar mi agradecimiento a mi director Gonzalo Olmo y a Diego Rubiera. Entre ellos y yo hemos escrito los artículos sobre los que trata esta tesis. Ha sido una experiencia muy bonita, donde he aprendido muchas cosas nuevas, y también he visto la física desde una perspectiva diferente a la que tenía antes de comenzar. También les agradezco haber leído mi tesis cuando estaba en una fase preliminar y haber sugerido un montón de cambios, que mejoraron notablemente la redacción. También quiero destacar los buenos ratos que he pasado compartiendo despacho con Gonzalo, y con Diego cuando estaba por Valencia; y decir que me alegro mucho de haber tomado la decisión de haber cambiado de línea investigadora para trabajar con ellos.

Me gustaría también agradecer al lector y a todo aquel que haya leído alguna parte de esta tesis. En especial al comité de expertos y el tribunal, José Navarro Salas, Prado Martín Moruno, Franciso Lobo, Victor I. Afonso,  Alessandro Fabbri  y Maria Antonia Lledó, que además han tenido que lidiar con la burocracia y los imprevistos que genera. Esta tesis ha sido fruto de una parte de mi vida, y me alegra que alguien dedique parte de su tiempo en ella.

Para llegar hasta este punto ha pasado un viaje educativo muy largo, y me gustaría agradecer a los profesores de física y matemáticas que he tenido a lo largo de estos años, en especial a Don Jesús en los Salesianos, y a Marc Mars, Miguel Ángel Vazquez-Mozo y Antonio López Almorox en la universidad de Salamanca. También me gustaría agradecer a Francisco Botella y a Miguel Nebot el darme la primera oportunidad en este viaje del doctorado, que si bien no salió como pensaba, aprendí mucho en ella y fueron muy comprensivos conmigo.

También me gustaría agradecer a todos los que en este tiempo me han dado apoyo moral y mantenido mi salud mental, bien tomando unas cervezas por ahí, montando rutas de senderismo y excursiones a calas, viendo al Barça, organizando timbas de póker, quejándose de sus respectivas tesis/oposiciones (y haciéndome ver que no estoy solo este viaje), compartiendo sus comics / dibujos / podcasts / relatos / actuaciones / ..., participando en el concursong, jugando a squash / frontón / futbito / fútbol 7 / baskets / ..., montando paellas y barbacoas, sugiriéndome todo tipo de nueva música (desde la infernal a la tranquilita), haciendo trucos de magia, preguntando y animándome por whatsapp, yendo al circo del sol, poniendo vídeos chorras, jugando a juegos de mesa, siendo conejillos de indias de mi cerveza casera, explorando cuevas llenas de murciélagos, hablando sobre extraterrestres, montando un comunio o una fantasy de la NBA... Sois mucha gente, y me sabría mal hacer un listado sin decir algo más personal de cada uno de vosotros. Y si bien después de dedicarle todo el tiempo que le he dedicado a la tesis, bien mereciera haber escrito unos agradecimientos en condiciones, como es costumbre he dejado esta tarea para el último momento y veo como el tiempo se echa encima para mandar la tesis a imprimir si quiero tener todo a tiempo (soy un dejado, lo sé...). Baste decir que en Salamanca, en Irlanda, en el Eramus y en Valencia he conocido a gente fantástica y quería daros las gracias por los buenos momentos en esta etapa de mi vida.

Por último me gustaría agradecer a mis padres y a mis hermanos, que me han apoyado en todo momento, y que si pasaba una mala racha sabía que estaban ahí para todo lo que necesitase. Tomar decisiones es siempre difícil, y uno nunca termina de saber si lo que elige es lo correcto o no, pero es mucho más llevadero cuando tu familia esta detrás.

\bibliographystyle{unsrt}
\pagestyle{fancy}
\bibliography{BBTesis} 

\end{document}